\documentclass[a4paper,12pt,customfont,custombib,index]{class/PhDThesisPSnPDF}

\ifsetCustomMargin
  \RequirePackage[left=37mm,right=30mm,top=35mm,bottom=30mm]{geometry}
  \setFancyHdr \fi

\ifsetCustomFont
\fi

\RequirePackage[labelsep=space,tableposition=top]{caption}

\usepackage{subcaption}

\usepackage{booktabs} \usepackage{multirow}

\usepackage{siunitx}

\ifuseCustomBib

\RequirePackage[backend=biber, style=alphabetic, sorting=nty, natbib=true, maxnames=99, minnames=99, maxalphanames=99, minalphanames=4, maxalphanames=4]{biblatex}
\bibliography{References/thesis-bib} \bibliography{References/software-etc} 
\renewbibmacro{in:}{}

\AtEveryBibitem{\ifentrytype{article}{\clearfield{url}\clearfield{URL}\clearfield{urldate}}{}}

\fi

\usepackage{sty/mmacells}
\usepackage[shortlabels]{enumitem}

\usepackage{underoverlap} \usepackage{import}
\usepackage{amsthm}
\usepackage{cleveref}
\usepackage{placeins} \usepackage{thm-restate}
\usepackage{tikz}
\usetikzlibrary{patterns}
\usepackage{tikz-3dplot}
\usepackage{tikz-cd}
\usepackage{pgfplots}
\usepackage{tikzscale} \usepackage{mathtools}
\usepackage{xparse,everysel} \usepackage{multirow}
\usepackage{lscape} 
\usepackage{pbox}
\usepackage{adjustbox} 
\newcommand{\inv}{^{-1}}
\DeclareMathOperator{\Ad}{Ad}
\DeclareMathOperator{\EB}{EB}
\DeclareMathOperator{\EEB}{EEB}
\DeclareMathOperator{\AEB}{AEB}
\DeclareMathOperator{\AES}{AES}
\DeclareMathOperator{\ES}{ES}
\DeclareMathOperator{\PPT}{PPT}
\DeclareMathOperator{\spr}{spr}
\newcommand{\iu}{\mathrm{i}\mkern1mu}

\newcommand{\bZ}{\mathbb{Z}}
\newcommand{\bR}{\mathbb{R}}
\newcommand{\bN}{\mathbb{N}}
\newcommand{\NN}{\mathbb{N}}
\newcommand{\bC}{\mathbb{C}}
\newcommand{\R}{\mathbb{R}}

\newcommand{\cT}{\mathcal{T}}
\newcommand{\cS}{\mathcal{S}}
\newcommand{\cX}{\mathcal{X}}
\newcommand{\cY}{\mathcal{Y}}
\newcommand{\cK}{\mathcal{K}}

\newcommand{\cD}{\mathcal{D}}
\newcommand{\D}{\mathcal{D}}

\newcommand{\cN}{\mathcal{N}}
\newcommand{\cB}{\mathcal{B}}
\newcommand{\cF}{\mathcal{F}}
\newcommand{\cM}{\mathcal{M}}
\newcommand{\cH}{\mathcal{H}}
\renewcommand{\H}{\mathcal{H}}
\newcommand{\cA}{\mathcal{A}}
\newcommand{\cL}{\mathcal{L}}
\newcommand{\cP}{\mathcal{P}}

\newcommand{\e}{\mathrm{e}}
\newcommand{\one}{\mathbf{1}}
\newcommand{\id}{\operatorname{id}}
\newcommand{\tr}{\operatorname{tr}}
\renewcommand{\d}{\operatorname{d\!}}

\newcommand{\dom}{\operatorname{dom}}
\newcommand{\gap}{\operatorname{gap}}

\DeclareMathOperator{\SEP}{SEP}
\newcommand{\linspan}{\operatorname{span}}
\newcommand{\eps}{\varepsilon}

\usepackage{braket}

\DeclareMathOperator*{\argmin}{argmin}
\DeclareMathOperator*{\argmax}{argmax}
\DeclareMathOperator{\diag}{diag}

\DeclareMathOperator{\conv}{conv}
\newcommand{\Be}{B_{\eps}}
\DeclareMathOperator{\rank}{rank}
\DeclareMathOperator{\ord}{ord}

\renewcommand{\subset}{\subseteq}
\newcommand{\mmm}{\mathcal{M}} \newcommand{\ketbra}[2]{\ket{#1}\!\bra{#2}}
\newcommand{\sa}{_\textnormal{sa}}

\newcommand{\Bsa}{\cB\sa}
\newcommand{\Bp}{\cB_+}
\newcommand{\Bpp}{\cB_{++}}

\newcommand{\HS}{_\textnormal{HS}}
\newcommand{\ip}{\braket{\,\cdot\,,\,\cdot\,}}

\newcommand{\majmax}{_{*,\eps}}
\newcommand{\majmin}{_{\eps}^*}

\DeclareMathOperator{\Eig}{Eig}
\DeclareMathOperator{\pow}{pow}
\newcommand{\Bep}{B_\varepsilon^+}
\DeclareMathOperator{\spec}{spec}

\newcommand{\Dpure}{\D_\text{pure}}
\newcommand{\RR}{\bR}

\newcommand{\infmaj}{\operatorname{inf}_\prec}
\newcommand{\minmaj}{\operatorname{min}_\prec}
\newcommand{\supmaj}{\operatorname{sup}_\prec}

\usepackage{accents} \newcommand{\ubar}[1]{\underaccent{\bar}{#1}}

\DeclareMathAlphabet{\mathpzc}{OT1}{pzc}{m}{it}
\newcommand{\cs}{\mathpzc{s}}

\DeclareMathOperator{\re}{Re}

\DeclareMathOperator{\TV}{TV}

\newcommand{\gen}{\mathcal{L}}
\renewcommand{\i}{\operatorname{i}}

\newcommand{\unorm}[1]{{\left\vert\kern-0.25ex\left\vert\kern-0.25ex\left\vert #1 
    \right\vert\kern-0.25ex\right\vert\kern-0.25ex\right\vert}}

\newcommand{\typeone}{Concave-Type}
\newcommand{\typetwo}{Convex-Type}

\DeclareMathOperator{\bin}{bin}

\newcommand{\bE}{\mathbb{E}}

\DeclarePairedDelimiter\ceil{\lceil}{\rceil}
\DeclarePairedDelimiter\floor{\lfloor}{\rfloor}

\renewcommand{\neg}{\operatorname{neg}}

\newcommand{\sys}{\mathcal{S}}
\newcommand{\env}{\mathcal{E}}
\newcommand{\init}{^\textnormal{i}}
\renewcommand{\Re}{\operatorname{Re}}
\renewcommand{\Im}{\operatorname{Im}}

\newcommand{\ealpha}{^{(\alpha)}}

	\def\inte{\operatorname{int}}
	\def\clos{\operatorname{cl}}

\newcommand{\invar}{^\textnormal{inv}}
\newcommand{\fin}{^\textnormal{f}}
\DeclareDocumentCommand \envstate { o } {\IfNoValueTF {#1} {\xi }{\xi_{#1}}}
\DeclareDocumentCommand \sysstate { o } {\IfNoValueTF {#1} {\rho }{\rho_{#1}}}

\newcommand{\ham}{h}
\newcommand{\hsys}{\ham_{\sys}}
	\DeclareDocumentCommand \henv { o } {\IfNoValueTF {#1} {\ham_{\mathcal{E}}}{\ham_{\mathcal{E}_{#1}}}}
\newcommand{\Exp}[1]{\mathrm{e}^{#1}}
\newcommand{\numset}[1]{\mathbb{#1}}

\newcommand{\Z}{\numset{Z}}
\newcommand{\N}{\numset{N}}
\def\cc{\numset{C}}
\def\rr{\numset{R}}

\def\nn{\numset{N}}

\newcommand{\E}{\mathbb{E}}
\newcommand{\ee}{\mathbb{E}}

\newcommand{\sP}{\mathcal{P}}

\newcommand{\bP}{\mathbb{P}}
\newcommand{\pp}{\mathbb{P}}

\newcommand{\Ai}{{A\init}}
\newcommand{\Af}{{A\fin}}
\newcommand{\ai}{{a\init}}
\newcommand{\af}{{a\fin}}
\newcommand{\rvY}{\Delta y^{\text{tot}}_{T}}
\newcommand{\rvZ}{{-\Delta a_T + \Delta y^{\text{tot}}_{T}}}
\newcommand{\rvW}{\Delta a_T}
\newcommand{\oldsigmaT}{\sigma^\textnormal{tot}_{T}}
\newcommand{\rhoadiab}{\rho_\textnormal{adiab}}
\newcommand{\invalpha}{\mathrm{I}\ealpha}

\makeatletter
\newcommand{\labitem}[2]{\def\@itemlabel{\textbf{#1}}
	\item
	\def\@currentlabel{#1}\label{#2}}
\makeatother \newtheorem{theorem}{Theorem}[section]

\newtheorem{lemma}[theorem]{Lemma}
\newtheorem{corollary}[theorem]{Corollary}
\newtheorem{proposition}[theorem]{Proposition}

\theoremstyle{definition}
\newtheorem{definition}[theorem]{Definition}
\theoremstyle{definition}
\newtheorem{example}[theorem]{Example}
\theoremstyle{definition}

\theoremstyle{remark}
\newtheorem*{remark}{Remark}
\newtheorem*{remarks}{Remarks} 
\allowdisplaybreaks 
\title{Entropic Continuity Bounds \& Eventually Entanglement-Breaking Channels}

\author{Eric P. Hanson}

\dept{Department of Applied Mathematics and Theoretical Physics}

\university{University of Cambridge}
\crest{\includegraphics[width=0.2\textwidth]{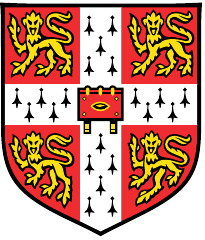}}

\degreetitle{Doctor of Philosophy}

\college{Churchill College}

\degreedate{April 2020} 

\subject{LaTeX} \keywords{{LaTeX} {PhD Thesis} {Engineering} {University of
Cambridge}}

\ifdefineAbstract
\fi

\ifdefineChapter
\fi

\begin{document}

\frontmatter

\maketitle

\clearpage{}

\begin{dedication} 

To my family, and to Hang Lu.

\end{dedication}\clearpage{}
\clearpage{}

\begin{declaration}

This dissertation is the result of my own work and includes nothing which is the
outcome of work done in collaboration except as declared in the Preface and
specified in the text. It is not substantially the same as any that I have
submitted, or, is being concurrently submitted for a degree or diploma or other
qualification at the University of Cambridge or any other University or similar
institution except as declared in the Preface and specified in the text. I
further state that no substantial part of my dissertation has already been
submitted, or, is being concurrently submitted for any such degree, diploma or
other qualification at the University of Cambridge or any other University of
similar institution except as declared in the Preface and specified in the text.

\end{declaration}\clearpage{}
\clearpage{}

\begin{acknowledgements}

I would like to sincerely thank Nilanjana Datta and Yan Pautrat for the excellent supervision over the past few years. Nilanjana always made time to discuss things with me and I benefited from many blackboard discussions with her. And while I could not meet Yan as frequently due to the distance to Paris, it was a pleasure whenever we met.

Over the course of my PhD, I have benefited tremendously from collaborations with Koenraad Audenaert, Andreas Bluhm, Hao-Chung Cheng, Nilanjana Datta, Daniel Stilck Fran\c{c}a, Min-Hsiu Hsieh, Alain Joye, Vishal Katariya, Yan Pautrat, Renaud Raqu\'epas, Cambyse Rouz\'e,  Mark Wilde, and Jiazheng Zhu. I hope to work with you all again.
I have also been lucky to have been able to have interesting mathematical discussions with many people, including Simon Andr\'eys, Koji Azuma, Ivan Bardet, Johannes Bausch, \'Angela Capel Cuevas, Michele Dall'Arno, Kun Fang, Hamza Fawzi, Lisa H\"{a}nggli, Christoph Hirche, Joel Klassen, Robert K\"onig, Felix Leditzky, M\=aris Ozols, Sam Power, and many others. I'd also like to thank Hamza Fawzi and Mark Wilde for serving as my thesis examiners and finding many typos in the first version of the document (any remaining mistakes should certainly be attributed to me!).

I would like to acknowledge my research group, the Centre for Quantum Information and Foundations, which has hosted many interesting seminars and discussions. In particular, I am thankful for Sathyawageeswar Subramanian and Mithuna Yoganathan who started their PhDs at the same time as me, and have been great company on this journey.

I would also like to thank the Cantab Capital Institute for the Mathematics of Information (CCIMI) for the funding to undertake this PhD, and in particular Carola Sch\"onlieb for her leadership and Rachel Furner and Josh Stevens for organizing everything. I'd also like to thank my cohort of students at the CCIMI: Edward Ayers, Andrew Celsus, Sam Power, Ferdia Sherry, and Sven Wang; it has been nice to share reading groups, offices, and houses with you. I feel like I have found a nice community, with others in Cambridge such as Lars Mennen, Federico Pasqualotto, and Sam Thomas, and with those further away, such as Peter Tsimiklis, Marshall Sumwalt, and Michail and Sarah Yasonik.

I also greatly appreciate the Julia programming language community, and in particular Alex Arslan, Beno\^it Legat, Harsha Nagarajan, and David P. Sanders. 
I am thankful to my masters supervisor Vojkan Jak\v{s}i\'c who suggested to me that I apply to work with Nilanjana and Yan, and taught me many things.
I also thank the administrative and facilities staff at the Centre for Mathematical Sciences and at Churchill College.

Lastly, I very much appreciate the support of my family, immediate and extended, and my partner, Hang Lu Su.

\end{acknowledgements}
\clearpage{}
\clearpage{}\begin{abstract}

This thesis combines two parallel research directions: an exploration into the
continuity properties of certain entropic quantities, and an investigation
into a simple class of physical systems whose time evolution
is given by the repeated application of a quantum channel.

In the first part of the thesis, we present a general technique for
establishing local and uniform continuity bounds for Schur concave functions;
that is, for real-valued functions which are decreasing in the majorization
pre-order. Continuity bounds provide a quantitative measure of robustness,
addressing the following question: If there is some uncertainty or error in
the input, how much uncertainty is there in the output? Our technique uses a
particular relationship between majorization and the trace distance between
quantum states (or total variation distance, in the case of probability
distributions). Namely,  the majorization pre-order attains a maximum and a
minimum over $\eps$-balls in this distance. By tracing the path of the
majorization-minimizer as a function of the distance $\eps$, we obtain the
path of  ``majorization flow’’. An analysis of the derivatives of Schur
concave functions along this path immediately yields tight continuity bounds
for such functions.

In this way, we find a new proof of the Audenaert-Fannes continuity bound for
the von Neumann entropy, and the necessary and sufficient conditions for its
saturation, in a universal framework which extends to the other functions,
including the R\'enyi and Tsallis entropies. In particular, we prove a novel
uniform continuity bound for the $\alpha$-R\'enyi entropy with $\alpha>1$ with
much improved dependence on the dimension of the underlying system and the
parameter $\alpha$ compared to previously known bounds. We show that this
framework can also be used to provide continuity bounds for other Schur
concave functions, such as the number of connected components of a certain
random graph model as a function of the underlying probability distribution,
and the number of distinct realizations of a random variable in some fixed
number of independent trials as a function of the underlying probability mass
function. The former has been used in modeling the spread of epidemics, while
the latter has been studied in the context of estimating measures of
biodiversity from observations; in these contexts, our continuity bounds
provide quantitative estimates of robustness to noise or data collection
errors.

In the second part, we consider repeated interaction systems, in which a
system of interest interacts with a sequence of probes, i.e.\@~environmental
systems, one at a time. The state of the system after each interaction is
related to the state of the system before the interaction by the so-called
reduced dynamics, which is described by the action of a quantum channel. When
each probe and the way it interacts with the system is identical, the reduced
dynamics at each step is identical. In this scenario, under the additional
assumption that the reduced dynamics satisfies a faithfulness property, we
characterize which repeated interaction systems break any initially-present
entanglement between the system and an untouched reference, after finitely
many steps. In this case, the reduced dynamics is said to be \emph{eventually
entanglement-breaking}. This investigation helps improve our
understanding of which kinds of noisy time evolution destroy entanglement.

When the probes and their interactions with the system are slowly-varying
(i.e.\@~adiabatic), we analyze the saturation of Landauer's bound, an inequality
between the entropy change of the system and the energy change of the probes,
in the limit in which the number of steps tends to infinity and both the 
difference between consecutive probes and the difference between their
interactions vanishes. This analysis proceeds at a fine-grained level by means
of a two-time measurement protocol, in which the energy of the probes is
measured before and after each interaction. The quantities of interest are
then studied as random variables on the space of outcomes of the energy
measurements of the probes, providing a deeper insight into the interrelations
between energy and entropy in this setting.

\end{abstract}
\clearpage{}
\clearpage{}\chapter*{Summary of contributions}

Part I of this thesis combines the results of \cite{HD18,HD19,AHD19,HKDW20}, along with some as-of-yet unpublished work, namely  \Cref{sec:motivate-from-inf,sec:other-p-norms,sec:symm-concave-cx,sec:LOCC,sec:guesswork_no_side_info}. \Cref{sec:geometry_trace_ball}  and \Cref{sec:general_optimality_conditions} are largely the results of \cite{HD18}, with is joint work with Nilanjana Datta, \Cref{chap:majflow_ctybounds} and many of the continuity bounds of \Cref{sec:applications} are the results of \cite{HD19}, which is also joint work with Nilanjana Datta, \Cref{sec:random-graph} is the result of \cite{AHD19}, which is joint work with Koenraad Audenaert and Nilanjana Datta, while \Cref{sec:guesswork} contains many of the results of \cite{HKDW20}, which is joint work with Vishal Katariya, Nilanjana Datta, and Mark Wilde. Part II of this thesis combines some of the results of \cite{HRS20}, which is joint work with Cambyse Rouz\'e and Daniel Stilck Fran\c{c}a, in \Cref{chap:char_EEB} and some of the results of \cite{HJPR18}, which is joint work with Alain Joye, Yan Pautrat, and Renaud Raqu{\'e}pas, in \Cref{chap:RIS}.

Thus, almost all of the results of this thesis were developed in collaboration with others. The proofs included in Chapters 1--11, however, are my own work, written with advice and suggestions from my supervisors Dr.~Datta and Dr.~Pautrat, and my other coauthors, with the exception of the proof of \Cref{thm:random-graph-cty-bound}, which was largely constructed by Dr.~Audenaert\footnote{My contributions to that proof consist of the numerical bounds, reducing the task to bounding \eqref{eq:EC-bound-1}, and some small fixes.}. Chapter 12 provides a summary of the paper \cite{HJPR18}; the few proofs included in that chapter were written collaboratively with my coauthors. 

The papers \cite{CHDH18,CHDH18b}, which are joint work with Hao-Chung Cheng, Nilanjana Datta, and Min-Hsiu Hsieh, were written during my PhD, but were omitted from this thesis in order to focus on the two themes of entropic continuity bounds and eventually entanglement-breaking channels.

\begin{refsegment}
  \nocite{HD18,HD19,AHD19,HKDW20,HRS20,HJPR18,CHDH18,CHDH18b}
  \printbibliography[segment=\therefsegment,heading=none]
\end{refsegment}

\clearpage{}

\setcounter{tocdepth}{2}
\tableofcontents

\printnomenclature

\mainmatter

\part{Entropic continuity bounds}
\graphicspath{{PartI-ECB/figures}}

\chapter{Introduction} \label{sec:ECB_intro}

Entropies play a fundamental role in classical and quantum information theory as characterizations of the optimal rates of information theoretic tasks, and as measures of uncertainty. The mathematical properties of entropic functions therefore have important physical implications. The von Neumann entropy $S$, given by
\[
 S(\rho) := - \tr[\rho\log \rho]
 \]
is strictly concave and continuous as a function of $d$-dimensional quantum states, and is bounded by $\log d$, where the logarithm is taken to be base~2. As the von Neumann entropy characterizes the optimal rate of data compression for a memoryless quantum information source \cite{Schumacher}, continuity of the von Neumann entropy, for example, implies that the quantum data compression limit is continuous in the source state. Likewise, for probability distributions $p\in \cP_d$, where $\cP_d$ is the set of probability distributions over $\{1,\dotsc,d\}$, the Shannon entropy, $S(p) := - \sum_{i=1}^d p_i \log p_i$ characterizes the data compression limit of a classical i.i.d.~source $X\sim p$. Note that the von Neumann entropy of a quantum state is given by the Shannon entropy of its eigenvalues. The Shannon entropy is commonly denoted by $H(p)$, but in this thesis we will reserve $H$ for a generic single-partite entropic quantity.

The von Neumann entropy satisfies the following tight continuity bound: Given $\eps > 0$ and $\rho,\sigma\in \cD(\cH)$, where $\cD(\cH)$ is the set of density matrices over a Hilbert space $\cH$ with dimension $d$, with trace distance $T(\rho,\sigma) := \frac{1}{2}\tr|\rho-\sigma|\leq \eps$, the following inequality holds:
 \begin{equation}
|S(\rho) - S(\sigma) | \leq \begin{cases}
\epsilon \log (d-1) + h(\epsilon) & \text{if } \epsilon < 1 - \tfrac{1}{d} \\
 \log d & \text{if } \epsilon \geq 1 - \tfrac{1}{d}
\end{cases} \label{eq:Audenaert-Fannes_bound}
\end{equation}
where $h(\eps) := - \eps \log \eps - (1-\eps) \log (1-\eps)$ denotes the binary entropy. The same bound holds in the classical case for $p,q\in \cP_d$, with total variation distance $\TV(p,q) := \frac{1}{2}\sum_{i=1}^d |p_i - q_i|\leq \eps$.

This inequality is known in the quantum information theory literature as the Audenaert-Fannes bound, which is a strengthened version of the Fannes bound \cite{Fan73} and was established in \cite{Aud07} by a direct optimization argument. The bound was also proven via a coupling argument in \cite[Theorem 3.8]{PetzQITbook}, with credit to Csisz\'ar. In the classical case, it was also proven by \cite[Cor.~1]{Zha07} by a coupling argument, and by Ho and Yeung \cite[Theorem 6]{HY10} via an optimization over local continuity bounds, similar in spirit to the techniques used later in this thesis (although with an analysis specific to the Shannon entropy, as opposed to more general families of entropies). The coupling argument was revisited in \cite[Theorem 3]{Sas13} and \cite[Lemma 1]{Winter16}. In each case (except for \cite[Theorem 3.8]{PetzQITbook}), it was shown that equality occurs if one state is pure, and the other state has the spectrum $\{1-\eps,\frac{\eps}{d-1},\dotsc,\frac{\eps}{d-1}\}$. These conditions were shown to be necessary in \cite{HD17} by an analysis of the coupling argument. 

The results established in Part I of this thesis can largely be seen as investigations into, and generalizations of, the inequality \eqref{eq:Audenaert-Fannes_bound} by interpreting it as a consequence of a relationship between trace distance and majorization.

Majorization \cite{LHP} (see also \cite{Schur} and \cite{Horn1954}) is an ordering of vectors (technically, a \emph{preorder}) which provides a means to describe one vector as being more disordered, less disordered, or incomparable with another vector. Majorization has natural connections to entropies, which quantify disorder or randomness of probability distributions or quantum states, and thus provide a measure of information encoded in a random variable or a quantum state. In fact, almost every entropy considered in the information-theoretic literature is Schur concave, meaning that if one vector is more disordered than another, according to the majorization order, then it has higher entropy than the other. Perhaps as a consequence, majorization has proven to be a very useful tool in classical and quantum information theory.

Majorization can be extended from vectors to quantum states by simply defining that a quantum state $\rho$ majorizes another quantum state $\sigma$ if the vector of eigenvalues of $\rho$ (counted with multiplicity) majorizes the vector of eigenvalues of $\sigma$. In quantum information theory, majorization plays a key role in the theory of bipartite pure state entanglement due to Nielsen's theorem \cite{Nielsen-LOCC}. This result states that given two pure states $\psi_{AB}$ and $\phi_{AB}$ of a bipartite system $AB$, the state $\psi_{AB}$ can be transformed into $\phi_{AB}$ via local operations and classical communication if and only if the reduced state of $\psi_{AB}$ for a subsystem ($A$ or $B$) is majorized by the corresponding reduced state of $\phi_{AB}$.

In \Cref{sec:geometry_trace_ball} it is shown that the majorization order on probability vectors behaves well with respect to the total variation distance, in the sense that given any probability vector $p$ and $\eps > 0$, there exists a {\em{minimal}} and {\em{maximal probability vector}} in majorization order in the $\eps$-ball around $p$ with respect to the total variation distance. The existence of such majorization extrema has recently appeared in several forms. Both these probability vectors were independently found in \cite{HOS18}, and in fact the minimal vector already appears in \cite{HY10}. The work \cite{Kog13} developed related ideas in the context of balls in total-variation distance of subnormalized distributions centered at a normalized distribution, and the works \cite{Remco} and \cite{vNW17} consider a generalization of majorization called thermo-majorization, and find thermo-majorization extrema over balls in trace distance.

In the quantum setting, the existence of these states provides a natural connection between majorization and trace distance (and between majorization and the total variation distance in the classical setting). In \Cref{chap:majflow_ctybounds}, we develop a notion of \emph{majorization flow}, which traces out the path of the minimizer in majorization order over the $\eps$-ball as $\eps$ is changed infinitesimally.

Majorization flow turns out to be a very useful tool for obtaining remarkably simple and universal proofs of continuity bounds for numerous well-known families of entropies, including R\'enyi entropies \cite{Ren61}, Tsallis entropies \cite{Tsallis1988}, the so-called unified entropies \cite{RT91}, entropies induced by $f$-divergences \cite{Pet85}, and the concurrence \cite{Woo01}. In particular, it allows us to establish Lipschitz continuity bounds for the $\alpha$-R\'enyi entropy with $\alpha > 1$. This resolves an open problem, left open since Audenaert's 2007 paper \cite{Aud07}, which was also presented at the Open Problem Session of the workshop \emph{2017 Beyond I.I.D. in Information Theory} (held in Singapore), and provides a substantial improvement over previously known bounds.

\paragraph{Overview of Part I}
In the following, let $H$ be a Schur concave function (such as the von Neumann entropy). Let us describe each chapter with an eye to how it provides insight into \eqref{eq:Audenaert-Fannes_bound}, or helps establish a generalization thereof.
\begin{itemize}
	\item \Cref{sec:notation_and_def} introduces basic notation and definitions.
	\item \Cref{sec:geometry_trace_ball} establishes the existence of a majorization-minimizer and -maximizer over the $\eps$-ball in trace distance (or total variation distance, in the classical case). Hence the Schur concavity of $H$ completely dictates its local continuity properties with respect to the trace distance, as manifested in \Cref{cor:local_cont_bound_VN_Renyi}.
	\item \Cref{chap:majflow_ctybounds} considers the path of the majorization-minimizer over the $\eps$-ball as $\eps$ varies, and shows that properties of the derivative $\Gamma_H$ of $H$ along the path provides insight into the global continuity properties of $H$. More specifically, a class of entropic quantities, which includes $S$, obtain a derivative $\Gamma_H$ which itself is Schur convex. This implies that the value of ``how much $H$ can locally increase'' itself respects the majorization order, and is maximized at pure states (or in the classical case, $\delta$-distributions such as $p= (1,0,\dotsc,0)$). Since a state with spectrum $(1-\eps, \frac{\eps}{d-1},\dotsc, \frac{\eps}{d-1})$ (with multiplicity) obtains the maximum entropy over a trace-ball of radius $\eps$ centered at a pure state---unless the completely mixed state lies in this ball, in which case it obtains the maximum entropy over the ball---\eqref{eq:Audenaert-Fannes_bound} follows.
	\item \Cref{sec:applications} applies this technique, and the related technique of bounding $\Gamma_H$ to obtain a Lipschitz continuity bound, to obtain uniform continuity bounds for a range of other quantities: other entropies, including the R\'enyi and Tsallis entropies, the number of distinct observations from $N$ independent trials, and the expected number of connected components of a particular model of random graph.
	\item \Cref{sec:guesswork} investigates the continuity of a quantity related to the Shannon entropy called the \emph{guesswork}, with and without quantum side information. The guesswork without side information is not differentiable on all of $\cP_d$ but nevertheless a tight uniform continuity bound can be obtained with a slight generalization to the techniques previously addressed (although the quantity is simple enough that other techniques would likely suffice). The guesswork with side information, similarly to the conditional entropy, is not Schur concave, and the techniques described so far do not provide a uniform continuity bound. However, by analyzing a semidefinite optimization problem formulation of the quantity, a Lipschitz continuity bound can be obtained which is tight up to at most a multiplicative factor of 2 and an additive factor of 1.
	\item \Cref{sec:general_optimality_conditions} calculates first-order optimality conditions for a quantum state to maximize a differentiable concave function over the $\eps$-ball in trace distance around a given state. These conditions dictate the form of the majorization-minimizer discussed in \Cref{sec:geometry_trace_ball} and provide necessary and sufficient conditions for a given state to maximize the quantum conditional entropy over the trace ball.
\end{itemize}

\chapter{Notation and definitions  \label{sec:notation_and_def}}

We define the set of probability vectors of length $d$ as $\cP_d \subset\R^d$,
\[
\cP_d := \left\{ p = (p_1,\dotsc,p_d)\in\R^d : p_i \geq 0 \text{ for }i=1,\dotsc,d,\, \sum_{i=1}^d p_i = 1 \right\}.
\]
We will often suppress the dependence on the dimension, $\cP\equiv \cP_d$.

The set of quantum states $\cD(\cH)$ where $\cH$ is a $d$-dimensional Hilbert space, $d<\infty$, is given by
\[
\cD(\cH) := \left\{ \rho \in \cB(\cH): \rho \geq 0, \,\tr(\rho) = 1 \right\},
\]
where $\cB(\cH)$ is the set of (bounded\footnote{In this thesis, we will only consider finite-dimensional Hilbert spaces, and hence all linear operators are bounded. We will thus slightly abuse notation and use $\cB(\cH)$ to denote the set of linear operators, the set of bounded operators, and the set of trace-class operators on $\cH$, since all three sets coincide.}) linear operators on $\cH$. The set $\Bsa(\cH)\subset \cB(\cH)$ is the set of self-adjoint operators on $\cH$. The extremal elements of $\cP$ are permutations of the probability vector $p := (1,0,\dotsc,0)$. The extremal elements of $\cD(\cH)$ are rank-1 projections, and are called \emph{pure states}. The identity matrix is denoted $\one\equiv \one_\cH \in \cB(\cH)$, where the dependence on $\cH$ in the notation may be suppressed.

We denote the completely mixed state by $\tau := \frac{\one}{d} \in \cD(\cH)$, and the analogous uniform distribution by $u := (\frac{1}{d},\dotsc, \frac{1}{d})\in \cP$. A pure state is a rank-1 density matrix; we denote the set of pure states by $\Dpure(\cH)$. For two quantum states $\rho,\sigma \in \D(\cH)$, the \emph{trace distance} between them 
is given by
\[
T(\rho,\sigma) := \frac{1}{2}\|\rho-\sigma\|_1
\]
where $\|A\|_1 = \tr|A| = \sqrt{A^\dagger A}$ for $A \in \cB(\cH)$.
The trace distance $T(\rho,\sigma)$ has an operational interpretation in terms of the optimal success probability $p$ in distinguishing between two quantum states $\rho$ and $\sigma$ by a 2-outcome POVM, namely
\[
p = \frac{1}{2}(1 + T(\rho, \sigma)).
\]
Hence, the trace distance can be seen as a measure of distinguishability between $\rho$ and $\sigma$. Analogously, the total variation distance between $p, q\in \cP$ is defined as
\[
\TV(p,q) := \frac{1}{2}\|p-q\|_1
\]
and is analogously endowed with an interpretation in terms of distinguishability.

A function $F: \cD(\cH)\to \mathbb{R}$ is $k$-\emph{Lipschitz} (with respect to the trace distance) if for all $\rho,\sigma \in \cD(\cH)$,
\[
|F(\rho) - F(\sigma)| \leq k\, T(\rho,\sigma),
\]
and likewise  $F: \cP\to\bR$ is $k$-Lipschitz if for all $p,q\in\cP$, $|F(p) - F(q)| \leq k\,\TV(p,q)$.
The smallest $k>0$ such that $F$ is $k$-Lipschitz is called the \emph{optimal Lipschitz constant} for $F$. The function $F$ is said to be \emph{Lipschitz} continuous if it is $k$-Lipschitz for some $k>0$.

For $\eps > 0$, we define the $\eps$-ball (in trace distance) around $\sigma\in \D(\cH)$ as the set
\begin{equation}\label{eq:eps-ball}
\Be(\sigma) := \{ \omega\in \D(\cH): T(\omega,\sigma) \leq \eps \}, 
\end{equation}
and likewise the $\eps$-ball (in total variation distance) around a probability vector $p\in \cP$ as the set
\begin{equation} \label{eq:eps-ball-TV}
\Be(p) := \{ q\in \cP: \TV(p,q) \leq \eps \}.
\end{equation}
For any $A\in \Bsa(\cH)$, let $\lambda_+ (A)$ and $\lambda_-(A)$ denote the maximum and minimum eigenvalue of $A$, respectively, and $k_+(A)$ and $k_-(A)$ denote their multiplicities. Let $\lambda_j^\downarrow(A)$ denote the $j$th largest eigenvalue, counting multiplicity; that is, the $j^{th}$ element of the ordering
\[
\lambda_1^\downarrow(A)\geq \lambda_2^\downarrow(A) \geq \dotsm \geq \lambda_d^\downarrow(A).
\]
We set $\vec \lambda^\downarrow (A) := (\lambda_i^\downarrow(A))_{i=1}^d \in\R^d$ and denote the spectrum of $A\in \Bsa(\cH)$ (i.e.\@~its set of eigenvalues) by $\spec A \subset \R$.

The set of probability vectors with strictly positive entries is denoted $\cP_+$. For a vector $r\in \R^d$, $r_+$ denotes its largest entry, and $r_-$ denotes its smallest entry, and $k_+$ the multiplicity of the largest entry $r_+$ in $r$, and likewise $k_-$ the multiplicity of the smallest entry. For $r\in \mathbb{R}^d$, we denote $r_i^\downarrow$ as the $i$th largest entry of $r$ (counting multiplicity), and let $r^\downarrow := (r_i^\downarrow)_{i=1}^d$. Analogously, $r^\uparrow_i$ denotes the $i$th smallest entry of $r$, counting multiplicity, and $r^\uparrow := (r_i^\uparrow)_{i=1}^d$. We set
\[
\cP^\downarrow \equiv \cP_d^\downarrow := \left\{ q^\downarrow: q\in \cP \right\} = \left\{ q\in \cP : q_1\geq q_2 \geq \dotsm \geq q_d \right\}.
\]

We use $\log x$ to denote the base-2 logarithm of $x$ and $\ln x$ to denote the natural logarithm of $x$.

\paragraph{Majorization of vectors}
For $x,y\in \R^d$, we say $x$ \emph{majorizes} $y$, written $x \succ y$, if 
	\begin{equation} \label{def:majorize}
	 \sum_{j=1}^k x^\downarrow_j \geq \sum_{j=1}^k y^\downarrow_j \quad \forall k=1,\dotsc,d-1, \quad \text{and}\quad \sum_{j=1}^d x^\downarrow_j = \sum_{j=1}^d y^\downarrow_j.
	 \end{equation}
We have that $x \succ y$ if and only if
\begin{equation}\label{eq:DS-char-maj}
y = M x
\end{equation}
for some doubly-stochastic matrix $M$ \cite[Section 2, B.2]{MOA11}.
We say a function $\varphi: \cP \to \R$ is Schur convex on a set $S\subset \cP$ if  for $p,q\in S$, $p\prec q\implies \varphi(p) \leq \varphi(q)$. If $S = \cP$, we simply say $\varphi$ is Schur convex. We say $\varphi$ is \emph{Schur concave} on $S$ if $-\varphi$ is Schur convex on $S$, and likewise, $\varphi$ is Schur concave if $-\varphi$ is Schur convex. One useful characterization of Schur convex functions is if $\varphi : \cP \to \R$ is  symmetric and continuous, and differentiable on $\cP_+$, then it is Schur convex if and only if
\begin{equation} \label{eq:S-convex-condition}
(p_i - p_j) \left[ \partial_{p_i}\varphi(p) - \partial_{p_j} \varphi(p) \right] \geq 0 \qquad \forall i,j
\end{equation}
for each $p \in \cP$ \cite[Section 3, A.4.a]{MOA11}.

\paragraph{Majorization of quantum states}
Given two quantum states $\rho,\sigma\in \cD(\cH)$, we say $\sigma$ majorizes $\rho$, written $\rho\prec \sigma$ if $\vec\lambda^\downarrow(\rho) \prec \vec\lambda^\downarrow(\sigma)$. We say that $\varphi: \D(\cH) \to \R$ is \emph{Schur convex} if $\varphi(\rho)\leq \varphi(\sigma)$ for any $\rho,\sigma\in \cD(\cH)$ with $\rho \prec \sigma$. If $\varphi(\rho) < \varphi(\sigma)$ for any $\rho,\sigma\in \cD(\cH)$ such that  $\rho \prec \sigma$, and $\rho$ is not unitarily equivalent to $\sigma$, then $\varphi$ is \emph{strictly Schur convex}. We say $\varphi$ is Schur concave (resp.~strictly Schur concave) if $(-\varphi)$ is Schur convex (resp.~strictly Schur convex).

\chapter{Relationship between majorization and the 1-norm ball} \label{sec:geometry_trace_ball}

This chapter is concerned with an interesting relationship between the majorization pre-order and the 1-norm: namely, that balls in 1-norm admit a majorization-\emph{maximum} and -\emph{minimum}, despite not all elements of the ball being pairwise comparable. More precisely, let $r\in \cP$ be a probability vector, and consider the total-variation ball
\[
B_\eps(r) := \left\{ q \in \cP: \TV(r, q) := \frac{1}{2}\|r-q\|_1 \leq \eps \right\}
\]
for $\eps >0$. There exists $r\majmin, r\majmax \in B_\eps(r)$ such that
\begin{equation} \label{eq:intro-maj-relation}
r\majmin \prec q \prec r\majmax \quad \forall q \in B_\eps(r).
\end{equation}
We call $r\majmin$ the \emph{majorization-minimizer} over $B_\eps(r)$, and $r\majmax$ the \emph{majorization-maximizer} over $B_\eps(r)$. We place the upper star on the majorization-minimizer as we derived the states in \cite{HD18} in the context of maximizing Schur concave functions.

In the following, we will discuss some of the recent history of \eqref{eq:intro-maj-relation}, and how that history connects with the focus of this chapter. The first steps towards establishing \eqref{eq:intro-maj-relation} begin in \Cref{sec:q-to-c}.

This relationship between majorization and total variation distance, or key parts thereof, was independently found on at least three occasions: by Ho and Yeung in 2010~\cite{HY10}, by my supervisor and I in 2018~\cite{HD18}, and by Horodecki, Oppenhiem, and Sparaciari in 2018 \cite{HOS18}\footnote{Horodecki and Oppenhiem had known about this relationship for several years already, however~\cite{MHJO}.}. A generalization of these states were also constructed in the context of thermal majorization by van der Meer and Wehner~\cite{Remco}, and subsequently used to study approximate state transitions in \cite{vNW17}.

As an immediate consequence of \eqref{eq:intro-maj-relation}, if $H$ is Schur concave, then $r\majmin$ maximizes $H$ over $B_\eps(r)$:
\[
r\majmin \in \argmax_{q\in B_\eps(r)} H(q)
\]
This is used in \Cref{sec:smoothed-entropies} for calculating smoothed entropies, and provides a method for obtaining $r\majmin$ by optimization.

In this chapter, we present three ways to obtain $r\majmin$. The first, based on \cite{HD18} and presented in \Cref{sec:motivate-from-optimality}, uses necessary and sufficient conditions for a given state to be a maximizer of some concave function $H$ (i.e.\@~not necessarily unitarily invariant) over a trace distance ball in $\cD(\cH)$ (established in \Cref{sec:general_optimality_conditions}) by specializing to the case in which $H$ is the von Neumann entropy. The general conditions reduce to ones suggestive of a particular form, which turns out to be the form of $\diag(r\majmin)$. In fact, these ``reduced conditions'' can be obtained directly from a waterfilling argument applied to maximizing the Shannon entropy over a total-variation distance ball as shown in \Cref{sec:motivate-waterfilling}. The state can also be obtained by calculating the majorization-infimum over the ball and checking that it indeed resides inside the ball itself. This is discussed in \Cref{sec:motivate-from-inf}.

To obtain the majorization-maximizer $r\majmax$, which turns out to have a very simple construction, one simply uses that $B_\eps(r)$ is a polytope, and $r\majmax$ is necessarily one of its vertices (see \Cref{sec:first-look} for more).

\cite{HY10} constructed the majorization-minimizer $r\majmin$, and \cite{HOS18} constructed both the minimizer and maximizer, but neither discuss in detail how they were obtained; possibly, a waterfilling argument like the one sketched in \Cref{sec:motivate-waterfilling} was used.

The relation \eqref{eq:intro-maj-relation} immediately provides \emph{local} continuity bounds, in the sense that if $H: \cP \to \R$ is a Schur concave function, and $r\in \cP$ is given, then the variation in $H$ near $r$ can be bounded as
\begin{equation} \label{eq:intro-local-bounds}
|H(q)- H(r)| \leq \max \left\{ H(r) - H(r\majmax), H(r\majmin) - H(r) \right\} \qquad \forall q\in B_\eps(r).
\end{equation}
These bounds, the relation \eqref{eq:intro-maj-relation}, and the derivation of optimality conditions for maximizing a concave function over a trace-ball, were the focus of \cite{HD18}. In that work, we also proved a semi-group property, namely for $r\in \cP$ and $\eps_1,\eps_2 >0$,
\begin{equation}\label{eq:intro-semigroup}
r^*_{\eps_1 + \eps_2} = (r^*_{\eps_1})^*_{\eps_2}.
\end{equation}
This property plays a central role in \Cref{chap:majflow_ctybounds}.

The work \cite{HY10} instead focussed on establishing a \emph{uniform} continuity bound (in contrast to a local one), for the Shannon entropy. In fact, likely unbeknownst to the authors, the optimal uniform continuity bound had already been established in 2007 by Audenaert \cite{Aud07} and independently by \cite{Zha07a}, as discussed in \Cref{sec:ECB_intro}. \cite{HY10} derive this bound in a different manner, with a proof prominently featuring majorization, but also using heavily the specific expression for the Shannon entropy (as opposed to a more general Schur concave function).

\cite{HOS18} established the existence of the majorization-minimizer and -maximizer states and used them to define a ``majorization distance'' between probability distributions, which is used in \Cref{sec:LOCC}. They also proved that if $p \prec q$, then $p\majmin \prec q\majmin$ for all $\eps >0$, which will prove useful in establishing uniform continuity bounds (and appears in \Cref{prop:properties_of_Lambda_eps}).

In \cite{HD17}, my supervisor and I showed that to establish a uniform continuity bound for a certain class of Schur concave functions, one only needs to establish the bound for $\eps$ small enough, and then use the semigroup property finitely many times to establish the bound for all $\eps>0$. While this provided a new technique for establishing uniform continuity bounds, most of the applicable entropic quantities already had known tight uniform continuity bounds. The biggest limitation of the technique is that it only applied to concave functions. However, one of the most widely used entropies without a known tight uniform continuity bound was the $\alpha$-R\'enyi entropy
\[
S_\alpha(r) = \frac{1}{1-\alpha}\log \sum_i r_i^\alpha
\]
for $\alpha > 1$, which is neither concave nor convex (but is Schur concave). In trying to establish a uniform continuity bound for $S_\alpha$, eventually it became apparent that an infinitesimal approach provided a much simpler way to prove the result. The semigroup property naturally allows statements about $r\majmin$ to be derived in terms of $\left. \partial_\eps r\majmin\right|_{\eps=0}$, leading to the notion of a \emph{path of majorization flow}, described in \Cref{chap:majflow_ctybounds} (and the preprint \cite{HD19}).

 \section{Quantum-to-classical reduction} \label{sec:q-to-c}

In this chapter, we are interested in the relationship between the majorization of quantum states and the trace distance, with an interest in developing continuity bounds for Schur concave functions on the set of density matrices. This task reduces to a ``classical'' one: understanding the relationship between the majorization of probability vectors and the total variation distance, and applying that to continuity bounds for Schur concave functions on the probability simplex. 

First, the reduction for majorization is trivial: $\rho \prec \sigma$ if and only if $\vec \lambda^\downarrow(\rho) \prec \vec \lambda^\downarrow(\rho)$, by definition.
Next, let us consider the trace distance. Let $A\in \Bsa(\cH)$. Then let $\Eig^\downarrow(A) := \diag (\vec \lambda^\downarrow(A))$ be the diagonal matrix whose diagonal entries are the sorted eigenvalues of $A$, counted with multiplicity. By \cite[IV.62]{Bha97}, we have for $A,B\in \Bsa(\cH)$
\begin{equation}\label{eq:unorm-q-to-c}
\unorm{\Eig^\downarrow(A)- \Eig^\downarrow(B)} \leq \unorm{A-B} 
\end{equation}
for any unitarily invariant norm $\unorm{\cdot}$. In the case of the trace distance,
\begin{equation*}T(\Eig^\downarrow(A), \Eig^\downarrow(B)) \equiv \frac{1}{2}\|\Eig^\downarrow(A)- \Eig^\downarrow(B)\|_1 = \frac{1}{2}\|\vec \lambda^\downarrow(A) - \vec \lambda^\downarrow(B)\|_1 \equiv \TV(\vec \lambda^\downarrow(A), \vec \lambda^\downarrow(B))
\end{equation*}
is the total variation distance between the sorted vectors of eigenvalues of $A$ and $B$,  and \eqref{eq:unorm-q-to-c} yields
\begin{equation}\label{eq:TV-bound-by-T}
 \TV(\vec \lambda^\downarrow(A), \vec \lambda^\downarrow(B)) \leq T(A, B).
\end{equation}
 I first encountered this reduction in \cite{Aud07}. 

Lastly, let $H: \cD(\cH)\subset \Bsa(\cH) \to \R$ be a unitarily invariant function. Then $H(A) = H(\Eig^\downarrow(A))$ is a function of the eigenvalues of $A$ alone: $H(A) = H_{\operatorname{cl}}(\vec \lambda^\downarrow(A))$ for some function $H_{\operatorname{cl}} : \cP \to \R$ (namely $H_{\operatorname{cl}}(r) := H(\diag(r))$ for $r\in \cP$, where $\diag(r)$ is the diagonal matrix with the entries of $r$ upon its main diagonal).

We can see that these reductions interact favorably for establishing continuity bounds, in the following sense. If $\frac{1}{2}\|\rho-\sigma\|_1 \leq \eps$, then $\frac{1}{2}\|\vec \lambda^\downarrow(\rho) - \vec \lambda^\downarrow(\sigma)\|_1\leq \eps$ as well, and
\[
|H(\rho) - H(\sigma)| = |H_{\operatorname{cl}}(\vec \lambda^\downarrow(\rho)) - H_{\operatorname{cl}}(\vec \lambda^\downarrow(\sigma))|.
\]
Hence, it remains to bound this difference in terms of $\eps$, using $\frac{1}{2}\|\vec \lambda^\downarrow(\rho) - \vec \lambda^\downarrow(\sigma)\|_1\leq \eps$. This is precisely the task of establishing a  continuity bound on $\cP$ with respect to the total variation distance.

Moreover, this reduction preserves tightness, as shown by the following simple argument. Assume we have established a bound of the form
\[
|H_{\operatorname{cl}}(p)-H_{\operatorname{cl}}(q)|\leq f(\eps)
\]
for all $p,q \in \cP$ such that $\TV(p,q)\leq \eps$, for some function $f$. Then by the above reduction, we know that for all $\rho,\sigma \in \cD(\cH)$ with $T(\rho,\sigma)\leq \eps$,
\[
|H(\rho)-H(\sigma)| \leq f(\eps).
\]
If two vectors $p^*,q^*\in \cP$ with $\TV(p^*,q^*)\leq \eps$ achieve $|H_{\operatorname{cl}}(p^*) - H_{\operatorname{cl}}(q^*)| = f(\eps)$, then $\rho^* := \diag(p^*)$ and $\sigma^* := \diag(q^*)$ satisfy
\[
|H(\rho^*)-H(\sigma^*)| = f(\eps).
\]
That is, a tight bound on the classical level yields a tight bound on the quantum level.

 \section{A first look at majorization over the 1-norm ball}\label{sec:first-look}

Fix $\eps>0$ and a state $\rho\in \D(\H)$, with $d:=\dim \H$. The main result of this chapter is that the $\eps$-ball around $\rho$, the $\eps$-ball in trace distance, $\Be(\rho)$, admits a minimum and maximum in the majorization order. Note that since majorization is a partial order, a priori one does not know that there are states in $\Be(\rho)$ comparable to every other state in $\Be(\rho)$.
	\begin{theorem}\label{thm:max-min-ball}
	Let $\rho\in \cD$ and $\epsilon > 0$. Then there exist two states $\rho\majmin$ and $\rho\majmax$ in $\Be(\rho)$ (which are defined by \Cref{eq:def_majmin_state_quantum} in \Cref{sec:construct_majmin} and \Cref{eq:def_rho_eps_lowerstar} in \Cref{sec:construct_majmax} respectively) both of which commute with $\rho$ and satisfy 
	\begin{equation}
	\rho\majmin \prec \omega \prec \rho\majmax, \qquad \forall \omega \in \Be(\rho). \label{eq:min-max-maj-cond-quantum}
	\end{equation}
	Moreover, $\rho\majmin$ is the unique state in $\Be(\rho)$ satisfying the left-hand relation of \eqref{eq:min-max-maj-cond-quantum}, and $\rho\majmax$ is the unique state in $\Be(\rho)$ satisfying the right-hand relation of \eqref{eq:min-max-maj-cond-quantum} up to unitary equivalence. 

	This result can be specialized to the case of probability vectors. Let $p\in \cP_d$ and $\epsilon > 0$. Then there exist two probability vectors $p\majmin$ and $p\majmax$ in $\Be(p)$ (which are defined by \Cref{def:majmin-state} in \Cref{sec:construct_majmin} and \Cref{eq:def_pjstar} in \Cref{sec:construct_majmax} respectively) which satisfy 
	\begin{equation}
	p\majmin \prec q \prec p\majmax, \qquad \forall q \in \Be(p). \label{eq:min-max-maj-cond}
	\end{equation}
	Moreover, $p\majmin$ is the unique element of $\Be(p)$ satisfying the left-hand relation of \eqref{eq:min-max-maj-cond}, and $p\majmax$ is the unique element of $\Be(p)$ satisfying the right-hand relation of \eqref{eq:min-max-maj-cond} up to permutations. 
	\end{theorem}

\begin{figure}[ht]
\centering
\includegraphics{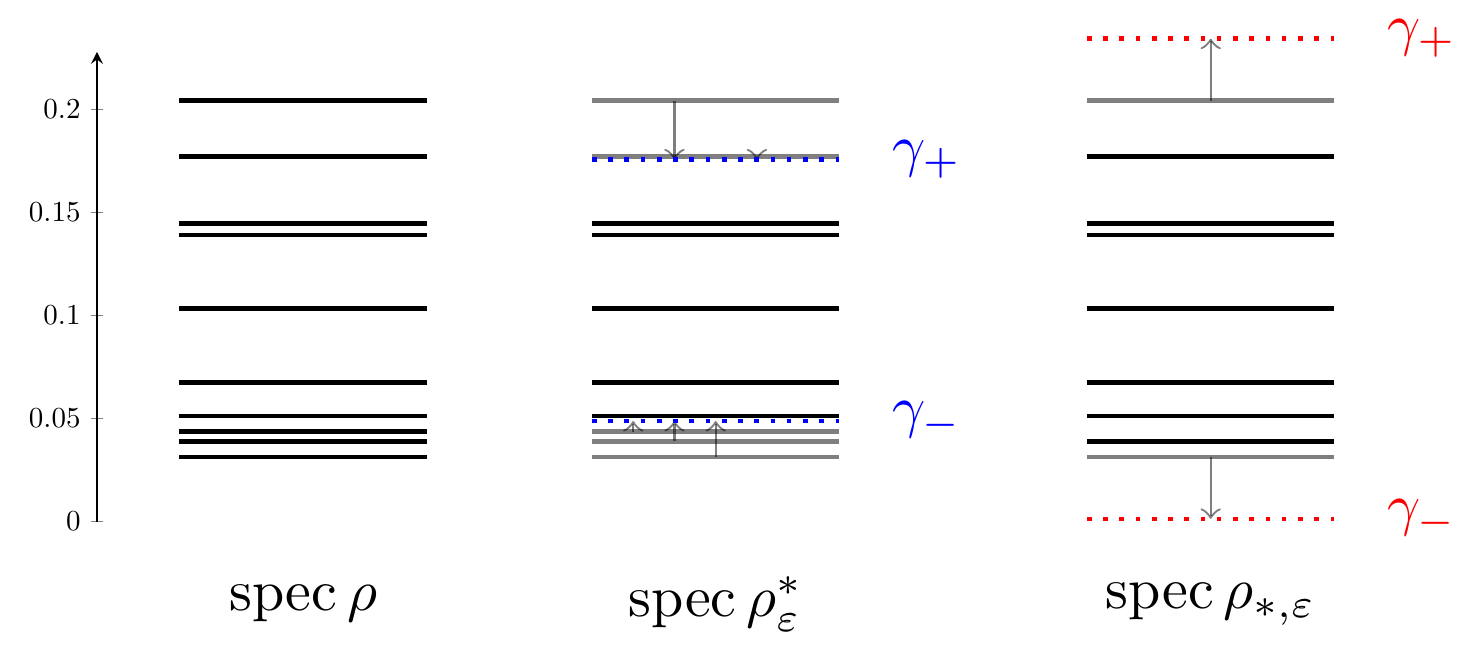}
\caption{An example of the majorization-minimizer and majorization-maximizer with $d=2$. The spectrum of a quantum state $\rho$ is on the left, the spectra of $\rho\majmin$ in the center and $\rho\majmax$ on the right, with $\eps = 0.03$. The states are constructed in \Cref{sec:construct_majmin} and \Cref{sec:construct_majmax} respectively, and have the structure of raising or lowering some eigenvalues of $\rho$ according to a prescribed formula.} \label{fig:example-rhomaxeps-rhomineps}
\end{figure}

\begin{remark}
See \Cref{fig:example-rhomaxeps-rhomineps} for an example of $\rho\majmin$ and $\rho\majmax$ for a particular state $\rho$.
\end{remark}

In this section, we consider the $\eps$-ball $\Be(\rho)$ around a state $\rho \in \D(\H)$, and motivate the construction of maximal and minimal states in the majorization order given in (\ref{eq:min-max-maj-cond-quantum}). As discussed in \Cref{sec:q-to-c}, we prove \Cref{thm:max-min-ball} by reducing it to the classical case of discrete probability distributions on $d$ symbols, and then constructing explicit states $\rho\majmin$ (in \Cref{sec:construct_majmin}), and $\rho\majmax$ (in \Cref{sec:construct_majmax}), whose eigenvalues are respectively given by the probability distributions which are minimal and maximal in majorization order.

\begin{figure}[ht]
\centering
\begin{minipage}[b]{0.28\textwidth}
\centering \includegraphics{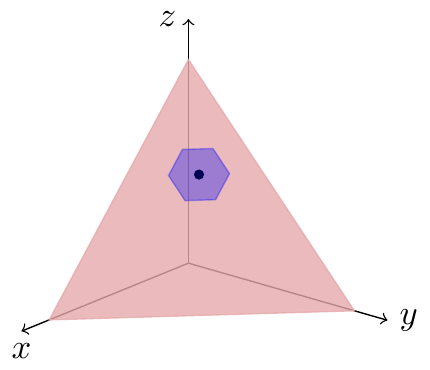}
\subcaption{$\Be(p)$, with $p$ in black}\label{fig:Mwa}
\end{minipage}\qquad
\begin{minipage}[b]{0.30\textwidth}
\centering \includegraphics{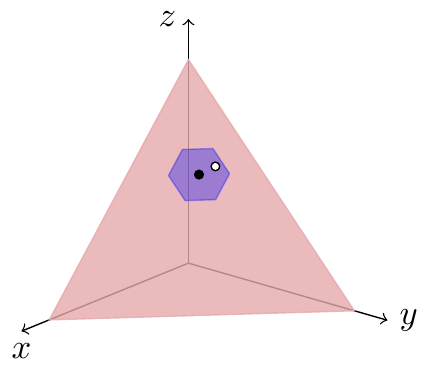}
\subcaption{Some $w\in \Be(p)$ (white).}\label{fig:Mwb}
\end{minipage}\qquad
\begin{minipage}[b]{0.28\textwidth}
\centering \includegraphics{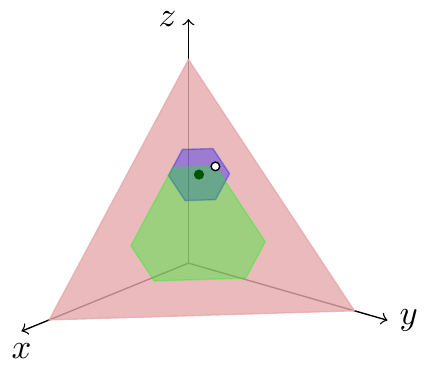}
\subcaption{The set $M_w$ (green).}\label{fig:Mwc}
\end{minipage}\caption{In dimension $d=3$, the simplex, $\cP_d$, of probability vectors is the shaded triangle shown in (a), along with the ball $\Be(p)$ which is the hexagon shown in blue, centered at $p = \left(0.21, 0.24, 0.55\right)$ (depicted by a black dot) with  $\epsilon = 0.1$. In (b), a point $w = \left(0.14, 0.28, 0.58\right) $ is depicted in white, and in (c), the set $M_w$ is shown in green. \label{fig:Mw}}
\end{figure}

As discussed in \Cref{sec:q-to-c}, it suffices to consider the simplex of probability vectors $\cP_d$, equipped with the total variation distance, instead of the set of density operators $\cD(\cH)$ equipped with the trace distance. Note that $\cP_d$ is the polytope (i.e.\@~the convex hull of finitely many points) generated by $(1,0,\dotsc,0)$ and its permutations. Recall the total variation ball,
\[
\Be(p) = \{w = (w_i)_{i=1}^d\in \cP_d:  \frac{1}{2}\|w-p\|_1 \leq \eps\}.
\] 
The set $\{ x \in \R^d: \|x\|_1 \leq 1\}$ can be written
\[
\{ x \in \R^d: \|x\|_1 \leq 1\} = \conv\{e_1,-e_1,\dotsc,e_d,-e_d\},
\]
where  $\conv(A)$ denotes the convex hull of a set $A$, and $e_1,\dotsc,e_d$ are the vectors of the standard basis (e.g.\@ $e_j = (0,\dotsc,0,1,0,\dotsc,0)$ with $1$ in the $j$th entry), and is therefore a polytope, called the \emph{$d$-dimensional cross-polytope} (see e.g.\@~\cite[p.~82]{matousek2002lectures}). As a translation and scaling of the $d$-dimensional cross-polytope, the set $\{w \in \R^d: \frac{1}{2}\|w -p\|_1 \leq \eps\}$ is a polytope as well. As $\Be(p)$ is the intersection of this set and $\cP_d$, it too is a polytope. See \Cref{fig:Mwa} for an illustration of $\Be(p)$ in a particular example.

The existence of $\rho\majmin$ and $\rho\majmax$  in $\Be(\rho)$ satisfying \eqref{eq:min-max-maj-cond} is equivalent to $p\majmin$ and $p\majmax$ in $\Be(p)$ satisfying
\begin{equation}\label{class-maj}
p\majmin \prec w \prec p\majmax 
\end{equation}
for all $w\in \Be(p)$. Using Birkhoff's Theorem (e.g.\@~\cite[Theorem 12.12]{NC}), the set of vectors majorized by a point $w \in \cP_d$ can be shown to be given by 
	\begin{equation}
	M_{w}:=\{p\in \cP_d: w \succ p\} = \conv \{ \pi (w): \pi \in S_d\}, \label{eq:def_Mw_char_conv_hull_of_perm}
	  \end{equation} where $S_d$ is the symmetric group on $d$ letters (see \cite[p.~34]{Marshall2011}). Let us illustrate this with an example in $d=3$. Let us choose $p = \left(0.21, 0.24, 0.55\right)$ and $\epsilon = 0.1$. The simplex $\cP_d$ and ball $\Be(p)$ are depicted in \Cref{fig:Mwa}. A point $w = \left(0.14, 0.28, 0.58\right) \in \Be(p)$ is shown in \Cref{fig:Mwb}, and the set $M_w$ in \Cref{fig:Mwc}. 

The geometric characterization \eqref{eq:def_Mw_char_conv_hull_of_perm}, depicted in \Cref{fig:Mw}, requires that $\Be(p) \subseteq M_{p\majmax}$, and conversely, $p\majmin \in M_p$ for each $p\in \Be(p)$.  \Cref{fig:Mwc} shows that for the point $w$, $\Be(p) \not \subseteq M_w$, implying that $w\neq p\majmax$. Moreover, one can check that that e.g.\@ $w \not \in M_p$, and hence $w\neq p\majmin$.

Next we consider Schur concave functions on $\cP_d$, in order to gain insight into the probability distributions $p\majmax$ and $p\majmin$ which arise
in the majorization order (\ref{class-maj}). In particular, let us consider Shannon entropy $S(w):= -\sum_{i=1}^d w_i \log w_i$ of a probability distribution $w = (w_i)_{i=1}^d$. It is known to be strictly Schur concave. Hence, if $p\majmax \in \Be(p)$ satisfying \eqref{class-maj} exists, it must satisfy
\[
S(p\majmax) \leq S(w)
\]
for any $w\in \Be(p)$. Thus, $p\majmax$ must be a minimizer of $S$, which is a concave function, over $\Be(p)$, a convex set. Similarly, $p\majmin$ must be a maximizer of $S$ over $\Be(p)$. Properties of maximizers of  concave functions over a convex sets are  well-understood; in particular, any local maximizer is a global maximizer.

The task of minimizing a concave function over a convex set is a priori more difficult; in particular, local minima need not be global minima. There is, however, a {\em{minimum principle}} which asserts that the minimum occurs on the boundary of the set; this is formulated more precisely in e.g.\@ \cite[Chapter 32]{Rockafellar_book}.  Since $\Be(p)$ is a polytope,  $S$ is minimized on one of the finitely many vertices of $\Be(p)$. This fact yields a simple solution to the problem of minimizing $S$ over $\Be(p)$, as described below by example, and in generality in \Cref{sec:construct_majmax}.

Let us return to the example of \Cref{fig:Mw}. We see $\Be(p)$ has six vertices; these are $\{ p + \pi((\epsilon,- \epsilon,0)): \pi \in S_d\}$. 
\begin{figure}[ht]
\centering
\begin{minipage}[b]{.46\linewidth}
\centering \includegraphics{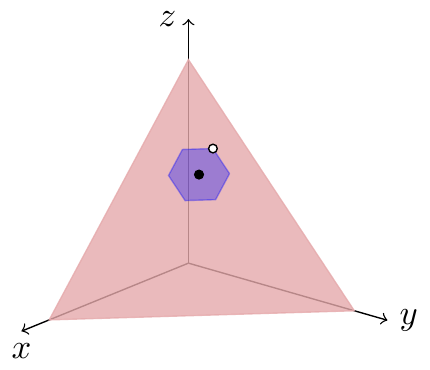}
\subcaption{A minimum $v$ of $S$ over $\Be(p)$, in white.}\label{fig:min-max-examplea}
\end{minipage}\qquad
\begin{minipage}[b]{.46\linewidth}
\centering \includegraphics{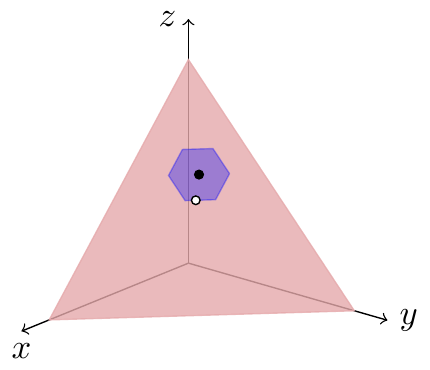}
\subcaption{Maximum $m$ of $S$ over $\Be(p)$, in white.}\label{fig:min-max-exampleb}
\end{minipage}\caption{For the example of \Cref{fig:Mw}, the (unique) maximum and minimum of the Shannon entropy $S$ over $\Be(p)$ are shown. Both $v$ and $m$ occur on the boundary of $\Be(p)$.
 }\label{fig:min-max-example}
\end{figure}
The vertex which minimizes $S$ is $v:=\left(0.21- \epsilon, 0.24, 0.55+ \epsilon\right)$, where the smallest entry is decreased and the largest entry is increased, as shown in \Cref{fig:min-max-exampleb}. Moreover, one can check that $w \prec v$ for any $w\in \Be(p)$.  This leads us to the conjecture that the vertex corresponding to decreasing the smallest entry and increasing the largest entry will yield $p\majmax$ satisfying \eqref{class-maj} in general. We see in \Cref{sec:construct_majmax} that this is indeed true, although in some cases more than one entry needs to be decreased.

On the other hand, finding the probability distribution $p\majmin$ in $\Be(p)$ which satisfies \eqref{class-maj} is more than a matter of checking the vertices of $\Be(p)$, as shown by \Cref{fig:min-max-exampleb}: in this example, $p\majmin$ is not a vertex of $\Be(p)$. Interestingly, useful insight into this probability distribution can be obtained by using results from convex optimization theory. This is discussed in the following section. 
\section{Constructing the majorization-minimizer \eqref{eq:min-max-maj-cond}}\label{sec:construct_majmin}

Let $r\in \cP_d$ and $\eps > 0$. If $u \in B_\eps(r)$, then $r_\eps^* = u$, where $u := (1/d,\dotsc,1/d)$ denotes the uniform distribution, so instead consider the case in which $u \not \in B_\eps(r)$. We first define a quantity $\gamma_+^{(m)}\equiv \gamma_+^{(m)}(r,\eps)$, for $m\in \{1,\dotsc,d-1\}$, as 
\[
    \gamma_+^{(m)} := \frac{1}{m}\left(\sum_{i=1}^{m} r_i^\downarrow  - \eps\right).
\]
Similarly, define $\gamma_-^{(m)} \equiv \gamma_-^{(m)}(r,\eps) $ by
\[
    \gamma_-^{(m)}:= \frac{1}{m}\left( \sum_{i=1}^m r_i^\uparrow + \eps \right) = \frac{1}{m}\left(\sum_{i=d-m+1}^d r_i^\downarrow  + \eps\right).
\]
Then we define $m_+ := m_+(r,\eps)$ as the unique solution to the following inequalities:
\begin{equation}\label{eq:m_is_sol_to_this}
    r^\downarrow_{m+1} \leq \gamma_+^{(m)} < r^\downarrow_m, \qquad m\in \{1,\dotsc,d-1\}
\end{equation}
and we set $m_+(u,\eps) = 0$.
Similarly, for $r\neq u$, we define $m_- := m_-(r,\eps)$ as the unique solution to the inequalities:
\begin{equation} 
    r^\uparrow_{m}< \gamma_-^{(m)}\leq r^\uparrow_{m+1}, \qquad  m\in \{1,\dotsc,d-1\}
\end{equation}
or equivalently, the inequalities
\begin{equation}\label{eq:n_is_sol_to_this}
    r^\downarrow_{d-m+1}< \gamma_-^{(m)}\leq r^\downarrow_{d-m}, \qquad  m\in \{1,\dotsc,d-1\}.
\end{equation}
Lastly, we set $\gamma_+ = \gamma_+(r,\eps) := \gamma_+^{(m_+)}$ and  $\gamma_-=\gamma_-(r,\eps) := \gamma_-^{(m_-)}$. The proof that the above equations have unique solutions follows from \Cref{lem:uniqueness_for_majmin}, which is postponed to the end of this section. In the following, we use Dirac notation where for $i=1,\dotsc,d$, the symbol $\ket{i}$ denotes the vector $e_i = (0,\dotsc,0,1,0,\dotsc,0)$, with the $1$ in the $i$th place. Then, given $r = \sum_{i=1}^d r_i^\downarrow \ket{\pi(i)}$ for some permutation $\pi \in S_d$ and where $\{\ket{i}\}_{i=1}^d$ is the standard basis of $\mathbb{R}^d$, we define
\begin{equation} \label{def:majmin-state}
    r_\eps^* := \sum_{i=1}^{m_+} \gamma_+\ket{\pi(i)} + \sum_{i=m_++1}^{d-m_-} r_i^\downarrow \ket{\pi(i)} + \sum_{i=d-m_-+1}^d \gamma_-\ket{\pi(i)}.
\end{equation}
Note that for any permutation matrix $P$ we have
\begin{equation}\label{eq:majmin-perm-covariance}
(Pr)\majmin = P(r\majmin).
\end{equation}
We check that $r\majmin\in \Be(r)$ as follows. From its definition, its entries lie in the interval $(0,1]$ and $r\majmin$ has sum 1. Additionally, if $\TV(r,u) \leq \eps$, then $r\majmin = u \in \Be(r)$. Otherwise,
\begin{equation}
    \|r\majmin-r\|_1 = \sum_{i=1}^{m_+}|\gamma_- - r_i^\downarrow | + \sum_{i= d- m_- + 1}^d |\gamma_+ - r_i^\downarrow | = \eps+\eps = 2 \eps; \label{eq:td_rho-eps-sigma}
\end{equation}
so for any $\epsilon\in[0,1]$, we have $r\majmin\in \Bep(r)$.

\begin{figure}[ht]
    \centering
    \includegraphics{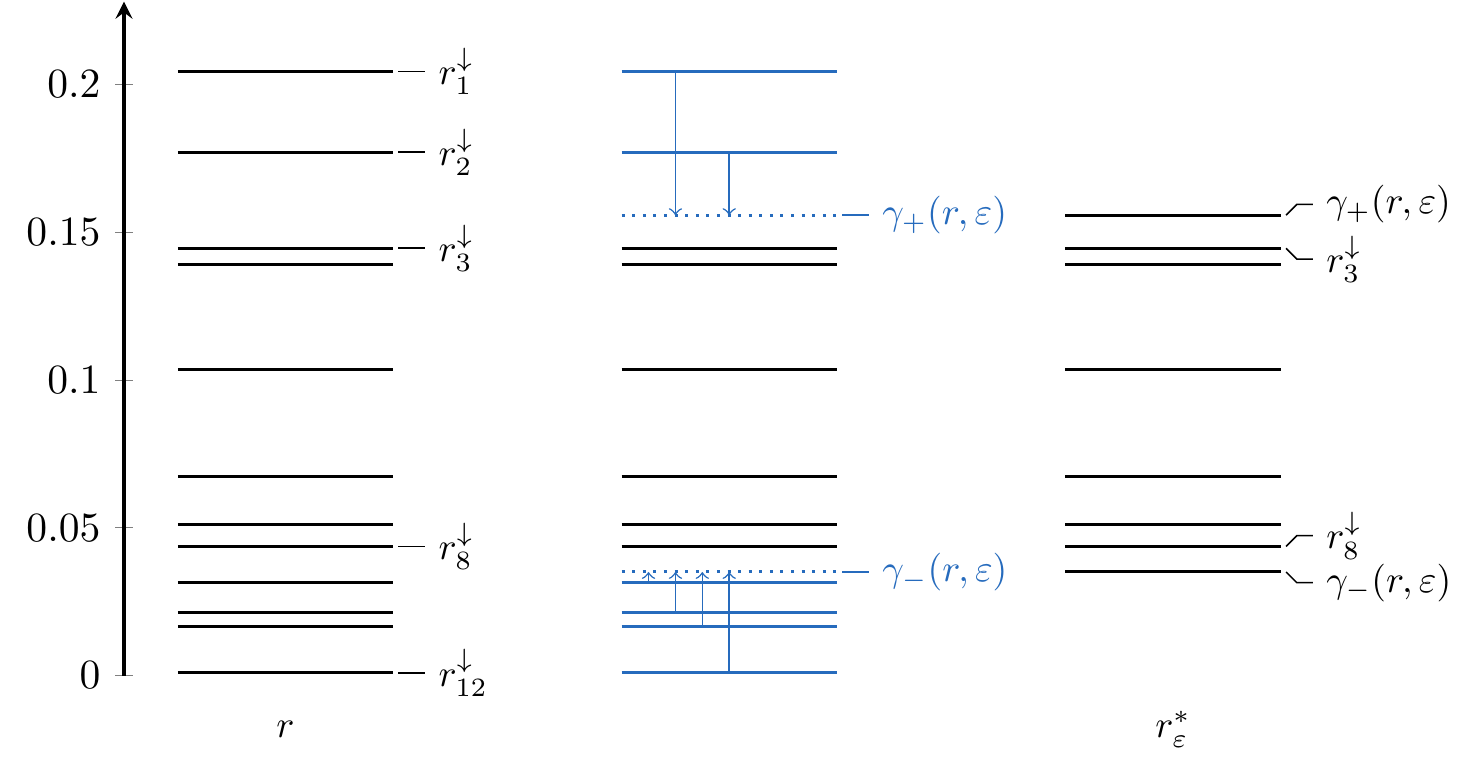}
    \caption{We choose $d=12$, a probability vector $r \in \cP_d$, and $\eps=0.07$, for which $m_+=2$ and $m_-=4$. Left: the entries $r^\downarrow_1 \geq r^\downarrow_2\geq \dotsc \geq r^\downarrow_d$ of $r$ are plotted. Center: the smallest four entries of $r$ are increased to $\gamma_- = \frac{1}{4}[r^\downarrow_d + \dotsm + r^\downarrow_{d+4}+  \eps]$, and the largest two entries of $r$ decreased to $\gamma_+ = \frac{1}{2}[r^\downarrow_1+r^\downarrow_2 - \eps]$. Right: the entries of $r_\eps^*$ are $\gamma_+$ with multiplicity two, $r^\downarrow_3,r^\downarrow_4,\dotsc,r^\downarrow_{d-4}$, and $\gamma_-$ with multiplicity four. \label{fig:rho-levels}}
\end{figure}

To summarize, we construct $r_\eps^*$ as follows: we decrease the $m_+$ largest entries of $r$ by setting them to $\gamma_+$ (where $m_+$ and $\gamma_+$ are related by \cref{eq:m_is_sol_to_this}), increase the $m_-$ smallest entries of $r$ by setting them to $\gamma_-$ (where $m_-$ and $\gamma_-$ are related by \cref{eq:n_is_sol_to_this}), and we keep the other entries of $r$ unchanged. This is illustrated in \Cref{fig:rho-levels}, for a state $r\in \cD(\cH)$ with $\eps = 0.07$ and $d=12$.

In the case of a quantum state $\rho\in \cD(\cH)$, let the sorted eigendecomposition of $\rho$ be given as $\rho =  \sum_{i=1}^d r_i^\downarrow\, U \ketbra{i}{i} U^*$, so $(r_i^\downarrow)_{i=1}^d \in \cP^\downarrow$ is the sorted collection of eigenvalues of $\rho$, counted with multiplicity (note $\bra{i} = e_i^T$ can be seen as a row vector). Then simply define
\begin{equation}\label{eq:def_majmin_state_quantum}
\rho\majmin := \sum_{i=1}^d (r\majmin)_i^\downarrow \, U \ketbra{i}{i} U^*.
\end{equation}
This choice ensures that $\rho\majmin \in B_\eps(\rho)$ and
\begin{equation}
(r^\downarrow)\majmin \prec q \quad \forall q \in B_\eps(r^\downarrow) \qquad\implies\qquad \rho\majmin \prec \omega \quad \forall \omega \in B_\eps(\rho)
\end{equation}
as discussed in \Cref{sec:q-to-c}.

In the next section, we provide three ways to motivate this construction. Then in \Cref{sec:max_of_pmaxeps}, we prove that $r\majmin$ satisfies the majorization relation \eqref{eq:min-max-maj-cond}.
 \section{Three ways to find the majorization-minimizer}\label{sec:motivate}

In this section, we present three ways to derive the form of the state given in \eqref{def:majmin-state}.

\subsection{From optimality conditions} \label{sec:motivate-from-optimality}

    One way to obtain $\rho\majmin$ is from the following theorem, which is a condensed form of \Cref{thm:optimality_conditions} in \Cref{sec:general_optimality_conditions}.
    Given a suitable function $\varphi:\cD(\cH) \to \R$,
    and any state $\rho \in \cD(\cH)$, the theorem provides a necessary and sufficient condition under which a state maximizes $\varphi$ in the $\eps$-ball (of positive definite states), $\Bep(\rho)$, of the state $\rho$. These are simply the first-order optimality conditions provided by Fermat's theorem; see \Cref{sec:general_optimality_conditions} for more details.

    \begin{theorem} \label{thm:short_optimality_conditions}
    Let $\rho\in \cD(\cH)$, $\epsilon > 0$, and $\varphi:\cD(\cH) \to \RR$ be a concave, continuous function which is G\^ateaux-differentiable\footnote{For the definition of Gateaux-differentiability and Gateaux gradient see \Cref{tools-convex}.} on $\cD_+(\cH)$.  A state $\xi\in \Bep(\rho)$, satisfies $\xi\in \argmax_{\Be(\rho)} \varphi$ if and only if \emph{both} of the following conditions are satisfied. Here $L_\xi := \nabla \varphi(\xi)$ denotes the G\^ateaux-gradient of $\varphi$ at $\xi$.
        \begin{enumerate}
        \item Either $\frac{1}{2}\|\xi-\rho\|_1= \eps$ or $L_\xi = \lambda \one$ for some $\lambda \in \R$, and
        \item  we have 
        \begin{align}\label{eq-eval}
        \pi_\pm L_\xi \pi_\pm &= \lambda_\pm(L_\xi) \pi_\pm,
        \end{align}
where $\pi_\pm$ is the projection onto the support of $\Delta_\pm$, and where $\Delta= \Delta_+ - \Delta_-$ is the Jordan decomposition of $\Delta:= \xi-\rho$, i.e.\@~$\Delta_\pm \geq 0$.
    \end{enumerate}
\end{theorem}

As shown in \Cref{sec:general_optimality_conditions}, the von Neumann entropy $S$ satisfies the requirements of the function $\varphi$ of the theorem, with $L_\xi = - \log \xi - \tfrac{1}{\log_{\mathrm{e}}(2)}\one$. Using \Cref{thm:short_optimality_conditions}, we deduce properties of and the form of a maximizer of $S$ in the $\eps$-ball.

Since $S$ is continuous and $\Be(\rho)$ is compact, $S$ achieves a maximum over $\Be(\rho)$. Moreover, since $S$ is strictly concave, the maximum is unique; otherwise, if $\xi_1,\xi_2\in \Be(\rho)$ were maximizers, $\xi = \frac{1}{2}\xi_1 + \frac{1}{2}\xi_2\in \Be(\rho)$ would have strictly higher entropy.

Now, let $\xi$ be the unique maximizer. Condition 1 of \Cref{thm:short_optimality_conditions} yields that either $\log \xi \propto \one$, so $\xi = \tau := \frac{\one}{d}$, or else $T(\xi,\rho) = \epsilon$. Since $\tau$ is the global maximizer of $S$ over $\cD$, we have $\xi = \tau$ whenever $\tau\in \Be(\rho)$. If $\tau\not\in \Be(\rho)$, then this condition yields the first piece of information about the maximizer: it is on the boundary of $\Be(\rho)$, in that $T(\xi,\rho) = \epsilon$, just as shown in \Cref{fig:min-max-exampleb} in the classical setup.

By \eqref{eq:unorm-q-to-c}, working in the basis in which $\rho = \Eig^\downarrow(\rho)$, we have
\[
    T(\Eig^\downarrow(\xi) , \rho) \leq T(\xi,\rho) \leq \epsilon
\]
and therefore $\Eig^\downarrow(\xi) \in \Be(\rho)$. Since $S$ is unitarily invariant, $S(\Eig^\downarrow(\xi)) = S(\xi)$, and by uniqueness of the maximizer, we have $\xi = \Eig^\downarrow(\xi)$. Hence, the maximizer $\xi$ commutes with $\rho$, and hence with $\Delta$, and the sums of its eigenprojections $\pi_\pm$. Since $[L_\xi,\xi]=0$ as well, \Cref{thm:short_optimality_conditions} yields
\begin{equation}
    L_\xi \pi_\pm = \lambda_\pm(L_\xi) \pi_\pm \label{eq:VN_L_rho_motivation}
\end{equation}
For any $\psi \in \pi_\pm \cH$, we have $L_\xi \psi = \lambda_\pm(L_\xi) \psi$, so \eqref{eq:VN_L_rho_motivation} is an eigenvalue equation for $L_\xi$. Since $\xi = \exp\left(-(L_\xi +\one)\right)$ is a function of $L_\xi$, it shares the same eigenprojections. In particular,
\[
    \xi \pi_\pm = \exp\left(- (\lambda_{\pm}(L_\xi) +1) \right)\pi_\pm
\] serves as an eigenvalue equation for $\xi$. Since $[\xi,\rho]=0$, we can discuss how each acts on each (shared) eigenspace. By definition, $\xi$ and $\rho$ act the same on $\ker \Delta = \ker(\xi-\rho)$. On the other hand, on the subspaces where the eigenvalues of $\xi$ are greater than those of $\rho$, i.e.\@ on $\pi_+\cH$, we see that $\xi$ has the constant eigenvalue $\gamma_-:= \exp\left(- (\lambda_{+}(L_\xi) +1)\right)$, and on $\pi_-\cH$, $\gamma_+:= \exp\left(- (\lambda_{-}(L_\xi) +1)\right)$. Note that since $x\mapsto - \log x - 1$ is monotone decreasing on $\R$,  the subspace $\pi_+\cH$ where $L_\xi$ has its largest eigenvalue, $\xi$ has its smallest eigenvalue, and vice-versa. That is, $\lambda_+(\xi) =\gamma_+$ and occurs on the subspace $\pi_-\cH$, and $\lambda_-(\xi) = \gamma_-$, and occurs on $\pi_+\cH$. Let us briefly remark on the notation: $\gamma_-$, for example, has the subscript $-$ as $\gamma_-$ is the smallest eigenvalue of $\xi$, whereas  $\pi_+$ is the projector onto  the subspace on which eigenvalues of $\rho$ \emph{increase} to form the eigenvalues of $\xi$ (which are, in fact, $\gamma_-$).

Let us summarize the above observations. In $\ker \Delta$, the maximizer $\xi$ has the same eigenvalues as $\rho$. On the subspace $\pi_-\cH$, the state $\xi$ has the constant eigenvalue $\gamma_+$, which is the largest eigenvalue of $\xi$, and $\xi \pi_- = \gamma_+ \pi_- \leq \rho \pi_-$. In the subspace  $\pi_+\cH$, $\xi$ has the constant eigenvalue $\gamma_-$, which is its smallest eigenvalue, and $\xi \pi_+ = \gamma_- \pi_+ \geq \rho \pi_+$. It remains to choose subspaces corresponding to $\pi_\pm$. The associated eigenvalues $\gamma_-$ and $\gamma_+$ are then determined by $\tr[(\xi-\rho)_+]=\tr[(\xi-\rho)_-] = \epsilon$, using that  $T(\xi,\rho) = \epsilon$.

These values, of course, turn out to be the same $\gamma_\pm$ from \Cref{sec:construct_majmin}. How might one guess $\pi_\pm$, however?
As the entropy is minimized on pure states and maximized on the completely mixed state $\tau$, one can guess that to increase the entropy, one should raise the small eigenvalues of $\rho$, and lower the large eigenvalues of $\rho$; moreover, from the non-linearity of $-x\log x$, one might guess to maximize the entropy it helps more to raise the \emph{smallest} eigenvalues of $\rho$ and lower the \emph{largest} eigenvalues. That is, $\pi_+$ should correspond to the eigenspaces of the $m_-$ smallest eigenvalues of $\rho$, and $\pi_-$ should correspond to the eigenspaces of the $m_+$ largest eigenvalues of $\rho$, for some $m_-,m_+\in \{1,\dotsc,d-1\}$. This turns out to be correct, as described in \Cref{sec:construct_majmin}. Moreover, as $\gamma_+$ is the largest eigenvalue of $\xi$, and $\gamma_-$ is the smallest one, we must have $\gamma_- \leq \mu \leq \gamma_+$ for any eigenvalue $\mu$ of $\rho$ with corresponding eigenspace which is a subspace of $\ker \Delta$. In \Cref{lem:uniqueness_for_majmin}, we prove there exists unique $\gamma_-,\gamma_+,m_-$ and $m_+$ which respect these considerations. Following the notation of \Cref{sec:construct_majmin}, we call the resulting state $\rho\majmin$ (instead of $\xi$).

\subsection{Waterfilling argument}\label{sec:motivate-waterfilling}
The discussion in \Cref{sec:motivate-from-optimality} uses the tools of \Cref{sec:general_optimality_conditions} which are applicable to any differentiable concave function over $B_\eps(\rho)$. However, in the specific case of maximizing the Shannon entropy over the total-variation distance ball, a simpler argument known as water-filling suffices to motivate the form of \eqref{def:majmin-state}. For the sake of brevity, we will only sketch the original argument and its applicability to motivating the form of \eqref{def:majmin-state} here, rather than including all the details.

As described by Boyd and Vandenberghe in \cite[Example 5.2]{bv2004convex}, this argument goes as follows. One considers $n$ classical communication channels, each with capacity $C(\vec x) = - \log(c_i + x_i)$ for $i=1,\dotsc,n$ with constants $c_i>0$, and variables $x_i \geq 0$ representing the percentage of power allocated to channel $i$, subject to the constraint $\sum_{i=1}^n x_i = 1$. By checking the Karush-Kuhn-Tucker (KKT) conditions from the theory of differentiable convex optimization, one finds that $C(\vec x)$ is maximized by $\vec x^* = (x_i^*)$ where $x_i^* = \max\{0, z^* - c_i\}$ with $z^*$ uniquely determined by the constraint $\sum_{i=1}^n x_i^* = 1$. This solution has the interpretation of waterfilling: one imagines the ground composed of $n$ patches, with the ground at height $c_i$ over patch $i$. One floods the ground to a depth $z$, using the total amount of water $\sum_{i=1}^n \max\{0,z - c_i\}$, and continues pouring water until a total amount $1$ of water is used. Then the amount of water added to patch $i$ is the optimal value $x_i^*$.

\begin{figure}[ht]
\centering
\includegraphics{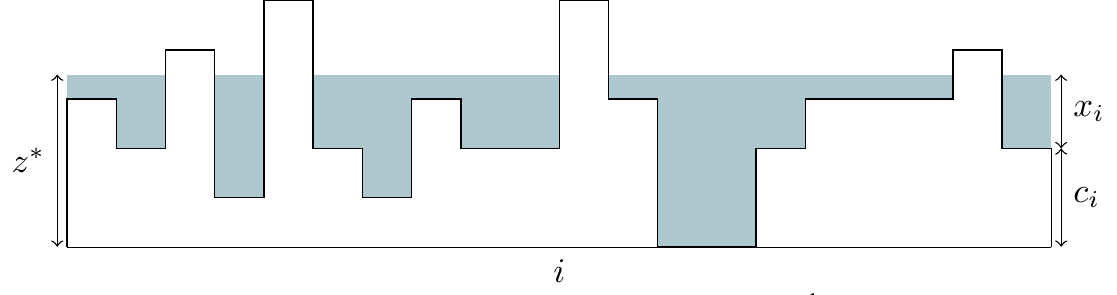}
\caption{Depiction of the waterfilling solution. The amount of water at site $i$ is given by $x_i^* = \max\{0,z^* - c_i\}$ where the depth $z^*$ is determined by the constraint $\sum_{i=1}^n x_i^* = 1$. Based on \cite[Figure 5.7]{bv2004convex}. }
\end{figure}

In fact, one finds the same solution if the objective function is changed to $\tilde C(\vec x) = \sum_{i=1}^n - (c_i + x_i) \log(c_i+x_i)$ subject to the constraint $\sum_{i=1}^n x_i = \eps$, namely $x_i^* = \max\{0, z^* - c_i\}$ with $z^*$ determined by $\sum_{i=1}^n x_i^* = \eps$.

Now, let us return to optimizing the Shannon entropy over the $\eps$-ball, and set the unique optimizer as $q \in \Be(p)$. If we fix $I_- \subset \{1,\dotsc,n\}$ as the set of indices such that $p_i \geq q_i$, we have $\sum_{i\in I_-} (p_i - q_i) = \eps$, and determining $x_i := p_i - q_i$ with $i\in I_-$ is exactly the problem of optimizing $\tilde C$ with the constraints $x_i\geq 0$, $\sum_{i\in I_-} x_i = \eps$, yielding a constant ``water depth'' $z_-$. By following the analogous argument with the set $I_+ = \{i: p_i< q_i\}$, yielding water depth $z_+$, one may see that   $p_i \in \{q_i, z_-, z_+\}$, depending on whether $p_i > q_i$, $p_i = q_i$, or $p_i < q_i$. The difficulty, then, is in determining the index sets $I_\pm$. This is exactly the task of determining the subspaces $\pi_\pm$ discussed in the previous section.

\subsection{From majorization-infimum} \label{sec:motivate-from-inf}

A third way to establish to find $r\majmin$ is via the majorization-infimum. This method has the advantage of providing an algorithm to exactly calculate $r\majmin$, although it's not immediate to derive the form \eqref{def:majmin-state} from it\footnote{In the case that $\eps$ is small (i.e.\@~$\eps< \delta(r)$, in the notation of \Cref{sec:mmm-small-eps}), it is straightforward, however.}. Beyond providing a means to construct $r\majmin$, by investigating the connection to the majorization-infimum, we find more insight into an apparent discrepancy between the majorization-minimizer and the majorization-maximizer: namely, that the majorization-maximizer of $B_\eps(p)$ lies on a vertex of the polytope, while the majorization-minimizer in general does not (see \Cref{sec:first-look}). As we will see in \Cref{prop:majmin-on-vertex-Epi}, the majorization-minimizer does lie on a vertex of one of a particular family of polytopes which make up $B_\eps(p)$.

Majorization, as defined in \eqref{def:majorize}, is a pre-order, meaning it is reflexive ($p\prec p$) and transitive ($p\prec q \prec r \implies p \prec r$), and on the set of sorted probability vectors, 
\[
 \cP^\downarrow = \left\{ p \in \R^d : p_1 \geq \dotsm \geq p_d \geq 0, \sum_{i=1}^d p_i = 1 \right\},
\]
majorization is a partial order, meaning it is also antisymmetric ($p\prec q$ and $q\prec p$ implies $p=q$). It also satisfies the so-called {\em{lattice}} property \cite{CV02}, meaning for any pair $p, q \in \cP^\downarrow$, there is a unique greatest lower bound $\infmaj(p,q)\in \cP^\downarrow$, which satisfies
\begin{itemize}
    \item $\infmaj(p,q) \prec p$ and $\infmaj(p,q) \prec q$
    \item if $r \in \cP^\downarrow$ is any other lower bound, meaning $r \prec p$ and $r\prec q$, then $r \prec \infmaj(p,q)$.
\end{itemize}
 Likewise, there is a unique least upper bound $\supmaj(p,q) \in \cP^\downarrow$ such that $p \prec \supmaj(p,q)$, $q \prec \supmaj(p,q)$, and  $\supmaj(p,q) \prec r$ for any $r\in \cP^\downarrow$ satisfying the relations $p \prec r$ and $q\prec r$. \cite{CV02} provides an explicit algorithms to construct $\infmaj(p,q)$ and $\supmaj(p,q)$.

In fact, majorization satisfies the stronger \emph{complete lattice} property, meaning the infimum and supremum of an arbitrary subset $\cS$ of $\cP^\downarrow$ exist and are unique, and can be obtained by an explicit algorithm \cite{YG19} (see also \cite{BBH+19}). For a possibly non-sorted set $\cS\subseteq \cP$, the supremum and infimum exist but are non-unique, since any permutation of a supremum (resp. infimum) is another supremum (resp. infimum). In this case, we will let $\infmaj \cS$ be the collection of majorization infima of $\cS$, and likewise $\supmaj \cS$ be the collection of majorization suprema of $\cS$.

As shown by \cite{YG19} and \cite{BBH+19}, in the case of a general set $\cS \subseteq \cP$, one can construct a majorization-infimum (which lies in $\cP$ but possibly not $\cS$) as follows. Let
\begin{equation}\label{eq:def_sk}
s_k := y_k - y_{k-1}
\end{equation}
where $y_0 := 0$ and $y_k := \inf_{q \in \cS} \sum_{i=1}^k q_i^\downarrow$. Then $s \in \infmaj \cS$.  Note that $s \in \cP^\downarrow$ by construction.

Since \Cref{thm:max-min-ball} shows that $r_\eps^* \in \infmaj B_\eps(r)$, equation \eqref{eq:def_sk} provides another means to construct it. However, \Cref{thm:max-min-ball} shows something stronger: not only is $r_\eps^* \in \infmaj B_\eps(r)$, but $r_\eps^* \in B_\eps(r)$ too (it is a minimum, not just an infimum).

Note that for a convex set $\cS\subseteq \cP$, if it has a majorization-minimum, then that minimum is unique. To see this, simply note that there exist functions $\cP\to \bR$ which are both strictly concave and Schur concave, such as the Shannon entropy $S$. If $r_1,r_2\in \cS$ were distinct majorization-minima, then $S$ would attain its maximum over $\cS$ at each of them, by Schur concavity. However, if $\cS$ is convex, then $r := \frac{1}{2}r_1 + \frac{1}{2}r_2 \in \cS$ too, and by strict convexity, then $S(r) > \frac{1}{2}S(r_1) + \frac{1}{2}S(r_2)$, a contradiction.

The following proposition shows that \eqref{eq:def_sk} can be used to establish a necessary and sufficient condition for the majorization-infimum to be a minimum.

\begin{proposition}\label{prop:majmin-inf-iff}
Let $\cS\subset \cP$ be an arbitrary set of probability vectors. $\cS$ admits a majorization minimum if and only if for some $q^* \in \cS$ the infimum
\begin{equation}\label{eq:prop-q-inf}
\inf_{q \in \cS} \sum_{i=1}^k q_i^\downarrow
\end{equation}
is achieved at $q^*$ for all $k=1,2,\dotsc,d$. In the latter case, $q^*$ is a majorization-minimum of $\cS$.
\end{proposition}
\begin{proof}
In the following, we use the notation of \eqref{eq:def_sk} and that $s\in \cP^\downarrow$, defined in that equation, is a majorization-infimum of $\cS$.
If the infimum in \eqref{eq:prop-q-inf} is achieved at $q^*$ for each $k=1,\dotsc,d$, then
\[
s_k = y_k - y_{k-1} = \sum_{i=1}^k (q_i^*)^\downarrow - \sum_{i=1}^{k-1} (q_i^*)^\downarrow = (q_k^*)^\downarrow
\]
and hence $s = (q^*)^\downarrow$. As a permutation of $s$, therefore, $q^* \in \infmaj \cS$. Since $q^* \in \cS$, $q^*$ is a majorization minimum of $\cS$.

On the other hand, assume that $\cS$ admits a majorization minimum $q^* \in \cS$. Since $q^*\in \infmaj \cS$, $q^* \prec q$ for all $q \in \cS$. By definition, this means that
\[
\sum_{i=1}^k (q_i^*)^\downarrow \leq \sum_{i=1}^k q_i^\downarrow \qquad \forall q\in \cS
\]
and hence infimum in \eqref{eq:prop-q-inf} is achieved on $q^*$ for each $k=1,\dotsc,d$.
\end{proof}
\begin{remark}
\Cref{prop:majmin-inf-iff} is not immediately helpful for establishing computationally that a given set $\cS$ admits no majorization-minima, because the optimization in \eqref{eq:prop-q-inf} could have multiple solutions. Hence, solving the optimization problem in \eqref{eq:prop-q-inf} for two different choices of $k$ and obtaining two different optima is not sufficient to prove that $\cS$ does not admit a majorization-minimum. Instead, in \Cref{sec:other-p-norms}, to establish that balls in $p$-norm with $1<p<\infty$ do not admit a majorization-minimizer, the majorization-infimum is constructed via \eqref{eq:def_sk} and shown to lie outside of the ball itself.
\end{remark}

Before moving to the following section, in which we prove the construction of $r_\eps^*$ given in \Cref{sec:construct_majmin} truly yields the majorization minimum, let us take brief detour to investigate the majorization-infimum further.

First, a word on the computational aspect of calculating the $y_k$ in \eqref{eq:def_sk}. Since the objective $\sum_{i=1}^k q_i^\downarrow$ in the optimization in $y_k$ is semidefinite representable, when membership in $\cS$ can be described by linear or semidefinite constraints, the optimization in $y_k$ can be represented as a semidefinite program.
In fact, when $\cS\subset \cP$ can be described by a set of linear constraints in the sense that
\[
 \cS = \left\{ q: A q \geq \alpha, B q = \beta \right\}
 \] 
for some matrices $A$ and $B$ and vectors $\alpha$ and $\beta$, \cite{YG19} shows that the majorization-infimum can be computed by a linear program.
One way to see\footnote{This point of view was shared with me by Xiao-Dong Yu in private communication.} the construction of $r\majmin$ in \Cref{sec:construct_majmin} then is as providing a closed-form solution for the majorization-infimum in the special case in which $\cS= B_\eps(r)$.

\paragraph{The majorization-minimizer of polytopes}

As discussed in \cite[Lemma 1]{BBH+19}, for a polytope $Q \subset \cP^\downarrow$, the optimization in the definition of $y_k$ (see \eqref{eq:def_sk}) may be restricted to the vertices of the polytope; this follows from the fact that when $Q\subseteq \cP^\downarrow$, $y_k$ constitutes a linear program.  Since $B_\eps(r)$ is a polytope (recalling the discussion in \Cref{sec:first-look}), the case of polytopes is quite relevant to us.

There are two apparent difficulties, however, to applying this result to the majorization-minimizer of polytopes. First, \Cref{prop:majmin-inf-iff} shows that the existence of a majorization-minimizer is equivalent to the optimization in the definition of $y_k$ being achieved on the same element for each $k$, and as noted in the following remark, the optimization in the $y_k$ may obtain multiple solutions. In that case, while one of the solutions may be a vertex of the polytope, the common solution, a priori, could lie in the interior. This is ruled out (in the more general case of the union of polytopes) by the following result, \Cref{prop:majmin-polytope-vertex}.

\begin{lemma}
Let $V \subseteq \cP^\downarrow$ be a finite set of sorted probability vectors. Let $\conv V := \left\{ \sum_i \lambda_i v_i : \lambda_i \geq 0, \sum_i \lambda_i = 1, v_i \in V \right\}$ be its convex hull. Then
\begin{equation}\label{eq:reduce-to-V}
p \prec V \iff p \prec \conv(V)
\end{equation}
where we use the notation $p \prec \cS$ for a set $\cS \subseteq \cP$ to mean $p \prec s$ for each $s \in \cS$.
Moreover if $p \prec V$ and $p \in \conv V$ then $p\in V$.
\end{lemma}
\begin{proof}    
The implication ``$\Leftarrow$'' of \eqref{eq:reduce-to-V} is trivial since $V \subset \conv V$. On the other hand, assume $p \prec V$ and $q\in \conv V$ is given by $q = \sum_{j}\lambda_j v^{(j)}$ for a probability distribution $\{\lambda_i\}_i$ and $v^{(j)}\in V$. Since for each $k$ and $j$,
\[
\sum_{i=1}^k p_i \leq \sum_{i=1}^k v^{(j)}_i
\]
we have
\[
\sum_{i=1}^k p_i \leq \sum_{j}\lambda_j \sum_{i=1}^k v^{(j)}_i = \sum_{i=1}^k q_i.
\]
As each vector lies in $\cP^\downarrow$, we conclude that $p \prec q$.

Next, assume $p\prec V$ and $p\in \conv V$ is given by $p = \sum_{j}\lambda_j v^{(j)}$ for a probability distribution $\{\lambda_i\}_i$ and $v^{(j)}\in V$. Then let $S_k^{(j)} = \sum_{i=1}^k v^{(j)}_i$. Set $J := \{ j : \lambda_j > 0\}$. We have that for each $k=1,\dotsc,d$,
\[
\sum_{i=1}^k q_i = \sum_{i=1}^k\sum_{j\in J}\lambda_j v^{(j)}_i =  \sum_{j\in J}\lambda_j S_k^{(j)} \leq \sum_{i=1}^k v^{(j)}_k = S_k^{(j)} \text{ for all }j\in J.
\]
Hence, $\sum_{j\in J} \lambda_j S_k^{(j)} \leq \min_{j \in J} S_k^{(j)}$. For the average to be smaller than the minimum, the distribution must be constant; in other words, we must have $S_k^{(j)} \equiv c_k :=\min_{j \in J} S_k^{(j)}$ for each $j\in J$. Thus, for $j \in J$,
\[
\sum_{i=1}^k v^{(j)}_i = c_k \qquad\implies\qquad v^{(j)}_k = c_{k} - c_{k-1} \,\,\, \forall k
\]
where we set $c_0=0$. That is, there can only be one vector $v^{(j)}$ such that $j\in J$, since $j\in J$ completely fixes the elements of $v^{(j)}$. Hence, $q = v^{(j)}$ for some $j$, and therefore $q\in V$ as desired.
\end{proof}
This lemma can be used to show that majorization-minimizers lie on vertices, as follows.
\begin{proposition}\label{prop:majmin-polytope-vertex}
Let $Q\subseteq \cP^\downarrow$ be a union of polytopes, $Q = \bigcup_i Q_i$. Let $V(Q_i)$ be the vertices of $Q_i$, and $V(Q) = \bigcup_i V(Q_i)$. Then $q^*\in \cP^\downarrow$ is a majorization-minimizer of $V(Q)$ if and only if $q^*$ is a majorization-minimizer of $Q$.
\end{proposition}
\begin{proof} 
If $V(Q)$ admits a majorization-minimizer $q^*$, then $q^* \prec V(Q)$. In particular, for each $i$, $q^* \prec V(Q_i)$ and hence $q^* \prec Q_i$ by the lemma. Thus, $q^* \prec Q$; since $q^* \in V(Q)\subset Q$, $q^*$ is a majorization-minimizer of $Q$.

On the other hand, if $Q$ admits a majorization-minimizer $q^*$, then $q^* \in Q_i$ for some $i$. Hence $q^* \prec V(Q_i)$ and $q^* \in \conv V(Q_i)$, so by the lemma, $q^* \in V(Q_i)\subset V(Q)$. Since $q^* \prec V(Q)$, $q^*$ is a majorization minimizer of $V(Q)$.
\end{proof}

\begin{figure}
\centering
\includegraphics{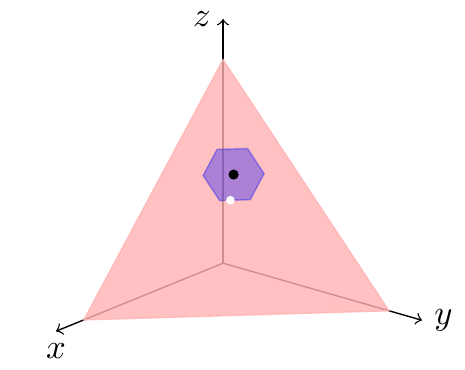}
\caption{The example of \Cref{fig:min-max-example}, again: In dimension $d=3$, the simplex, $\Delta$, of probability vectors is the shaded triangle, along with the ball $\Be(p)$ which is the hexagon shown in blue, centered at $p = \left(0.21, 0.24, 0.55\right)$ (depicted by a black dot) with  $\epsilon = 0.1$. The majorization-minimizer $p_\eps^*$ is drawn in white.} \label{fig:polytope-1}
\end{figure}

The second difficulty that is that for a polytope $Q\subset \cP$ which may not be sorted, the majorization-minimum may not lie on a vertex of $Q$. For example, see \Cref{fig:polytope-1} for an example in which the majorization-minimizer $p_\eps^*$ does not lie on a vertex of $B_\eps(p)$.

\begin{figure}[ht]
\centering
\begin{minipage}[b]{.46\linewidth}
\centering \includegraphics{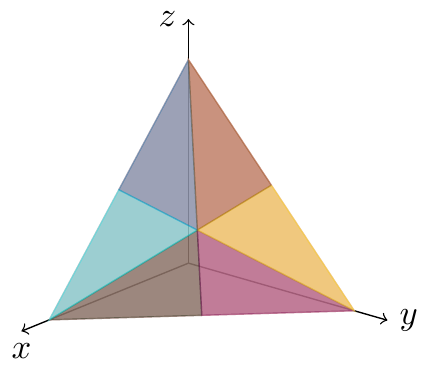}
\subcaption{Each set $E_\pi$ is highlighted in a different color.}
\end{minipage}\qquad
\begin{minipage}[b]{.46\linewidth}
\centering \includegraphics{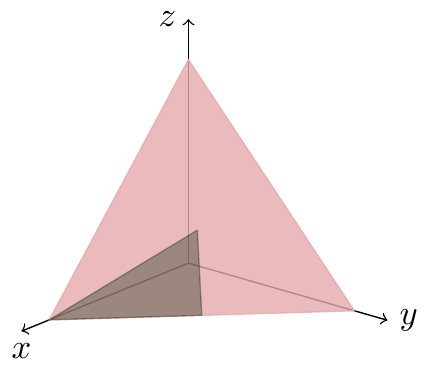}
\subcaption{Only $\cP^\downarrow = E_{(1,2,3)}$ is highlighted in brown.}
\end{minipage}\caption{The simplex probability $\cP\subseteq \mathbb{R}^3$ with $d=3$, decomposed into the six components $\cP = \bigcup_{\pi \in S_3} E_\pi$.
 } \label{fig:Epi}
\end{figure}

However, if $\cS\subset \cP$ is a polytope, then $\cS^\downarrow$ can be seen as a union of polytopes, as follows. For a permutation $\pi \in S_d$, let
\begin{equation}\label{eq:def_Epi}
E_\pi := \left\{ q \in \cP : q_{\pi(1)} \geq q_{\pi(2)} \geq \dotsm \geq q_{\pi(d)} \right\}
\end{equation}
be the part of the simplex $\cP$ whose elements are sorted according to $\pi$. Note $\cP^\downarrow = E_{(1,2,3)}$. As $E_\pi$ can be written as $\cP$ intersected with $d-1$ halfspaces, with each halfspace enforcing an inequality $q_{\pi(i)} \geq q_{\pi(i+1)}$ for $i = 1,\dotsc,d-1$, we have that $E_\pi$ is a polytope. The sets $E_\pi$ are shown for $d=3$ in \Cref{fig:Epi}.

\begin{figure}[ht]
\centering
\begin{minipage}[b]{.46\linewidth}
\centering \includegraphics{figs/fig3_qball}
\subcaption{The ball $B_\eps(p)$}\label{fig:qball_pieces_a}
\end{minipage}\qquad
\begin{minipage}[b]{.46\linewidth}
\centering \includegraphics{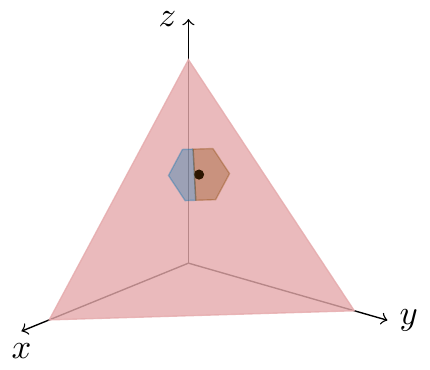}
\subcaption{$B_\eps(p) = B_1 \cup B_2$.} \label{fig:qball_pieces_b}
\end{minipage}\caption{Continuing the example of \Cref{fig:polytope-1}, we depict $B_\eps(p)$ decomposed according to the $\{E_\pi\}_{\pi \in S_d}$. In this example, only two of the components are non-zero, $B_1 := E_{(3,1,2)} \cap B_\eps(p)$ and $B_2 := E_{(3,2,1)} \cap B_\eps(p)$.} \label{fig:qball_pieces}
\end{figure}

We have that $\bigcup_{\pi \in S_d} E_\pi = \cP$. Moreover, for any polytope $\cS$ and $\pi \in S_d$, the intersection $\cS_\pi := \cS \cap E_\pi$ is a polytope. We can see $B_\eps(q)$ decomposed in this way in \Cref{fig:qball_pieces}, using the same example as from \Cref{sec:first-look}. 

\begin{figure}[ht]
\centering \includegraphics{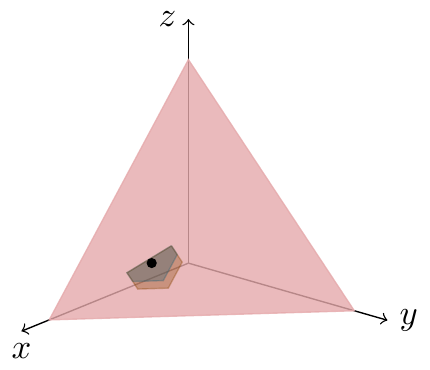}
 \caption{Using the notation of \Cref{fig:qball_pieces}, we find the sorted set $(B_\eps(p))^\downarrow = B_1^\downarrow \cup B_2^\downarrow$ by sorting each components $B_1$ and $B_2$. } \label{fig:q_pieces_sorted}
\end{figure}

In fact, $\cS_\pi^\downarrow$ is a polytope too, as it is just a rotation of $\cS_\pi$ (in contrast to $\cS^\downarrow$, in general). In particular, if $\cS_\pi = \conv\{q_1,\dotsc,q_n\}$, then $\cS_\pi^\downarrow = \conv\{q_1^\downarrow,\dotsc,q_n^\downarrow\}$. So we can obtain $\cS^\downarrow = \bigcup_{\pi\in S_d} \cS_\pi^\downarrow$ in this way. This is illustrated in \Cref{fig:q_pieces_sorted}.

Lastly, let us return to the discussion of the majorization minimizer.

\begin{proposition} \label{prop:majmin-on-vertex-Epi}
If $Q\subseteq \cP$ is a polytope (such as a total-variation ball) and admits a majorization-minimizer, then it lies on one of the vertices of the polytope $Q \cap E_\pi$ for some $\pi \in S_d$, where $E_\pi$ is defined in \eqref{eq:def_Epi}.
\end{proposition}
\begin{proof}    
As discussed in the text above, $Q^\downarrow = \bigcup_\pi ( Q \cap E_\pi )^\downarrow$. Since $Q$ admits a majorization-minimizer, $Q^\downarrow$ does too. Hence \Cref{prop:majmin-polytope-vertex} shows that there is a majorization-minimizer of $Q^\downarrow$ such that $(q^*)^\downarrow \in V( ( Q \cap E_\pi )^\downarrow)$ for some $\pi$. Then there is a permutation $q^*$ of $(q^*)^\downarrow$ that lies in $Q\cap E_\pi$, and $q^*$ is a majorization-minimizer of $Q$.
\end{proof}

\begin{figure}[ht]
\centering
\begin{minipage}[b]{.46\linewidth}
\centering \includegraphics{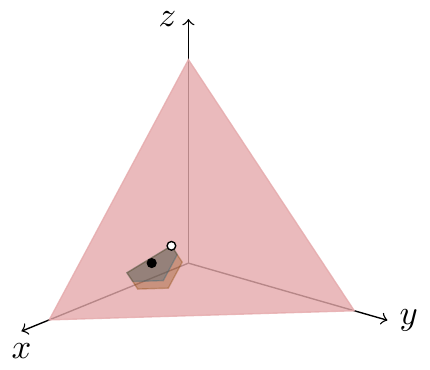}
\subcaption{\Cref{fig:q_pieces_sorted} with $(p_\eps^*)^\downarrow$ marked in white. }
\end{minipage}\qquad
\begin{minipage}[b]{.46\linewidth}
\centering \includegraphics{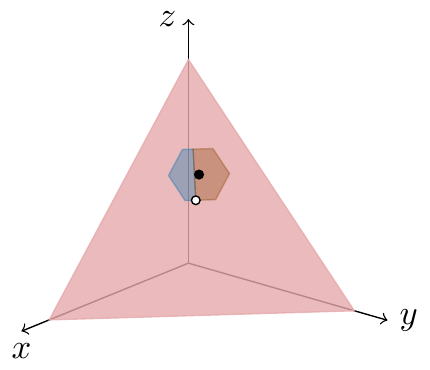}
\subcaption{\Cref{fig:qball_pieces_b} with $(p_\eps^*)^\downarrow$ marked in white.} \label{fig:majmin-on-vertex-Epi-b}
\end{minipage}\caption{Continuing the example of \Cref{fig:polytope-1}, we see that in agreement with \Cref{prop:majmin-polytope-vertex}, $(B_\eps(p))^\downarrow$ admits its majorization-minimizer on a vertex. Returning to the original, unsorted, form in \Cref{fig:majmin-on-vertex-Epi-b}, we see that indeed, $p_\eps^*$ lies on a vertex of $B_\eps(p) \cap E_\pi$ for some $\pi$ (in this case, either $\pi = (3,1,2)$ or $\pi = (3,2,1)$). } \label{fig:majmin-on-vertex-Epi}
\end{figure}

We can see \Cref{prop:majmin-on-vertex-Epi} illustrated in our running example in \Cref{fig:majmin-on-vertex-Epi}.

\begin{remark}
In the case of the particular polytope given by the total-variation ball $B_\eps(r)$, one in fact does not need to consider a union of polytopes, since \eqref{eq:unorm-q-to-c} shows that if $r\in E_\pi$ for some $\pi$, then for any $q\in B_\eps(r)$, for some permutation $\pi'\in S_d$, we have $\pi'(q) \in E_\pi$. In particular, we can take $r\majmin \in B_\eps(r)\cap E_\pi$. Then $(B_\eps(r) \cap E_\pi)^\downarrow$ itself is a sorted polytope, for which \Cref{prop:majmin-polytope-vertex} yields that the majorization-minimizer lies on a vertex.
\end{remark}
 
\section{Minimality of \texorpdfstring{$r\majmin$}{r*} in the  majorization order \eqref{eq:min-max-maj-cond}}\label{sec:max_of_pmaxeps}
Given $r\in \cP^\downarrow_d$. Let $w \in \Be(r)$. We aim to show $r\majmin \prec w$; equivalently, $r\majmin \prec w^\downarrow$. Since $\|r^\downarrow - w^\downarrow\|_1 \leq \|r - w\|_1$, using \eqref{eq:unorm-q-to-c} (by promoting vectors to diagonal matrices), we have $w^\downarrow \in \Be(r)$ as well. Hence, without loss of generality, we may consider $w = w^\downarrow$, with entries
\[
    w_1\geq \dotsm \geq w_d.
\]
Let $r_1\geq \dotsm \geq r_d$ be the entries of $r$, and $r^*_1\geq \dotsm \geq r^*_d$ be the entries of $r_{\epsilon}^*$. In the following, we show $r_{\epsilon}^* \prec w$.

\begin{enumerate}
    \item First, we establish that $r^*_1 \leq w_1$.

          To prove this, let us assume the contrary: $r^*_1 > w_1$.
          Then, since $r^*_1=r^*_2=\dotsm = r^*_{m_+} = \gamma_+$,
          \begin{align*}
              m_+\gamma_+  = \sum_{i=1}^{m_+}r^*_i > m_+ w_1 \geq \sum_{i=1}^{m_+}w_i.
          \end{align*}

          We conclude this step with the following lemma.
          \begin{lemma} \label{lem:n2sum_contradiction}
              If $m_+\gamma_+ > \sum_{i=1}^{m_+}w_i$, then $w\not\in \Be(r)$.
          \end{lemma}
          \begin{proof}
              Multiplying each side by $-1$ and adding $\sum_{i=1}^{m_+} r_i$, we have
              \begin{equation}
                  \sum_{i=1}^{m_+}(r_i - r^*_i) < \sum_{i=1}^{m_+} (r_i - w_i) \leq \sum_{i=1}^{m_+} (r_i - w_i)_+ \leq \sum_{i=1}^{d} (r_i - w_i)_+ = \frac{1}{2}\|r - w\|_1. \label{eq:LB_norm_dist}
              \end{equation}
              Using $\gamma_+ = \frac{1}{m_+}\left(\sum_{i=1}^{m_+} r_i - \epsilon\right)$, the far left-hand side is
              \[
                  \sum_{i=1}^{m_+}r_i - m_+\gamma_+= \sum_{i=1}^{m_+}r_i  -\left(\sum_{i=1}^{m_+}r_i  - \epsilon\right) =\epsilon.
              \]
              Then \eqref{eq:LB_norm_dist} becomes
              \[
                  \epsilon <  \frac{1}{2}\|r - w\|_1,
              \]
              contradicting that $w\in \Be(r)$.
          \end{proof}

    \item Next, for $k\in \{1,\dotsc,m_+\}$, we have $\sum_{i=1}^k r_i^* \leq \sum_{i=1}^k w_i$.

          We prove this recursively: assume the property holds for some $k\in \{1,\dotsc,m_+-1\}$ but not for $k+1$. Note we have proven the base case of $k=1$ in the previous step. Then

          \begin{equation}
              \sum_{i=1}^{k} r_i^* \leq \sum_{i=1}^{k}w_i, \quad \text{and} \quad \sum_{i=1}^{k+1}r_i^* > \sum_{i=1}^{k+1} w_i. \label{eq:n2-assump}
          \end{equation}
          Subtracting the first inequality in (\ref{eq:n2-assump}) from the second, we have
          \[
              r_{k+1}^* > w_{k+1}
          \]
          yielding $\gamma_+> w_{k+1} \geq w_{k} \geq \dotsm \geq w_{m_+}$. Summing the inequalities $r_{k+\ell}^*=\gamma_+ > w_{k+\ell}$ for $\ell=2,3,\dotsc,m_+-k$, we have
          \[
              \sum_{j=k+2}^{m_+} r_j^* > \sum_{j=k+2}^{m_+} w_j.
          \]
          Adding this to the second inequality of \eqref{eq:n2-assump}, we have
          \[
              \sum_{i=1}^{m_+}r_i^* > \sum_{i=1}^{m_+} w_i.
          \]

          We thus conclude by \Cref{lem:n2sum_contradiction}.

    \item Next, let $k\in\{m_++1,\dotsc,d-m_-\}$. Assume $\sum_{i=1}^k r_i^* > \sum_{i=1}^k w_i$.
          Then
          \[
              \sum_{i=1}^k r_i - r_i^* < \sum_{i=1}^k r_i - w_i
          \]
          However, the left-hand side is
          \[
              \sum_{i=1}^k r_i - r_i^* = \sum_{i=1}^{m_+} r_i - m_+ \gamma_+ = \epsilon.
          \]
          Hence,
          \[
              \epsilon < \sum_{i=1}^k r_i - w_i \leq \sum_{i=1}^d (r_i-w_i)_+ = \frac{1}{2}\|w-r\|_1
          \]
          which is a contradiction.

    \item Finally, we finish with a recursive proof similar to step 2. Assume the property holds for  $k \in \{d-m_-,\dotsc,d-1\}$, but not for $k+1$. For this case too we have proven the base case $k=d-m_-$ in the previous step. We therefore assume
          \begin{equation}
              \sum_{i=1}^{k} r_i^* \leq \sum_{i=1}^{k} w_i, \qquad \text{and} \qquad \sum_{i=1}^{k+1} r_i^* > \sum_{i=1}^{k+1} w_i. \label{eq:contradiction>n1}
          \end{equation}
          Subtracting the two equations, we have
          \[
              r_{k+1}^* > w_{k+1}.
          \]
          Since $\gamma_- = r_{k+1}^*$ we have  $\gamma_- > w_{k+1}\geq  w_{k+1} \geq \dotsm \geq w_d$. Summing $r_{k+\ell}^* = \gamma_- > w_{k+\ell}$ for $\ell=2,3,\dotsm,d$, we have
          \[
              \sum_{i=k+2}^d r_i^* > \sum_{i=k+2}^d w_i.
          \]
          Adding to the second inequality of \eqref{eq:contradiction>n1}, we find
          \[
              1 = \sum_{i=1}^d r_i^* > \sum_{i=1}^d w_i
          \]
          which contradicts the assumption that $r\in \Delta$.\qed
\end{enumerate}

\subsection{Uniqueness of construction for \texorpdfstring{$r\majmin$}{r*} \label{sec:proofs_VN}}

The uniqueness claims in the construction of $r\majmin$ in \Cref{sec:construct_majmin} follow from the following lemma. These properties prove useful in establishing the so-called semi-group property in \Cref{prop:VN_semigroup_property}, and are not present in \cite{HOS18,HY10}.
\begin{lemma} \label{lem:uniqueness_for_majmin}
  Let $r\in \cP_d$ have sorted entries $r^\downarrow = (r^\downarrow_1,\dotsc,r^\downarrow_d)$.
  Let $0 < \eps < \TV(r,u)$, where $u = (1/d,\dotsc,1/d)$, i.e.\@~$u \not\in B_\eps(r)$. There is a unique pair $(\gamma_-,m_-) \in [0,1]\times \{1,\dotsc,d-1\}$ such that
  \begin{equation}
    \sum_{i=d-m_-+1}^{d} | \gamma_- - r^\downarrow_i| = \eps \quad\text{and}\quad r^\downarrow_{d-m_-+1} < \gamma_- \leq r^\downarrow_{d-m_-}, \label{eq:def_alpha1_n1}
  \end{equation}
  and similarly,  a unique pair $(\gamma_+,m_+) \in [0,1]\times \{1,\dotsc,d-1\}$ with
  \begin{gather}
    \sum_{j = 1}^{m_+} |\gamma_+ - r^\downarrow_i| = \eps \quad\text{and}\quad  r^\downarrow_{m_++1} \leq \gamma_+  < r^\downarrow_{m_+}. \label{eq:def_alpha2_m}
  \end{gather}
  We define the index sets
  \begin{equation} \label{eq:def_IH_IM_IL}
    \begin{aligned}
      I_H & = \{1,\dotsc,m_+ \}, & I_M & =\{m_++1,\dotsc,d-m_- \}, & I_L & = \{d-m_-+1,\dotsc,d\}
    \end{aligned}
  \end{equation}
  corresponding to the ``highest'' $m_+$ entries, ``middle'' $d-m_--m_+$ entries, and ``lowest'' $m_-$ entries of $r$, respectively.

  We have the following properties. The numbers $\gamma_-$ and $\gamma_+$ are such that $\gamma_- < \frac{1}{d}< \gamma_+$, and we have the following characterizations of the pairs $(\gamma_-,m_-)$ and $(\gamma_+,m_+)$: For any $m \in {\mathbb{N}}$, defining the functions
  \begin{align}\label{eq-alpha}
    \gamma_-^{(m)} :=\frac{1}{m}\left(  \sum_{j=d-m+1}^d r^\downarrow_j + \eps \right), & \qquad \gamma_+^{(m)}:= \frac{1}{m}\left( \sum_{j=1}^m r^\downarrow_j - \eps \right),
  \end{align}
  we have $\gamma_- = \gamma_-^{(m_-)}$ and $\gamma_+ = \gamma_+^{(m_+)}$, where $m_-$ and $m_+$ are, respectively, the unique solutions of the following:
  \begin{gather}
    r^\downarrow_{d-m_-'+1} < \gamma_- \leq r^\downarrow_{d-m_-'}\quad:\quad m_-' \in \{1,\dotsc,d-1\}, \label{eq:n1_as_alpha1(n1)_inequality} \\
    r^\downarrow_{m_+'+1} \leq \gamma_+  < r^\downarrow_{m_+'} \quad:\quad m_+' \in \{1,\dotsc,d-1\}, \nonumber
  \end{gather}
  and satisfy
  \begin{gather}
    m_- = \min \{ m'\in \{1,\dotsc,d-\ell\}: \gamma_-^{(m')} \leq r^\downarrow_{d-m_-'} \}, \label{eq:n1_as_min}\\
    m_+ = \min \{ m'\in \{1,\dotsc,\ell\}: \gamma_+^{(m')} \geq r^\downarrow_{m'+1} \}, \nonumber
  \end{gather}
  where $\ell$ is defined by
  \begin{align}\label{def-ell}
    r^\downarrow_{\ell+1} < \frac{1}{d}\leq r^\downarrow_{\ell}.
  \end{align}
\end{lemma}

\paragraph{Proof of Lemma \ref{lem:uniqueness_for_majmin}}
Here we prove the results pertaining to the pair $(\gamma_+,m_+)$. The results for the pair $(\gamma_-,m_-)$ can be obtained analogously. Note that if any pair $(\gamma_+,m_+) \in [0,1]\times \{1,\dotsc,d-1\}$ satisfies \eqref{eq:def_alpha1_n1} then we have
\begin{equation*}
  \eps = \sum_{i=1}^{m_+}|\gamma_+ - r^\downarrow_i| =  \sum_{i=1}^{m_+}( r^\downarrow_i-\gamma_+ ) = \sum_{i=1}^{m_+}r^\downarrow_i-m_+ \gamma_+ ,
\end{equation*}
implying $\gamma_+ = \frac{1}{m_+}\left( \sum_{i=1}^{m_+}r^\downarrow_i - \eps \right) = \gamma_+^{(m_+)}$.
Conversely, if for some $m_+'\in \{1,\dotsc,d-1\}$, the corresponding value $\gamma_+^{(m_+')}$ satisfies $r^\downarrow_{m_+'+1} \leq \gamma_+^{(m_+')} < r^\downarrow_{m_+'}$, then
\[
  \eps = \sum_{i=1}^{m_+'}r^\downarrow_i- m_+' \gamma_+^{(m_+')} = \sum_{i=1}^{m_+'}( r^\downarrow_i-\gamma_+^{(m_+')})=\sum_{i=1}^{m_+'}|\gamma_+^{(m_+')}- r^\downarrow_i|
\]
and in particular
\[
  \sum_{i=1}^{m_+'}|\gamma_+^{(m_+')}- r^\downarrow_i| = \eps.
\]
Hence, the existence and uniqueness of $(\gamma_+,m_+)$ satisfying \eqref{eq:def_alpha1_n1} is equivalent to the existence and uniqueness of $m_+\in \{1,\dotsc,d-1\}$ such that $\gamma_+^{(m_+)}$ satisfies $r^\downarrow_{m_++1} \leq \gamma_+^{(m_+)} < r^\downarrow_{m_+}$.

Next, we show that
\[
  m_+=\min \{m_+'\in \{1,\dotsc,\ell\}: \gamma_+^{(m_+')}\geq r^\downarrow_{m_+'+1} \}
\]
by checking that the minimum exists and uniquely solves $r^\downarrow_{m_++1} \leq \gamma_+^{(m_+)} < r^\downarrow_{m_+}$. The proof is then completed by showing that we must have $\gamma_+^{(m_+)} > \frac{1}{d}$.
The steps of the construction are elucidated below.

~\begin{enumerate}[label={Step }\arabic*.]

  \item $ \{ m_+'\in \{1,\dotsc,\ell\}:\gamma_+^{(m_+')}\geq r^\downarrow_{m_+'+1}\}\neq \emptyset$.

        Let us assume the contrary. Then, in particular, that $\gamma_+^{(\ell)} < r^\downarrow_{\ell+1}$. By substituting $\gamma_+^{(\ell)} = \frac{1}{\ell}\sum_{i=1}^\ell r^\downarrow_i - \frac{1}{\ell}\eps$ in this inequality, we have
        \begin{equation} \label{eq:lem_proofalpha1n1_contradiction1}
          \sum_{i=1}^\ell r^\downarrow_i< \ell r^\downarrow_{\ell+1} + \eps
        \end{equation}
        Using $\sum_{i=1}^d(r_i-u_i) = 0$, we have $\frac{1}{2}\|r-u\|_1 = \sum_{i=1}^d \neg( u_i-r_i)$, where $\neg(x) := \min(0,x)$. Using the definition of $\ell$, \cref{def-ell}, this can
        be written as
        \begin{equation}
          \frac{1}{2}\|r-u\|_1 = \sum_{i=1}^\ell (r^\downarrow_i-\frac{1}{d}) = \sum_{i=1}^\ell r^\downarrow_i-\frac{\ell}{d } \label{eq:tracedist_sigma-pi}.
        \end{equation}
        Employing \eqref{eq:lem_proofalpha1n1_contradiction1}, and using the definition of $\ell$, we find that
        \begin{align}
          \frac{1}{2}\|r-u\|_1 <  \ell r^\downarrow_{\ell+1} + \eps -\frac{\ell}{d}  =  \eps + \ell ( r^\downarrow_{\ell+1} - \frac 1d) <  \eps.
        \end{align}
        That is, $u\in B_\eps(r)$, which contradicts our assumption.

  \item The value $x := \min \{ m_+'\in \{1,\dotsc,\ell\}: \gamma_+^{(m_+')} \geq r^\downarrow_{m_+'+1}\}$ solves $r^\downarrow_{x+1} \leq \gamma_+^{(x)} <  r^\downarrow_{x}$.

        If $x=1$, then using \cref{eq-alpha} we see that $r^\downarrow_1 > r^\downarrow_1 - \eps = \gamma_+^{(1)}$, and hence $r^\downarrow_{x+1} \leq \gamma_+^{(x)} <  r^\downarrow_{x}$.
        Otherwise, by minimality of $x$, we have $\gamma_+^{(x-1)} < r^\downarrow_x$.
        We first establish  that for $m_+'\in \{2,3,\dotsc,d\}$,
        \begin{equation}
          \label{eq:alpha_id}
          \gamma_+^{(m_+'-1)} < r^\downarrow_n \iff \gamma_+^{(m_+')}  < r^\downarrow_n
        \end{equation}
        and the result follows by taking $m_+'=x$. To prove \eqref{eq:alpha_id}, we write
        \begin{align*}
          \gamma_+^{(m_+'-1)}< r^\downarrow_n & \iff\frac{1}{m_+'-1}\Big[\sum_{j=1}^{m_+'-1}r^\downarrow_j - \eps\Big]< r^\downarrow_n             \\
                                         & \iff \sum_{j=1}^{m_+'-1}r^\downarrow_j - \eps< (m_+'-1)r^\downarrow_n = m_+' r^\downarrow_n - r^\downarrow_n \\
                                         & \iff \sum_{j=1}^{m_+'}r^\downarrow_j - \eps<  m_+' r^\downarrow_n                                  \\
                                         & \iff \frac{1}{m_+'}\Big[\sum_{j=1}^{m_+'}r^\downarrow_j - \eps\Big]<  r^\downarrow_n               \\
                                         & \iff \gamma_+^{(m_+')} < r^\downarrow_n.
        \end{align*}

  \item Uniqueness of $m_+$ satisfying $r^\downarrow_{m_++1} \leq \gamma_+^{(m_+)}< r^\downarrow_{m_+}$.

        By substituting the inequality $r^\downarrow_{m_+'-1} \geq r^\downarrow_n$ into \eqref{eq:alpha_id}, we find for $m_+'\in \{ 2,3,\dotsc,d\}$,
        \begin{equation}
          \label{eq:alpha_id2}
          r^\downarrow_n > \gamma_+^{(m_+')}  \implies  r^\downarrow_{m_+'-1} > \gamma_+^{(m_+'-1)}.
        \end{equation}
        Now, assume that there exists a $y$ satisfying $y>x>0$ for which $r^\downarrow_y > \gamma_+^{(y)} \geq r^\downarrow_{y+1}$. By applying the implication \eqref{eq:alpha_id2} a total of $(y-x-1)$ times, we see that
        \[
          r^\downarrow_y > \gamma_+^{(y)} \implies r^\downarrow_{y-1} > \gamma_+^{(y-1)} \implies \dotsm \implies r^\downarrow_{x+1} > \gamma_+^{(x+1)}
        \]
        which by \eqref{eq:alpha_id} is equivalent to $\gamma_+^{(x)} < r^\downarrow_{x+1}$.
        This contradicts the assumption that $\gamma_+^{(x)} \geq r^\downarrow_{x+1}$. Hence, such a $y$ cannot exist.

  \item $\gamma_+^{(m_+)} > \frac{1}{d}$.

        We prove this by showing that if $\gamma_+^{(m_+)} \leq \frac{1}{d}$ then we obtain a contradiction to the assumption $\frac{1}{2}\|r-u\|_1 >  \eps$
        of \Cref{lem:uniqueness_for_majmin}.

        Assume  $\gamma_+^{(m_+)} \leq \frac{1}{d}$. Then,
        \begin{align*}
          \gamma_+^{(m_+)} = \frac{1}{m_+}\sum_{j=1}^{m_+} r^\downarrow_j - \frac{\eps}{m_+}  \leq \frac{1}{d}  \iff
          \sum_{j=1}^{m_+} \left(r^\downarrow_j- \frac{1}{d} \right) \leq \eps.
        \end{align*}
        Now, since $m_+\leq \ell$ by \eqref{eq:n1_as_min}, $r^\downarrow_j -\frac{1}{d}\geq 0$ for each $j\in \{1,\dotsc,m_+\}$.

        If $m_+=\ell$, then
        \begin{equation}
          \sum_{j=1}^{\ell}  \left(r^\downarrow_j- \frac{1}{d} \right)\leq \eps. \label{eq:pos_part_leq_eps}
        \end{equation}
        On the other hand, if $m_+ <\ell$, then $m_++1 \leq \ell$. Then, using the assumption $\gamma_+^{(m_+)}\leq \frac{1}{d}$,
        \[
          \frac{1}{d} \geq \gamma^{(m_+)}\geq r^\downarrow_{m_++1} \geq \dotsm \geq r^\downarrow_\ell \geq \frac{1}{d},
        \]
        so $r^\downarrow_{m_++1} = r^\downarrow_{m_++2} =\dotsm = r^\downarrow_\ell = \frac{1}{d}$. Then,
        \eqref{eq:pos_part_leq_eps}
        holds in this case as well. Then by \eqref{eq:tracedist_sigma-pi},
        \[
          \frac{1}{2}\|r - u\|_1 =  \sum_{j=1}^\ell \left(r^\downarrow_j- \frac{1}{d} \right)\leq \eps,
        \]
        which contradicts the assumption that $u \not\in B_\eps(r)$.\qed
\end{enumerate}

\section{Constructing the maximal state in the majorization order \eqref{eq:min-max-maj-cond}\label{sec:construct_majmax}}
	
\begin{figure}[ht]
	\centering
	\includegraphics{figs/fig0_Mw}\qquad \includegraphics{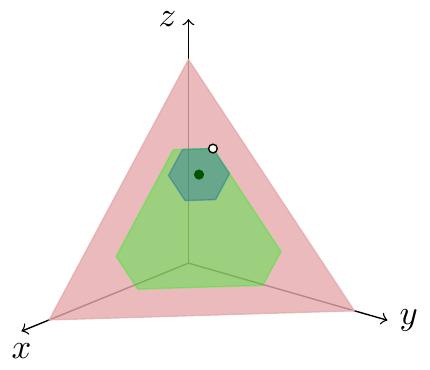}
	\caption{Left: The set $M_w$ for the point $w\in \Be(p)$ shown in white. Right: The set $M_{p\majmax}$ for $p\majmax$ the minimizer of $H$ over $\Be(p)$.\label{fig:simplex_Mqmin_Mp}}
	\end{figure}

As mentioned earlier, the minimum principle tells us that the entropy is minimized on a vertex of $\Be(p)$. In the example of \Cref{fig:Mw} in $d=3$, with $p=  ( 0.21, 0.24,0.55)$ and $\epsilon = 0.1$, we considered the set $M_w$ of points of $\cP$ majorized by $w\in \Be(p)$. In the same example in \Cref{fig:min-max-example}, the minimum of the entropy over $\Be(p)$ was found to be $p\majmax =( 0.21 - \epsilon, 0.24,0.55+ \epsilon )$. In \Cref{fig:simplex_Mqmin_Mp}, we see that in fact $\Be(p) \subseteq M_{p\majmax}$. That is, $w \prec p\majmax$ for any $w\in \Be(p)$. In this section, we provide a general construction of $p\majmax$, and show that this property holds. As in the example, the construction will proceed by forming $p\majmax$ by decreasing the smallest entries of $p$ and increasing the largest entry. Intuitively, one ``spreads out'' the entries of $p$ to form $p\majmax$ so that $M_{p\majmax}$ covers the most area, in order to cover $\Be(p)$.

	Let $\eps > 0$. We construct a probability vector $p\majmax$ which we show has $M_{p\majmax} \supseteq \Be(p)$ by using the definition of majorization given in \eqref{def:majorize}.
	\begin{definition}[$p\majmax$]
	\label{def:pstarepsq}
	If $p_d^\downarrow > \epsilon$, let $p\majmax^\downarrow := (p_1^\downarrow+\epsilon,p_2^\downarrow,\dotsc,p_d^\downarrow- \epsilon)$. Otherwise, let $\ell \in \{1,\dotsc,d-1\}$ be the largest index such that the sum of the $\ell$ smallest entries $Q_\ell := \sum_{j=d-\ell+1}^d p_j^\downarrow$ has $Q_\ell \leq \eps$. If $\ell=d-1$, set $p\majmax^\downarrow := (1,0,\dotsc,0)$.
	 Otherwise, choose $p\majmax^\downarrow := (p_{j*}^\downarrow)_{j=1}^d$ for
	\begin{equation} \label{eq:def_pjstar}
	p_{j*}^\downarrow := \begin{cases}
	p_1^\downarrow + \epsilon & j=1 \\
	p_j^\downarrow & 2\leq j \leq d-\ell-1\\
	p_{d-\ell+1}^\downarrow - (\epsilon - Q_\ell) & j=d-\ell\\
	0 & j\geq d-\ell+1.
	 \end{cases}
	\end{equation}
	\end{definition}
	Note that the condition $p_d^\downarrow > \epsilon$ is equivalent to every vector $w\in \Be(p)$ having strictly positive entries. This is the case of \Cref{fig:min-max-example}, and $p\majmax^\downarrow$ reduces to $(p_1^\downarrow+\epsilon,p_2^\downarrow,\dotsc,p_d^\downarrow- \epsilon)$.

	Then we set $p\majmax := \pi(p\majmax^\downarrow)$ for the permutation $\pi \in S_d$ such that $p = \pi(p^\downarrow)$.
	Using \Cref{def:pstarepsq}, we verify that $p\majmax\in \Be(p)$, as follows. First, if $p_d^\downarrow > \epsilon$, then
	\[
	\frac{1}{2}\|p\majmax - p\| = \frac{1}{2}\|p\majmax^\downarrow - p^\downarrow\| = \frac{1}{2}(|p_1^\downarrow+\epsilon - p_1^\downarrow| + |p_d^\downarrow - \epsilon - p_d^\downarrow|) = \epsilon.
	\]
	 If $\ell = d-1$, then $p\majmax^\downarrow = (1,0,\dotsc,0)$ and
	\[
	 \frac{1}{2}\|p\majmax - p\| = \frac{1}{2}\|p\majmax^\downarrow - p^\downarrow\| = \sum_{j=1}^d (p_j^\downarrow-p_{j*}^\downarrow)_+ = \sum_{j=2}^{d} p_j^\downarrow = Q_\ell \leq \eps
	 \]
	 yielding $p\majmax\in \Be(p)$. Otherwise, $\left(p_{j*}^\downarrow\right)_{j=1}^d$ is defined via \eqref{eq:def_pjstar}.
	For $j\neq d-\ell$, that $p_{j*}^\downarrow \geq 0$ is immediate, and by maximality of $\ell$, we have $p_{(d-\ell)*}^\downarrow = Q_{\ell+1} - \eps > 0$. Additionally, $\sum_{j=1}^d p_{j*}^\downarrow =   \sum_{j=1}^{d-\ell-1} p_j^\downarrow + Q_\ell -\eps + \eps = \sum_{j=1}^d p_j^\downarrow = 1$. Furthermore,
	\begin{align*}	
	\sum_{j=1}^d |p_{j*}^\downarrow - p_j^\downarrow| 
	&= |p_1^\downarrow+\eps-p_1| +\sum_{j=2}^{d-\ell-1} |p_j^\downarrow - p_j^\downarrow| + |p_{d-\ell}^\downarrow - (\eps - Q_\ell) - p_{d-\ell}^\downarrow|  + \sum_{j=d-\ell+1}^d |p_j^\downarrow- 0| \\
	&= Q_\ell + (\eps - Q_\ell) + \eps = 2 \eps,
	\end{align*}
	so $p\majmax \in \Be(p)$.

	The following lemma shows that $p\majmax$ is indeed the maximal distribution in the majorization order \eqref{class-maj}.
	\begin{lemma}  \label{lem:pstar_succ_p}
	We have that $p\majmax \succ w$ for any $w\in \Be(p)$.
	\end{lemma}
	\begin{proof} As in \Cref{sec:max_of_pmaxeps}, without loss of generality we may take $p= p^\downarrow$ and $w=w^\downarrow$.
	If $p\majmax = (1,0,\dotsc,0)$, then the result is immediate. If $p\majmax = (p_1+\epsilon,\dotsc,p_d - \epsilon)$, then consider $\ell=0$ and $Q_0 = 0$ in the following. Now, $p\majmax = (p_{j*})$ for $p_{j*}$ defined via \eqref{eq:def_pjstar}. Our task is to show that for any $w=\left(w_{j}\right)_{j=1}^d \in \Delta $,
	\begin{equation} \label{eq:proof_def_maj}
	\sum_{j=1}^k w_j \leq \sum_{j=1}^k p_{j*}
	\end{equation}
	holds for each $k\in\{1,\dotsc,d-1\}$. Equality in (\ref{eq:proof_def_maj}) obviously holds for $k=d$ since $w,p\majmax \in \Delta$.

	Since $\sum_{j=1}^d (w_j - p_j)_+ = \sum_{j=1}^d (w_j - p_j)_- \leq \eps$, in particular $\sum_{j=1}^k (w_j - p_j) \leq \eps$
	and therefore
	\[
	\sum_{j=1}^k w_j \leq \eps +\sum_{j=1}^k p_j.
	\]
	For $k \leq d-\ell-1$, we have $\sum_{j=1}^k p_{j*} = \eps +\sum_{j=1}^k p_j$, yielding \eqref{eq:proof_def_maj} in this case.
	On the other hand, for $k \geq d-\ell$ we have $\sum_{j=1}^k p_{j*} = \sum_{j=1}^d p_{j*} = 1$. Since $\sum_{j=1}^k w_j \leq 1$, this completes the proof.
	\end{proof}

	Given $\epsilon > 0$ and a quantum state $\rho\in \cD$, with eigen-decomposition $\rho =  \sum_{i=1}^d p_i\ketbra{i}{i}$ in the sorted eigenbasis for which $\rho = \Eig^\downarrow(\rho)$, we define
	\begin{equation} \label{eq:def_rho_eps_lowerstar}
	\rho\majmax = \sum_{i=1}^d p_{i*} \ketbra{i}{i}
	\end{equation}
	where $p\majmax$ is defined via \Cref{def:pstarepsq} where $p = \vec \lambda^\downarrow(\rho)$. \Cref{lem:pstar_succ_p} therefore proves $\rho\majmax \succ \omega$ for any $\omega\in \Be(\rho)$, proving the second majorization relation of \Cref{thm:max-min-ball}. The state $\rho\majmax$ is unique up to unitary equivalence as follows. If another state $\tilde \rho\in \Be(\rho)$ had $\tilde \rho \succ \omega$ for all $\omega \in \Be(\rho)$, then in particular, $\tilde \rho \succ \rho\majmax \succ \tilde \rho$, which implies that $\tilde \rho$ and $\rho\majmax$ are unitarily equivalent.
	 
\section{Other $p$-norms} \label{sec:other-p-norms}

We have established that balls in 1-norm admit a majorization-minimizer and majorization-maximizer. Is that true for other norms?

In the case $d=2$, all compact sets, and in particular $\eps$-balls induced by norms, admit a majorization-minimizer. To see this, note that for $p,q\in \cP_2$, i.e.\@~distributions of the form $(x,1-x)$ for $x\in [0,1]$, we have that $p \prec q$ if and only if $\max(p) \leq \max(q)$. On any compact set $C \subset \cP_2$, the continuous function $p \mapsto \max(p)$ must attain a minimum which is therefore a majorization-minimizer of $C$.

What about for $d > 2$? We show that in dimension $d=3$, for $p$-norms, i.e.\@~
\[
 \|r\|_p := \left( \sum_{i=1}^d r_i^p \right)^{1/p}
 \]
 with $1<p<\infty$, the answer is ``no''. 
To establish this, we will simply compute the majorization-infimum over a ball in $p$-norm via \eqref{eq:def_sk} and show that it lies outside the ball itself. 

We denote the $p$-norm ball as
\[
\Be^p(r) := \left\{ q \in \cP : \|r-q\|_p \leq \eps \right\}.
\]

\begin{proposition}
For any $1<p<\infty$ there exists (an uncountable collection of) $r\in \cP_3$ and $\eps > 0$ such that $\Be^p(r)$ does not admit a majorization-minimizer.
\end{proposition}
\begin{remark}
In the case $p=\infty$, the ball $\Be^p(r)$ does appear to admit a majorization-minimizer for any $d > 2$, $r\in \cP_d$ and $\eps > 0$ (ongoing work).
\end{remark}
\begin{proof}	
We will work in arbitrary dimensions $d > 2$ for now, and only specialize to $d=3$ near the end of the proof.
For $\eps > 0$ and $r\in \cP$, and recall a majorization-infimum $s$ of $B_\eps^p(r)$ (which is unique up to permutations) is given by $s_k = y_{k} - y_{k-1}$ for
\begin{equation}\label{eq:yk-p-proof}
 \begin{aligned}
y_k :=\text{minimize} \quad& \sum_{i=1}^k q_i^\downarrow\\
\text{subject to} \quad& q\in \Be^p(r)
\end{aligned}
\end{equation}
with $k =1,\dotsc,d$ and $y_0 :=0$.
Define
\[
\gap(r) := \min \{ |r_i - r_j| : i\neq j, \,\,i, j = 1,\dotsc,d\}.
\]
Note that $\gap(r) = 0$ if $r_i=r_j$ for some $i\neq j$. We have that if $r\in \cP^\downarrow$ and $\eps < \frac{\gap(r)}{2}$, then $\Be^p(r) \subseteq \cP^\downarrow$. To see this, note that if $q\in \Be^p(r)$, then
\[
\|q-r\|_\infty \leq \|q-r\|_p \leq \eps < \frac{\gap(r)}{2}
\]
and hence each entry of $q$ must not vary by more than $\frac{\gap(r)}{2}$, and thus the entries do not cross and instead remain sorted.

We choose $r\in \cP^\downarrow$ and $\eps < \frac{\gap(r)}{2}$. Then \eqref{eq:yk-p-proof} becomes
\[
 \begin{aligned}
y_k =\text{minimize} \quad& \sum_{i=1}^k q_i\\
\text{subject to} \quad& q\in \Be^p(r)
\end{aligned}
\]
or equivalently,
\[
 \begin{aligned}
y_k =\text{minimize} \quad&\sum_{i=1}^k (r_i + \delta_i)\\
\text{subject to} \quad& \delta \in \mathbb{R}^n\\
& \|\delta\|_p \leq \eps\\
& \sum_{i=1}^d \delta_i = 0.
\end{aligned}
\]
Moreover, while majorization-infimum is non-unique, $s$ defined via $s_k = y_{k}-y{k-1}$ satisfies $s \in \cP^\downarrow$. Then for any permutation $\pi \in S_d$,
\[
\|s-r\|_p = \|s^\downarrow - r^\downarrow\|_p \leq \| \pi(s) - r\|_p
\]
by \eqref{eq:unorm-q-to-c} since $\|\cdot\|_p$ is permutation invariant. Thus, if $s\not\in \Be^p(r)$, then no majorization-infimum of $\Be^p(r)$ lies in the ball. It remains to check that $\|s-r\|_p > \eps$.

Next, by substituting $\delta \to \eps \delta$, we obtain for $k=1,\dotsc,d$
\[
y_k =\sum_{i=1}^k r_i +\eps \gamma_{k,p,d}
\]
where
\begin{equation}\label{eq:def_gamma}
 \begin{aligned}
\gamma_{k,p,d} := \text{minimize} \quad& \sum_{i=1}^k \delta_i\\
\text{subject to} \quad& \delta \in \mathbb{R}^d\\
& \|\delta\|_p \leq 1\\
& \sum_{i=1}^d \delta_i = 0
\end{aligned}
\end{equation}
for $1\leq k<d$, and $\gamma_{d,p,d} := 0$.
Note that the optimization problem only depends on $p$, $k$, and $d$, not $\eps$ or $r$. Then the majorization-infimum over $\Be^p(r)$ is given by $s\in \cP^\downarrow$ with
\[
s_k := y_k - y_{k-1} =r_k + \eps ( \gamma_{k,p,d} - \gamma_{k-1,p,d})
\]
using the definitions $y_0 := 0$ and $\gamma_{0,p,d} = 0$. Then
\begin{equation}\label{eq:norm-as-gamma}
\frac{1}{\eps^p}\|r - s\|_p^p = |\gamma_{1,p,d}|^p + \sum_{k=2}^{d-1} |\gamma_{k,p,d} - \gamma_{k-1,p,d}|^p + |\gamma_{d-1,p,d}|^p.
\end{equation}
The right-hand side only depends on $p$ and $d$; if it can be shown to exceed $1$, then it demonstrates the existence of a large family of counterexamples (i.e.\@ corresponding to any $r\in \cP^\downarrow_d$ and $\eps < \frac{\gap(r)}{2}$).

We will compute the dual problem to \eqref{eq:def_gamma}. The following is standard; see e.g.\@ \cite[Chapter 5]{BV04}. The Lagrangian is given by
\[
L(\delta, \lambda, \nu) = \sum_{i=1}^k \delta_i + \lambda (\|\delta\|_p - 1) + \nu \sum_{i=1}^d \delta_i.
\]
for $\delta \in \bR^d$, with dual variables $\lambda \geq 0$ and $\nu \in \bR$. Let
\[
\Gamma := (\underbrace{\nu + 1, \dotsc, \nu + 1}_{k \text{ times}}, \nu,\dotsc, \nu).
\]
Then we can write
\[
L(\delta,\lambda,\nu) = -\lambda + \braket{\Gamma, \delta} + \lambda \|\delta\|_p
\]
where $\braket{\Gamma,\delta} := \sum_{i=1}^d\Gamma_i \delta_i$. Then the dual problem is given by
\[
\begin{aligned}
\text{maximize} \quad & g(\lambda, \nu)\\
\text{subject to} \quad & \lambda \geq 0\\
& \nu \in \bR
\end{aligned}
\]
for
\[
g(\lambda,\nu) := \min_{\delta \in \bR^d} L(\delta,\lambda,\nu).
\]
Since there exists a strictly feasible point to the primal problem (e.g.\@ $\delta = 0_{\bR^d}$), strong duality holds, and we have $\gamma_{k,p,d}= \max_{\lambda\geq 0, \nu\in \bR} g(\lambda,\nu)$.

We have that
\begin{equation*}	
 g(\lambda,\nu) =  -\lambda + \min_{\delta \in \bR^d} \left[ \braket{\Gamma, \delta} + \lambda \|\delta\|_p\right].
\end{equation*}
If $\lambda = 0$, then $g(0,\nu) = -\infty$. Otherwise, we may change variables $\delta \to \lambda \delta$, yielding
\[
g(\lambda,\nu) = -\lambda + \min_{\delta \in \bR^d} \left[ \braket{\tilde \Gamma, \delta} + \|\delta\|_p\right].
\]
for $\tilde \Gamma = \frac{1}{\lambda}\Gamma$. Then
\begin{align*}	
g(\lambda,\nu) &= -\lambda - \max_{\delta \in \bR^d} \left[ \braket{-\tilde \Gamma, \delta} - \|\delta\|_p\right]\\
&= \lambda - f^*(-\tilde \Gamma)
\end{align*}
where $f(x) = \|x\|_p$ and $f^*$ is its conjugate function,
\[
f^*(y) := \max_{x \in \bR^d} \braket{y, x} - f(x).
\]
Since $f$ is a norm, its conjugate is the indicator function of the dual norm (see e.g.\@ \cite[Example 3.26]{BV04}),
\[
f^*(\tilde \Gamma) = \begin{cases}
0 & \|\tilde \Gamma\|_q \leq 1\\
\infty & \text{otherwise}
\end{cases}
\]
where $q > 1$ is the H\"older conjugate given by
\[
\frac{1}{q}+ \frac{1}{p}=1.
\]
Thus,
\begin{align*}	
g(\lambda, \nu)  &= \begin{cases}
-\lambda & \lambda > 0 \text{ and }\|\tilde \Gamma\|_q \leq 1\\
-\infty & \text{otherwise,}
\end{cases}\\
&= \begin{cases}
-\lambda & \lambda > 0 \text{ and }\|\Gamma\|_q \leq \lambda\\
-\infty & \text{otherwise.}
\end{cases}
\end{align*}
Since $\|\Gamma\|_q = \left(k \cdot |\nu + 1|^q + (d-k) \cdot |\nu|^q\right)^{1/q}$ does not depend on $\lambda$,
we have that the maximal $\lambda$ corresponds to $\lambda = \|\Gamma\|_q$, and
\[
\gamma_{k,p,d} = \max_{\lambda \geq 0, \nu\in \bR} g(\lambda,\nu) = - \min_{\nu \in \bR}\left(k \cdot |\nu + 1|^q + (d-k) \cdot |\nu|^q\right)^{1/q}.
\]
The function
\[
h(\nu) = k \cdot |\nu + 1|^q + (d-k) \cdot |\nu|^q
\]
is differentiable on $(-\infty,-1)\cup(-1,0)\cup(0,\infty)$. Since $h$ diverges as $\nu \to \pm \infty$, its minimal value occurs either at $h(0) = k$, $h(-1) = d-k$, or at some $\nu \in (-\infty,-1)\cup(-1,0)\cup(0,\infty)$ with $h'(\nu) = 0$. Since
\[
h'(\nu) = \begin{cases}
- kq |\nu+1|^{q-1} - (d-k)q|\nu|^{q-1} & \nu < -1 \\
 kq |\nu+1|^{q-1} - (d-k)q|\nu|^{q-1} & -1 < \nu < 0 \\
 kq |\nu+1|^{q-1} + (d-k)q|\nu|^{q-1} & \nu > 0
\end{cases}
\]
we have $h'(\nu) \neq 0$ for $\nu \not \in (-1,0)$. Moreover, we find
\[
h(\nu^*) = 0 \quad \iff \quad \nu^* = \frac{-\beta}{\beta + 1}
\]
where $\beta := \left( \frac{k}{d-k} \right)^{\frac{1}{q-1}} >0$. Hence,
\[
h(\nu^*) = k \left( \frac{1}{1 + \beta} \right)^q + (d-k) \left( \frac{1}{1 + \beta\inv} \right)^q.
\]
 Next, we show $h(\nu^*) \leq \min(k,d-k)$ to establish that $h(\nu^*)$ is the minimum of $h$. Note $ \frac{1}{1 + \beta} +  \frac{1}{1 + \beta\inv} = 1$. Let $t = \frac{1}{1+\beta}$. We have the following sequence of equivalences
\begin{gather*}
k t^q + (d-k)(1-t)^q \leq d -k \\
\beta^{q-1} t^q + (1-t)^q \leq 1 \\
\beta^{q-1} + (t\inv - 1)^q \leq t^{-q}\\
\beta^{q-1} + \beta^q \leq (1+\beta)^q\\
\beta^q (\beta + 1) \leq \beta ( 1 +\beta)^q\\
q \log \beta + \log(\beta + 1) \leq \log \beta + q \log (1+\beta)\\
(q-1) \log \beta \leq (q-1) \log(\beta + 1)\\
\beta \leq \beta + 1
\end{gather*}
using $q>1$, which shows that $h(\nu^*) \leq d-k$. Likewise we have the equivalences
\begin{gather*}
k t^q + (d-k) (1-t)^q \leq k\\
t^q + \beta^{1-q} (1-t)^q \leq 1\\
1 + \beta^{1-q}(t\inv - 1)^q \leq t^{-q}\\
1 + \beta \leq (\beta + 1)^q
\end{gather*}
where the last line holds since $q>1$. Thus, we have
\begin{equation}\label{eq:gamma-formula}
\gamma_{k,p,d} = - h(\nu^*)^{1/q} = -\left(k \left( \frac{1}{1 + \beta} \right)^q + (d-k) \left( \frac{1}{1 + \beta\inv} \right)^q\right)^{1/q}
\end{equation}
for $\beta = \left( \frac{k}{d-k} \right)^{\frac{1}{q-1}} = \left(  \frac{k}{d-k}  \right)^{p-1}$, where $q$ is the H\"older conjugate to $p$.

Together, the relation \eqref{eq:norm-as-gamma} and the formula \eqref{eq:gamma-formula} can be used to prove whether or not the majorization-minimizer exists for a given pair $(p,d)$ with $1<p<\infty$ and $d>2$. Let us specialize to the case $d=3$. For $k=1$, we have $\beta=2^{-\frac{1}{(q-1)}}$ and for $k=2$ we have $\beta = 2^{\frac{1}{q-1}}$. Thus,
\[
\gamma_{1,p,3} = \gamma_{2,p,3} = - \left[ \left( 1 +  2^{1/(q-1)}\right)^{-q} + 2\left( 1 +  2^{1/(1-q)}\right)^{-q}  \right]^{1/q}
\]
Substituting for $p$ and simplifying, we obtain
\[
|\gamma_{1,p,3}|^p = \frac{2^p}{2 + 2^p}.
\]
Thus, the right-hand side of \eqref{eq:norm-as-gamma} becomes $2\frac{2^p}{2 + 2^p}$. Since
\[
2\frac{2^p}{2 + 2^p} > 1 \iff 2^p > 2 \iff p > 1
\]
we have that indeed, the right-hand side of \eqref{eq:norm-as-gamma}  exceeds $1$ for $p\in (1,\infty)$. Note that while $\lim_{p\to\infty}2\frac{2^p}{2 + 2^p} = 2 > 1$, the assumption $q>1$ was used earlier in the proof, requiring $p<\infty$.
\end{proof}
\begin{remark}
If we specialize to the case $p=2$, which implies $q=2$, then \eqref{eq:gamma-formula} gives
\[
\gamma_{k,2,d} = - \sqrt{\frac{k(d-k)}{d}}.
\]
Hence,
\[
|\gamma_{1,2,d}|^2 + |\gamma_{d-1,2,d}|^2 = 2 \frac{(d-1)}{d} > 1 \iff d > 2.
\]
Thus, the right-hand side of \eqref{eq:norm-as-gamma} is strictly larger than $1$ for all $d>2$, when $p=2$.
\end{remark}

\chapter{Majorization flow and continuity bounds}\label{chap:majflow_ctybounds}
In \Cref{sec:local-bounds}, we show how \Cref{thm:max-min-ball} immediately yields \emph{local} continuity bounds for any Schur concave function. Then in \Cref{sec:prop-majmin} we introduce the so-called majorization-minimizer map and prove some of its fundamental properties. These are used to establish a notion of ``majorization flow'' in \Cref{sec:majflow}, which in turn is used to obtain uniform continuity bounds in \Cref{sec:uniform-bounds-majflow}. This chapter concerns generic techniques applicable to large classes of functions. These techniques and others are used in \Cref{sec:applications} to provide continuity bounds for specific functions and families of functions.

\section{Local bounds}\label{sec:local-bounds}

As discussed in \Cref{sec:geometry_trace_ball}, \Cref{thm:max-min-ball} immediately yields local continuity bounds, as formalized in the following proposition.
\begin{proposition}[Local continuity bounds] \label{cor:local_cont_bound_VN_Renyi}
     Let $f$ be Schur concave on $\cP$. Let $p\in \cP$ and $\eps\in [0,1]$. Then for any $q \in \Be(p)$,
    \begin{equation}
        |f(q) - f(p)| \leq \max \{  f(p\majmin) - f(p), f(p) - f(p\majmax) \}\label{eq:classical_local_cont_bound_Schur-concave}.
    \end{equation}
    This result can be restated in the language of quantum states as follows.
    Let $f$ be Schur concave on $\cD(\cH)$. Let $\rho\in \D$ and $\eps\in [0,1]$. Then for any state $\omega \in \Be(\rho)$,
    \begin{equation}
        |f(\omega) - f(\rho)| \leq \max \{  f(\rho\majmin) - f(\rho), f(\rho) - f(\rho\majmax) \}\label{eq:local_cont_bound_Schur-concave}.
    \end{equation}
\end{proposition}
\begin{remarks}
   \begin{enumerate}
       \item  Note that this inequality indeed gives a local continuity bound: the right-hand side of \eqref{eq:classical_local_cont_bound_Schur-concave} only depends on $p$ and $\eps$. Moreover, the value of the right-hand side may be computed via the definitions of $p\majmin$ and $p\majmax$.
       \item This bound can be used in quite general circumstances because it only requires ``oracle-access'' to $f$, in the sense that one does not need to know the inner workings of $f$, but rather only needs three evaluations of it (at $p$, $p\majmax$, and $p\majmin$), and the promise that $f$ is Schur concave. For example, if $f$ describes the result of an optimization problem, it may be relatively easy to evaluate it (one just needs to solve the problem), but it may not be easy to compute e.g.\@ the derivative of $f$ along the path of majorization flow, $\Gamma_f$, which is introduced in \Cref{sec:majflow} and is used subsequently to obtain uniform continuity bounds. Note also that $p\majmin$ and $p\majmax$ can be computed in time $O(d\log d)$, where the time is dominated by sorting the entries of $p$. In the quantum case, one additionally needs to compute the eigenvalues of $\rho$.
   \end{enumerate}
   
\end{remarks}
 
\section{The majorization-minimizer map}\label{sec:prop-majmin}

By regarding the transformation from $r\in \cP_d$ to the minimal state in majorization order in the $\eps$-ball around $r$, namely $r\majmin$, as a map, we can obtain additional information about the behavior of the majorization order on the set of probability vectors.
Define the map
\begin{equation} \label{eq:def_mmm-eps}
  \begin{aligned}
    \mmm_\eps : \qquad \cP_d & \to  \cP_d        \\
    r                        & \mapsto r_\eps^*.
  \end{aligned}
\end{equation}
Note that $\mmm_\eps$ is a \emph{non-linear} map on the set of states, and therefore does not represent a  physical time evolution. Nevertheless, it has useful mathematical properties.

\begin{theorem}[Properties of $\mmm_\eps$ on $\cP_d$] \label{prop:properties_of_Lambda_eps}
  The (non-linear) map $\mmm_\eps$ has following properties, for any $\eps > 0$. Let $r\in \cP_d$.
  \begin{enumerate}[label*=\alph*.,ref=(\alph*)]
    \item \label{item:Lambda_epsstates-to-states} Maps probability vectors to probability vectors: $\mmm_\eps : \cP_d \to \cP_d$.
    \item  \label{item:Lambda_eps-semigroup} Semi-group property:
          if $\eps_1,\eps_2>0$, then $\mmm_{\eps_1+\eps_2}(r) = \mmm_{\eps_1} \circ \mmm_{\eps_2}(r)$.
    \item \label{item:Lambda_eps_preserves_sorting} Covariance with permutations: If $P$ is a permutation matrix, then $\mmm_\eps(Pr) = P\mmm_\eps(r)$. In addition, $\mmm_\eps(r^\downarrow) \in \cP^\downarrow_d$.
    \item \label{item:Lambda_eps_min_maj_order} Minimal in majorization order:   $\mmm_\eps(r)\in \Be(r)$ and for any $\omega\in \Be(r)$, we have $\mmm_\eps(r) \prec \omega$.
    \item \label{item:Lambda_eps-maj-preserving} Majorization-preserving: let $q\in \cP_d$ be such that $r \prec q$. Then $\mmm_\eps(r) \prec \mmm_\eps(q)$.
    \item \label{item:Lambda_eps_fixed_point} $u := (1/d,\dotsc,1/d)$ is the unique fixed point of $\mmm_\eps$, i.e.\@~the unique solution to $r = \mmm_\eps(r)$ for $r \in \cP_d$.
    \item \label{item:mmm-near-tau}For any state $r \in \Be(u)$, we have $\mmm_\eps(r) = u$. If $r\not\in \Be(u)$, then $\mmm_\eps(r)$ is on the boundary of $\Be(r)$.
    \item \label{item:mmm-on-psi} For $p = (1,0,\dotsc,0)$, the probability vector $\mmm_\eps(p)$ is given by
          \begin{equation}
            \mmm_\eps(p) = \begin{cases}
              (1- \eps, \frac{\eps}{d-1},\dotsc \frac{\eps}{d-1}) & \eps < 1 - \frac{1}{d}     \\
              u                       & \eps \geq 1 - \frac{1}{d}.
            \end{cases} \label{eq:mmm-pure}
          \end{equation}
    \item  \label{item:mmm-saturates-triangle} Saturates triangle inequality with $u$:
          \begin{equation}
            \frac{1}{2}\|r-u\|_1 =\frac{1}{2}\| u -  \mmm_\eps(r) \|_1+\frac{1}{2}\|  \mmm_\eps(r) - r\|_1  .
          \end{equation}
  \end{enumerate}
\end{theorem}

\begin{proof}
  The properties \ref{item:Lambda_epsstates-to-states}, \ref{item:Lambda_eps_preserves_sorting}, \ref{item:mmm-near-tau} and \ref{item:mmm-on-psi} follow from the construction given in \Cref{sec:construct_majmin}.  Note in particular, that \cref{eq:td_rho-eps-sigma} establishes the statement that $r\majmin$ lies on the boundary of $\Be(r)$ when $r\majmin\neq u$. Property \ref{item:Lambda_eps-maj-preserving} can be found in Lemma 2 of \cite{HO17}. Property \ref{item:Lambda_eps-semigroup} is shown by \Cref{prop:VN_semigroup_property} below. Property \ref{item:Lambda_eps_min_maj_order} is due to \Cref{thm:max-min-ball}, and is proven in \Cref{sec:construct_majmin}.  The property \ref{item:Lambda_eps_fixed_point} follows from the fact that the entries of $\mmm_\eps(r)$ differ from $r$ if and only if $r\neq u$. Lastly, Property \ref{item:mmm-saturates-triangle} is shown by \Cref{prop:VN_opt_saturates_triangle} below.
\end{proof}

Each of these properties has an exact analog in the case of quantum states, which follows from the definition \eqref{eq:def_majmin_state_quantum}. For the sake of completeness, these are listed below.
\begin{theorem}[Properties of $\mmm_\eps$ on $\cD(\cH)$] \label{prop:properties_of_Lambda_eps_quantum}
  The (non-linear) map $\mmm_\eps$ has following properties, for any $\eps>0$. Let $\rho\in \cD(\cH)$.
  \begin{enumerate}[label*=\alph*.,ref=(\alph*)]
    \item Maps states to states: $\mmm_\eps : \D(\cH) \to \D(\cH)$.
    \item Semi-group property:
          if $\eps_1,\eps_2>0$, we have $\mmm_{\eps_1+\eps_2}(\rho) = \mmm_{\eps_1} \circ \mmm_{\eps_2}(\rho)$.
    \item Covariance with unitaries: If $U$ is a unitary matrix, then $\mmm_\eps(U \rho U^*) = U\mmm_\eps(\rho)U^*$. In addition, $\mmm_\eps(\Eig^\downarrow(\rho))$ is diagonal and sorted.
    \item Minimal in majorization order:   $\mmm_\eps(\rho)\in \Be(\rho)$ and for any $\omega\in \Be(\rho)$, we have $\mmm_\eps(\rho) \prec \omega$.
    \item Majorization-preserving: let $\sigma\in \D(\cH)$ such that $\rho \prec \sigma$. Then $\mmm_\eps(\rho) \prec \mmm_\eps(\sigma)$.
    \item $\tau = \frac{\one}{d}$ is the unique fixed point of $\mmm_\eps$, i.e.\@~the unique solution to $\rho = \mmm_\eps(\rho)$ for $\rho \in \D(\cH)$. Moreover, for any $\rho \neq \tau$, $\mmm_\eps(\rho)$ is not unitarily equivalent to $\rho$.
    \item For any state $\rho \in \Be(\tau)$, we have $\mmm_\eps(\rho) = \tau$. If $\rho\not\in \Be(\tau)$, then $\mmm_\eps(\rho)$ is on the boundary of $\Be(\rho)$.
    \item For any pure state $\psi\in \Dpure(\cH)$, we may write $\psi = \diag(1,0,\dotsc,0)$ in an eigenbasis. In the same basis, the state $\mmm_\eps(\psi)$ is given by
          \begin{equation}
            \mmm_\eps(\psi) = \begin{cases}
              \diag(1- \eps, \frac{\eps}{d-1},\dotsc \frac{\eps}{d-1}) & \eps < 1 - \frac{1}{d}     \\
              \tau := \frac{\one}{d}                                   & \eps \geq 1 - \frac{1}{d}.
            \end{cases} \label{eq:mmm-pure-quantum}
          \end{equation}
    \item Saturates triangle inequality with $\tau := \tfrac{\one}{d}$:
          \begin{equation}
            \frac{1}{2}\|\rho-\tau\|_1 =\frac{1}{2}\| \tau -  \mmm_\eps(\rho) \|_1+\frac{1}{2}\|  \mmm_\eps(\rho) - \rho\|_1  .
          \end{equation}
  \end{enumerate}
\end{theorem}

Two of the properties, namely saturating the triangle inequality with $u$ and the semi-group property, are proven below. The first, the saturation of the triangle inequality, follows directly from the construction of $r\majmin$, and proves useful in establishing the second.

\begin{proposition} \label{prop:VN_opt_saturates_triangle} The state $r\majmin$ saturates the triangle inequality for $u := (1/d,\dotsc,1/d)$ and $r$, in that
  \[
    \frac{1}{2}\|r-u\|_1 =\frac{1}{2}\| u -  r\majmin \|_1+\frac{1}{2}\| r\majmin - r\|_1  .
  \]
\end{proposition}
\begin{proof}
  Note \[
    \frac{1}{2}\| r\majmin - r\|_1 = \min\left(\eps, \,\frac{1}{2}\|r-u\|_1\right),
  \]
  from the construction of $r\majmin$.  Now, if  $r\majmin=u$, we have the result immediately. Otherwise,
  \begin{align*}
    \|u -r_{ \eps}\|_1
     & =  m_-\left( \frac{1}{d} - \gamma_- \right) +   m_+\left( \gamma_+-\frac{1}{d}  \right) + \sum_{j=m_++1}^{d-m_-} | \tfrac{1}{d}-r^\downarrow_j|                    \\
     & = \frac{m_-}{d} - \sum_{j=1}^{m_+} r^\downarrow_j -  \eps +\sum_{j=d-m_-+1}^d r^\downarrow_j - \frac{m_+}{d} -  \eps + \sum_{j=m_++1}^{d-m_-} | \tfrac{1}{d}-r^\downarrow_j| \\
     & = \sum_{j=1}^{d} | \tfrac{1}{d}-r^\downarrow_j| - 2 \eps                                                                                                           \\
     & = \| u - r\|_1 - 2  \eps = \| u - r\|_1 - \|r\majmin- r\|_1 . \qedhere
  \end{align*}
\end{proof}

Next, we establish the semi-group property. This plays a crucial role in the rest of the chapter, and consequently underpins the continuity bounds established in \Cref{sec:applications}. We highlight that the proof depends heavily on the uniqueness results established in \Cref{lem:uniqueness_for_majmin}.

\begin{proposition} \label{prop:VN_semigroup_property}
  If $\eps_1,\eps_2 > 0$ and $r\in\cP_d$, we have
  \[
    \mmm_{\eps_1 + \eps_2}(r) = \mmm_{\eps_1} ( \mmm_{\eps_2}(r)).
  \]
\end{proposition}
\begin{proof}
  Without loss of generality, we may consider $r = r^\downarrow \in \cP^\downarrow$. Then $\mmm_{\eps_1+\eps_2}(r), \mmm_{\eps_1} \circ \mmm_{\eps_2}(r), \mmm_{\eps_2}(r)\in \cP^\downarrow$ too.

We first treat the cases in which $\mmm_{\eps_1+ \eps_2}(r) = u$; that is, when $\frac{1}{2}\|r-u\|_1 \leq \eps_1+\eps_2$.
If $\frac{1}{2}\|r-u\|_1 \leq \eps_2$, then  $ \mmm_{\eps_2}(r) = u$ as well; since $\mmm_{\eps_1}(u)=u$, we have $\mmm_{\eps_1 + \eps_2}(r) = \mmm_{\eps_1}(\mmm_{\eps_2}(r))$. Next, consider the case  $\eps_2 < \frac{1}{2}\|r-u\|_1 \leq \eps_1 +\eps_2$.  To show $ \mmm_{\eps_1} ( \mmm_{\eps_2}(r)) = u$, we need $\frac{1}{2}\|\mmm_{\eps_2}(r) - u\|_1 \leq \eps_1$.
By Proposition \ref{prop:VN_opt_saturates_triangle},
\[
 \frac{1}{2}\|u - \mmm_{\eps_2}(r)\|_1  =   \frac{1}{2}\|r-u\|_1 - \frac{1}{2}\| \mmm_{\eps_2}(r) - r\|_1  \leq \eps_1 + \eps_2 -  \frac{1}{2}\| \mmm_{\eps_2}(r) - r\|_1
\]
but since $ \frac{1}{2}\| \mmm_{\eps_2}(r) - r\|_1 = \eps_2$, using that $\mmm_{\eps_2}(r)\neq u$, we have $ \frac{1}{2}\|u - \mmm_{\eps_2}(r)\|_1  \leq \eps_1$ as required. This completes the proof of the cases for which $\mmm_{\eps_1+\eps_2}(r)=u$.

For the remainder of the proof, we will consider the construction of the states generated by $\mmm_\eps$ via their characterization in \Cref{lem:uniqueness_for_majmin}. The uniqueness results of that lemma will allow us to show that $\mmm_{\eps_1+ \eps_2}(r)$ indeed is the same state as $\mmm_{\eps_1} \circ \mmm_{\eps_2}(r)$.

Assume that $\frac{1}{2}\|r-u\|_1 > \eps_1 + \eps_2$. Let $r = \sum_{i=1}^d r_i \ket{i}$ and write
\[
\mmm_{\eps_2}(r) = \sum_{i\in I_L} \alpha_- \ket{i} + \sum_{i\in I_M} r_i \ket{i} + \sum_{i\in I_H}\alpha_+ \ket{i}
\]
for $I_L = \{d-m_-+1,\dotsc,d\}$, $I_M = \{m_++1,\dotsc,d-m_-\}$, and $I_H = \{1,\dotsc,m_+\}$, and  $(\alpha_-,m_-)$ and $(\alpha_+,m_+)$ are determined by $r$ and $\eps_2$ via \Cref{lem:uniqueness_for_majmin}.

Now, consider
\begin{align}\label{rho-1-2}
\mmm_{\eps_1}( \mmm_{\eps_2}(r) ) =\sum_{i\in I_{L'}} \beta_- \ket{i} + \sum_{i\in I_{M'}} r_i \ket{i} + \sum_{i\in I_{H'}} \beta_+ \ket{i} 
\end{align}
with $I_{L'} = \{d-m_-'+1,\dotsc,d\}$, $I_{M'} = \{m_+'+1,\dotsc,d-m_-'\}$, and $I_{H'} = \{1,\dotsc,m_+'\}$, and where $(\beta_-,m_-')$ and $(\beta_+,m_+)$ are determined by $ \mmm_{\eps_2}(r)$ and $\eps_1$ via \Cref{lem:uniqueness_for_majmin}.

We aim to compare the expression (\ref{rho-1-2}) of $\mmm_{\eps_1}( \mmm_{\eps_2}(r) ) $ to the following:
\[
\mmm_{\eps_1 + \eps_2} (r) = \sum_{i\in I_{L''}} \gamma_- \ket{i} + \sum_{i\in I_{M''}}r_i \ket{i} + \sum_{i\in I_{H''}} \gamma_+ \ket{i}
\]
with $I_{L''} = \{d-m_-''+1,\dotsc,d\}$, $I_{M''} = \{m_+''+1,\dotsc,d-m_-''\}$, and $I_{H''} = \{1,\dotsc,m_+''\}$, and where $(\gamma_-,m_-'')$ and $(\gamma_+,m_+'')$ are determined by $(r,\eps_1+\eps_2)$ via \Cref{lem:uniqueness_for_majmin}. That is, we wish to show $(\beta_-,m_-') = (\gamma_-,m_-'')$, and $(\beta_+,m_+') = (\gamma_+,m_+'')$. We only consider the first equality here; the second is very similar.

 Let $(\nu)_{i=1}^d$ be the entries of  $\mmm_{\eps_2}(r)$. That is, $\nu_i = \alpha_-$ for $i\in I_L$, $\nu_i = r_i$ for $i \in I_M$, and $\nu_i = \alpha_+$ for $i\in I_H$. Equation~\eqref{eq:n1_as_alpha1(n1)_inequality} for $\mmm_{\eps_1}(\mmm_{\eps_2}(r))$ in Lemma~\ref{lem:uniqueness_for_majmin} yields
\begin{equation}
\nu_{m_+'} > \beta_+ = \frac{1}{m_+'}\left(\sum_{i\in I_{H'}} \nu_i - \eps_1\right) \geq  \nu_{m_+'+1}. \label{eq:defd_beta1_(proof)}
\end{equation}
Thus,
\begin{itemize}
	\item $m_+' \geq m_+$. Otherwise, $\beta_+ = \alpha_+ - \frac{\eps_1}{m_+'} < \alpha_+ = \nu_{m_+'+1}$.
	\item $m_+' \leq d- m_-$. Otherwise, $\beta_+ < \nu_{m_+'} = \alpha_- < \frac{1}{d}$, contradicting that $\beta_+ > \frac{1}{d}$ by Lemma~\ref{lem:uniqueness_for_majmin}.
\end{itemize}
Hence,
\begin{align*}	
\beta_+ &= \frac{1}{m_+'}\left(\sum_{i\in I_H} \nu_i + \sum_{m_++1}^{m_+'} \nu_i - \eps_1 \right) =\frac{1}{m_+'}\left(m_+\alpha_+ + \sum_{m_++1}^{m_+'} r_i - \eps_1 \right)\\
&=\frac{1}{m_+'}\left(\sum_{i\in I_H}r_i - \eps_2 + \sum_{m_++1}^{m_+'} r_i - \eps_1 \right) = \frac{1}{m_+'} \left( \sum_{i\in I_{H'}} r_i - \eps_1 -\eps_2 \right).
\end{align*}
That is, $\beta_+ = \gamma_+(m_+')$. It remains to show $m_+' = m_+''$.

\begin{itemize}
\item If $m_+'=m_+$, then $\nu_{m_+'} = \alpha_+ < r_{m_+'}$ by equation \eqref{eq:def_alpha2_m} for $\mmm_{\eps_2}(r)$.
\item If $m_+' > m_+$, then $ \nu_{m_+'} =r_{m_+'}$. 
\end{itemize}

In either case, $\nu_{m_+'} \leq r_{m_+'}$. Hence, \eqref{eq:defd_beta1_(proof)} becomes
\[
r_{m_+} \geq \nu_{m_+'}  > \beta_+ \geq \nu_{m_++1}.
\]
Either $\nu_{m_++1} = r_{m_++1}$, or $\nu_{m_++1} = \alpha_- > r_{m_++1}$; in either case, $\beta_+ \geq r_{m_++1}$. Thus, writing $\beta_+ = \gamma_+(m_+')$, we have
\[
r_{m_+'} > \gamma_+(m_+') \geq r_{m_+'+1}.
\]
Equation~\eqref{eq:n1_as_alpha1(n1)_inequality} in \Cref{lem:uniqueness_for_majmin} defines $m_+''$ as the unique solution of 
\[
 r_{m_+''} >  \gamma_+(m_+'') \geq r_{m_+''+1} 
\]
and therefore $m_+'' = m_+'$. \end{proof}

\section{Majorization flow}\label{sec:majflow}
In this section, we consider the infinitesimal action of $\mmm_\eps$. This action is very simple to describe because, as shown in \Cref{sec:mmm-small-eps}, the action of $\mmm_\eps$ on $\cP$ is piecewise linear for $\eps$ small enough. Then in \Cref{sec:path-majflow}, we use the semigroup property to derive integral formulas to track the variation of a quantity along the path traced out by $(\mmm_\eps(r))_{\eps \geq 0}$.

\subsection{The majorization-minimizer for small \texorpdfstring{$\eps$}{epsilon}} \label{sec:mmm-small-eps}

\begin{definition}[$\delta(r)$] \label{def:delta-rho-sigma}Let $r \in \cP$. If $r = u \equiv (1/d,\dotsc,1/d)$, then set $\delta(r) := 0$. 
Otherwise, let $\mu_1 > \mu_2 > \dotsm > \mu_\ell$ denote the distinct ordered entries of $r$, let $k_+$ be the multiplicity of the largest entry of $r$ and let $k_-$ the multiplicity of the smallest entry.  Define
  \begin{equation}\label{eq:delta_rho}
    \delta(r) := \begin{cases}
      \frac{k_+ k_-}{k_+ + k_-}(\mu_1 - \mu_2)
                                                                           & \ell = 2 \\
      \min \{k_+ (\mu_1-\mu_2), k_-(\mu_{\ell-1}-\mu_\ell) \} & \ell > 2.
    \end{cases}
  \end{equation}
\end{definition}
Let us briefly comment on the form of $\delta(r)$ in the case $\ell=2$. The form used in the definition shows that the value of $\delta(r)$ is proportional to a gap between consecutive values of $r$. For $\ell=2$, however, it can also be written as follows.
We have that
\[
  \TV(r,u) = k_+\left( r_+ - \frac{1}{d}\right) = k_+\left( r_+ - \frac{r_+k_+ + r_- k_-}{k_++ k_-}\right)
\]
using $k_+ + k_- = d$ and $r_+k_+ + r_- k_- =1$. Hence,
\[
  \TV(r,u) = \frac{k_+ r_+ (k_+ + k_-) - r_+k_+^2 - r_- k_+k_-}{k_+ + k_-} = \delta(r).
\]

\smallskip

\begin{figure}[ht]
  \centering
  \includegraphics[width=.75\textwidth]{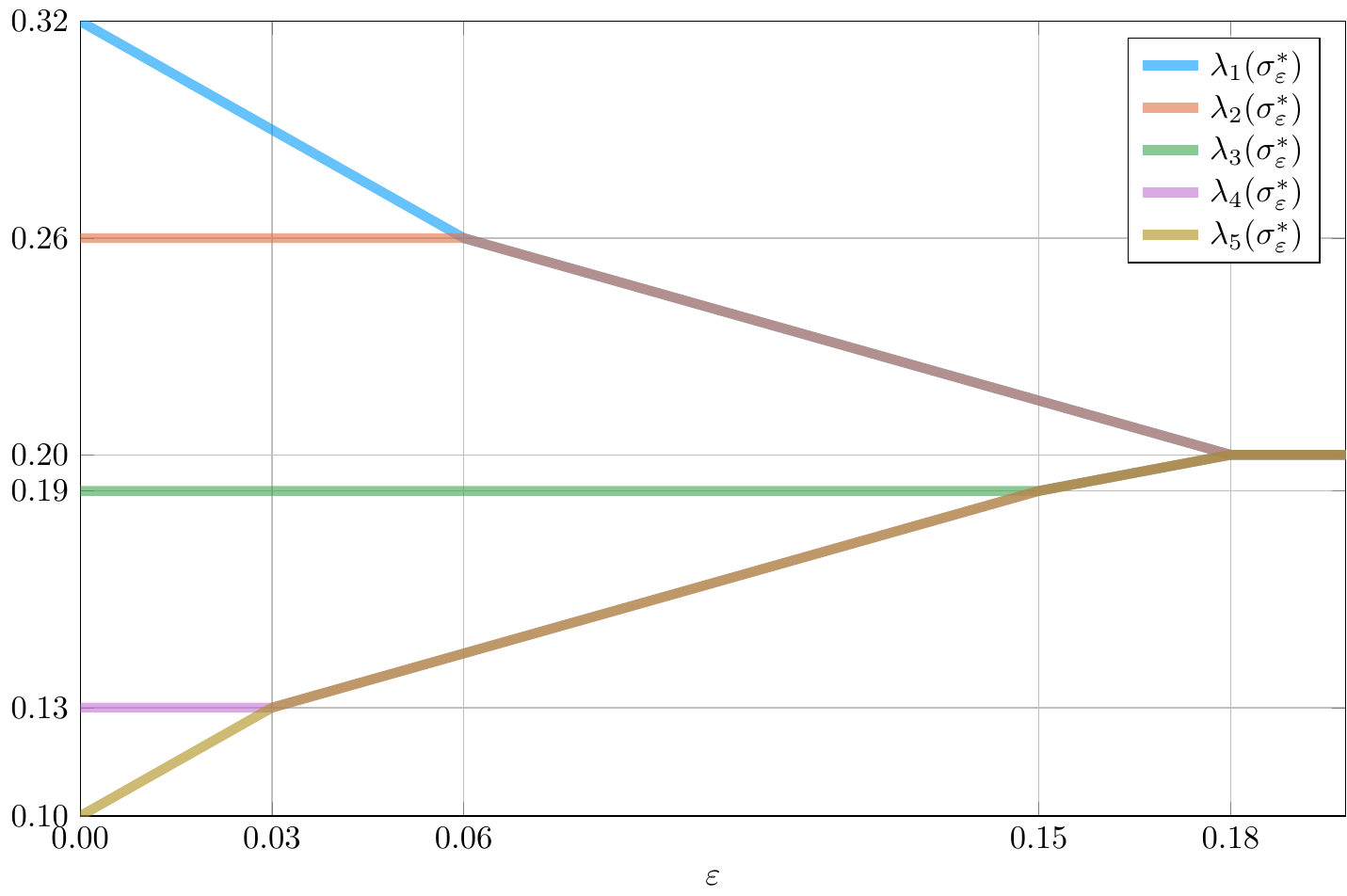}
  \caption{For the 5-dimensional state $\sigma = \diag(0.32, 0.26, 0.19, 0.13, 0.10)$, the spectrum of $\sigma_\eps^* = \mmm_\eps(\sigma)$ is plotted as a function of $\eps$. This plot is a continuous (in $\eps$) analog to the type of plot shown in \Cref{fig:rho-levels}, which shows the spectrum of $\sigma_\eps^*$ at two discrete points, $\eps = 0$ and $\eps= 0.07$, in a different example.
    Here, at $\eps=0$, the five lines correspond to the five eigenvalues of $\sigma$, each with multiplicity one. For $\eps \leq 0.03$, $\sigma_\eps^* = \diag(0.32 -\eps, 0.26, 0.19, 0.13, 0.10+\eps)$ and differs from $\sigma$ only in the smallest and largest eigenvalue. When $\eps$ reaches $0.03$, the multiplicity of the smallest eigenvalue of $\sigma_\eps^*$ increases to 2. Between $\eps=  0.03$ and $\eps = 0.06$, again only the smallest and largest eigenvalues change, but the smallest eigenvalue has multiplicity 2. This process continues until every eigenvalue reaches $\frac{1}{d}=0.2$ at $T(\sigma,\tau) = 0.18$.} \label{fig:mmm}
\end{figure}

For any $\eps \leq \delta(r)$, the map $\mmm_\eps$ only ``moves'' the largest and smallest entries of $r$, as shown by the following result and illustrated through an example in \Cref{fig:mmm}.

\begin{lemma} \label{lem:linear-perturbation}
  Let $r\in \cP$. Then $r_\eps^*$ is a linear perturbation of $r$ for $\eps$ small enough, in the sense that
  \begin{equation} \label{eq:linear-perturbation}
    r_\eps^* = r + \eps \gen(r) \qquad  \forall \eps \leq \delta(r)
  \end{equation}
  where $\delta(r) \geq 0$ is defined in \Cref{def:delta-rho-sigma} and $\gen(r) \in \R^d$ is defined as follows. If $r = u$ is the uniform distribution, set $\cL(r) := 0$. For $r\neq u$, denote the largest entry of $r$ as $r_+$, and its multiplicity by $k_+$, and likewise the smallest entry of $r$ as $r_-$ and its multiplicity by $k_-$.
  Then, $\gen(r) \in \R^n$ is a vector defined by
  \begin{equation} \label{eq:def_X}
    \gen(r)_i = \begin{cases}
      -\frac{1}{k_+} & r_i = r_+    \\
      \frac{1}{k_-}  & r_i = r_-    \\
      0              & \text{else,}
    \end{cases}
  \end{equation}
  for $i=1,\dotsc, d$. Moreover, for any $r\in \cP$,
  \begin{equation}\label{eq:constant-speed}
    \frac{1}{2}\|\gen(r)\|_1 = \begin{cases}
      0 & r = u,            \\
      1 & \text{otherwise},
    \end{cases}
  \end{equation}
  and either
  \begin{itemize}
    \item $r + \delta(r) \cL(r) = u$,
    \item $k_+( r + \delta(r)\cL(r) ) > k_+(r)$, or
    \item $k_- ( r + \delta(r) \cL(r)) > k_-(r)$.
  \end{itemize}
\end{lemma}
\begin{remark}
The last statement shows that the range of validity of \Cref{eq:linear-perturbation} is exactly until the degeneracy of $r$ is increased. Since a vector $r\in \cP_d$ has $d$ entries, the degeneracy can only increase at most $d-1$ times, and hence, using the semigroup property, the action of $\mmm_\eps$ on $r\in \cP$ can be described as the composition of at most $d$ linear maps (note that exactly \emph{which} maps depends on $r$; $\mmm_\eps$ is not linear on all of $\cP$). We used this decomposition in \cite{HD17} to establish uniform continuity bounds for various entropies (the so-called \typeone{} $(h,\phi)$-entropies in the language of \Cref{sec:entropic-cty-from-majflow}). However, it has proven useful to consider the action of $\mmm_\eps$ fully infinitesimally, dividing it into infinitely many pieces instead of finitely many, as is done in \Cref{sec:path-majflow}. This technique allows us to consider the derivatives of various quantities along the path traced out by $(\mmm_\eps(r))_{\eps \geq 0}$ in $\cP$ and subsequently allows us to formulate uniform continuity bounds for a wider class of quantities (e.g.\@~the so-called \typetwo{} $(h,\phi)$-entropies of \Cref{sec:entropic-cty-from-majflow}, the guesswork without side information considered in \Cref{sec:guesswork_no_side_info}, and the expected number of connected components of a particular model of random graph considered in \Cref{sec:random-graph}.)
\end{remark}
\begin{proof}
Note that \eqref{eq:constant-speed} follows directly from the definition \eqref{eq:def_X}. It remains to prove \eqref{eq:linear-perturbation} and the final statements.
  We use the notation of \Cref{def:delta-rho-sigma}. We consider two cases: either $\eps = \TV(r,u)$, or $\eps  < \TV(r,u)$.

  Let us consider the case $\eps = \TV(r,u)$. First, we establish the inequality $\delta(r)\leq \TV(r,u)$. If $\ell=2$, we have that $\delta(r) = \TV(r,u)$ as discussed after \Cref{def:delta-rho-sigma}. If $\ell > 2$, one cannot have both $\mu_{\ell-1} > \frac{1}{d}$ and $\mu_2 < \frac{1}{d}$. In the case $\mu_2 \geq \frac{1}{d}$, we have
  \begin{equation}\label{eq:gen-proof-case1}
    \delta(r) \leq k_+ (\mu_1 - \mu_2) = k_+ \left(\mu_1 - \frac{1}{d}\right) + k_+ \left( \frac{1}{d}-\mu_2 \right) \leq k_+ \left(\mu_1 - \frac{1}{d}\right)  \leq \TV(r,u)
  \end{equation}
  and likewise if $\mu_{\ell-1}\leq \frac{1}{d}$,
  \begin{equation} \label{eq:gen-proof-case2}
    \delta(r) \leq k_- (\mu_{\ell-1}-\mu_\ell) = k_- \left( \mu_{\ell-1} - \frac{1}{d} \right) + k_- \left(\frac{1}{d} -  \mu_{\ell}  \right) \leq k_- \left(\frac{1}{d} -  \mu_{\ell}  \right) \leq \TV(r,u).
  \end{equation}

  Next, let us establish the result in the case $\eps=\TV(r,u)$.
  Note that $\eps = \TV(r,u)$ necessitates $\eps = \delta(r) = \TV(r,u)$. If $\ell = 2$, then we  directly check that in this case, for $\eps = \TV(r,u)$, we have $r + \eps \cL = \mmm_\eps(r) = u$, as desired.
  If $\ell > 2$ and $\eps = \TV(r,u)$, then $\delta(r) = \TV(r,u)$. Let us inspect when equality can occur in \eqref{eq:gen-proof-case1} and \eqref{eq:gen-proof-case2}.
  We find that in either case, if $\delta(r) = \TV(u,r)$ then $\ell = 3$ and $\mu_2 = \mu_{\ell-1} = \frac{1}{d}$, and consequently, $r + \delta(r) \cL(r) = u$. For example, assume $\mu_2 \geq \frac{1}{d}$ and equality holds everywhere in \eqref{eq:gen-proof-case2} (the case $\mu_{\ell-1} \leq \frac{1}{d}$ follows similarly). Then we must have
  \begin{equation}\label{eq:gen-proof-kp}
    k_+(\mu_1 - \mu_2) \leq k_-(\mu_{\ell-1}-\mu_\ell)
  \end{equation}
  and $\mu_2 = \frac{1}{d}$. Hence, either $\ell-1 = 2$ and $\mu_{\ell-1} = \frac{1}{d}$, or else $\ell > 3$ and $\mu_{\ell-1} < \frac{1}{d}$. In the latter situation, the right-hand side of \eqref{eq:gen-proof-kp} is strictly less than $\TV(r,u)$ by following \eqref{eq:gen-proof-case2}, while the left-hand side of \eqref{eq:gen-proof-kp} equals $\TV(r,u)$, which is a contradiction.

  Next, it remains to consider the case $\eps < \TV(r,u)$. If $\ell =2$, then since $\frac{k_+k_-}{k_+ + k_-}\leq \min(k_+,k_-)$, we also have
  \[
    \eps \leq \delta(r) \leq \min( k_+ (\mu_1-\mu_2), k_-(\mu_1-\mu_2)) =  \min( k_+ (\mu_1-\mu_2), k_-(\mu_{\ell-1}-\mu_\ell)).
  \]
  The inequality
  \begin{equation}\label{eq:gen-proof-delta-ieq}
    \eps \leq \min( k_+ (\mu_1-\mu_2), k_-(\mu_{\ell-1}-\mu_\ell)).
  \end{equation}
  therefore holds for any $\ell\geq 2$ (since it holds by definition of $\delta(r)$ for $\ell > 2$).
  Hence, for any $\ell \geq 2$, it suffices to prove \eqref{eq:linear-perturbation} (and that either $k_+(r + \delta(r)\gen(r)) > k_+(r)$ or $k_-(r + \delta(r)\gen(r)) > k_-(r)$) under the assumptions that $\eps < \TV(r,u)$ and \eqref{eq:gen-proof-delta-ieq}. Here and in the following, we write make the dependence on $r$ in $k_\pm$ explicit, writing $k_\pm \equiv k_\pm(r)$.

 To establish \eqref{eq:linear-perturbation}, we simply check that the choice $m=k_+(r)$ satisfies the definition of $m_+(r,\eps)$, namely that the choice $m=k_+(r)$ solves \eqref{eq:m_is_sol_to_this}, and likewise for $k_-$ with $m_-$. To see this, recall that $\gamma_+^{(m)}(r,\eps) = \frac{1}{m}\left(\sum_{i=1}^{m} r_i^\downarrow  - \eps\right)$.
  Then, by taking $m = k_+(r)$ we find
  \[
    r_{k_+(r)+1}^\downarrow = \mu_2 \leq \frac{1}{k_+(r)}\left(\sum_{i=1}^{k_+(r)} r_i^\downarrow  - \eps\right) = \mu_1 - \frac{\eps}{k_+(r)}
  \]
  since $\frac{\eps}{k_+(r)} \leq \frac{1}{k_+(r)}\delta(r) \leq \mu_1-\mu_2$. Additionally, $ \mu_1 - \frac{\eps}{k_+(r)} < \mu_1 = r_{k_+(r)}^\downarrow$. Therefore, $m = k_+(r)$ solves \eqref{eq:m_is_sol_to_this}, hence $m_+(r,\eps) = k_+(r)$.
  Proving that $m_-(r,\eps) = k_-(r)$ is analogous.

  Next, consider $\eps = \delta(r)$. If $\ell = 2$, then $\eps = \TV(r,u)$ which has already been considered. So consider $\ell > 3$ and $\eps < \TV(r,u)$ in the following.
  Without loss of generality, assume $\delta(r) = k_+(r) (\mu_1-\mu_2)$. We show that $k_+(\mmm_\eps(r)) > k_+(r)$. By the above, $m_+(r,\eps) = k_+(r)$, and therefore
  \[
    \gamma_+(r,\eps)= \mu_1 + \frac{\eps}{k_+(r)} =  \mu_1 +(\mu_1-\mu_2) = \mu_2.
  \]
  Hence, $k_+(r + \delta(r) \gen(r)) \geq k_+(r) + k_2 > k_+(r)$ where $k_2$ is the multiplicity of $\mu_2$ in $r$.
  Similarly, if $\delta(r) = k_-(r)(\mu_{\ell-1}-\mu_{\ell})$ then $k_-( \mmm_{\eps}(r)) > k_-(r)$.
\end{proof}
Let us highlight that the above proof also makes use of the uniqueness results provided by \Cref{lem:uniqueness_for_majmin}.

Together, the semigroup property \Cref{prop:VN_semigroup_property} and \Cref{lem:linear-perturbation} provide a simple and effective characterization of $\mmm_\eps$ via $\gen$, which is explored in the next section.

\subsection{The path of majorization flow} \label{sec:path-majflow}
For $r\in \cP$, define the path $\gamma(t) = \mmm_t(r)$. We call the path $(\gamma(t))_{t=0}^1 \subseteq \cP$ as the \emph{path of majorization flow} starting from $r$, with respect to the total variation distance. Intuitively speaking, along this path the probability vector decreases in majorization order as quickly as possible, while changing at constant speed in $1$-norm. Note that for $t\geq 1 - \frac{1}{d}$, one has $\gamma(t) \equiv u$, the uniform distribution, and we take the final time as $t=1$ for simplicity.

Recall the quantity $\gen$ given in \Cref{lem:linear-perturbation}. Then for $r\in \cP$,
\begin{equation}
    \gen(r) = \left.\partial_s^+ \mmm_s(r)\right|_{s=0}
\end{equation}
where $\partial_s^+$ indicates the one-sided derivative from above. This immediately follows from \eqref{eq:linear-perturbation}. We consequently call $\gen$ the \emph{generator of majorization flow}.
The semigroup property established in \Cref{prop:VN_semigroup_property} leads to the following integral formulas.

\begin{theorem}\label{thm:integral-formulas}
    Let $\eps > 0$ and $(\mmm_s)_{s\geq 0}$ be as defined above. Then
    \begin{equation} \label{eq:mmm-integral-formula}
        \mmm_\eps(r) = r + \int_0^\eps \gen(\mmm_s(r)) \d s.
    \end{equation}
    Additionally, for $H: \cP \to \R$ which is continuously differentiable on $\cP_+$, the quantity $\Gamma_H(r) := \partial_s^+ H(\mmm_s(r))|_{s=0}$ exists and satisfies
    \begin{equation} \label{eq:H-integral-formula}
        H(\mmm_\eps(r))  = H(r) + \int_0^\eps \Gamma_H(\mmm_s(r)) \d s.
    \end{equation}
\end{theorem}

\begin{remark}
    The relations \eqref{eq:mmm-integral-formula} and \eqref{eq:H-integral-formula} are of particular importance. Equation \eqref{eq:mmm-integral-formula} allows one to determine properties of $\mmm_s(r)$ by simply analyzing $\gen$, while \eqref{eq:H-integral-formula} allows one to analyze continuity properties of $H$ using $\Gamma_H$.
\end{remark}

\begin{proof}
    To establish \eqref{eq:mmm-integral-formula}, note that equation \eqref{eq:linear-perturbation} immediately yields
    \[
        \left.\partial_\eps^+ \mmm_\eps(r)\right|_{\eps=0} = \gen(r)
    \]
    where $\partial_\eps^+$ indicates the one-sided derivative in $\eps$ from above. Then
    \[
        \left.\partial_s^+ \mmm_s(r)\right|_{s=t} = \left.\partial_t^+ \mmm_{s+t}(r)\right|_{t= 0}= \left.\partial_t^+ \mmm_{t}(\mmm_s(r))\right|_{t= 0} = \gen(\mmm_s(r))
    \]
    follows using the semigroup property. As the path $s\mapsto \mmm_s(r)$ is piecewise affine with at most $d$ pieces, for each fixed $r\in \cP$, the two-sided derivative $\partial_s \mmm_s(r)$ exists for all but at most $d$ elements $s\in [0,1]$. Then, using the fundamental theorem of calculus,
    \[
        \mmm_\eps(r) = r + \int_0^\eps \gen(\mmm_s(r)) \d s.
    \]
    This establishes \eqref{eq:mmm-integral-formula}. \Cref{eq:H-integral-formula} follows in the same manner by considering $H \circ \mmm_\eps(r)$ instead of $\mmm_\eps(r)$, and using that $H$ is continuously differentiable on $\cP_+$.
\end{proof}

\begin{remark}
    Let us show the power of \eqref{eq:mmm-integral-formula} with a quick application.
    Since $\frac{1}{2}\|\gen(r)\|_1 = 1$ for all $r\in \cP\setminus\{u\}$ as shown in \Cref{lem:linear-perturbation}, and $\gen(u) = 0$, the triangle inequality
    \[
        \left\|\int_0^\eps \gen(\mmm_s(r)) \d s \right\|_1 \leq \int_0^\eps \|\gen(\mmm_s(r))\|_1 \d s
    \]
    immediately yields
    \[
        \frac{1}{2}\|\mmm_\eps(r) - r\|_1 \leq \eps.
    \]
    This, of course, is not new information; $\mmm_\eps(r)$ was constructed in \Cref{sec:construct_majmin} to have this property. However, we can see that the full construction detailed in that section is no longer needed to derive properties of $\mmm_\eps$, in this case at least. Instead, the simpler construction of $\gen$ (detailed in \Cref{lem:linear-perturbation}) and the semigroup property (as manifested in \eqref{eq:mmm-integral-formula}) is enough.
\end{remark}

Recall the notation that $r_\pm$ denotes the largest and smallest elements of $r \in \cP$.
\begin{proposition}\label{prop:Gamma-for-symmetric-H}
    Let $H: \cP\to \mathbb{R}$ be \emph{symmetric}\footnote{Note that symmetry is implied by Schur concavity.}, meaning $H(r) = H(\pi(r))$ for any permutation $\pi$, and differentiable on $\cP_+$. Then for $r\in \cP_+$,
    \begin{equation}\label{eq:Gamma-for-symmetric-H}
        \Gamma_H(r) = H_-(r) - H_+(r)
    \end{equation}
    where $H_+ = \partial_{r_i} H(r_1,\dotsc,r_d)$ for any index $i$ such that $r_i = r_+$, and similarly for $H_-$. Note, by symmetry, the value of $H_+$ does not depend on which index $i$ is chosen, as long as $r_i = r_+$.

    Moreover, if $H$ is Schur concave and continuous on $\cP$, then \eqref{eq:Gamma-for-symmetric-H} shows that
    \begin{equation}\label{eq:Gamma_H_pos}
        \Gamma_H(r) \geq 0 \qquad \forall r \in \cP
    \end{equation}
    using \eqref{eq:S-convex-condition}, and likewise if $H$ is Schur convex, then $\Gamma_H(r) \leq 0$ for all $r \in \cP$.
\end{proposition}
\begin{proof}
    We have
    \begin{align*}
        \Gamma_H(r) & =  \left.\frac{\d}{\d y} H(\mmm_y(r))\right|_{y=0}                                                                      \\
                    & = \sum_{i=1}^d \gen(r)_i H_i(r)= \sum_{i : r_i = r_+} \frac{-1}{k_+} H_i(r) + \sum_{i : r_i = r_-} \frac{1}{k_-} H_i(r)
    \end{align*}
    where $H_i(r) = \partial_{r_i}H(r)$. Since $H$ is symmetric, $H_i(r) = H_j(r)$ if $r_i = r_j$. Hence, we have \eqref{eq:Gamma-for-symmetric-H}.
\end{proof}
That is, $\Gamma_H(r)$ is simply the difference between two partial derivatives of $H$, evaluated at $r$.

\begin{example} \label{ex:non-diff}
Assume $H: \cP\to \R$ is symmetric.
If $H$ is continuously differentiable on $\cP^\downarrow$ but not on an open set containing $\cP^\downarrow$, then \eqref{eq:H-integral-formula} may not hold. For example, define $H(r) = \sum_{j=1}^d j \, r_j^\downarrow$; this in fact is the \emph{guesswork}, discussed in \Cref{sec:guesswork}. Then $H$ is Schur concave, and for $r\in \cP^\downarrow$,  we have the simple formula $H(r) = \sum_{j=1}^d j\, r_j$. Hence, $H$ differentiable on $\cP^\downarrow$ with $\Gamma_H(r) = d-1$ for $r\neq u$ and $\Gamma_H(u) =0$. However, setting $p = (1,0,\dotsc,0)$, and $0<\eps<1- \frac{1}{d}$, we have
\[
H(p_\eps^*) =(1-\eps) + \frac{\eps}{d-1} \sum_{i=2}^d i = 1 + \eps \frac{d}{2},
\]
while $H(p) + \int_0^\eps \Gamma_H(\mmm_s(p)) \d s = 1 + \eps (d-1)$.
\end{example}

To ameliorate that, we can formulate a slight generalization of \Cref{thm:integral-formulas} as follows.
\begin{corollary} \label{cor:non-diff-integral}
Let $H: \cP\to \mathbb{R}$ be symmetric, and let $F = \left.H\right|_{\cP^\downarrow}$ have a (possibly non-symmetric) extension $\tilde F : \cP \to \mathbb{R}$ which is differentiable on $\cP_+$. Then for all $\eps > 0$,
\begin{equation}
H(\mmm_\eps(r)) = H(r) + \int_0^\eps \Gamma_{\tilde F}(\mmm_s(r^\downarrow)) \d s.
\end{equation}
\end{corollary}
\begin{proof}    
We have that by \Cref{thm:integral-formulas},
\begin{equation}\label{eq:proof-tildeF}
\tilde F(\mmm_\eps(r)) = \tilde F(r) + \int_0^\eps \Gamma_{\tilde F}(\mmm_s(r)) \d s.
\end{equation}
Next, since $H$ is symmetric, we have that
\[
 H(\mmm_\eps(r)) = H(\mmm_\eps(r^\downarrow)) = F(\mmm_\eps(r^\downarrow)) = \tilde F(\mmm_\eps(r^\downarrow))
\]
and similarly $H(r) = \tilde F(r^\downarrow)$. Hence \eqref{eq:proof-tildeF} with $r$ replaced by $r^\downarrow$ yields the result.
\end{proof}

Let us see how this repairs \Cref{ex:non-diff}. We may choose $\tilde F(r) = \sum_{j=1}^d j r_j$ as a differentiable extension of $\left. H \right|_{\cP^\downarrow}$. Then $\Gamma_{\tilde F}(r) = \frac{1}{k_-} \sum_{j: r_j = r_-} j - \frac{1}{k_+}\sum_{j: r_j = r_+} j$, as in the proof of \Cref{prop:Gamma-for-symmetric-H}. Then for $0<s<1-\frac{1}{d}$ and $p^\downarrow = p=(1,0,\dotsc,0)$,
\[
\Gamma_{\tilde F} (\mmm_s(p^\downarrow)) = \Gamma_{\tilde F} (1-s, \frac{s}{d-1}, \dotsc, \frac{s}{d-1}) = \frac{1}{d-1}\sum_{j=2}^d j - 1 = \frac{d}{2}.
\]
Hence,
\[
H(p) + \int_0^\eps \Gamma_{\tilde F}(\mmm_s(p^\downarrow)) \d s = 1 + \eps \frac{d}{2}
\]
matching $H(\mmm_\eps(p))$.

\paragraph{Comparison to other flows}
The notion of a flow arises naturally in various branches of physics, mathematics, and engineering.  It is interesting to compare
and contrast the notion of majorization flow that we have introduced to the notion of gradient flow that arises in optimal transport and differential geometry, and has been applied to study open quantum systems.

The gradient flow induced by a function $F$ in a metric space can be loosely interpreted as the flow that decreases $F$ as quickly as possible \cite[p.~645]{Vil09}; in a similar sense, the majorization flow decreases the majorization order as quickly as possible (while following a unit speed path in total variation distance). However, there are several complications that prevent making the connection between majorization flow and gradient flow more precise:
\begin{itemize}
    \item Perhaps the most obvious one: decreasing in majorization order requires non-increasingness of all the partial sums given in \eqref{def:majorize}, while decreasing a function only requires decreasing a scalar value.
    \item The theory of gradient flow is well-developed on continuous spaces, such as probability measures on $\R^d$ equipped with a Wasserstein metric (see, e.g.\@ \cite{Vil09}), but is less well-developed in the discrete case considered here (probability measures on $\{1,\dotsc,d\}$).
    \item Here we consider the total variation distance, which can be seen as the $1$-Wasserstein distance induced by the Hamming distance on the set $\{1,\dotsc,d\}$. Almost all of the literature in discrete or continuous space takes the metric to be the $p$-Wasserstein distance for $p>1$ (for smoothness reasons).
\end{itemize}

\section{Uniform continuity bounds from majorization flow}\label{sec:uniform-bounds-majflow}

For $\eps >0$ and Schur concave $H$ (meaning $H(p) \geq H(q)$ if $q \prec p$), we have $H(\mmm_\eps(r)) = \max_{p \in B_\eps(r)} H(p)$ since $\mmm_\eps(r) = \minmaj( B_\eps(r))$. In this case, if $H$ is continuously differentiable on $\cP_+$, then by using \eqref{eq:H-integral-formula} we obtain
\begin{equation} \label{eq:Delta-eps-by-Gamma}
\Delta_\eps^H(r):= \max_{p \in B_\eps(r)} H(p) - H(r) = \int_0^\eps \Gamma_H(\mmm_s(r)) \d s.
\end{equation}
In other words, the amount $H$ can locally increase near $r$ (quantified by $\Delta_\eps^H(r)$) is determined by $\Gamma_H$. Moreover, the global continuity properties of $H$ are determined by $\sup_{r\in \cP}\Delta_\eps^H(r)$. To see this, note that if $p,q \in \cP$ satisfy $\TV(p,q)\leq \eps$, then
\[
H(p) - H(q) \leq H(q\majmin) - H(q) = \Delta_\eps^H(q)
\]
and
\[
H(q) - H(p) \leq H(p\majmin) - H(p) = \Delta_\eps^H(p)
\]
so we have
\begin{equation}
|H(p) - H(q)| \leq \max\{\Delta_\eps^H(p), \Delta_\eps^H(q)\} \leq \sup_{r\in \cP} \Delta_\eps^H(r) \label{eq:uniform-bound-from-Delta-eps}
\end{equation}
This fact and \eqref{eq:Delta-eps-by-Gamma}  have two immediate consequences for the continuity properties of $H$:
\begin{enumerate}
    \item If $\Gamma_H$ is Schur convex, then by \eqref{eq:Delta-eps-by-Gamma}, $\Delta_\eps^H$ is Schur convex too, as $\mmm_s$ is majorization preserving for all $s\in [0,1]$. This provides the upper bound $\Delta_\eps^H(r) \leq\Delta_\eps^H(\psi)$ for $\psi = (1,0,\dotsc,0)$, (since $\psi$ majorizes every $r \prec \psi$ for every $r \in \cP$) which yields a tight uniform continuity bound by \eqref{eq:uniform-bound-from-Delta-eps}.
    \item If $\Gamma_H$ can be upper bounded by $k>0$ on $\cP_+$, then by \eqref{eq:Delta-eps-by-Gamma},  $\Delta_\eps^H(r)\leq \eps k$. This immediately yields a Lipschitz continuity bound for $H$ by \eqref{eq:uniform-bound-from-Delta-eps}.
\end{enumerate}
\begin{remark}
In case $H$ is not continuously differentiable on $\cP_+$ but satisfies the conditions of \Cref{cor:non-diff-integral}, the above two points hold with $\Gamma_H$ replaced by $\Gamma_{\tilde F}$, in the notation of that corollary.
\end{remark}
The second point can be strengthened to the following corollary to \Cref{thm:integral-formulas}.
\begin{corollary}\label{cor:opt-Lip}
Let $H: \cP\to \R$ be a Schur concave function which is continuously differentiable on $\cP_+$. We write $H(r_1,\dotsc,r_d) \equiv H(r)$ for $r\in \cP$. Next, for $r \in \cP$, let $i_+ \in \{1,\dotsc,d\}$ be an index such that $r_+ = r_{i_+}$, and similarly $i_- \in \{1,\dotsc,d\}$ such that $r_- = r_{i_-}$.
Then $\Gamma_H$ is given by
\begin{equation}
\begin{aligned}
\Gamma_H : \quad \cP_+ &\to \R\\
 r &\mapsto (\partial_{r_{i_+}} - \partial_{r_{i_-}})H(r_1,\dotsc, r_d).
\end{aligned}
\end{equation}
Note that this expression does not depend on the choice of $i_\pm$ since $H$ is permutation invariant. Moreover, $H$ is Lipschitz continuous if and only if
\[
k := \sup_{r \in \cP_+} \Gamma_H(r)
\]
satisfies $k < \infty$. Moreover, in the latter case $k$ is the optimal Lipschitz constant for $H$.
\end{corollary}

\begin{proof}[Proof of \Cref{cor:opt-Lip}]
    In the case that $\sup_{r\in \cP_+}\Gamma_H(r) = \infty$, for each $n \in \mathbb{N}$, there exists $r^{(n)} \in \cP_+$ such that
    \[
        \lim_{s\downarrow 0}\frac{H(\mmm_s(r^{(n)})) - H(r^{(n)})}{s} > n.
    \]
    Additionally, $r^{(n)}\neq u$, otherwise the numerator would be zero. Hence, for any $\eps > 0$ there exists $0 < s_n < \TV(r^{(n)}, u)$ such that
    \[
        \frac{H(\mmm_{s_n}(r^{(n)})) - H(r^{(n)})}{s_n} > n - \eps.
    \]
    Since $s_n < \TV(r^{(n)}, u)$, we have $s_n = \TV(\mmm_{s_n}(r^{(n)}), r^{(n)} )$, and hence
    \[
        \frac{H(\mmm_{s_n}(r^{(n)})) - H(r^{(n)})}{ \TV(\mmm_{s_n}(r^{(n)}), r^{(n)} )} > n - \eps.
    \]
    Any Lipschitz constant $k$ must be larger than the left-hand side, for any $n$, and hence must be $\infty$.

    Next, consider the case in which $\sup_{r\in \cP_+}\Gamma_H(r)$ is finite. \Cref{eq:Gamma-for-symmetric-H} shows that the quantity defined in \Cref{cor:opt-Lip} is indeed the $\Gamma_H$ of \Cref{thm:integral-formulas}. Hence, in the case that $\sup_{r\in \cP_+}\Gamma_H(r) < \infty$ holds, \eqref{eq:Delta-eps-by-Gamma} and \eqref{eq:uniform-bound-from-Delta-eps} show that $\sup_{r\in \cP_+}\Gamma_H(r)$ is indeed a Lipschitz constant for $H$. It remains to show this constant is optimal.

    Assume there is some Lipschitz constant $k' < \sup_{r\in \cP_+}\Gamma_H(r)$.
    For each $\eps > 0$, let $r^{(\eps)} \in \cP_+$ satisfy $\Gamma_H(r^{(\eps)}) > \sup_{r\in \cP_+}\Gamma_H(r) - \eps$. Then
    \[
        k' \geq \frac{H(\mmm_s(r^{(\eps)})) - H(r^{(\eps)})}{\frac{1}{2}\|\mmm_s(r^{(\eps)}) -r^{(\eps)}\|_1 } \geq \frac{H(\mmm_s(r^{(\eps)})) - H(r^{(\eps)})}{s}
    \]
    for each $s > 0$. Taking the limit $s\to 0$ yields
    \[
        k' \geq \Gamma_H(r^{(\eps)}) > \sup_{r\in \cP_+}\Gamma_H(r) - \eps.
    \]
    Taking $\eps \to 0$ shows $k' \geq \sup_{r\in \cP_+}\Gamma_H(r)$, which contradicts $k' < \sup_{r\in \cP_+}\Gamma_H(r)$. Hence, the quantity $\sup_{r\in \cP_+}\Gamma_H(r)$ is indeed the optimal Lipschitz constant.
\end{proof} 

\chapter{Applications} \label{sec:applications}

This section applies the techniques developed in the previous chapters to various tasks (primarily continuity bounds).
\begin{itemize}
	\item In \Cref{sec:entropic-cty-from-majflow}, tight uniform continuity bounds and optimal Lipschitz constants are obtained for a class of entropies known as $(h,\phi)$-entropies.
	\item \Cref{sec:tsallis-renyi} specializes these continuity bounds to the case of the R\'enyi and Tsallis entropies. In \Cref{sec:Renyi_opt_scaling}, it is shown that the continuity bound for the Shannon entropy (or von Neumann entropy) obtains the best dimensional scaling out of all the $\alpha$-R\'enyi entropies. In \Cref{sec:discuss-prev-bounds}, majorization flow is used to show why previous uniform continuity bounds for the $\alpha$-R\'enyi entropy with $\alpha>1$ achieve poor scaling with $d$ and $\alpha$. Lastly, in \Cref{sec:connect-thermo}, the connections between R\'enyi entropies and thermodynamic free energies introduced by \cite{Bae11} are discussed, and the continuity bounds for the R\'enyi entropies are interpreted in this framework.
	\item In \Cref{sec:more-cty-bounds}, continuity bounds are obtained for several other families of entropies.
	\item In \Cref{sec:butterflies}, continuity bounds are obtained for the number $K$ of distinct realizations of $N$ i.i.d.~discrete random variables $X_i\sim p$, as a function of $p \in \cP_M$ (for some number of possible outcomes $M$).
	\item In \Cref{sec:random-graph}, a continuity bound is obtained for the expected number of connected components of a particular model of random graphs.
	\item In \Cref{sec:smoothed-entropies}, the majorization-minimizer and majorization-maximizer are used to provide explicit formulas for smoothed Schur concave and Schur convex functions. The maximum-entropy principle (MaxEnt) is also briefly discussed.
	\item In \Cref{sec:LOCC}, the majorization-minimizer and majorization-maximizer are used to obtain bounds on approximately transforming bipartite pure states via local operations and classical communication.
	\item In \Cref{sec:cty-bound-sigma}, a simple continuity bound is established for the function $\rho \mapsto D(\rho\|\sigma)$ on quantum states.
\end{itemize}

Many of the results established in this chapter are Lipschitz continuity bounds, in which a function $F$ on $\cP_d$ is shown to satisfy
\[
|F(p) - F(q)| \leq k \TV(p,q)\qquad\forall \, p,q\in \cP_d
\]
for some constant $k$, or a function $F$ on $\cD(\cH)$ with $\dim(\cH)=d$ is likewise shown to satisfy
\[
|F(\rho) - F(\sigma)|\leq k  T(\rho,\sigma)\qquad\forall \, \rho,\sigma\in \cD(\cH)
\]
for some constant $k$. The Lipschitz smallest $k$ is the optimal Lipschitz constant, and \Cref{tab:Lipschitz-constants} summarizes the values or bounds on $k$ established for various quantities in this chapter.
\begin{table}[ht]
\centering
\begin{tabular}{ c c c c }
Name &  Parameter Regime  & Optimal Lipschitz constant $k$ & Reference \\ \toprule
Tsallis entropy $T_\alpha$  & $0 < \alpha < 1$ & $k = \infty$ & Prop.~\ref{cor:Tsallis-Lipschitz} \\ 
  & $\alpha > 1$ & $k=\frac{\alpha}{\alpha-1}$ & Prop.~\ref{cor:Tsallis-Lipschitz} \\\addlinespace[0.75em]
R\'enyi entropy $S_\alpha$  & $0< \alpha < 1$ & $k = \infty$ & Prop.~\ref{cor:Renyi-lipschitz} \\
 & $\alpha > 1$ & $\frac{\alpha}{\alpha-1}\frac{(d-2)^{1 - 1/\alpha}}{2\ln(2)} \leq k\leq \frac{d\alpha}{\alpha-1} \frac{1}{\ln(2)}$ & Prop.~\ref{cor:Renyi-lipschitz} \\
 & $\alpha = 2$ & $k=\begin{cases}
 \frac{2}{\ln(2)} & d = 2\\ 
 \frac{d-2}{\sqrt{d-1}-1}\frac{1}{\ln(2)} & d > 2,
 \end{cases}$  & Prop.~\ref{cor:Renyi-lipschitz} \\
 & $\alpha = \infty$ & $k=\frac{d}{\ln(2)}$ & Prop.~\ref{cor:Renyi-lipschitz} \\\addlinespace[0.75em]
von Neumann entropy $S$ &  & $k = \infty$ & above \eqref{eq:smoothed-entropy-Lipschitz} \\
\begin{tabular}{@{}c@{}}
$\delta$-smoothed ($S^\delta$)
\end{tabular} & $0<\delta<1$ & $k = \log(\delta\inv - 1) + \log(d-1)$ &  \eqref{eq:smoothed-entropy-Lipschitz} \\\addlinespace[0.75em]
Unified entropies $E_\alpha^s$ & $\alpha > 1$, $s\geq 1$ & $k \leq \frac{\alpha}{\alpha-1}$ & \cite{HY06} \\
 & $\alpha > 1$, $s \leq 1$ & $k \leq \begin{cases}
\frac{\alpha}{\alpha-1}d^{1-\alpha s} &s \alpha < 1\\
\frac{\alpha}{\alpha-1} & s\alpha \geq 1.
\end{cases}$ & \eqref{eq:unified-Lipschitz-alpha_gt_1_s_lt_1} \\
 & $\alpha\in (0,1)$  & $k = \infty$ & above \eqref{eq:unified-Lipschitz-alpha_gt_1_s_lt_1}\\\addlinespace[0.75em]
$S_f(\rho) = - \tr[f(\rho)]$ & \begin{tabular}{@{}c@{}}$f$ strictly convex\\ $f(0)=f(1)=0$\end{tabular} & $k = f'(1) - f'(0)$ & \eqref{eq:Sf-Lipschitz}\\\addlinespace[0.75em]
Concurrence $C$ &  & $k = \infty$ & \eqref{eq:concurrence_GCB} \\\addlinespace[0.75em]
\begin{tabular}{@{}c@{}}
Expected number of\\
connected components\\
$E_C$
\end{tabular} & & $k\leq 3 + 0.35 \sqrt{d-2}$  & \eqref{eq:Lip-upper-bound} \\\addlinespace[0.75em]
Guesswork $G_{\vec c}(X)$ & & $k = c_d - c_1$ & Prop.~\ref{prop:guesswork-Lipschitz-nonconditional}\\
with QSI ($G_{\vec c}(X|B)$) & $c_{d}<\infty$  & $k \leq 2c_{d}$ & Prop.~\ref{prop:guesswork_Lipschitz}\\\addlinespace[0.75em]
$\bE[K]$ & $M < \infty$ & $k=N$ & \eqref{eq:trials-Lip} \\\addlinespace[0.75em]
\end{tabular}
\caption{Optimal Lipschitz constants $k$ for various functions considered in this chapter, as a function of the underlying dimension $d$ and parameters of the function. $k=\infty$ means that the function is not Lipschitz continuous.}\label{tab:Lipschitz-constants}
\end{table}

\FloatBarrier

\section{\texorpdfstring{$(h,\phi)$}{(h, phi)}-entropies}\label{sec:entropic-cty-from-majflow}

There are several families of single-partite entropies: $\alpha$-R\'enyi entropies, Tsallis entropies, unified entropies, and so forth (definitions and references for these can be found in \Cref{sec:tsallis-renyi,sec:more-cty-bounds}). These various entropies have a lot of similarities, in particular sharing a functional form: for $r \in \cP_d$,
\begin{equation}\label{eq:functional-form-h-phi}
H(r) = h\left( \sum_{i=1}^d \phi(r_i)\right)
\end{equation}
for a pair of scalar functions $h$ and $\phi$. However, they exhibit different mathematical properties, most notably concavity or lack thereof. For example, the $\alpha$-R\'enyi entropies, defined by
\[
 S_\alpha(r)= \frac{1}{1-\alpha}\log \left( \sum_{i=1}^d r_i^\alpha \right)
 \] are concave for $\alpha \in (0,1)$, but are neither concave nor convex for $\alpha > 1$. Certain properties of the latter have proven particularly difficult to study, perhaps as a consequence of the lack of concavity.

While Audenaert proved a tight uniform continuity bound for the $\alpha$-R\'enyi entropies for $\alpha \in (0,1)$ in 2007 (\cite{Aud07}; see \eqref{eq:uniform_Renyi_bound} below), a uniform continuity bound on the $\alpha$-R\'enyi entropies for $\alpha > 1$ was not established until 2011. At that time, Rastegin \cite{Ras2011} proved the bound
\begin{equation}
|S_\alpha(p) - S_\alpha(q)| \leq  \frac{d^{2(\alpha-1)}}{\alpha-1}[  1  - (1-\eps)^\alpha  -\eps^\alpha (d-1)^{1-\alpha} ] \label{eq:bound-Rastegin}
\end{equation}
where $\eps = \TV(p,q)$, for $\alpha > 1$.
This bound, however, suffers from an exponential dependence on $\alpha$ (and for fixed $\alpha$, polynomial dependence on $d$), while scaling linearly with $\eps$, as $1  - (1 - \eps)^\alpha \approx \alpha \eps - \frac{1}{2}\alpha(\alpha-1) \eps^2 + O(\eps^3)$. Since the inequality $|S_\alpha(p) - S_\alpha(q)| \leq 2 \log d$ holds trivially, for even moderately large $\alpha$, \eqref{eq:bound-Rastegin} provides a non-trivial bound for a very small range of $\eps$.

In 2017, Chen et al \cite{Renyi-CMNF} improved upon this bound, showing that for $\alpha > 1$,
\begin{equation} \label{eq:Chen-renyi-bound}
|S_\alpha(p) - S_\alpha(q)| \leq \frac{d^{\alpha-1}}{\alpha-1}[  1  - (1-\eps)^\alpha  -\eps^\alpha (d-1)^{1-\alpha} ] 
\end{equation}
However, this bound still suffers from exponential dependence on $\alpha$.
The proof of both bounds proceeds by reducing to the case of the Tsallis entropy,
\[
T_\alpha(p) = \frac{1}{1-\alpha}\left(\Big(\sum_{i=1}^d p^\alpha\Big) - 1\right),
 \]
picking up an exponential prefactor along the way.

In this chapter, we prove that the dimensional dependence is at most linear\footnote{This is also established in-effect by \cite[Theorem 7, (2)]{WH19}, which was developed independently and posted slightly later than \cite{HD19} in which we established the linear bound discussed here.}, and in fact
\begin{equation} \label{eq:intro-Renyi-Lip}
|S_\alpha(p) - S_\alpha(q)| \leq  \frac{d\alpha}{\alpha-1} \frac{1}{\ln(2)} \eps
\end{equation}
for $\TV(p,q) \leq \eps$ and $\alpha > 1$. In fact, we prove that the Tsallis entropy increases the fastest near the corners of the probability simplex (i.e.\@ the extremal points which are permutations of $(1,0,\dotsc,0)$) and the slowest near the center of the simplex (see \Cref{thm:Delta-eps-Schur-convex}), while the $\alpha$-R\'enyi entropy increases the fastest close to the center of the probability simplex. This mismatch shows why bounding the difference of R\'enyi entropies of two probability vectors by the difference of Tsallis entropies of the same two distributions does not work well: a large prefactor is needed to bound the rapidly-changing R\'enyi entropies near the center of the simplex by the Tsallis entropies which change the slowest there. With the benefit of hindsight (and our proof techniques), we can find that indeed, a linear prefactor suffices to compare the maximum differences in R\'enyi entropies between two probability vectors which are at a fixed total variation distance apart, and the maximum difference in Tsallis entropies between two distributions at the same distance apart; however, these two maximum differences occur at very different parts of the probability simplex. These two effects can be seen quantitatively in \Cref{prop:compare-Renyi-Tsallis}.

To prove the bound \eqref{eq:intro-Renyi-Lip}, as well as determine where each entropy increases the fastest, we take a unified approach to establishing  entropic continuity bounds. While concavity only holds for certain entropies, we exploit the fact that all the above entropies $H$ are Schur concave.

We consider a class of entropic functionals called $(h,\phi)$-entropies, which were introduced by \cite{SMMP93}. These are defined by the formula
\[
 H_{(h,\phi)}(r) := h\left(\sum_{i=1}^d \phi(r_i)\right)
\]
for $r \in \cP_d$, and likewise by $H_{(h,\phi)}(\rho) = h(\tr(\phi(\rho)))$ for $\rho \in \cD(\cH)$, using the functional calculus (see \eqref{eq:functional-calculus}). The concept of $(h,\phi)$-entropies for quantum states was introduced by \cite{Bosyk2016}. In other words, the $(h,\phi)$-entropy of a quantum state is defined by the $(h,\phi)$-entropy of the probability vector given by its eigenvalues.

We will consider two classes of $(h,\phi)$-entropies which capture almost all single-partite entropies considered in the literature.
\begin{itemize}
	\item We say a function $H: \cP\to \R$ is a  \emph{\typeone{} $(h,\phi)$-entropy} if $H = H_{(h,\phi)}$ for some  $\phi: [0,1] \to \R$ which is continuously differentiable on $(0,1]$ and continuous on $[0,1]$ and $h: [\phi(1), \phi(\tfrac{1}{d})d] \subset \R \to \R$ which is continuously differentiable on $(\phi(1), \phi(\tfrac{1}{d})d]$ and continuous on $[\phi(1), \phi(\tfrac{1}{d})d]$, such that $\phi(0) = 0$ and $h(\phi(1)) = 0$, with $h$  strictly increasing and (not necessarily strictly) concave, and $\phi$  strictly concave.
	\item We say a function $H: \cP\to \R$ is a  \emph{\typetwo{}  $(h,\phi)$-entropy} if $H = H_{(h,\phi)}$ for some  $\phi: [0,1] \to \R$ which is continuously differentiable on $(0,1]$ and continuous on $[0,1]$ and $h:  [\phi(\tfrac{1}{d})d, \phi(1)] \subset \R \to \R$ which is continuously differentiable on $[\phi(\tfrac{1}{d})d, \phi(1))$ and continuous on $[\phi(\tfrac{1}{d})d, \phi(1)]$, such that $\phi(0) = 0$ and $h(\phi(1)) = 0$, with $h$  strictly decreasing and (not necessarily strictly) convex, and $\phi$  strictly convex.
\end{itemize}
Both classes of $(h,\phi)$-entropies are strictly Schur concave; this follows immediately from the fact that symmetric and strictly convex (resp.~strictly concave) functions are strictly Schur convex (resp.~strictly Schur concave), and that the composition with a strictly increasing function preserves strict Schur convexity and strict Schur concavity. In contrast, composition with a strictly decreasing function swaps strict Schur convexity and strict Schur concavity. Likewise, both classes of $(h,\phi)$-entropies are continuously differentiable on $\cP_+$ and continuous on $\cP$.

\paragraph{Examples of $(h,\phi)$-entropies}
\begin{itemize}
    \item The $\alpha$-R\'enyi entropy for $\alpha <1$, the $\alpha$-Tsallis entropy for $\alpha>0$, the von Neumann entropy (or Shannon entropy in the classical case), the $(s,\alpha)$-unified entropies for $\alpha\in(0,1)$ and $s\leq 1$, and entropies induced by $f$ divergences with strictly convex $f$ are all \typeone{} $(h,\phi)$-entropies. See also \Cref{sec:butterflies} for an example of a different flavor: there it is shown that $p\mapsto \bE_p[K]-1$ constitutes a \typeone{} $(h,\phi)$-entropy, where $K$ denotes the number of distinct realizations of $N$ i.i.d. random variables $X_i \sim p$ with outcomes in $\{1,\dotsc,M\}$.
    \item The $\alpha$-R\'enyi entropy for $\alpha >1$ and the $(s,\alpha)$-unified entropies for $\alpha > 1$ and $s\leq 1$ are  \typetwo{} $(h,\phi)$-entropies.
\end{itemize}
The R\'enyi and Tsallis entropies are discussed in more detail in \Cref{sec:tsallis-renyi}, and the other entropies are discussed in \Cref{sec:more-cty-bounds}.

While both \typeone{} and \typetwo{} $(h,\phi)$-entropies are Schur concave, \typeone{} $(h,\phi)$-entropies are additionally concave, as the composition of a concave increasing function with a concave function. On the other hand, in general \typetwo{} $(h,\phi)$-entropies are neither convex nor concave. In this chapter, we investigate the continuity properties of these two classes of entropies.

Let $H=H_{(h,\phi)}$ be an $(h,\phi)$-entropy (of either type). By \eqref{eq:Gamma-for-symmetric-H}, we have immediately that
\begin{equation}
\Gamma_H(r) = \left. \frac{\d}{\d t}  H_{(h,\phi)}(\mmm_{t}(r))\right|_{t=0} = h'( {\textstyle \sum_i} \phi(r_i) ) (  \phi'(r_-) - \phi'(r_+) ) \label{eq:Gamma_hphi}
\end{equation}
where $r_+$ denotes the largest element of $r$, and $r_-$ the smallest. Note that $r \mapsto r_-$ is Schur concave, while $r \mapsto r_+$ is Schur convex. 

\paragraph{\typeone{} $(h,\phi)$-entropies}
We see that for a \typeone{} $(h,\phi)$-entropy $H \equiv H_{(h,\phi)}$,
\[
r \mapsto (  \phi'(r_-) - \phi'(r_+) )
\]
is Schur convex and strictly positive for $r$ non-uniform, and likewise
\[
r \mapsto h'( {\textstyle \sum_i} \phi(r_i) )
\]
is strictly positive and strictly Schur convex. Thus, $\Gamma_H$ is strictly Schur convex on $\cP_+$, and following the discussion of \Cref{sec:uniform-bounds-majflow}, we immediately obtain the following results.

\begin{theorem} \label{thm:Delta-eps-Schur-convex}
Let $H_{(h,\phi)}$ be a \typeone{} $(h,\phi)$-entropy. Then $\Gamma_H$ is strictly Schur convex on $\cP_+$, and
\[
r \mapsto \sup_{p \in B_\eps(r)} H_{(h,\phi)}(p) - H_{(h,\phi)}(r)
\]
is strictly Schur convex on $\cP$.
\end{theorem}

\begin{corollary}[Tight uniform continuity bounds for \typeone{} $(h,\phi)$-entropies] \label{cor:tight-uniform-bounds-concave-type}
For $\eps > 0$ and any states $p,q\in \cP$  such that $\TV(p,q)\leq \eps$, we have
 \begin{equation} 
 | H_{(h,\phi)}(p) - H_{(h,\phi)}(q) | \leq g(\eps) \label{eq:hphi-uniform-bound} 
 \end{equation}
 where
 \begin{equation} \label{eq:def_g}
 g(\eps) := \begin{cases}
h( \phi(1-\eps) + (d-1) \phi( \frac{\eps}{d-1})) & \eps < 1-\frac{1}{d}\\
h(d\phi(\frac{1}{d})) & \eps \geq 1 - \frac{1}{d}
\end{cases} 
 \end{equation}
and $d$ is the dimension. Moreover, equality in~\eqref{eq:hphi-uniform-bound} occurs if and only if one of the two distributions (say, $q$) is extremal (i.e.\@~a permutation of $(1,0,\dotsc,0)$), and either
\begin{enumerate}
	\item $\eps < 1 - \frac{1}{d}$ and $p = \pi( 1- \eps, \frac{\eps}{d-1},\dotsc, \dotsc, \frac{\eps}{d-1})$ for some permutation $\pi$, or
	\item  $\eps \geq 1- \frac{1}{d}$, and $p = u$ is uniform.
\end{enumerate}\label{thm:hphi-GCB}

\end{corollary}
This provides a tight uniform continuity bound for the Tsallis entropies, the $\alpha$-R\'enyi entropies for $\alpha \in (0,1)$, the Shannon entropy, the $(s,\alpha)$-unified entropies with $\alpha\in (0,1)$ and $s\leq 1$, and any entropy induced by an $f$-divergence or maximal $f$-divergence with strictly convex $f$. See \Cref{sec:more-cty-bounds} for more details and references.

Given an $(h,\phi)$-entropy, we may also consider its smoothed variant,
\begin{equation} \label{eq:def-H-smoothed}
H_{(h,\phi)}^\delta(p) := \max_{q \in B_\delta(p)} H_{(h,\phi)}(q) = H_{(h,\phi)}\circ \mmm_\delta(p)
\end{equation}
for $\delta\in[0,1]$. If $H_{(h,\phi)}$ is \typeone{}, we can simply establish Lipschitz continuity bounds for any $\delta > 0$ by using the Schur concavity of $\Gamma_{H_{(h,\phi)}}$ and \Cref{cor:opt-Lip}.

\begin{corollary} \label{prop:typ1-Lip}
Let $H_{(h,\phi)}^\delta$ be the smoothed variant of a \typeone{} $(h,\phi)$-entropy, for $\delta \in [0, 1]$ (as defined in \eqref{eq:def-H-smoothed}). Then $H_{(h,\phi)}^\delta$ is Lipschitz continuous on $\cP$ if and only if
\begin{equation}\label{eq:opt-Lip-smoothed}
k := \lim_{\eps \to 0} \frac{g(\eps+\delta)-g(\delta)}{\eps} < \infty
\end{equation}
where $g(\eps) := H_{(h,\phi)}(\psi_\eps^*)$ for $\psi_\eps^* = \diag(1-\eps, \frac{\eps}{d-1},\dotsc, \frac{\eps}{d-1})$ is given in \eqref{eq:def_g}. Moreover, if $k$ is finite, then it is the optimal Lipschitz constant for $H_{(h,\phi)}^\delta$. In particular, if $\delta >0$, then
\[
k = g'(\delta) = h'(\phi(1-\delta) + (d-1) \phi(\tfrac{\delta}{d-1}))(\phi'(\tfrac{\delta}{d-1}) - \phi'(1-\delta))
\]
and $H_{(h,\phi)}^\delta$ is Lipschitz continuous.
\end{corollary}
\begin{proof}
If $\delta = 0$, the result follows from \Cref{cor:tight-uniform-bounds-concave-type}. Hence, consider the case $\delta > 0$. For $r\in \cP_+$,
\begin{align*}	
\Gamma_{H_{(h,\phi)}^\delta}(r) &= \lim_{s\downarrow 0} \frac{1}{s}\left(H_{(h,\phi)}^\delta(\mmm_s(r)) - H_{(h,\phi)}^\delta(r)\right)\\
&= \lim_{s\downarrow 0} \frac{1}{s}\left(H_{(h,\phi)}(\mmm_{s+\delta}(r)) - H_{(h,\phi)}(\mmm_\delta(r))\right)\\
&= \lim_{s\downarrow 0} \frac{1}{s}\left(H_{(h,\phi)}(\mmm_{s}(\mmm_\delta(r)) - H_{(h,\phi)}(\mmm_\delta(r))\right)\\
&= \Gamma_{H_{(h,\phi)}}(\mmm_\delta(r)).
\end{align*}
Since $\Gamma_{H_{(h,\phi)}}$ is strictly Schur convex on $\cP_+$ by \Cref{thm:Delta-eps-Schur-convex}, and for any $\delta > 0$, the map $\mmm_\delta: \cP\to \cP_+$ is majorization-preserving, we have that $r \mapsto \Gamma_{H_{(h,\phi)}}(\mmm_\delta(r))$ is strictly Schur convex on $\cP$. Hence, it is maximized at $r = (1,0,\dotsc,0)$ (or any permutation thereof). Invoking \Cref{cor:opt-Lip} completes the proof.
\end{proof}

\paragraph{\typetwo{} $(h,\phi)$-entropies}

For a \typetwo{} $(h,\phi)$-entropy $H \equiv H_{(h,\phi)}$, 
\[
r \mapsto (  \phi'(r_+) - \phi'(r_-) )
\]
is Schur convex and strictly positive for non-uniform $r$, while
\[
r \mapsto -h'( {\textstyle \sum_i} \phi(r_i) )
\]
is strictly Schur concave and strictly positive. Hence, $\Gamma_H$ is the product of a Schur convex and Schur concave function. The former only depends on the largest and smallest entries of $r$, however. For $d=2$, these are all the entries, and for $x\in (0, \frac{1}{2}]$,
\[
\Gamma_H(\left\{ x, 1-x \right\}) = h'(\phi(x) + \phi(1-x)) (\phi'(x) - \phi'(1-x)) = \bin_{(h,\phi)}'(x)
\]
is the derivative of the binary $(h,\phi)$-entropy, where
\[
\bin_{(h,\phi)}(x) := H_{(h,\phi)}(\left\{ x, 1-x \right\}).
\]
For $d > 2$, define
\[
\bar r = (r_1, \underbrace{z, \dotsc, z}_{d-2 \text{ times}}, r_d)
\]
where without loss of generality, $r_1 \geq r_2 \geq \dotsm \geq r_d$ are the sorted elements of $r$ and $z = \frac{1 - r_1 - r_d}{d-2}$.  Note that $\bar r = Mr$ for
\[
M := \begin{pmatrix}
1 &	0				& \dotsm &	0				&	0\\
0 &	\tfrac{1}{d-2}	& \dotsm &	\tfrac{1}{d-2}	&	0\\
0 &	\vdots			&		 &	\vdots			&	0\\
0 &	\tfrac{1}{d-2}	& \dotsm &	\tfrac{1}{d-2}	&	0\\
0 &	0				& \dotsm &	0				&	1	
\end{pmatrix}
\]
which is a doubly-stochastic matrix, and hence the characterization of majorization provided in \eqref{eq:DS-char-maj} yields that $\bar r \prec r$. Since $\bar r$ and $r$ additionally have the same largest and smallest elements, we obtain $\Gamma_H(r) \leq \Gamma_H(\bar r)$. This leads to the following result.

\begin{theorem} \label{thm:Convex-type-Lip}
Let $H_{(h,\phi)}$ be a \typetwo{} $(h,\phi)$-entropy.
If $d = 2$, then
\begin{equation}\label{eq:Convex-type-Lip-d-2}
\sup_{r\in \cP_+}\Gamma_H(r) = \sup_{0 < x \leq \frac{1}{2}} \bin'_{(h,\phi)}(x).
\end{equation}
If $d > 2$, then
\begin{equation} \label{eq:Convex-type-Lip-d-gt-2}
\sup_{r\in \cP_+}\Gamma_H(r) =\sup_{\substack{x, y: \\ 0 < x\leq \frac{1}{d} \leq y\\x \leq z \leq y}} -h'(\phi(y) + (d-2)\phi(z) + \phi(x))(\phi'(y) - \phi'(x)), \quad z := \frac{1-y-x}{d-2}
\end{equation}
In either case, if $\phi$ is differentiable at zero and $h$ is differentiable at $\phi(1)$, then $H_{(h,\phi)}$ is Lipschitz continuous, and the optimal Lipschitz constant is given by $\sup_{r\in \cP_+}\Gamma_H(r)$.
\end{theorem}
\begin{proof}	
The discussion before \Cref{thm:Convex-type-Lip} and \eqref{eq:Gamma_hphi} establishes the expressions for $\sup_{r\in \cP_+}\Gamma_H(r)$, and the proof concludes by \Cref{cor:opt-Lip}.
\end{proof}
\begin{remark}
While \eqref{eq:Convex-type-Lip-d-2} (resp. \eqref{eq:Convex-type-Lip-d-gt-2}) do not provide a closed-form expression for $\sup_{r \in \cP_+} \Gamma_H(r)$, they reduce the naively $d$-dimensional optimization problem to a 1- (resp.~2-) dimensional problem.
\end{remark}

\subsection{An aside on the necessity of the \texorpdfstring{$(h,\phi)$}{(h, phi)} form}\label{sec:symm-concave-cx}
\Cref{thm:Delta-eps-Schur-convex} shows \typeone{} $(h,\phi)$-entropies $H$, which are symmetric and concave, are such that their derivative along the path of majorization flow, $\Gamma_H$, is Schur convex. A natural question is whether or not the restiction to the form of an $(h,\phi)$-entropy is necessary; perhaps all symmetric and concave functions are such that their derivative along the path of majorization flow is Schur convex? This turns out to not be the case, as shown by \Cref{prop:symm-concave-cx} below. First, we establish a simple characterization of the Schur convexity of $\Gamma_f$.
\begin{lemma}
Let $f: \cP \to \R$ be symmetric and concave, and twice differentiable on $\cP_+$. Then $\Gamma_f$ is Schur convex on $\cP_+$ iff for each $x \in \cP^\downarrow$, the Hessian $H(x)$ of $f$ at $x$ satisfies
\begin{equation} \label{eq:Hess-condition}
H(x)_{i,d}+ H(x)_{j,1} - H(x)_{j,d} - H(x)_{i,1} \geq 0
\end{equation}
for each $i < j$.
\end{lemma}
\begin{proof}	
If $x\in \cP^\downarrow$, then $\Gamma_f(x) = (\partial_{x_d} - \partial_{x_1})f(x)$. \Cref{eq:S-convex-condition} then shows that Schur convexity of $\Gamma_f$ implies $(\partial_{x_i} - \partial_{x_j}) \Gamma_f(x) \geq 0$ for $i < j$ and $x\in \cP^\downarrow$.
To recover the reverse implication, one needs to show the assumption \eqref{eq:Hess-condition} yields $(x_i - x_j)(\partial_{x_i} - \partial_{x_j}) \Gamma_f(x) \geq 0$ for all $x\in \cP$ and $i,j$. We can use that if $f$ is symmetric, then $H$ satisfies
\[
 H(Px) = P^T H(x) P
 \]
 for any permutation matrix $P$. Given $x\in \cP$, let $P$ be a permutation matrix such that $x^\downarrow = Px$. Then  applying \eqref{eq:Hess-condition} to $H(x^\downarrow) = P^T H(x) P$ yields the result.
\end{proof}
Next, we use this characterization to establish a counter-example.
\begin{proposition}\label{prop:symm-concave-cx}
There exists a function $f:\cP \to \R$ which is symmetric, concave, and smooth, but is such that $\Gamma_f$ is not Schur convex.
\end{proposition}
\begin{proof}	
Consider $d=4$, $y = (1,2,3,4)$, and $z = (4,3,2,1)$, and define
\begin{equation*}	
f(x) = - \frac{1}{|S_4|} \sum_{\pi \in S_4}\log\left(\e^{\braket{\pi(x), y}} + \e^{\braket{\pi(x), z}}\right)
\end{equation*}
Note that $-f(x)$ is the symmetrization of \[
g(x) = \log\left(\e^{\braket{x, y}} + \e^{\braket{x, z}}\right).
\]
Let $Y$ be a vector-valued random variable which with probability $1/2$ takes on the value $2y$ and with probability $\frac{1}{2}$ takes on the value $2z$. Then $g(x)$ is the cumulant-generating function of $Y$, and hence is convex. Thus, $f(x)$ is concave and symmetric.

Next, we simply check \eqref{eq:Hess-condition} with $i=2$, $j=3$, and $x = (0.5, 0.3, 0.2, 0) \in \cP^\downarrow$. We find\footnote{with some symbolic manipulation by Mathematica; see Appendix~\ref{sec:cx_mathematica} for the code} the left-hand side of \eqref{eq:Hess-condition} is given by
\[
L:= -\frac{N\left(e^{1/5}-1\right)^2 e^{1/5}}{D \left(1+e^{1/5}\right)^2 \left(1+e^{2/5}\right)^2 \left(1-e^{1/5}+e^{2/5}\right)^2 \left(1+e^{4/5}\right)^2   }
\]
where
\begin{align*}
N :\!&= 3-4 \sqrt[5]{e}+13 e^{2/5}-13 e^{3/5}+36 e^{4/5}-32 e+84 e^{6/5}-82 e^{7/5}+157 e^{8/5}-127 e^{9/5}\\
&\qquad +235 e^2-202 e^{11/5}+382 e^{12/5}-290 e^{13/5}+500 e^{14/5}-382 e^3+685 e^{16/5}-513 e^{17/5}\\
&\qquad +817 e^{18/5}-574 e^{19/5}+970 e^4-680 e^{21/5}+1052 e^{22/5}-691 e^{23/5}+1114 e^{24/5}-756 e^5\\
&\qquad +1114 e^{26/5}-691 e^{27/5}+1052 e^{28/5}-680 e^{29/5}+970 e^6-574 e^{31/5}+817 e^{32/5}-513 e^{33/5}\\
&\qquad +685 e^{34/5}-382 e^7+500 e^{36/5}-290 e^{37/5} +382 e^{38/5}-202 e^{39/5}+235 e^8-127 e^{41/5}\\
&\qquad +157 e^{42/5}-82 e^{43/5}+84 e^{44/5}-32 e^9+36 e^{46/5}-13 e^{47/5}+13 e^{48/5}-4 e^{49/5}+3 e^{10}\\
D :\!&= 3\left(1-e^{1/5}+e^{2/5}-e^{3/5}+e^{4/5}-e+e^{6/5}\right)^2\left(1+e^{8/5}\right)^2\left(1-e^{2/5}+e^{4/5}\right)^2
\end{align*}
which yields $L \approx -0.15621966760667033050$ and in particular, $L < 0$. Hence, $\Gamma_f$ is not Schur convex.
\end{proof}
\begin{remark}
The example used in the proof is for dimension $d=4$. One can likely construct examples in any dimension $d \geq 4$, possibly by embedding this example in a larger one.
Numerically, I was not able to find a counterexample for $d=3$ or lower. One difficulty in constructing such a counterexample is that necessarily two of the indices $(1,i,j,d)$ must coincide. Perhaps in dimensions $d=2$ or $d=3$, symmetric, concave, and differentiable functions $f$ yield Schur convex $\Gamma_f$.
\end{remark} 

\section{R\'enyi and Tsallis entropies} \label{sec:tsallis-renyi}
Let us apply the results of \Cref{thm:hphi-GCB} and \Cref{thm:Convex-type-Lip} to two cases of particular interest: R\'enyi and Tsallis entropies. In addition to establishing much sharper continuity bounds for the $\alpha$-R\'enyi entropy for $\alpha >1$, we will also see why previous continuity bounds for the quantity, which were derived from continuity bounds for Tsallis entropies, scale poorly with the dimension.

\paragraph{R\'enyi entropies}
The $\alpha$-R\'enyi entropies $S_\alpha$ are parametrized by $\alpha \in (0,1)\cup (1,\infty)$, and are a generalization of the von Neumann entropy in the sense that $\lim_{\alpha\to 1} S_\alpha = S$. The $\alpha$-R\'enyi entropy has been used to bound the quantum communication complexity of distributed information-theoretic tasks \cite{vDH02}, can be interpreted in terms of the free energy of a quantum or classical system \cite{Baez11}, and is the fundamental quantity defining the entanglement $\alpha$-R\'enyi entropy \cite{WMVF16}.

The $\alpha$-R\'enyi entropy \cite{Ren61}  for $\alpha \in (0,1)\cup(1,\infty)$, of a quantum state $\rho \in \cD(\cH)$ is defined by
\[
S_\alpha(\rho) := \frac{1}{1-\alpha} \log \tr \left(\rho^\alpha\right).
\]
$S_\alpha$ is the $(h,\phi)$-entropy with $h(x) = \frac{1}{1-\alpha}\log x$ for $x\in \RR$ and $\phi(x) = x^\alpha$ for $x\in [0,1]$. For $\alpha\in (0,1)$, $h$ is concave and strictly increasing and $\phi$ is strictly concave. For $\alpha >1$, $h$ is convex and strictly decreasing, and $\phi$ is strictly convex. Hence, $S_\alpha$ is a \typeone{} $(h,\phi)$-entropy for $\alpha \in (0,1)$, and is a \typetwo{} $(h,\phi)$-entropy for $\alpha > 1$. It is known that $\lim_{\alpha \to 1}S_\alpha(\rho) = S(\rho)$, and $\lim_{\alpha\to\infty}S_\alpha(\rho) = S_\infty(\rho) := - \log \lambda_{\max}(\rho)$, where $\lambda_{\max}(\rho)$ denotes the largest eigenvalue of $\rho$. The R\'enyi entropies, like the von Neumann (or Shannon) entropy, are additive under tensor products: $S_\alpha(\rho\otimes \sigma) = S_\alpha(\rho) + S_\alpha(\sigma)$. In thermodynamic contexts, this property is often called \emph{extensivity}.

For a probability distribution $p \in \cP$, the above quantity reduces to
\[
S_\alpha(p) = \frac{1}{1-\alpha} \log \left(\sum_{i=1}^d p_i^\alpha\right),
\]
and $S_\infty(p) = - \log \max_{1\leq i \leq d} p_i$. 

\paragraph{Tsallis entropies}

The $\alpha$-Tsallis entropy \cite{Tsallis1988} for $\alpha\in(0,1)\cup(1,\infty)$ of a quantum state $\rho \in \cD(\cH)$ is defined by
\[
T_\alpha(\rho) := \frac{1}{1-\alpha}[\tr(\rho^\alpha)-1],
\]
and in the case of a probability vector $p \in \cP$,
\[
T_\alpha(p) = \frac{1}{1-\alpha}\left[\sum_{i=1}^d p_i^\alpha - 1\right].
\]
The Tsallis entropy can be seen as a version of the R\'enyi entropy in which the logarithm has been linearized (up to a factor of $\ln 2$), using the first-order Taylor series $\log x \approx \frac{1}{\ln 2} (x - 1)$. The Tsallis entropy is not additive under tensor products (it is \emph{nonextensive}) and instead satisfies the relation
\begin{equation}\label{eq:Talpha_non_subadd}
T_\alpha(\rho\otimes \sigma) = T_\alpha(\rho) + T_\alpha(\sigma) + (1-\alpha)T_\alpha(\rho)T_\alpha(\sigma)
\end{equation}
for $\alpha \in (0,1)\cup(1,\infty)$ and $\rho,\sigma\in \cD(\cH)$, as can be verified by direct computation.

We have that $T_\alpha = H_{(h,\phi)}$ for $h(x) = x$ and $\phi(x) = \frac{x^\alpha-x}{1-\alpha}$ and hence is a \typeone $(h,\phi)$-entropy. 

\paragraph{Previously known continuity bounds for Tsallis entropies}

Raggio \cite[Lemma 2]{Rag95} showed that $T_\alpha$ is Lipschitz continuous for $\alpha>1$:
\begin{equation} \label{eq:Tsallis-Lip}
|T_\alpha(\rho) - T_\alpha(\sigma)| \leq \frac{2\alpha}{\alpha-1} \eps
\end{equation}
if $T(\rho,\sigma) \leq \eps$,
while Zhang \cite[Theorem 1]{Zha07a} proved that if $T(\rho,\sigma)\leq \eps$ and $\alpha>1$, then
\begin{equation}\label{eq:Tsallis-bound}
|T_\alpha(\rho) - T_\alpha(\sigma)| \leq \begin{cases}
 \frac{1}{1-\alpha}(\eps^\alpha (d-1)^{1-\alpha} +  (1-\eps)^\alpha -1)  & \eps < 1 - \frac{1}{d}\\
 \frac{d^{1-\alpha} - 1}{(1-\alpha)} & \eps \geq 1 - \frac{1}{d}
\end{cases}
\end{equation}
using a coupling technique\footnote{In fact, \cite[Theorem 1]{Zha07a} considers the case $T(\rho,\sigma)=\eps$; in \eqref{eq:Tsallis-bound}, their bound has been monotonized to hold for $T(\rho,\sigma)\leq \eps$.}. In fact, \eqref{eq:Tsallis-bound} also holds in the case  $0<\alpha<1$ as was shown by \cite[(A.2)]{Aud07} via a direct optimization method (adapting the proof of \eqref{eq:Audenaert-Fannes_bound}). This bound for all $\alpha\in (0,1)\cup(1,\infty)$ also appears as \cite[Lemma 1.2]{Renyi-CMNF}, whose proof appears to follow the same direct optimization method as Audenaert. Zhang  \cite[Remark 4]{Zha07a} also derived \eqref{eq:Tsallis-Lip} from \eqref{eq:Tsallis-bound}, and \eqref{eq:Audenaert-Fannes_bound} from the limit $\alpha\to 1$ of \eqref{eq:Tsallis-bound}.

 \cite[Theorem 2.4]{FYK07} showed that if $\alpha\in [0,2]$ and $p,r\in \cP$ such that $\TV(p,r)= \eps \leq \alpha^{1/(1-\alpha)}$, then
\[
|T_\alpha(p) - T_\alpha(r)| \leq (2\eps)^\alpha \ln_\alpha(d) + \eta_\alpha(2\eps)
\]
where $\eta_\alpha(x) = - x^\alpha \ln_\alpha(x)$ and $\ln_\alpha(x) = \frac{x^{1-\alpha}-1}{1-\alpha}$. This bound is less tight than \eqref{eq:Tsallis-bound}, however.

\paragraph{Previously known continuity bounds for R\'enyi entropies}

Audenaert \cite[Appendix A]{Aud07} proved a tight uniform continuity bound for the $\alpha$-R\'enyi entropies for $\alpha \in (0,1)$ in 2007, namely
for  $\eps\in [0,1]$ and $\rho,\sigma\in \mathcal{D}$ with $T(\rho,\sigma)\leq \eps$,
	\begin{equation}\label{eq:uniform_Renyi_bound}
 | S_\alpha(\rho) - S_\alpha(\sigma) | \leq \begin{cases}
\frac{1}{1-\alpha} \log ( (1-\eps)^\alpha + (d-1)^{1-\alpha} \eps^\alpha) & \eps < 1-\frac{1}{d}\\
\log d& \eps \geq 1 - \frac{1}{d},
\end{cases} 
 \end{equation}

See \eqref{eq:bound-Rastegin} and \eqref{eq:Chen-renyi-bound} of \Cref{sec:entropic-cty-from-majflow} for the previously known continuity bounds in the case $\alpha > 1$. Note also \cite[Theorem 7]{WH19} provides continuity bounds for the $\alpha$-R\'enyi-entropy, although in the case $\alpha \in (0,1)$ bounds are not optimal (in contrast to \eqref{eq:uniform_Renyi_bound}), and in the case $\alpha>1$, are not as tight as the bounds presented here.

\paragraph{New continuity bounds}

\Cref{thm:hphi-GCB} provides an alternate proof of \eqref{eq:Tsallis-bound} for any $\alpha>0$ and of \eqref{eq:uniform_Renyi_bound} for $\alpha\in(0,1)$ and establishes that in either case for equality to occur, it is necessary and sufficient for one state to be pure, and the other state to have spectrum $\{1-\eps, \frac{\eps}{d-1}, \dotsc, \frac{\eps}{d-1})$ if $\eps < 1-\frac{1}{d}$, or $\{\frac{1}{d},\dotsc,\frac{1}{d}\}$ if $\eps \geq 1 - \frac{1}{d}$.

\begin{proposition} \label{cor:Tsallis-Lipschitz}
The $\alpha$-Tsallis entropies are Lipschitz continuous for $\alpha > 1$, with optimal Lipschitz constant $\frac{\alpha}{\alpha-1}$.
\end{proposition}
\begin{proof}	
\[
\Gamma_{T_\alpha}(r) = \frac{\alpha}{1-\alpha} (r_-^{\alpha-1} -r_+^{\alpha-1}) = \frac{\alpha}{\alpha-1} (r_+^{\alpha-1} -r_-^{\alpha-1}) \leq \frac{\alpha}{\alpha-1}
\]
with equality achieved by $r=  (1,0,\dotsc,0)$.
\end{proof}
\begin{remark}
This improves upon \eqref{eq:Tsallis-Lip} by a factor of 2, but can also be derived directly from \eqref{eq:Tsallis-bound}.
\end{remark}

The following is a corollary of \Cref{thm:Convex-type-Lip}.
\begin{proposition} \label{cor:Renyi-lipschitz}
The $\alpha$-R\'enyi entropy is Lipschitz continuous if and only if  $\alpha> 1$. In the latter case, the
optimal Lipschitz constant $k_\alpha$ satisfies
\begin{equation} \label{eq:Lip-estimate-Renyi}
 \frac{\alpha}{\alpha-1}(d-2)^{1 - 1/\alpha}\frac{1}{2\ln(2)} \leq k_\alpha \leq \frac{d\alpha}{\alpha-1}\frac{1}{\ln(2)}.
\end{equation}
For certain values of $\alpha$, we compute $k_\alpha$ exactly or provide tighter bounds. We have
\begin{align}
k_\infty &= \frac{1}{\ln(2)} d,\\
k_2 &= \begin{cases}
 \frac{2}{\ln(2)} & d = 2\\ 
 \frac{d-2}{\sqrt{d-1}-1}\frac{1}{\ln(2)} & d > 2,
 \end{cases}
\end{align}
and for $\alpha\in(1,2)$,
\[
k_\alpha \leq \frac{\alpha}{\alpha-1} \frac{d^{\alpha-1}}{\ln(2)}< \frac{d\alpha}{\alpha-1}\frac{1}{\ln(2)}.
\]

Additionally, for $\alpha \in (0,1)$ and any $\delta> 0$, the smoothed entropy $S_\alpha^\delta$ is Lipschitz continuous, with optimal Lipschitz constant
\[
\frac{\alpha}{1-\alpha} \frac{1}{\ln(2)} \frac{ (\tfrac{\delta}{d-1})^{\alpha-1} - (1-\delta)^{\alpha-1} }{(1-\delta)^\alpha + (d-1)^{\alpha-1}\delta^\alpha}.
\]
\end{proposition}
\begin{remark}
\cite[Theorem 7 (2)]{WH19} can be used to establish the bound $k_\alpha \leq 2\frac{d\alpha}{\alpha-1}\frac{1}{\ln(2)}$.
\end{remark}
\begin{proof}	
The fact that the $\alpha$-R\'enyi entropies are not Lipshitz for $\alpha \leq 1$ follows from the fact that $g(\eps)$ defined in \eqref{eq:def_g} has $\lim_{\eps \to 0}\frac{g(\eps)}{\eps}=+\infty$.

Let us prove \eqref{eq:Lip-estimate-Renyi}; consider $\alpha > 1$. First, let's prove the upper bound. We have $h(x) = \frac{1}{1-\alpha}\log x$ and $\phi(x) =x^\alpha$. Then for $\alpha > 1$, we have $\phi(x)\geq 0$ and $\phi'(x)\geq 0$ for $x\in [0,1]$. Hence, since $-h'$ is strictly decreasing,
\begin{align*}	
 -h'(\phi(y) + (d-2)\phi(z) + \phi(x))(\phi'(y) - \phi'(x)) &\leq - h'(\phi(y))(\phi'(y) - \phi'(x)) \\
 &\leq - h'(\phi(y))\phi'(y)  \\
 &= - (h\circ \phi)'(y).
\end{align*}
Then
\[
 - (h\circ \phi)'(y) = \frac{\alpha}{\alpha-1}\frac{1}{\ln(2)}\frac{1}{y} \leq \frac{\alpha}{\alpha-1}\frac{1}{\ln(2)} d
\]
since $y \geq \frac{1}{d}$. Next, consider the lower bound.

Let $x=0$, $y = \frac{1}{(d-2)^{1 - 1/\alpha}}=\frac{1}{(d-2)^{(\alpha-1)/\alpha}}$, $z = \frac{1-y}{d-2}$, and let $r = (x,z,\dotsc,z,y) \in \cP^\downarrow$. Then
\begin{align*}	
\Gamma_{S_\alpha}(r) &= \frac{\alpha}{\alpha-1} \frac{1}{\ln(2)}\frac{y^{\alpha-1} - x^{\alpha-1}}{x^\alpha + y^\alpha + (d-2) z^\alpha} \\
&= \frac{\alpha}{\alpha-1} \frac{1}{\ln(2)}\frac{y^{\alpha-1}}{y^\alpha + (d-2) z^\alpha} \\
&= \frac{\alpha}{\alpha-1} \frac{1}{\ln(2)} \frac{y^{\alpha-1}}{y^\alpha + (1-y)^\alpha (d-2)^{1-\alpha}} \\
&= \frac{\alpha}{\alpha-1} \frac{1}{\ln(2)}\frac{1}{y} \frac{1}{1 + (1/y-1)^\alpha (d-2)^{1-\alpha}} \\
&\geq \frac{\alpha}{\alpha-1} \frac{1}{\ln(2)}\frac{1}{y} \frac{1}{1 + y^{-\alpha} (d-2)^{1-\alpha}} \\
&\geq \frac{\alpha}{\alpha-1} \frac{1}{\ln(2)}\frac{1}{y} \frac{1}{1 + 1} \\
&= \frac{\alpha}{\alpha-1} \frac{1}{2\ln(2)}(d-2)^{1 - 1/\alpha}.
\end{align*}
The proof for $\alpha=\infty$ is in the next proposition.
The proof for $\alpha \in(1,2)$ follows from \Cref{cor:Tsallis-Lipschitz} and \eqref{eq:bound-Renyi-by-Tsallis}. Next, for $\alpha < 1$ and $\delta > 0$, we find
\[
\lim_{\eps \to 0}\frac{g(\eps + \delta)}{\eps} = \frac{\alpha}{1-\alpha} \frac{1}{\log_\mathrm{e}(2)} \frac{ (\tfrac{\delta}{d-1})^{\alpha-1} - (1-\delta)^{\alpha-1} }{(1-\delta)^\alpha + (d-1)^{\alpha-1}\delta^\alpha}.
\]
from which the optimal Lipschitz constant follows by \Cref{prop:typ1-Lip}.

Lastly, we prove establish the optimal Lipschitz constant in the case $\alpha=2$.
It suffices to calculate the right-hand side of \eqref{eq:Convex-type-Lip-d-gt-2} in the case that $h(x) = -\log x$ and $\phi(x) = x^2$. We first establish the result in the case $d=2$, and then in the case $d>2$. For $d=2$ and $\alpha\in (1,2]$, we aim to maximize the function 
\[
h(x,y) = \frac{y^{\alpha-1} - x^{\alpha-1}}{x^\alpha + y^\alpha}
\] 
where $y=1-x$. For $\alpha\in (1,2]$, we have the bound
\begin{align*}	
h(x,y) \leq 1 &\iff y^{\alpha-1} - x^{\alpha-1} \leq y^\alpha + x^\alpha\\
&\iff y^{\alpha-1} - y^{\alpha} \leq x^{\alpha-1} + x^\alpha\\
&\iff y^{\alpha-1}(1-y) \leq x^{\alpha-1}(1-x)\\
&\iff y^{\alpha-1}x \leq x^{\alpha-1}y\\
&\iff y^{\alpha-2}\leq x^{\alpha-2}
\end{align*}
using $y=1-x$ and $x=1-y$. We find $h(x,y) \leq 1$ since $x\leq y$ and $\alpha\in (1,2]$. On the other hand, $h(0,1) = 1$ for all $\alpha$. 

For $d>2$, and $\alpha = 2$, we consider
\[
h(x,y) = \frac{y - x}{x^2 + y^2 + (d-2)z^2}
\]
where $z \equiv z(x,y) = \frac{1-x-y}{d-2}$.
Then for $D=x^2 + y^2 + (d-2)z^2$,
\begin{align*}	
D^2 \frac{\d}{\d x} h(x,y) &= -(x^2 + y^2 + (d-2)z^2) - (y-x)(2x - 2z)\\ 
&=-(x^2 + y^2 + (d-2)z^2) +2 (y-x)(z-x)
\end{align*}
so $\frac{\d}{\d x} h(x,y)\leq 0$ iff
\begin{gather*}
2 (y-x)(z-x) \leq x^2 + y^2 + (d-2)z^2\\
2(yz - xz + x^2 - xy)\leq x^2 + y^2 + (d-2)z^2\\
2(yz - xz - xy)\leq -x^2 + y^2 + (d-2)z^2\\
\end{gather*}
We have $2yz \leq y^2 + d^2 \leq y^2 + (d-2)z^2$, so it remains to show $-2(xz + xy) \leq - x^2$. But that follows from
\[
x^2 \leq 4x^2 \leq 2(xz + xy)
\]
using $x\leq y$ and $x\leq z$. Thus, for any $y \geq x$, such that $x\leq z \leq y$, $h(x,y)$ is decreasing in $x$. 

Hence, we consider $x=0$. Then $z = \frac{1-y}{d-2}$, and $y$ is constrained only by $y\in[\frac{1}{d-1}, 1]$. Then
\[
h(0,y) = \frac{y}{y^2 + (d-2)\inv (1-y)^2} = \frac{y (d-2)}{ (d-2)y^2 + 1 - 2y + y^2} = \frac{(d-2)y}{(d-1)y^2 + 1 - 2y}.
\]
We have
\[
((d-1)y^2 + 1 - 2y)^2\partial_y h(0,y) = ((d-1)y^2 + 1 - 2y)(d-2) - (d-2)y ( 2(d-1)y - 2)
\]
so $\partial_y h(0,y) = 0$ if and only if
\begin{gather*}
(d-1)y^2 + 1 - 2y = y ( 2(d-1)y - 2)\\
(d-1)y^2 + 1 - 2y =  2(d-1)y^2 - 2y\\
 1  =  (d-1)y^2 \\
y = \frac{1}{\sqrt{d-1}}.
\end{gather*}
Note that with $x=0$, $y = \frac{1}{\sqrt{d-1}}$, we have $0\leq x\leq z\leq y\leq 1$ so the constraints are satisfied. It could be that this choice of $y$ yields only a local maximum. To rule this case out, since $y \in [ \frac{1}{d-1}, 1]$, we check $h(0, \frac{1}{d-1})  = h(0, 1) = 1$, while
\[
h( 0, (d-1)^{-1/2}) = \frac{d-2}{2\sqrt{d-1}-2} > 1
\]
for $d \geq 3$.
\end{proof}

For $\alpha=\infty$, we can obtain both a tight uniform continuity bound and the optimal Lipschitz continuity constant.
\begin{proposition} \label{prop:H-infinity}
For $\eps > 0$ and $p,q\in \cP_d$ such that $\TV(p,q)\leq \eps$, we have
$|S_\infty(p) - S_\infty(q)| \leq \log(1 + \eps d)$ if $\TV(p,q)\leq \eps$. In particular, $S_\infty$ has an optimal Lipschitz constant of $\frac{d}{\ln(2)}$.
\end{proposition}
\begin{remark}
\cite[Lemma 18]{WH19} provides a simple direct proof to establish the Lipschitz constant $\frac{d}{\ln(2)}$.
\end{remark}
\begin{proof}	
Since $S_\infty(r) = -\log r_+$, for $r\neq u$,  \eqref{eq:Gamma-for-symmetric-H} yields
\[
\Gamma_{S_\infty}(r) = (\partial_--\partial_+) S_\infty(r) = \partial_{+} \log r_+ = \frac{1}{\ln(2)}\frac{1}{r_+},
\]
whereas $\Gamma_{S_\infty}(u)=0$ as $\gen(u)=0$.
The optimal Lipschitz constant follows from the fact that $r_+ \geq \frac{1}{d}$ and for some probability vectors, $r_+ = \frac{1}{d}$. 

Moreover, $r \mapsto \frac{1}{\ln(2)}\frac{1}{r_+}$ is Schur concave (as the composition of a decreasing function and the Schur convex function $r\mapsto r_+$). Hence, $\Gamma_{S_\infty}$ is Schur concave on $\cP\setminus \{u\}$.
Since $\mmm_s$ is majorization-preserving for all $s\in[0,1]$, we have
\[
r\mapsto \Delta_\eps^{S_\infty} := S_\infty(\mmm_\eps(r)) -r = \int_0^\eps \Gamma_{S_\infty}(\mmm_s(r))\d s
\]
is Schur concave on $\cP\setminus B_\eps(u)$. This uses the fact that for $r\in \cP\setminus B_\eps(u)$, $\TV(r,u) > \eps$ and hence $\mmm_s(r)\neq u$ for all $s\in[0,\eps]$.
For any $r\in \cP\setminus B_\eps(u)$, $\mmm_t(r) \prec r$ for $t = \TV(r,u)$. Hence,
\[
\max_{r\in \cP}S_\infty(\mmm_\eps(r)) -r = \max_{r\in B_\eps(u)}S_\infty(\mmm_\eps(r)) -r = \max_{r\in B_\eps(u)} \log (d) - S_\infty(r)
\]
using that $\mmm_\eps(r) = u$ for $r\in B_\eps(u)$. Then
\[
\max_{r\in B_\eps(u)} \log (d) - S_\infty(r) = \max_{r\in B_\eps(u)} \log (d r_+).
\]
For $r\in B_\eps(u)$, $r_+ \leq \frac{1}{d}+\eps$, with equality for $r =  (\frac{1}{d}+\eps, \frac{1}{d},\dotsc, \frac{1}{d},\frac{1}{d}-\eps)$. Hence, putting it all together,
\[
\max_{r\in \cP}S_\infty(\mmm_\eps(r)) -r =  \log \left(d \left(\frac{1}{d}+\eps\right)\right) = \log (  1 + d\eps).\qedhere
\]
\end{proof}
\begin{remark}
The proof of \Cref{prop:H-infinity} shows the Schur concavity of $\Delta_\eps^{S_\infty}$ on $\cP\setminus B_\eps(u)$. This contrasts strongly with the Schur \emph{convexity} of $\Delta_\eps^H$ on $\cP$ for all \typeone{} $(h,\phi)$ entropies $H$ proven in \Cref{thm:Delta-eps-Schur-convex}.
\end{remark}

\subsection{R\'enyi entropy of parameter \texorpdfstring{$\alpha =1$}{alpha = 1} has optimal dimensional scaling in its continuity bound}\label{sec:Renyi_opt_scaling}

Note that the R\'enyi entropy of parameter $\alpha=1$ is the Shannon entropy (or von Neumann entropy, in the quantum case). There is a sense in which the continuity properties as a function of dimension $d$ of the R\'enyi entropy $S_\alpha$ are much improved at $\alpha=1$ compared to $\alpha \neq 1$. Let us introduce some notation. Define
\[
C_\alpha(d,\eps) := \sup_{\substack{p,q\in \cP_d\\ \TV(p,q)\leq \eps}} |S_\alpha(p) - S_\alpha(q)|
\]
as the optimal uniform continuity bound for $S_\alpha$ over $\cP_d$. Consider a sequence $(\eps_d)_{d\in \bN}$ such that $\eps_d \xrightarrow{d\to\infty} 0$. Clearly, if
\begin{equation}\label{eq:Calpha_to_zero}
\limsup_{d\to\infty} C_\alpha(d, \eps_d) =0
\end{equation}
then for any sequences of distributions $(p_d)_{d\in \bN}$ and $(q_d)_{d\in \bN}$ with $p_q,q_d\in \cP_d$ such that $\TV(p_d,q_d) \leq \eps_d$, we have
\[
\limsup_{d\to\infty} |S_\alpha(p_d) - S_\alpha(q_d)| =0.
\]
Thus, any $(\eps_d)_{d\in \bN}$ satisfying \eqref{eq:Calpha_to_zero} provides a dimensionally-aware notion of continuity for $S_\alpha$. Moreover, the slower $\eps_d$ converges to zero, the stronger the statement of continuity provided by \eqref{eq:Calpha_to_zero}. The following proposition therefore demonstrates that the case $\alpha=1$ is the ``most continuous'' in this sense.
\begin{proposition}
For any $s > 0$,
\begin{equation} \label{eq:C1_invpoly_zero}
C_1(d, d^{-s}) \xrightarrow{d\to\infty} 0.
\end{equation}
In fact, if $(\eps_d)_{d\in \bN}$ has $\eps_d\log(d) \to 0$, then $\lim_{d\to\infty}C_1(d, \eps_d) = 0$.
However, for $\alpha \in(0,1)\cup(1,\infty)$,
\begin{equation} \label{eq:Calpha_invpoly_nonzero}
\liminf_{d\to\infty}C_\alpha(d, d^{-\frac{|\alpha-1|}{\alpha}}) > 0.
\end{equation}
and likewise $\liminf_{d\to\infty}C_\infty(d, d\inv) > 0$.
\end{proposition}
\begin{remark}
Note that this result contrasts with the scenario of fixed dimension $d$. The fact that $S_\alpha$ is Lipschitz continuous on $\cP_d$ if and only if $\alpha > 1$ provides a notion in which $S_\alpha$ is ``more continuous'' for $\alpha > 1$ than for $\alpha > 0$. In other words, if $\alpha > 0$, the only constant $k$ satisfying $|S_\alpha(p) - S_\alpha(q) | \leq k \TV(p,q)$ is $k=\infty$, while for $\alpha > 1$, finite $k$ suffices. This notion is not ``dimensionally-aware'', however, in the sense that $d$ is fixed.

Additionally, the parameter $\alpha=1$ for the Tsallis entropy, which again coincides with the Shannon entropy, does not admit optimal scaling of the continuity bound out of the whole family of Tsallis entropies; in fact, \eqref{eq:Tsallis-bound} shows that the optimal bound actually decreases with dimension for $\alpha>1$. This is related to the nonextensivity of the Tsallis entropy; extensivity enforces the scaling
\[
 S_\alpha(p^{\otimes n}) - S_\alpha(q^{\otimes n}) = n(S_\alpha(p) - S_\alpha(q))
\]
and since
\[
\TV(p^{\otimes n}, q^{\otimes n}) \leq n\TV(p,q)
\]
we find that
\[
C_\alpha(d^n, n \eps_d) \geq n C_\alpha (d,\eps_d),
\]
or, for $d = 2^n$,
\[
C_\alpha(d, \eps_d\log(d)) \geq \log(d) C_\alpha(2,\eps_d).
\]
Hence for any extensive entropy, the optimal continuity bound must grow at least logarithmically with dimension (up to the modification $\eps_d \leadsto \log(d) \eps_d$).
\end{remark}
\begin{proof}	
For $0\leq \eps < 1 - \frac{1}{d}$, the Audenaert-Fannes bound \eqref{eq:Audenaert-Fannes_bound} gives
\[
C_1(d, \eps) = h_2(\eps) + \eps\log(d-1)
\]
where $h_2(\eps) = -\eps\log(\eps) - (1-\eps)\log(1-\eps)$ is the binary entropy. Hence, if $\eps_d \log(d) \to 0$, then $\eps_d \to 0$ and
\[
C_1(d, \eps_d) = h_2(\eps_d) + \eps_d \log(d-1) \to 0
\]
as well. Since for any $s>0$ we have $d^{-s} \log(d) \to 0$, we recover \eqref{eq:C1_invpoly_zero}.

Next, let us establish \eqref{eq:Calpha_invpoly_nonzero} for $0<\alpha<1$. In this case, for $0\leq \eps < 1 - \frac{1}{d}$ \eqref{eq:uniform_Renyi_bound} gives
\[
C_\alpha(d, \eps) = \frac{1}{1-\alpha} \log ( (1-\eps)^\alpha + (d-1)^{1-\alpha} \eps^\alpha).
\]
For $\eps_d = d^{-\frac{1-\alpha}{\alpha}}$, we have $(d-1)^{1-\alpha} \eps_d^\alpha \to 1$ while $(1-\eps_d)^\alpha\to 0$, and hence $C_\alpha(d,\eps_d) \to \frac{1}{1-\alpha} > 0$.

For $\alpha > 1$, we do not have an exact expression for $C_\alpha(d,\eps)$. However, since a Lipschitz bound provides a uniform continuity bound, we have
\[
C_\alpha(d, \eps) \geq \eps \sup_{r\in \cP_+} \Gamma_{S_\alpha}(r)
\]
using \Cref{cor:opt-Lip}. Then \Cref{cor:Renyi-lipschitz} shows that $k_\alpha = \sup_{r\in \cP_+} \Gamma_{S_\alpha}(r)$ satisfies
\[
k_\alpha \geq \frac{\alpha}{\alpha-1}(d-2)^{1 - 1/\alpha}\frac{1}{2\ln(2)}.
\]
Hence, if $\eps_d = d^{-\frac{\alpha-1}{\alpha}}$, then
\[
\eps_d(d-2)^{1 - 1/\alpha} = \eps_d (d-2)^{(1-\alpha)/\alpha} \to 1.
\]
Therefore $\liminf_{d\to\infty}  C_\alpha(d, \eps) \geq  \frac{\alpha}{\alpha-1}\frac{1}{2\ln(2)}>0$.
Likewise, for $\alpha=\infty$, the optimal Lipschitz constant is given in \Cref{prop:H-infinity} as $k_\infty = \frac{d}{\ln(2)}$, and hence $\liminf_{d\to\infty}C_\infty(d,d\inv) \geq \frac{1}{\ln(2)}>0$.
\end{proof}

\subsection{Discussion of previous continuity bounds for \texorpdfstring{$S_\alpha$}{the alpha-Renyi entropy} with \texorpdfstring{$\alpha > 1$}{alpha > 1}} \label{sec:discuss-prev-bounds}

As mentioned at the start of the \Cref{sec:entropic-cty-from-majflow}, continuity bounds on the $\alpha$-R\'enyi entropy for $\alpha > 1$ were not known until 2011, and the bounds known until this work have poor scaling $\sim d^{\alpha-1}$. Here, we use majorization flow as a tool to understand why the previous bounds performed poorly. The technique used to establish the previous bounds was to relate the difference in R\'enyi entropy of two distributions to the corresponding difference in Tsallis entropy \cite{Ras2011,Renyi-CMNF}. In the following proposition, we show that even a relaxed version of this pointwise comparison must necessarily yield a bound scaling as $d^{\alpha-1}$. In contrast, we show that the comparison between the \emph{maximum} difference of the two entropies exhibits much better scaling, and in fact can yield the Lipschitz continuity bound of \eqref{eq:Lip-estimate-Renyi}. This can be understood by the fact that the Tsallis entropy is a \typeone{} $(h,\phi)$-entropy, and hence it increases the slowest near the uniform distribution (in the sense that \Cref{thm:Delta-eps-Schur-convex} holds). On the other hand, for $\alpha>1$, the $\alpha$-R\'enyi entropy is a \typetwo{} $(h,\phi)$-entropy, and increases  quickest at a distribution of the form $(x,z,\dotsc,z,y)$ for $x\leq z \leq y$ as shown by \Cref{thm:Convex-type-Lip}, which can be close to uniform.

Recall the definition $\Delta_\eps^H(r) := \max_{q \in B_\eps(r)}(H(q) - H(r))$ for a function $H: \cP_d\to \bR$.
\begin{proposition} \label{prop:compare-Renyi-Tsallis}

Let $\alpha > 1$.
The smallest constant $c$ such that
\begin{equation} \label{eq:bound-Renyi-by-Tsallis-prointwise}
\Delta_\eps^{S_\alpha}(r) \leq c \, \Delta_\eps^{T_\alpha}(r)
\end{equation}
for all $r\in \cP$ and $\eps > 0$ is $c = \frac{d^{\alpha-1}}{\ln(2)}$. However, the smallest constant $\tilde c$ such that
\begin{equation} \label{eq:bound-Renyi-by-Tsallis}
\max_{r \in \cP}\Delta_\eps^{S_\alpha}(r) \leq \tilde c\, \max_{r' \in \cP}\Delta_\eps^{T_\alpha}(r)
\end{equation}
for all $\eps > 0$ satisfies $\tilde c \leq \frac{\alpha d}{\ln(2)}$.
\end{proposition}
\begin{remark}
\eqref{eq:bound-Renyi-by-Tsallis-prointwise} shows that at some points $r\in \cP$ the Tsallis entropy can locally increase much more than the R\'enyi entropy, while \eqref{eq:bound-Renyi-by-Tsallis} shows that their maximal local increases are not so different.
\end{remark}
\begin{proof}	
Since
\[
\Delta_\eps^{S_\alpha}(r) = \int_0^\eps \Gamma_{S_\alpha}(\mmm_\eps(r)) \d s
\]
and
\[
\Delta_\eps^{T_\alpha}(r) = \int_0^\eps \Gamma_{T_\alpha}(\mmm_\eps(r)) \d s
\]
it suffices to bound the ratio $\frac{\Gamma_{S_\alpha}(p)}{\Gamma_{T_\alpha}(p)}$ uniformly in $p\in \cP$ by $c$. On the other hand, for \eqref{eq:bound-Renyi-by-Tsallis} to hold for all $\eps > 0$ and $r\in \cP$, the same ratio must in fact be bounded by $c$.  We have
\[
\Gamma_{S_\alpha}(p) =\frac{1}{\ln(2)} \frac{\alpha}{\alpha-1} \frac{r_+^{\alpha-1} - r_-^{\alpha-1}}{\sum_{i=1}^d p_i^\alpha}, \qquad \Gamma_{T_\alpha}(p) = \frac{\alpha}{\alpha-1} (r_+^{\alpha-1} - r_-^{\alpha-1})
\]
and hence
\[
\frac{\Gamma_{S_\alpha}(p)}{\Gamma_{T_\alpha}(p)} =\frac{1}{\ln(2)} \left( \sum_{i=1}^d p_i^\alpha \right)^{-1}.
\]
Since $p\mapsto \sum_{i=1}^d p_i^\alpha$ is Schur convex, the above ratio is Schur concave, and hence maximized at the uniform distribution. Thus,
\[
\max_{p\in \cP}\frac{\Gamma_{S_\alpha}(p)}{\Gamma_{T_\alpha}(p)} =\frac{1}{\ln(2)} \left( d \left(\frac{1}{d}\right)^\alpha\right)^{-1} =\frac{1}{\ln(2)} d^{\alpha-1}.
\]
To estimate $\tilde c$, we simply rewrite the uniform continuity bound for $T_\alpha$ given in \Cref{eq:hphi-uniform-bound} as:
\[
 \max_{r' \in \cP} \Delta_\eps^{T_\alpha}(r') = \frac{1}{\alpha-1} ( 1 - (1-\eps)^\alpha - (d-1)^{1-\alpha}\eps^\alpha ),
\]
noting that the maximum is achieved at $r' = (1,0,\dotsc,0)$. We have $(1-\eps)^\alpha  \leq 1-\eps$ and hence $1-(1-\eps)^\alpha \geq \eps$. Then
\[
 \max_{r' \in \cP} \Delta_\eps^{T_\alpha}(r') \geq  \frac{ \eps}{\alpha-1}.
\]
On the other hand, \eqref{eq:Lip-estimate-Renyi} yields
\[
\max_{r \in \cP}\Delta_\eps^{S_\alpha}(r)  \leq \eps\frac{\alpha}{\alpha-1}\frac{d}{\ln(2)}\qedhere
\]
\end{proof}

\Cref{prop:compare-Renyi-Tsallis} helps show that the  notion of majorization flow can provide additional insight into the entropy landscape with respect to the TV distance, beyond the entropic bounds it can establish.

\subsection{Connection to thermodynamics} \label{sec:connect-thermo}
\cite{Baez11} introduced an interesting connection between the $\alpha$-R\'enyi entropy and free energies in thermodynamics. In this section, we recall this relationship, and remark on the resulting consequences of our continuity bounds for R\'enyi entropies of Gibbs states and their connection to changes in free energy.

The following holds in either a quantum or classical picture. We will work in quantum notation for consistency with \cite{Baez11}. Consider a Gibbs state
\[
\rho(T) = Z(T)\inv\e^{-H/T}
\]
where $H$ is the Hamiltonian, $T \geq 0$ the temperature, $Z(T) = \tr(\e^{-H/T})$ is the partition function, and we have set Boltzmann's constant $k_\text{B}\equiv 1$. We can define the free energy as
\[
F(T) = - T \ln Z(T).
\]
By direct calculation, we find that the $\alpha$-R\'enyi entropy $S_\alpha$ satisfies
\begin{equation} \label{eq:Renyi-entropy-as-free-energies}
    S_{\frac{T_0}{T}}(\rho(T_0)) = -\frac{F(T)- F(T_0)}{T-T_0}.
\end{equation}
for any $T > 0$ \cite[Equation (9)]{Baez11}. In the limit $T\to T_0$, we recover the thermodynamic relation
\[
 S(\rho(T_0)) = - \left.\frac{\d F}{\d T}\right|_{T_0}
 \]
 that the entropy is the derivative of the free energy with respect to temperature. Note that any full-rank state $\sigma$ can be seen as a Gibbs state at temperature $T$ associated to the Hamiltonian $H = -\frac{1}{T}\log \sigma$. This gives a physical interpretation to $S_\alpha(\sigma)$ for any full-rank state $\sigma$: consider $\sigma$ to be the Gibbs state at initial temperature $T_0=1$. Then $S_\alpha(\sigma)$ equals the negative of the ratio of the change in free energy to the change in temperature when the temperature is changed from $T_0$ to $\alpha\inv T_0$. This can be seen as the maximum amount of work the system, initially in thermal equilibrium at temperature $T_0$, can do when its temperature is suddenly changed from $T_0$ to $\alpha\inv T_0$ as it moves to the new thermal equilibrium, divided by the change in temperature \cite{Baez11}.

The Schur concavity of $S_\alpha$ for all $\alpha > 0$ can be interpreted through this physical picture as well. The relation $\rho \prec \sigma$ means that the distribution of the eigenvalues of $\rho$ is ``flatter'' and ``more disordered'' than those of $\sigma$; correspondingly, $H_\rho := - \frac{1}{T}\log \rho$ log-majorizes $H_\sigma := - \frac{1}{T}\log \sigma$, where log majorization is defined by $A \prec_{\log} B$ if $\log A \prec \log B$. By Schur concavity, if $\rho \prec \sigma$, then $S_\alpha(\rho)\geq S_\alpha(\sigma)$ for any $\alpha > 0$. Hence, the Schur concavity of the $\alpha$-R\'enyi entropy can be interpreted as a statement about how the distribution of energy levels of a Hamiltonian relates to the free-energy increase or decrease of the system (per unit change in temperature) under a sudden change in temperature.

Now, recall that the trace distance is endowed with an operational interpretation in terms of distinguishability under measurement. We say that $\rho$ and $\sigma$ are $\eps$-indistinguishable if $\frac{1}{2}\|\rho-\sigma\|_1 \leq \eps$. Consider an experiment in which the system is in a state $\sigma$ which is not known precisely, but is $\eps$-indistinguishable from a known state $\rho$, which is a Gibbs state, $\rho = \rho(T_0)$. We pose the following question:

\smallskip
If the temperature is abruptly changed from $T_0\to T$, can one bound the ratio of the change in free energy and the corresponding change in temperature?
\smallskip

This quantity is exactly the $\alpha$-R\'enyi entropy of order $\alpha = T_0/T$ by \eqref{eq:Renyi-entropy-as-free-energies}, and hence \eqref{eq:uniform_Renyi_bound} and \Cref{cor:Renyi-lipschitz} provide an answer in the affirmative. Moreover, \Cref{cor:Renyi-lipschitz} shows that the $\alpha$-R\'enyi entropy is Lipschitz continuous if and only if $\alpha > 1$. In other words, if $T_0 > T$, then there exists $k_{T_0/T} < \infty$ such that
\[
|S_{T_0/T}(\sigma) - S_{T_0/T}(\rho(T_0))| \leq \eps k_{T_0/T}.
\]
If $T_0 \leq T$, then no such linear bound can hold uniformly in $\sigma$ and $\rho(T_0)$, but \eqref{eq:hphi-uniform-bound} gives a tight uniform (nonlinear) bound. 
 
\section{Other entropies} \label{sec:more-cty-bounds}

\begin{itemize}
	\item The von Neumann entropy
	\[
	 S(\rho) = -\tr (\rho \log \rho).
	 \] 
	 satisfies $S = S_{(h,\phi)}$ for $h(x) = x$ and $\phi(x) = -x \log x$, and is a \typeone{} $(h,\phi)$-entropy. The Audenaert-Fannes bound, \eqref{eq:Audenaert-Fannes_bound}, provides a tight uniform continuity bound for the von Neumann (or Shannon) entropy.

\textbf{Our contribution} \Cref{thm:hphi-GCB} provides an alternate proof for \eqref{eq:Audenaert-Fannes_bound} and the necessary and sufficient conditions for equality as a consequence of the fact that the von Neumann entropy is a \typeone{} $(h,\phi)$-entropy, and \Cref{prop:typ1-Lip} shows that $S$ is not Lipschitz continuous with respect to the trace distance, but its smoothed variant $S^\delta$ is Lipschitz continuous for any $\delta \in (0,1)$, with an optimal Lipschitz constant of
\begin{equation}\label{eq:smoothed-entropy-Lipschitz}
\log ( \delta\inv - 1) + \log(d-1). 
\end{equation}

\item The $(s,\alpha)$-unified entropies,
	\[
	E_\alpha^s(\rho) = \frac{1}{s(1-\alpha)} (\tr[\rho^\alpha]^s - 1)
	\]
	for $\alpha\in (0,1)\cup (1,\infty)$, $s\in \mathbb{R}\setminus\{0\}$, were introduced in the quantum case by \cite{HY06} and in the classical case by \cite{RT91}. This family of entropies includes the Tsallis entropies in the case $s=1$, and the $\alpha$-R\'enyi entropies (up to a factor of $\ln(2)$) in the limit $s\to 0$. We have $E_\alpha^s = H_{(h,\phi)}$ for $\phi(x) = x^\alpha$, and $h(x) = \frac{1}{s(1-\alpha)}(x^s - 1)$, which satisfy $\phi(0)=0$, and $h(\phi(1))=0$. If $\alpha\in (0,1)$, $h$ is strictly increasing and $\phi$ is strictly concave, while if $\alpha >1$, $h$ is strictly decreasing, and $\phi$ is strictly convex. Additionally, $h$ is convex if ($s>1$ and $\alpha <1$) or if ($s<1$ and $\alpha >1$), and is concave otherwise. Thus, if $0<\alpha < 1$ and $s\leq 1$, $E_\alpha^s$ is a \typeone{} $(h,\phi)$-entropy, and if $\alpha > 1$ with $s\leq 1$, then $E_\alpha^s$ is a \typetwo{} $(h,\phi)$-entropy. 
	If $s>1$, then for any $\alpha\in (0,1)\cup(1,\infty)$,  $E_\alpha^s$ is an $(h,\phi)$-entropy in the sense defined by \cite{SMMP93}, but not of \typeone{} or \typetwo{}, and hence the results of \Cref{sec:entropic-cty-from-majflow} do not apply in that case.
	\begin{remark}
	 \cite[Proposition 5]{HY06} incorrectly claims that the unified entropies are not Schur concave. However, they are indeed strictly Schur concave for all $\alpha\in (0,1)\cup (1,\infty)$ and $s\in \mathbb{R}\setminus\{0\}$.
	\end{remark}
	Rastegin \cite{Ras2011} showed that for $0<\alpha<1$, $s\in (-\infty,-1]\cup[0,1]$, and $T(\rho,\sigma) \leq \frac{1}{2}\alpha^{\frac{1}{1-\alpha}}$, the bound
	\begin{equation}
	|E_\alpha^s(\rho) - E_\alpha^s(\sigma)| \leq (2 \eps)^\alpha \ln_\alpha d + n_\alpha(2\eps)
	\end{equation}
	holds, where $\ln_\alpha = \frac{x^{1-\alpha} - 1}{1-\alpha}$ and $n_\alpha(x) = \frac{x^\alpha - x }{1-\alpha}$.
    If $\alpha > 1$ and $s\in[-1,0]\cup [1,+\infty]$,
	\begin{equation}
	|E_\alpha^s(\rho) - E_\alpha^s(\sigma)| \leq \chi_s [ \eps^\alpha \ln_\alpha(d-1) + t_\alpha(\eps) ]
	\end{equation}
	where $t_\alpha(\eps) := T_\alpha(\{\eps, 1-\eps\})$ is the binary Tsallis entropy. In \cite[Proposition 6]{HY06}, the Lipschitz continuity bound
\begin{equation}
|E_\alpha^s(\rho) - E_\alpha^s(\sigma)| \leq \frac{\alpha}{\alpha-1} \eps
\end{equation}
for any $\rho,\sigma\in \mathcal{D}(\cH)$ with $T(\rho,\sigma)\leq \eps$ for $\alpha > 1$, and $s\geq 1$ was proven.

\textbf{Our contribution:}

For $\alpha\in (0,1)$ and $s\leq 1$, then $E_\alpha^s$ is not Lipschitz continuous on $\cD(\cH)$ by \Cref{prop:typ1-Lip}, but satisfies the following tight uniform continuity bound by \Cref{thm:hphi-GCB}: If $\eps\in [0,1]$ and $\rho,\sigma\in \mathcal{D}$ with $T(\rho,\sigma)\leq \eps$,
\begin{equation}
|E_\alpha^s(\rho) - E_\alpha^s(\sigma)| \leq \begin{cases}\frac{1}{s(1-\alpha)}
\left[\big( (1-\eps)^\alpha + (d-1)^{1-\alpha}\eps^\alpha \big)^s - 1\right] & \eps < 1-\frac{1}{d}\\
\frac{1}{s(1-\alpha)}\left[d^{s(1-\alpha)}-1\right] & \eps \geq 1 - \frac{1}{d}
\end{cases}
\end{equation}
 with equality if and only if one state is pure, and the other state has spectrum $\{1-\eps, \frac{\eps}{d-1}, \dotsc, \frac{\eps}{d-1}\}$ if $\eps < 1-\frac{1}{d}$, or $\{\frac{1}{d},\dotsc,\frac{1}{d}\}$ if $\eps \geq 1 - \frac{1}{d}$.

If $\alpha>1$ and $s\leq 1$, then $E_\alpha^s$ is Lipschitz continuous on $\cD(\cH)$ by \Cref{thm:Convex-type-Lip}, and the associated optimal Lipschitz constant $k_{\alpha}^s$ satisfies
\begin{equation} \label{eq:unified-Lipschitz-alpha_gt_1_s_lt_1}
k_\alpha^s \leq \begin{cases}
\frac{\alpha}{\alpha-1}d^{1-\alpha s} &s \alpha < 1\\
\frac{\alpha}{\alpha-1} & s\alpha \geq 1.
\end{cases}
\end{equation}

\item Entropies induced by divergences. Denoting left multiplication by an operator $A$ as $L_A$, and right multiplication by $A$ as $R_A$, one defines the $f$-divergence
\begin{equation}
S_f(\rho\|\sigma) := \tr[ \sigma^{1/2} f(L_\rho R_{\sigma\inv}) (\sigma^{1/2})]
\end{equation}
which was first introduced by Petz \cite{Pet85} (see \cite{HM17} for a useful overview). The \emph{maximal $f$-divergence} \cite{PR98} is given by
\begin{equation}
\hat S_f(\rho\|\sigma) := \tr[\sigma^{1/2} f(\sigma^{-1/2}\rho \sigma^{-1/2}) \sigma^{1/2}].
\end{equation}

From either divergence, one can define an associated entropy by evaluating at $\sigma=\one$ (and reversing the sign). The two entropies coincide, yielding
\begin{equation}
S_f(\rho) := - S_f(\rho\|\one) = - \hat S_f(\rho\||\one) = - \tr [ f(\rho)].
\end{equation}
For strictly convex $f$ with $f(0)=f(1)=0$ we can define $\phi = -f$ and $h(x) = x$, yielding a \typeone{} $(h,\phi)$-entropy.

\textbf{Our contribution:}
For strictly convex $f$ with $f(0)=f(1)=0$, \Cref{thm:hphi-GCB} gives that for $\eps\in [0,1]$ and $\rho,\sigma\in \cD(\cH)$ with $T(\rho,\sigma)\leq \eps$,
\begin{equation} \label{eq:unif-bound-f-div}
 | S_f(\rho) - S_f(\sigma) | \leq \begin{cases}
 -f(1-\eps) - (d-1) f( \frac{\eps}{d-1}) & \eps < 1-\frac{1}{d}\\
-d f(\frac{1}{d}) & \eps \geq 1 - \frac{1}{d}.
\end{cases} 
\end{equation}
\Cref{prop:typ1-Lip} shows that $S_f$ is Lipschitz continuous on $\cD(\cH)$ if and only if
\begin{equation} \label{eq:Sf-Lipschitz}
k := \lim_{\eps\to 0}-\frac{1}{\eps}f(1-\eps) - (d-1) \frac{1}{\eps}f( \frac{\eps}{d-1})
\end{equation}
is finite. In the latter case, $k$ is the optimal Lipschitz constant for $S_f$. Note that if $f$ is differentiable at $0$ and $1$, then $k = f'(1) - f'(0)$.
\item The \emph{concurrence} of a bipartite pure state $\psi_{AB}$ is an entanglement monotone defined as
\[
C(\psi_{AB}) = \sqrt{2(1 - \tr[\rho_{A}^2])}
\]
where $\rho^\psi = \tr_B[\psi_{AB}]$ is the reduced state on system $A$ \cite{Woo01,RBC+01}. Regarded as a function of the reduced state, the concurrence can be seen as \typeone{} $(h,\phi)$-entropy with $\phi(x) = -x^2$ and $h(x) = \sqrt{2(1+x)}$, and hence \Cref{thm:hphi-GCB} gives a tight uniform continuity bound in terms of the trace distance between the reduced states. If for some $\eps>0$ two bipartite pure states $\psi_{AB}$ and $\phi_{AB}$ satisfy $T(\psi_{AB},\phi_{AB})\leq \eps$, then by monotonicity of the trace distance under partial trace, $T(\rho^\psi, \rho^\phi) \leq \eps$ as well. Hence, \Cref{thm:hphi-GCB} yields
\begin{equation}\label{eq:concurrence_GCB}
 |C(\psi_{AB}) - C(\phi_{AB})| \leq \begin{cases}
 \sqrt{2( 1 - (1-\eps)^2 - (d-1)^{-1}\eps^2 )} & \eps < 1 - \frac{1}{d}\\
\sqrt{2(1-d\inv)} & \eps \geq 1 - \frac{1}{d}
 \end{cases}
\end{equation}
for all bipartite pure states $\psi_{AB}$ and $\phi_{AB}$ such that $T(\psi_{AB},\phi_{AB}) \leq \eps$. The concurrence is not Lipschitz continuous, by \Cref{prop:typ1-Lip}.
\end{itemize}

\section{Number of distinct realizations from $N$ i.i.d.~random variables} \label{sec:butterflies}
In this section, we consider another application of the majorization flow to continuity bounds.

Consider an experiment in which outcome $i \in \{1,\dotsc,M\}$ is observed with probability $p_i$, for some probability distribution $p\in \cP_M$. Repeat this experiment $N$ times, independently, and consider the random variable $K$ which denotes the number of distinct outcomes observed. In the following, we establish the Lipschitz continuity of each entry of the cumulative probability distribution (c.d.f.) of $K$, as a function of $p$, a tight uniform continuity bound for $\bE[K]$, and the optimal Lipschitz constant for $\bE[K]$, as functions of $p$. These results quickly follow from the results of \Cref{chap:majflow_ctybounds} and the expressions for the c.d.f.~of $K$ and  $\bE[K]$ given in \cite{WY73}. These bounds provide a notion of robustness in order to understand how much $K$ can vary given some uncertainty in $p$.

This experiment was introduced in the excellent review \cite{Arn07} as describing an experiment on an island in which there is some unknown (but fixed) number $M$ of butterfly species, and butterflies are sequentially trapped until $N$ individuals have been caught. In this context, $K$ is the number of distinct species of butterflies observed, and $p$ is an underlying ``catchability'' distribution which gives the probability $p_j>0$ that a butterfly of specifies $j$ will be caught on any given trial\footnote{The independence of the trials corresponds to an assumption that there is a large number of individuals of each species such that trapping one does not influence the probability of which species will be caught on the next trial.}. In this scenario, the continuity bounds for $K$ provide a notion of robustness by describing how much the number of distinct observed species can vary given some change in the catchability distribution.

In \cite{WY73} it was shown that each entry of the cumulative distribution of $K$, namely
\[
F_j \equiv F_j(p) := \Pr[ K \leq j],
\]
is a Schur convex function of $p$ (Theorem 4.1 of the above article), and has the expression
\[
F_j = \sum_{i=1}^j (-1)^{j-i}{M - i - 1 \choose j - i} \sum_{1 \leq l_1 < l_2 < \dotsm < l_i \leq M} (p_{l_1} + \dotsm + p_{l_i})^N
\]
which is given in \cite[Corollary 3.2]{WY73}. Taking the derivative, we obtain
\[
\frac{\partial F_j}{\partial p_1} - \frac{\partial F_j}{\partial p_2} = f_j(p_1) - f_j(p_2)
\]
for
\[
f_j(s)  =  N \sum_{i=1}^j (-1)^{j-i}{M - i - 1 \choose j - i} \sum_{\substack{1 < l_2 < \dotsm < l_i \leq M\\ l_2,\dotsc,l_i \neq 1,2}} (s + p_{l_2} + \dotsm + p_{l_i})^{N-1}
\]
as was calculated in \cite[Equation 11]{WY73}. To compute the optimal Lipschitz constant for $F_j$, it remains to maximize this difference over $p\in \cP_M$. We leave that for future work, and simply show that $F_j$ is a Lipschitz continuous function of $p\in \cP_M$ by showing that $f_j(p_1) - f_j(p_2) < \infty$ for any $p\in \cP_M$. We can use that the summand $(s + p_{l_2} + \dotsm + p_{l_i})$ is less than $1$ for $s\in \{p_1, p_2\}$, and that the number of elements in the second summation is ${M-2 \choose i}$ to find the simple bound
\[
|f_j(s)| \leq N \sum_{i=1}^j {M - i - 1 \choose j - i} {M-2 \choose i} < \infty
\]
for $s\in \{p_1, p_2\}$, which completes the proof.

Next, \cite[Corollary 3.3]{WY73} shows that the expected number of distinct elements, $\bE[K]$ satisfies
\[
\bE[K] \equiv \bE_p[K] = M- \sum_{i=1}^M (1 - p_i)^N.
\]
In fact, we can identify $\bE[K]-1$ as a \typeone{} $(h,\phi)$-entropy, with $h(x) = x-1$ and $\phi(x) = 1 - (1-x)^N$, as defined in \Cref{sec:entropic-cty-from-majflow}. Hence, \Cref{cor:tight-uniform-bounds-concave-type} shows that for $\eps >0$ if $p,q \in \cP_M$ satisfy $\TV(p,q) \leq \eps$, then
\[
|\bE_p[K] - \bE_q[K]| \leq \begin{cases}
\frac{(M-1)^N - (M-1-\varepsilon)^N}{(M-1)^{N-1}} - \varepsilon^N & \varepsilon \leq 1 - \frac{1}{M} \\ 
\frac{M^N - (M-1)^N}{M^{N-1}}-1 & \eps > 1 - \frac{1}{M}.
\end{cases}
\]
In particular, using \Cref{prop:typ1-Lip},
\begin{equation}\label{eq:trials-Lip}
|\bE_p[K] - \bE_q[K]| \leq \varepsilon N,
\end{equation}
and $N$ is the optimal Lipschitz constant. Note that \eqref{eq:trials-Lip} does not depend on $M$, but its derivation assumes $M < \infty$. 
\section{Number of connected components of a random graph}\label{sec:random-graph}

In this section, we examine the continuity of a quantity with a fairly different flavor: the expected number of connected components $E_C(p)$ of a particular random graph model which is parametrized by a probability distribution $p\in \cP$. This quantity has the two necessary features to apply the tools of \Cref{sec:uniform-bounds-majflow}: it is a Schur concave function of a probability vector, and it has a closed-form expression which we can differentiate. Moreover, the bounds we obtain are quite sharp: we obtain the optimal Lipschitz constant, which scales as $\sqrt{n-2}$ for $n$-dimensional distributions, up to an additive factor of at most five and a multiplicative factor of at most $\sqrt{2}$, for all $n\geq 5$.

We consider the following random graph $G$ with $n \in \mathbb{N}$ nodes (or vertices), which was constructed in \cite{Ros81}. Fix a probability distribution $p$ on $\{1,\dotsc,n\}$, and take $n$ independent and identically distributed (i.i.d.) random variables $X_1,\dotsc,X_n \sim p$ such that $\Pr[X_i = j] = p_j$ for all $i, j \in \{1,2, \ldots, n\}$. Then construct $n$ edges by connecting $i$ to $X_i$. The result is a graph with $n$ nodes and edges, such that every node has at least one edge.

\begin{definition}\label{compt}
A connected component of a graph $G$ is a subgraph $H$ such that for every pair of nodes $x,y \in H$, there is a path (made up of contiguous edges) between $x$ and $y$, and moreover there are no edges between nodes in $H$ and $G\setminus H$ (in other words, no edges leave the connected component).
\end{definition}

\begin{figure}[ht]
\centering
\includegraphics[width=.8\textwidth]{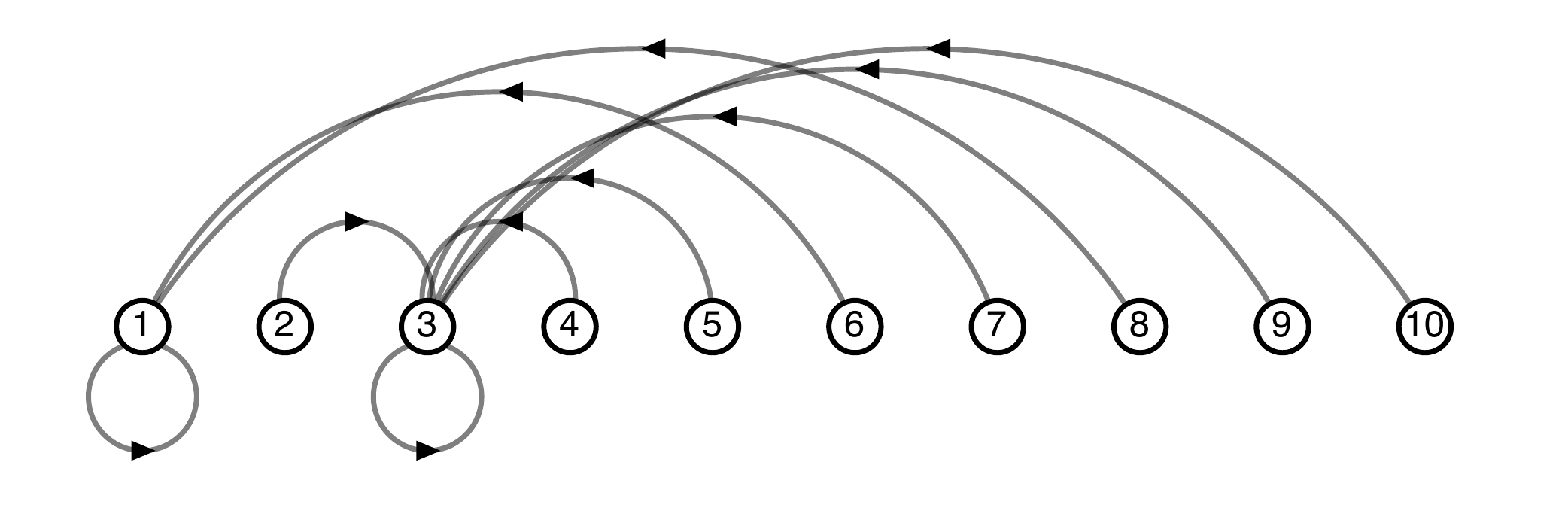}
\caption{\label{fig:onethird-twothirds} A realization of the random graph $G$ associated to the distribution $p = (1/3, 0, 2/3, 0,\dotsc, 0)$. A set of edges $\{(i, X_i): i = 1,\dotsc, 10\}$ is constructed by independently sampling each $X_i \sim p$. The edges shown here are directed, but for the purpose of calculating the number of connected components, we consider the associated undirected graph.}
\end{figure}

\begin{figure}[ht]
\centering
\includegraphics[width=.8\textwidth]{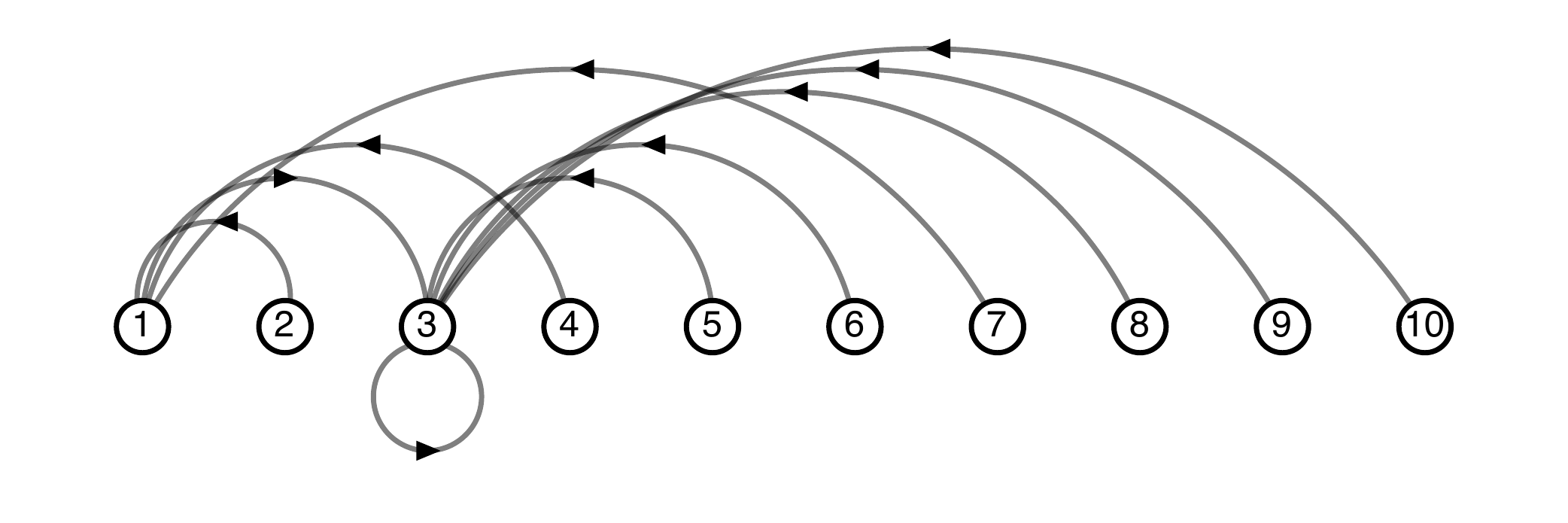}
\caption{\label{fig:onethird-twothirds-1c} Another realization of the graph $G$ from \Cref{fig:onethird-twothirds}, with underlying probability distribution $p = (1/3, 0, 2/3, 0,\dotsc, 0)$. This realization has only one connected component.}
\end{figure}

As a simple example, if $p = (1,0,\dotsc,0)$, then $1$ has a single self-edge, and every other node has a single edge touching it, which is connected to $1$, and hence the graph has one connected component. As another example, if $p = (1/3, 0, 2/3, 0,\dotsc, 0)$, then one realization of the random graph is shown in \Cref{fig:onethird-twothirds}, and another shown in \Cref{fig:onethird-twothirds-1c}.

\begin{figure}[ht]
\centering
\includegraphics[width=.8\textwidth]{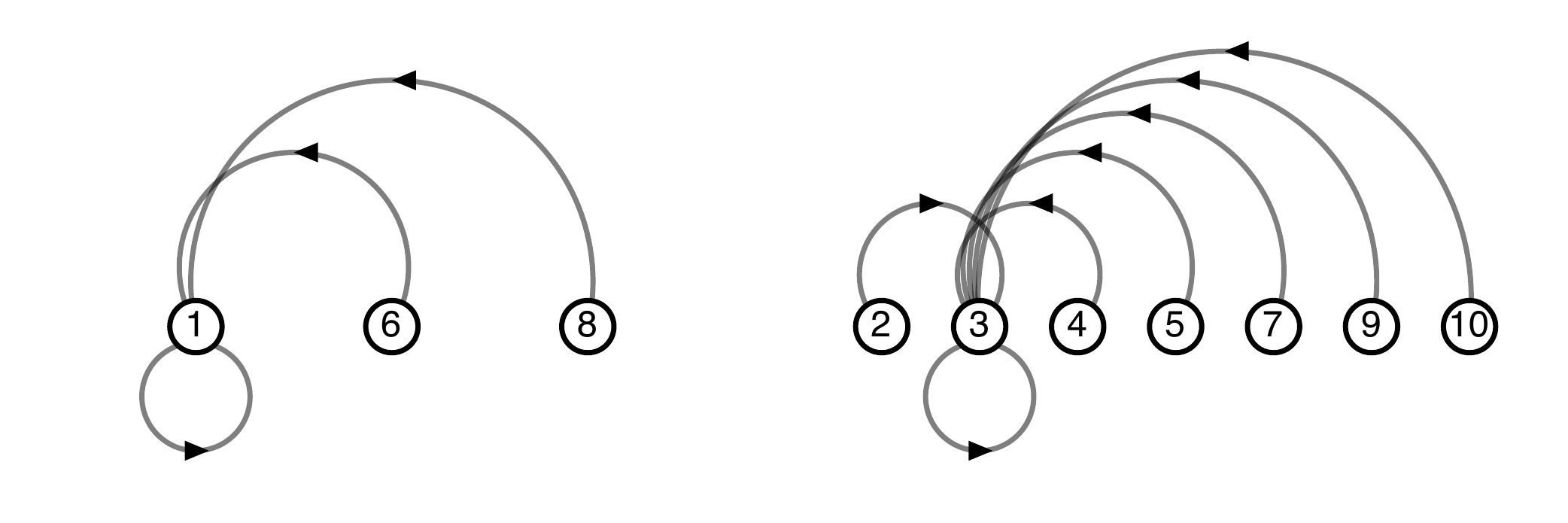}
\caption{\label{fig:onethird-twothirds-components} The two connected components of the same realization of the graph $G$ from \Cref{fig:onethird-twothirds}. These consist of elements which connect to $\{1\}$, and elements which connect to $\{3\}$, respectively. Note that we neglect directionality in computing connected components.}
\end{figure}

The number of connected components $C$ of $G$, the random graph constructed above, is a random variable. Its expectation satisfies
\begin{equation}
\bE[C] = \sum_{\substack{S\subset \{1,\dotsc, n\}\\ S \text{ nonempty}}} (|S| - 1)! \prod_{j\in S}p_j \label{eq:expected_num_connected_components}
\end{equation}
by \cite[Equation 3]{Ros81}. To make the dependence on $p$ explicit, let us write 
\begin{equation}\label{eq:def_EC}
E_C(p) \equiv \bE[C].
\end{equation}
Note that $E_C(p)$ is a polynomial in the entries of $p$, and hence is smooth. On the compact set $\cP_n$, $E_C$ is thus a Lipschitz function. In the following, we establish upper and lower bounds on the optimal Lipchitz constant.
These bounds involve the Mills ratio,
\begin{equation}\label{eq:def_Mills_ratio}
    M(x) := \frac{1-\Phi(x)}{\phi(x)}
\end{equation}
where $\phi$ and $\Phi$ are the probability mass function and cumulative density function, respectively, of the standard normal distribution.

\noindent
\begin{theorem} \label{thm:random-graph-cty-bound}
Let $E_C(p)$ and $E_C(q)$ denote the expected number of connected components of the random graph of $n \geq 3$ nodes, corresponding to probability distributions $p$ and $q$, respectively. Let $f(x) := x - x^2 M(x)$, and $\mu$ be its maximum value on the domain $x\geq 0$. Then the following bound holds:
\begin{equation}\label{eq:Lip-upper-bound}
|E_C(p) - E_C(q)| \leq  (3 + \mu \sqrt{n-2}) \TV(p, q).
    \end{equation}

Moreover, setting $x_0$ to be the unique maximizer of $f$ on $x\geq 0$, we have that any Lipschitz constant $\kappa$ satisfies
\begin{equation}\label{eq:Lip-lower-bound}
\kappa \geq \frac{\mu \sqrt{n-2} }{\sqrt{2}} - \frac{\mu x_0}{2} - \sqrt{\frac{2}{n-2}} x_0  -  \frac{x_0^2}{n-2}- \sqrt{n-2}\e^{-(n-2)}\frac{x_0\e^{x_0^2/2}}{\sqrt{2}}.
\end{equation}
Additionally, $x_0$ and $\mu$ satisfy the following explicit bounds\footnote{These bounds can be refined to arbitrary precision as shown in \cite{AHD19_code}.}:
\begin{align}\label{eq:bound-x0}
1.1615278892744612 &\leq x_0 \leq 1.1615278892744958,\\
0.346813047097384 &\leq \mu \leq 0.346813047097549. \label{eq:bound-mu}
\end{align}
\end{theorem}
\begin{remark}
For $n\geq 5$, the lower order terms in \eqref{eq:Lip-lower-bound} satisfy
\[
- \frac{\mu x_0}{2} - \sqrt{\frac{2}{n-2}} x_0  -  \frac{x_0^2}{n-2}- \sqrt{n-2}\e^{-(n-2)}\frac{x_0\e^{x_0^2/2}}{\sqrt{2}} \geq -2
\]
yielding the simpler lower bound $\kappa \geq \frac{\mu \sqrt{n-2} }{\sqrt{2}} - 2$ for $n\geq 5$.
\end{remark}

\paragraph{Connection to models of epidemics}

This kind of random graph model has been used for modelling epidemics; in fact, $G$ corresponds to the ``two-sided epidemic process'' of \cite{Ger77}. We can interpret the continuity bound given in \Cref{thm:random-graph-cty-bound} in this framework as follows. Let the nodes of the graph represent people (or, e.g.\@, groups of people), and let the edges represent pairwise interactions between people which can spread the infection. We can see the probability distribution $p$ governing the placement of edges as being associated to a model of interactions. Now, what is the minimum number of people who need to be infected initially in order for the whole population (all $n$ nodes) to become infected eventually? The answer is simply the number of connected components of the graph. This is because if one person per connected component is infected, that person can then spread the infection to the rest of the people in the connected component. On the other hand, any connected component lacking an infected person will never become infected.

In this interpretation, \eqref{eq:expected_num_connected_components} provides a formula for computing $E_C(p)$, the expected minimal number of initially infected people required to infect the whole population. However, if interactions are better modelled by some $p' \neq p$, then $E_C(p')$ provides a better estimate of the minimal number of people required to infect the whole population. \Cref{thm:random-graph-cty-bound} then provides a quantitative measure of the \emph{robustness} of $E_C$ by establishing a bound on the error $|E_C(p) - E_C(p')|$ as a function of the model error given by the total variation distance between $p$ and $p'$. There are, of course, many other measures of robustness that may be useful in this setting which are not considered here; e.g.\@~of the average against fluctuations, of the framework of the model itself, and so forth.

\paragraph{Overview of the remainder of the section}
The upper bound \eqref{eq:Lip-upper-bound} is proven in \Cref{sec:upper-bound}, and the lower bound \eqref{eq:Lip-lower-bound} is proven in \Cref{sec:lower-bound}. The uniqueness of the maximizer of $f(x) = x - x^2 M(x)$ and the bounds on the maximizer $x_0$ and the maximum value $\mu$ are established using interval arithmetic, as described in \Cref{sec:x0mu}.

\subsection{Proof of the upper bound \eqref{eq:Lip-upper-bound}}\label{sec:upper-bound}

In this section, we use \Cref{cor:opt-Lip} to establish \Cref{thm:random-graph-cty-bound}. Note that by \eqref{eq:expected_num_connected_components}, $E_C : \cP_n \to \R$ is a polynomial in the components of the probability vector $p$ and in particular is continuously differentiable.
In  \cite[Proposition 1]{Ros81}, the author proves that $p \mapsto E_C(p)$ is Schur concave using the criterion \eqref{eq:S-convex-condition}, by showing that
\[
\partial_{p_i}E_C(p) - \partial_{p_j} E_C(p) = (p_j - p_i) \sum_{S^*} (|S^*|+1)! \prod_{j\in S^*} p_j
\]
where $S^*$ ranges over nonempty sets of $\{1,\dotsc,n\}\setminus\{i,j\}$. Hence,
\[
 \Gamma_{E_C}(r) =  (r_+ - r_-) \sum_{S^*} (|S^*|+1)! \prod_{j\in S^*} r_j
\] 
where $S^*$ ranges over nonempty sets of $I :=\{1,\dotsc,n\}\setminus\{i_+,i_-\}$, where $i_\pm$ are indices such that $r_{i_\pm} = r_{\pm}$. We can use  the criterion \eqref{eq:S-convex-condition} again by repeating the proof of \cite[Proposition 1]{Ros81} to show that for
\[
S(\{r_i\}_{i\in I}) :=  \sum_{S^*} (|S^*|+1)! \prod_{j\in S^*} r_j,
\]
we have
\[
\partial_{r_i}S(\{r_i\}_{i\in I}) - \partial_{r_j} S(\{r_i\}_{i\in I}) =  (r_j - r_i) \sum_{S^*} (|S^*|+3)! \prod_{j\in S^*} r_j
\]
and hence $S$ is Schur concave on the set $\left\{ p \in \R^{n-2} : p_i \geq 0, \sum_i p_i = 1-r_+ - r_i \right\}$. For such $p$,
\[
S(p) \leq S\left( \left\{ \frac{1-r_- - r_+}{n-2} \right\}_{i \in I}\right)
\]
and thus
\begin{equation} \label{eq:EC-bound-1}
\Gamma_{E_C}(r) \leq (r_+ - r_-)\sum_{k=1}^{n-2} {n-2 \choose k} (k+1)! (1-r_- - r_+)^{k} (n-2)^{-k}.
\end{equation}
To obtain a Lipschitz bound on $E_C(r)$, it suffices to bound $\Gamma_{E_C}(r)$ independently of $r\in \cP$. We upper bound \eqref{eq:EC-bound-1} by taking $r_-=0$. For the simplicity of notation, let $s = r_+$ and $m = n-2$. Then we aim to bound
\begin{equation} \label{eq:def_Bns}
B_m(s) := s \sum_{k=1}^m{m \choose k} (k+1)! (1-s)^{k}m^{-k}
\end{equation}
for $s\in \left[\frac{1}{m+2}, 1\right]$, using that $r_+ \in \left[\frac{1}{n},1\right]$ which follows from $r \in \cP$. 
Let
$$
S_m(s) := \sum_{k=1}^m c_{k,m} (1-s)^{k-1}
$$
with
$$
c_{k,m} := {m \choose k} \frac{(k+1)!}{m^k} = (k+1)\prod_{j=1}^{k-1} \left(1-\frac{j}{m}\right),
$$
then 
\begin{equation}\label{eq:Bm-Sm-relationship}
B_m(s) = s(1-s) S_m(s).
\end{equation}

Applying the inequality $1-x\le\exp(-x)$ to every factor in $c_{k,m}$ gives the simple upper bound
\begin{align*}
c_{k,m}
&\leq  (k+1) \prod_{j=1}^{k-1} \exp(-j/m) \\
&=(k+1)\exp\left(-\sum_{j=1}^{k-1} j/m\right)
=(k+1)\exp\left(-\frac{(k-1)k}{2m}\right) \\
&\le (k+1)\exp\left(-\frac{(k-1)^2}{2m}\right).
\end{align*}

As $c_{k,m}\ge0$, we can also use the same inequality $1-s\le \exp(-s)$ in the formula for $S_m(s)$.
This gives as a first upper bound:
\begin{align*}
S_m(s) &\leq \sum_{k=1}^m  (k+1)\exp\left(-\frac{(k-1)^2}{2m} - (k-1)s\right) \\
&= \sum_{l=0}^{m-1}  (l+2)\exp\left(-\frac{l^2}{2m} - ls\right) \\
&= 2+\sum_{l=1}^{m-1}  (l+2)\exp\left(-\frac{l^2}{2m} - ls\right).
\end{align*}

We can interpret this sum as a lower Riemann sum for a certain Riemann integral.
Noting that the factor $l+2$ increases with $l$ and the factor $\exp\left(-\frac{l^2}{2m} -ls\right)$ decreases, we have
$$
(l+2)\exp\left(-\frac{l^2}{2m} -l s\right) 
\le \int_{l-1}^l (u+3) \exp\left(-\frac{u^2}{2m}-u s \right) \;\d u.
$$
Therefore,
\begin{align*}
S_m(s)&=2+\sum_{l=1}^{m-1}  (l+2)\exp\left(-\frac{l^2}{2m}-ls \right)\\
&\le 2+\sum_{l=1}^{m-1} \int_{l-1}^l (u+3) \exp\left(-\frac{u^2}{2m} -us\right) \;\d u \\
&= 2+\int_{0}^{m-1} (u+3) \exp\left(-\frac{u^2}{2m}-us \right) \;\d u \\
&\le 2+\int_{0}^{\infty} (u+3) \exp\left(-\frac{u^2}{2m}-us \right) \;\d u \\
&= 2+\exp(ms^2/2)  \int_{0}^{\infty} (u+3) \exp\left(-\frac{(u+ms)^2}{2m} \right) \;\d u.
\end{align*}
In terms of the probability density function $\phi(x)$ of the standard normal distribution, $\phi(x) = \exp(-x^2/2)/\sqrt{2\pi}$,
and making the substitution $v=(u+ms)/\sqrt{m}$,
this last expression can be written as
\begin{multline*}
\frac{1}{\phi(\sqrt{m}s)} \int_0^\infty (u+3) \phi\left(\frac{u+ms}{\sqrt{m}}\right)\;\d u\\
\begin{aligned}
&= \frac{\sqrt{m}}{\phi(\sqrt{m}s)} \int_{\sqrt{m}s}^\infty (\sqrt{m}v-ms+3) \phi(v)\;\d v \\
&= \frac{\sqrt{m}}{\phi(\sqrt{m}s)} \left(\sqrt{m} \int_{\sqrt{m}s}^\infty v \phi(v)\;\d v + (3-ms) \int_{\sqrt{m}s}^\infty \phi(v)\;\d v\right).
\end{aligned}
\end{multline*}
Exploiting the fact that $x\phi(x)=-\phi'(x)$, and with $\Phi(x)$ the cumulative density function of the standard normal distribution,
this last expression is equal to
$$
\frac{\sqrt{m}}{\phi(\sqrt{m}\;s)} \left(\sqrt{m}\; \phi(\sqrt{m}\;s) + (3-mx) (1-\Phi(\sqrt{m}\;s))\right)
=m+\sqrt{m}(3-ms)\frac{1-\Phi(\sqrt{m}\;s)}{\phi(\sqrt{m}\;s)},
$$
so that
$$
S_m(s) \le 2+m+\sqrt{m}(3-ms)\frac{1-\Phi(\sqrt{m}\;s)}{\phi(\sqrt{m}\;s)}.
$$
The function in the last factor,
$$M(x):=\frac{1-\Phi(x)}{\phi(x)},$$ is known as the Mills ratio, and several bounds are known for it.
A well-known upper bound valid for $x>0$ is $M(x) < 1/x$ \cite{Gor41,YC15}, which follows from the fact that $M'(x) = x M(x) - 1$ and that $M$ is a strictly decreasing function.
Therefore, 
$$
3\sqrt{m} M(\sqrt{m}\;s)\le 3/s,
$$
and
$$
S_m(s) \le 2+m+\frac{3}{s}-m^{3/2}s M(\sqrt{m}\;s).
$$
Setting $\mu$ to be as in \Cref{thm:random-graph-cty-bound}, we have
$$
-M(x)\le\frac{\mu-x}{x^2}.
$$
Therefore,
$$
S_m(s) \le 2+m+\frac{3}{s}+\sqrt{m}\frac{\mu-\sqrt{m}\;s}{s} = 2+\frac{3+\mu\sqrt{m}}{s},
$$
and
$$
B_m(s) \le (1-s)(2s+3+\mu\sqrt{m})=2(1-s)(1+s)+(1-s)(1+\mu\sqrt{m})\le 2+(1+\mu\sqrt{m}),
$$
over the interval $0\le s\le 1$.  \qed

\bigskip
Explicit numerical calculations of $B_m(s)$ for $m$ up to $10^6$ suggest that the maximal value of $B_m(s)$ is bounded below 
by $\mu\sqrt{m}$ and, hence, lies within a constant not exceeding 3 of our bound, which is remarkable. In the following, we prove a slightly weaker bound, which recovers the square-root scaling at leading order.

\subsection{Proof of the lower bound \eqref{eq:Lip-lower-bound}}\label{sec:lower-bound}
Let $r = \left(r_+, \frac{1 - r_+}{n-2}, \dotsc, \frac{1 - r_+}{n-2}, 0\right) \in \cP$ for some $r_+ \in \left[\frac{1}{n-1}, 1\right]$, so that $r$ is a probability distribution with largest element $r_+$. Then the start of \Cref{sec:upper-bound} establishes that
\[
\Gamma_{E_C}(r) = B_m(s)
\]
where $B_m(s)$ is defined in \eqref{eq:def_Bns}, and $s := r_+$, and $m := n-2$. By \Cref{cor:opt-Lip}, it remains to lower bound $B_m(s)$ for some $s \in \left[ \frac{1}{m+1}, 1 \right]$. As in \eqref{eq:Bm-Sm-relationship}, we write
\begin{equation}\label{eq:lb-decompose-B}
B_m(s) = s \sum_{k=1}^m c_{k, m} (1-s)^{k}, \qquad c_{k,m} := (k+1)\prod_{j=1}^{k-1}\left(1-\frac{j}{m}\right).
\end{equation}
Then
\[
\ln \frac{c_{k,m}}{k+1} = \sum_{j=1}^{k-1}\ln \left( 1 - \frac{j}{m} \right) = \sum_{j=0}^{k-1}\ln \left( 1 - \frac{j}{m} \right) \geq \int_0^k \ln \left( 1 - \frac{u}{m} \right) \d u
\]
using that since $j \mapsto \ln \left( 1 - \frac{j}{m} \right)$ is decreasing, the integral forms an underapproximation to the sum. By changing variables to $v = u/m$, we obtain
\[
\ln \frac{c_{k,m}}{k+1}\geq m \int_0^{k/m} \ln(1-v)\d v =  -k  - (m-k) \ln\left( 1 - \frac{k}{m} \right) \geq - \frac{k^2}{m}
\]
using that $\ln\left( 1 - \frac{k}{m} \right) \leq - \frac{k}{m}$. Hence,
\begin{equation}\label{eq:c_km-lower-bound}
c_{k,m} \geq (k+1) \exp\left( - \frac{k^2}{m}\right).
\end{equation}
From \eqref{eq:lb-decompose-B}, defining $c_{0,m} = 1$, we have
\begin{align*}
\frac{1}{s}B_m(s) &= \sum_{k=1}^{m} c_{k,m}(1-s)^k = \sum_{k=0}^{m} c_{k,m}(1-s)^k - 1\\
&\geq \sum_{k=0}^{m}(k+1) \exp\left( - \frac{k^2}{m} + k \ln(1-s) \right) - 1\\
&\geq \int_0^{m+1} u \exp \left(  - \frac{u^2}{m} + u \ln(1-s) \right) \d u - 1.
\end{align*}
using \eqref{eq:c_km-lower-bound} for the first inequality. For the second inequality, notice that the sum is of the form $\sum_{k=0}^m f(k+1) g(k)$ where $f(k)=k$ is monotone increasing, and $g(k) = \exp\left( - \frac{k^2}{m} + k \ln(1-s) \right)$ is monotone decreasing. Hence, we have $f(k+1)\geq \int_k^{k+1} f(u) \d u = \|\left.f\right|_{[k,k+1]}\|_1$ and $g(k) = \sup_{k \leq u \leq k+1} g(u) = \|\left.g\right|_{[k,k+1]}\|_\infty$, using that both functions are non-negative. H\"older's inequality gives
\[
\int_k^{k+1} f(u) g(u)\d u \leq  \|\left.f\right|_{[k,k+1]}\|_1\, \|\left.g\right|_{[k,k+1]}\|_\infty \leq f(k+1)g(k)
\]
and summing over $k$ yields the inequality. Next, since
\[
 - \frac{u^2}{m} + u \ln(1-s) = - \frac{1}{m} \left( \left( u - \frac{m \ln(1-s)}{2} \right)^2 - \left( \frac{m \ln(1-s)}{2} \right)^2\right),
\]
we obtain
\begin{align*}
\frac{1}{s}B_m(s) &\geq \frac{1}{\exp\left( - \frac{1}{2} \left(\sqrt{\frac{m}{2}}\frac{\ln(1-s)}{2}\right)^2\right)}\int_0^{m+1}u  \exp \left( - \frac{1}{2} \left( \sqrt{\frac{2}{m}}  \left(u - \frac{m}{2} \ln(1-s) \right)\right)^2\right) -1\\
&= \frac{1}{\phi(b)} \int_0^{m+1} u\phi(au -b) \d u-1
\end{align*}
for $a = \sqrt{\frac{2}{m}}$, $ b = \sqrt{\frac{m}{2}}\ln(1-s)$, and $\phi(x) := \frac{1}{\sqrt{2\pi}}\e^{-\frac{x^2}{2}}$ is the p.d.f.~of a standard normal distribution. Changing variables to $v = au - b$, we find
\begin{align*}
\frac{1}{s}B_m(s) &\geq  \frac{1}{\phi(b)a^2} \left[\int_{-b}^{a(m+1)-b} v \phi(v) \d v + b \int_{-b}^{a(m+1)-b} \phi(v) \d v\right]\\
&=  \frac{1}{\phi(b)a^2} \left[ \phi(-b) - \phi( a(m+1) - b) + b( \Phi(a(m+1)-b) - \Phi(-b))\right] - 1
\end{align*}
where $\Phi$ is the c.d.f~of the standard normal distribution. Since $a(m+1)-b \leq \sqrt{2m}$ and $\phi(x)$ is decreasing on $x > 0$, we have $\phi( a(m+1) - b) \leq \phi(\sqrt{2m})$. Using also that $\Phi( a(m+1)-b) \leq 1$, we obtain
\begin{equation*}
\frac{1}{s}B_m(s) \geq - \frac{\phi(\sqrt{2m})}{a^2 \phi(-b)} + \frac{1}{a^2} \left( 1 + b M(-b)\right) - 1 
\end{equation*}
where $M(x) = \frac{1 - \phi(x)}{\Phi(x)}$ is the Mills ratio. Substituting in $a$, we have
\begin{equation*}
\frac{1}{s}B_m(s) \geq - \frac{m\phi(\sqrt{2m})}{2\phi(-b)} + \frac{m}{2} \left( 1 + b M(-b)\right) - 1 .
\end{equation*}
Recalling the definition of $x_0$ and $\mu$ from \Cref{thm:random-graph-cty-bound}, we choose $s = 1 - \e^{- \sqrt{\frac{2}{m}}x_0}$ so that $b = - x_0$, and $\mu = x_0 - x_0^2 M(x_0)$. Substituting for $\mu$, we have
\[
B_m(s) \geq s\left(\frac{m\mu}{2x_0} - 1- \frac{m\phi(\sqrt{2m})}{2\phi(x_0)} \right) .
\]
Using the bound $\e^{-x} \leq 1 - x + \frac{x^2}{2}$ for $x\geq 0$, we have $s \geq \sqrt{\frac{2}{m}} x_0 - \frac{x_0^2}{m}$. Hence,
\[
B_m(s) \geq \frac{\mu \sqrt{m} }{\sqrt{2}} - \frac{\mu x_0}{2} - \sqrt{\frac{2}{m}} x_0  -  \frac{x_0^2}{m}- \sqrt{m}\e^{-m}\frac{x_0}{2\sqrt{\pi}\phi(x_0)}. \tag*{\qed}
\]

\subsection{Maximizing \texorpdfstring{$x - x^2 M(x)$}{x - x*x*M(x)}} \label{sec:x0mu}

The function $f(x) := x - x^2 M(x)$ on the domain $x \geq 0$ has a maximum value $\mu$ satisfying \eqref{eq:bound-mu} which occurs at a unique point $x_0$ satisfying \eqref{eq:bound-x0}. To prove this, we will use the tools of \emph{interval arithmetic}. Interval arithmetic is a method for rigorous calculation using finite-precision floating point numbers on a computer, as follows. Instead of considering a real number $x\in \R$, which may not be exactly representable with a particular finite precision arithmetic, a small interval $[a,b] \subseteq \R$ containing $x$  whose endpoints are exactly representable is used instead. Then to estimate e.g.\@ $f(x)$, an interval $[c,d] \subseteq \R$ is found such that $f(y) \in [c,d]$ for any $y \in [a,b]$. This yields rigorous bounds on $f(x)$ which are not subject to the ``roundoff error'' of usual floating point arithmetic. In addition, we will use the \emph{interval Newton's method}, a powerful extension of the iterative root-finding method which provides rigorous bounds on the zeros of a differentiable function and gives a sufficient condition to guarantee the function has a unique zero in a given interval \cite[Ch.~5]{Tuc11}.

First note that $f(0) = 0$ and $f(x) > 0$ using the simple upper bound $M(x) < \frac{1}{x}$ for $x>0$. Hence, any maximum of $f$ cannot occur at zero. Next, $f$ is smooth, with first derivative
\[
f'(x) = 1 + x^2 - x(2+x^2)M(x)
\]
and second derivative
\[
f''(x) =  x^3 + 4x - M(x)(2+5x^2+x^4).
\]
By using the interval Newton's method as implemented in the Julia  programming language \cite{BEKS17} package \texttt{IntervalRootFinding.jl} \cite{IntervalRootFinding.jl}, we can verify that for $x\in [0.0, 3.0]$, the equation $f'(x)=0$ has a unique solution $x_0$ which satisfies \eqref{eq:bound-x0}.
Moreover, bounding $f$ on the interval given by \eqref{eq:bound-x0} with interval arithmetic, as implemented in  \texttt{IntervalArithmetic.jl} \cite{IntervalArithmetic.jl} shows that $\mu :=f(x_0)$ satisfies  \eqref{eq:bound-mu}.
Lastly, we likewise find that
\[
-0.16730889431005824 \leq f''(x_0) \leq -0.16730889430876594,
\]
and hence $f''(x_0) < 0$ confirming that $x_0$ is a local maxima of $f$. The code used to establish these bounds can be found here: \cite{AHD19_code}. This code uses the MPFR library \cite{FHL+07} for a correctly-rounded implementation of the complementary error function, $1-\Phi(x)$.

Lastly, for $x > 0$, we use the lower bound $M(x) > \frac{x}{x^2+1}$ which holds for $x>0$ \cite{YC15}. This bound yields $f(x) < \frac{x}{1+x^2}$. The right-hand side is strictly monotone decreasing for $x > 1$, and evaluates to $0.3$ at $x=3$. Hence $f(x) < 0.3$ for $x > 3$. Thus, the local maximum at $x_0$ is in fact a global maximum.  

\section{Smoothed entropies and MaxEnt} \label{sec:smoothed-entropies}

\Cref{thm:max-min-ball} immediately yields maximizers and minimizers over $\Be(\rho)$ for any Schur concave function $\varphi$. Note that, as stated in the following corollary, the minimizer $\rho\majmin$ (resp.~the maximizer $\rho\majmax$) in the majorization order (\ref{eq:min-max-maj-cond}) is the maximizer (resp.~minimizer) of $\varphi$ over the $\eps$-ball $\Be(\rho)$.

\begin{corollary}\label{cor:max-min-ball-Schurconcave}
    Let $\varphi: \cP \to \R$ be Schur concave. Then for $p\in \cP$,
    \[
        p\majmin \in \argmax_{\Be(p)}\varphi, \qquad \text{and} \qquad p\majmax \in \argmin_{\Be(p)}\varphi.
    \]
    Additionally, if $\varphi$ is strictly Schur concave, any other state $p' \in \argmax_{\Be(p)} \varphi$ (resp.~ $p'\in \argmin_{\Be(p)}\varphi$) is unitarily equivalent to $p\majmin$ (resp.~$p\majmax$). If $\varphi$ is strictly concave, then $\argmax_{\Be(p)} \varphi = \{p\majmin\}$.

    This result can be equivalently stated in terms of quantum states by subsituting $\rho\in \cD(\cH)$ for $p$ in the above if $\varphi : \cD(\cH) \to \mathbb{R}$ is Schur concave.
\end{corollary}

In an experimental setup, one may not know a quantum state exactly, and need to determine an estimate of the state for mathematical analysis. Let us briefly investigate \Cref{cor:max-min-ball-Schurconcave}'s implications for this task. The so-called MaxEnt (or maximum-entropy) principle gives that an appropriate estimate of $\sigma$ is one which is compatible with the constraints on $\sigma$ and which has maximum entropy subject to those constraints \cite{BDDAW99,Jaynes57}.

Consider an experimental device which attempts to produce a given target pure state $\sigma_\text{target}$, and let $\sigma$ denote the actual state produced by the device.  One can estimate the fidelity between $\sigma$ and $\sigma_\text{target}$ efficiently, using few Pauli measurements of $\sigma$ \cite{FL11,dSLCP11}. Using the bound $\frac{1}{2}\|\sigma - \sigma_\text{target}\|_1 \leq \eps:=\sqrt{1 - F(\sigma,\sigma_\text{target})^2}$, one therefore obtains a bound on the trace distance between $\sigma$ and $\sigma_\text{target}$. \Cref{thm:max-min-ball} gives that the state with maximum entropy in $\Be(\sigma_\text{target})$ is $\rho\majmin(\sigma_\text{target})$. Using that $\sigma_\text{target}$ is a pure state, \Cref{prop:properties_of_Lambda_eps_quantum} yields
\begin{equation} \label{eq:def_rho_eps_sigma_target}
    (\sigma_\text{target})^*_\eps = \begin{cases}
        \diag(1-\eps, \frac{\eps}{d-1},\dotsc,\frac{\eps}{d-1}) & \text{ if } \eps \leq 1 - \frac{1}{d} \\
        \tau:=\frac{\one}{d}                                    & \text{else}
    \end{cases}
\end{equation}
in the basis in which $\sigma_\text{target} = \diag(1,0,\dotsc,0)$, where $d$ is the dimension of the underlying Hilbert space. The maximum-entropy principle therefore implies that $(\sigma_\text{target})^*_\eps$ defined by \eqref{eq:def_rho_eps_sigma_target} is the appropriate estimate of $\sigma$, when given only the condition that ${\frac{1}{2}\|\sigma - \sigma_\text{target}\|_1 \leq \eps}$.
\begin{remarks}
    ~\begin{itemize}
        \item it may be possible to determine additional constraints on $\sigma$ by the Pauli measurements performed to estimate $F(\sigma,\sigma_\text{target})$. In that case, MaxEnt gives that the appropriate estimate of $\sigma$ is the state with maximum entropy subject to these constraints as well, and not only the relation $\frac{1}{2}\|\sigma - \sigma_\text{target}\|_1 \leq \eps$.
        \item it may be possible to devise a measurement scheme to estimate $\frac{1}{2}\|\rho-\sigma_\text{target}\|_1$ directly, which could be more efficient than first estimating the fidelity and then employing the Fuchs-van de Graaf inequality to bound the trace distance.
    \end{itemize}
\end{remarks}

Moreover, \Cref{cor:max-min-ball-Schurconcave} yields maximizers and minimizers of any Schur concave function, not simply the von Neumann entropy. This allows computation of (trace-ball) ``smoothed'' Schur-concave functions. Given $\varphi: \cD\to \R$, we define
\begin{align}
    \bar\varphi^{(\epsilon)}(\sigma) & := \max_{\omega\in \Be(\rho)} \varphi(\sigma), \qquad \ubar \varphi^{(\epsilon)}(\sigma) := \min_{\omega\in \Be(\rho)} \varphi(\sigma).
    \label{eq-x}
\end{align}
By \Cref{cor:max-min-ball-Schurconcave}, we therefore obtain explicit formulas: $\bar \varphi^{(\epsilon)}(\sigma) = \varphi(\rho\majmin)$, and $\ubar \varphi^{(\epsilon)}(\sigma) = \varphi(\rho\majmax)$. In particular, this provides an exact version of Theorem 1 of \cite{Skorski16}, which formulates approximate maximizers for the smoothed $\alpha$-R\'enyi entropy $\bar S_\alpha^{(\eps)}$.

Note that by setting $\varphi = S_\text{min}$ or $S_\text{max}$, the min- and max-entropies, in (\ref{eq-x}), yields explicit expressions for the min- and max- entropies smoothed over the $\eps$-ball. These choices are of particular interest, due to their relevance in one-shot information theory (see e.g.\@~\cite{Renner2005a,Tom16} and references therein). In particular, let us briefly consider one-shot classical data compression. Given a source (random variable) $X$, one wishes to encode output from $X$ using codewords of a fixed length $\log m$, such that the original message may be recovered with probability of error at most $\eps$. It is known that the minimal value of $m$ at fixed $\eps$, denoted $m_*(\eps)$, satisfies
\begin{equation}
    \ubar S_\text{max}^{(\eps)}(X) \leq \log m_*(\eps) \leq \inf_{\delta\in(0,\eps)} [ \ubar S_\text{max}^{(\eps)}(X) + \log \frac{1}{\delta}]
\end{equation}
as shown by \cite{Tom16,RR12}. Equation~\eqref{eq-x} provides the means to explicitly evaluate the quantity $ \ubar S_\text{max}^{(\eps)}(X)$ in this bound.

\begin{remark}
    Although Part I of this thesis is concerned with the $\eps$-ball defined via the trace distance, these results have some implications for other distance measures. For example, the so-called \emph{sine distance} \cite{Ras02,Ras06} or \emph{purified distance} \cite{Tom-Thesis} is often used in defining smoothed entropies. The purified distance between two states $\rho$ and $\sigma$, denoted $P(\rho,\sigma)$, satisfies the inequality
    \begin{equation}
        T(\rho,\sigma) \leq P(\rho,\sigma)
    \end{equation}
    by Lemma 3.5 of \cite{Tom16}. Therefore, denoting $\tilde \Be(\rho)$ by the $\eps$-ball
    \begin{equation}
        \tilde \Be(\rho) := \left\{ \omega \in \cD(\cH): P(\omega,\sigma) \leq \eps \right\}
    \end{equation}
    one has $\tilde \Be(\rho)\subset \Be(\rho)$. Therefore, for any Schur concave function $f$,
    \begin{equation}
        f(\rho\majmax) \leq \min_{ \tilde \Be(\rho)} f \leq \max_{ \tilde \Be(\rho)} f \leq  f(\rho\majmin)
    \end{equation}
    by \Cref{cor:max-min-ball-Schurconcave}. We therefore obtain immediate bounds on the maxima and minima of Schur concave functions over $\tilde \Be(\rho)$.
\end{remark}

\section{Application to LOCC transformations}\label{sec:LOCC}
Nielsen's theorem \cite{Nie99} states that one bipartite pure state $\psi_{AB} \in \Dpure(\cH_A\otimes\cH_B)$ may be converted into another $\phi_{AB} \in  \Dpure(\cH_A\otimes\cH_B)$ by a protocol consisting of \emph{local operations and classical communication} (LOCC) if and only if the reduced states $\rho_A := \tr_B[\psi_{AB}]$ and $\sigma_A := \tr_B[\phi_{AB}]$ satisfy $\rho_A \prec \sigma_A$. We write $\psi_{AB}\to \phi_{AB}$ to denote that $\psi_{AB}$ can be transformed into $\phi_{AB}$ by an LOCC protocol, and call $\psi_{AB}$ the \emph{source state} and $\phi_{AB}$ the \emph{target state}. For the sake of brevity, we will omit the full mathematical definition of an LOCC protocol in this text; see e.g.\@~\cite{CLM+14} as a reference.

In the case that $\rho_A \not \prec \sigma_A$, can an LOCC protocol perform a  transformation of the source state to an approximation of the target state, $\psi_{AB} \to \omega_{AB} \approx \phi_{AB}$?  The question, in fact, has an optimal solution with respect to two different distance measures. 

\cite{VJN00} construct an optimal state $\chi^F_{AB} \in  \Dpure(\cH_A\otimes\cH_B)$ such that
\[
 \chi^F_{AB} \in \argmax_{\substack{\chi_{AB} \in  \Dpure(\cH_A\otimes\cH_B)\\ \psi_{AB}\to \chi_{AB}}} F(\chi_{AB}, \phi_{AB})^2
 \]
 where $F(\rho,\sigma) := \|\sqrt{\rho}\sqrt{\sigma}\|_1$ is the \emph{fidelity} of two states $\rho$ and $\sigma$. In the case of pure states $\chi$ and $\phi$, the fidelity can be expressed as $F(\chi,\phi) = |\braket{\chi|\phi}|$. The fidelity can be related to the trace distance via the Fuchs and van de Graaf inequalities,
 \begin{equation}\label{eq:Fuchs-and-van-de-Graaf}
 1 - F(\rho,\sigma)\leq T(\rho,\sigma) \leq \sqrt{1- F(\rho,\sigma)^2}.
 \end{equation}
In the case that $\rho$ and $\sigma$ are both pure states, the second inequality is an equality. Thus, maximizing the fidelity between pure states is equivalent to minimizing the trace distance, and the state $\chi^F_{AB}$ is likewise optimal with respect to the trace distance.

\cite{CGV13} define a metric $\operatorname{d}$ on $\cP_d$ (where $d := \dim \cH_A$) by
\[
\operatorname{d}(p,q) = H(p) + H(q) - 2 H( \supmaj(p,q) )
\]
where $\supmaj(p,q)$ is the majorization-supremum discussed in \Cref{sec:motivate-from-inf}. Let  $\vec \lambda(\rho) := (\lambda_i^\downarrow(\rho))_i \in \cP^\downarrow_d$  be the sorted eigenvalues of $\rho_A = \tr_B [\psi_{AB}]$  and likewise $\vec \lambda(\sigma) := (\lambda_i^\downarrow(\sigma))_i \in \cP^\downarrow_d$ be the sorted eigenvalues of $\sigma_A = \tr_B [\phi_{AB}]$. Define $x = \supmaj(\vec \lambda(\rho), \vec \lambda(\sigma))$, and let $\phi_{AB}$ have the Schmidt decomposition
\begin{equation}\label{eq:schimdt-phi}
\ket{\phi}_{AB} = \sum_{i} \sqrt{\lambda^\downarrow_i(\sigma)} \ket{i}_A \ket{i}_B
\end{equation}
for some basis $\{\ket{i}_A\}_i$ of $\cH_A$ and $\{\ket{i}_B\}_i$ of $\cH_B$. Then
\cite{BSF+17} shows that $\chi^\text{sup}_{AB} \in \Dpure(\cH_A\otimes \cH_B)$ given by
\begin{equation}\label{eq:def_chi_sup}
\ket{\chi^\text{sup}}_{AB} = \sum_i \sqrt{x_i}\ket{i}_A\otimes \ket{i}_B
\end{equation}
satisfies
\[
\chi^\text{sup}_{AB} \in \argmin_{\substack{\chi_{AB} \in  \Dpure(\cH_A\otimes\cH_B)\\ \psi_{AB}\to \chi_{AB}}} \operatorname{d}(\chi_{AB}, \phi_{AB}).
\]
In other words, taking the majorization-supremum of the Schmidt coefficients gives the closest target state to $\phi_{AB}$ according to the metric $\operatorname{d}$ that can be reached from $\psi_{AB}$ by an LOCC protocol.

Here, we provide a modest contribution along the same lines, although without establishing optimality. \cite{HOS18} showed that for $p,q \in \cP_d$ with $p\not \prec p$, the quantity
\[
\delta^*(p,q) := 2\max_{k\in \{1,\dotsc,d\}} \sum_{i=1}^k (q_i^\downarrow - p_i^\downarrow)
\]
satisfies
\begin{equation}\label{eq:def-deltastar}
\delta^*(p,q) = \min \{\delta_1 \geq 0 : q \prec p_{*,\delta_1} \} = \min \{\delta_2 \geq 0 : q^*_{\delta_2} \prec p\}
\end{equation}
where $p_{*,\delta_1}$ is the majorization-maximizer of $B_{\delta_1}(p)$ and $q^*_{\delta_2}$ is the majorization-minimizer of $B_{\delta_2}(q)$, as defined in \Cref{sec:geometry_trace_ball}.

Then, similarly to the construction \eqref{eq:def_chi_sup}, define
\begin{equation}\label{eq:def_chi_lowerstar}
\ket{\chi_*}_{AB} := \sum_i \sqrt{(q_{\delta,*})_i} \ket{i}_A\ket{i}_B
\end{equation}
where $\{\ket{i}_A\}_i$ and $\{\ket{i}_B\}_i$ are defined by the Schmidt decomposition of $\phi_{AB}$ in \eqref{eq:schimdt-phi}, and where $q := \vec \lambda(\sigma)$, and $\delta := \delta^*(\vec \lambda(\rho), \vec \lambda(\sigma))$.
Likewise, define
\begin{equation}\label{eq:def_chi_upperstar}
\ket{\chi^*}_{AB} := \sum_i \sqrt{(p_\delta^*)_i} \ket{f_i}_A\ket{f_i}_B
\end{equation}
where $\{\ket{f_i}_A\}_i$ and $\{\ket{f_i}_B\}_i$ are orthonormal bases defined by the Schmidt decomposition of $\psi_{AB}$, and $p = \vec \lambda(\rho)$.
Then, since $p \prec q_{\delta,*}$, we have $\psi_{AB} \to (\chi_*)_{AB}$ and since $p_{\delta}^* \prec q$, we have $(\chi^*)_{AB} \to \phi_{AB}$.

Thus, $(\chi_*)_{AB}$ serves as an approximation to the target state such that the LOCC transformation is possible, while likewise $(\chi^*)_{AB}$ serves as an approximation to the source state. We can quantify this approximation by bounding the trace distance between $(\chi^*)_{AB}$ and $\psi_{AB}$.
\begin{proposition}\label{prop:LOCC}
Let $\phi_{AB}\in \Dpure(\cH_A\otimes\cH_B)$ and $\psi_{AB} \in  \Dpure(\cH_A\otimes\cH_B)$ be pure states with reduced states given by $\rho_A := \tr_B[\psi_{AB}]$ and $\sigma_A := \tr_B[\phi_{AB}]$. The state $(\chi^*)_{AB}$ defined in \eqref{eq:def_chi_upperstar} satisfies $(\chi^*)_{AB} \to \phi_{AB}$ and the bound
\[
T((\chi^*)_{AB}, \psi_{AB}) \leq \sqrt{2\delta}
\]
while $(\chi_*)_{AB}$ defined in \eqref{eq:def_chi_lowerstar} satisfies $\psi_{AB} \to (\chi_*)_{AB}$ and the bound
\[
T((\chi_*)_{AB}, \phi_{AB}) \leq \sqrt{2\delta},
\]
where $\delta := \delta^*(\vec \lambda(\rho),\vec \lambda(\sigma))$, which is defined in \eqref{eq:def-deltastar}.
\end{proposition}
\begin{remark}
The state $(\chi_*)_{AB}$ is not the optimal approximation to $\phi_{AB}$ in trace distance that can be obtained from $\psi_{AB}$ by an LOCC protocol; that state is $(\chi^F)_{AB}$ constructed by \cite{VJN00}, as discussed above. By using the majorization-minimizer and maximizer, however, one can approximate either the source state and target state, as shown above.
\end{remark}
\begin{proof}
We will establish the bound on $T((\chi^*)_{AB}, \psi_{AB})$; the bound on $T((\chi_*)_{AB}, \phi_{AB})$ follows in the same manner.
We have that the fidelity between $(\chi^*)_{AB}$ and $\psi_{AB}$ satisfies
\[
F((\chi^*)_{AB}, \psi_{AB}) = |\braket{\chi^*| \psi}| = \sum_i \sqrt{(p_{\delta}^*)_i p_i}.
\]
This is precisely the fidelity between the reduced states
\[
\rho_{\delta}^* = \diag(p_{\delta}^*), \qquad \rho = \diag(p).
\]
The Fuchs and van de Graaf inequalities, \eqref{eq:Fuchs-and-van-de-Graaf}, give
\begin{equation}\label{eq:proof-F-bound}
F(\rho_{\delta}^*,\rho) \geq 1 - T(\rho_{\delta}^*, \rho)
\end{equation}
Since $(\chi^*)_{AB}$ and $\psi_{AB}$ are pure states, they achieve equality in the upper bound of Fuchs and van de Graaf, namely $T((\chi^*)_{AB},\psi_{AB}) = \sqrt{1 - F((\chi^*)_{AB}, \psi_{AB})^2}$. Thus,
\begin{align*}	
T((\chi^*)_{AB},\psi_{AB}) &= \sqrt{1 - F((\chi^*)_{AB}, \psi_{AB})^2} \\
&= \sqrt{1 - F(\rho_{\delta}^*,\rho)^2}\\
&\leq \sqrt{1 - (1 - T(\rho_{\delta}^*, \rho))^2}\\
&\leq \sqrt{2 T(\rho_{\delta}^*, \rho)}.
\end{align*}
using \eqref{eq:proof-F-bound} for the first inequality. The proof is completed by the observation that $T(\rho_{\delta}^*, \rho) \leq \delta$ since $\rho_{\delta}^*\in B_{\delta}(\rho)$.
\end{proof}
 
\section{A continuity bound for \texorpdfstring{$\rho \mapsto D(\rho\|\sigma)$}{rho -> D(rho, sigma)}} \label{sec:cty-bound-sigma}
The Audenaert-Fannes bound \eqref{eq:Audenaert-Fannes_bound} can be interpreted as a uniform continuity bound for $\rho \mapsto D(\rho \| \frac{1}{d}\one)$, using that $S(\rho) = \log d - D(\rho \| \frac{1}{d}\one)$, where $D(\rho\|\sigma) =  \tr[ \rho \log \rho - \rho\log \sigma]$ is the quantum relative entropy between two quantum states $\rho,\sigma \in \cD(\cH)$ with $d := \dim \cH$.

We can easily generalize this bound to allow arbitrary full-rank states $\omega$ in the second slot, instead of the maximally mixed state $\frac{1}{d}\one$, as follows. Note that this is not an application of majorization flow.

\begin{proposition}
For any $\rho,\sigma\in \cD(\cH)$ and $\omega > 0$, we have
\begin{equation}\label{eq:bound-diff-D}
|D(\rho\|\omega) - D(\sigma\|\omega)| \leq |S(\rho) - S(\sigma)| + T(\rho,\sigma) (\log \lambda_+(\omega) - \log \lambda_-(\omega)).
\end{equation}
Applying the Audenaert-Fannes bound to the first term yields that for $\eps \in [0, 1-\tfrac{1}{d}]$ and $\omega > 0$ the bound
\begin{equation}\label{eq:sup-bound-diff-D}
 \sup_{\substack{\rho,\sigma\in \cD(\cH)\\ T(\rho,\sigma) \leq \eps}} |D(\rho\|\omega) - D(\sigma\|\omega)| \leq \eps \log (d-1) + h(\eps) + \eps  (\log \lambda_+(\omega) - \log \lambda_-(\omega))
\end{equation}
holds. Moreover, this bound is tight in the sense that for any $\eps \in [0, 1-\tfrac{1}{d}]$ and any numbers $\lambda_\pm > 0$ (with $\lambda_+ \geq \lambda_-$) there exists a positive operator $\omega$ with smallest and largest eigenvalues given by $\lambda_\pm(\omega) \equiv \lambda_\pm$ and such that equality holds in \eqref{eq:sup-bound-diff-D}.
\end{proposition}

\begin{proof} 
We have that
\begin{align} 
D(\rho\|\omega) - D(\sigma\|\omega) &= \tr( \rho \log(\rho)) - \tr(\rho \log \omega) - \tr(\sigma \log \sigma) + \tr(\sigma \log \omega)\\
&= S(\sigma) - S(\rho) + \tr( (\sigma-\rho)\log \omega).
\end{align}
Let $(\sigma-\rho)_\pm$ denote the positive and negative parts of the operator $\sigma-\rho$ (sometimes called the Jordan decomposition). Then
\begin{equation}  
D(\rho\|\omega) - D(\sigma\|\omega) = S(\sigma) - S(\rho) + \tr( (\sigma-\rho)_+ \log \omega) - \tr( (\sigma-\rho)_- \log \omega).
\end{equation}
Since $\lambda_-(\omega) \one \leq \omega \leq \lambda_+(\omega) \one$, and $(\sigma-\rho)_\pm \geq 0$ in semi-definite order, we have
\begin{equation}\label{eq:proof-D-bound}
D(\rho\|\omega) - D(\sigma\|\omega) \leq S(\sigma) - S(\rho) + \tr( (\sigma-\rho)_+)\log \lambda_+(\omega) - \tr( (\sigma-\rho)_- )\log \lambda_-(\omega).
\end{equation}
Since $\tr(\sigma-\rho) = 0$, we have $\tr((\sigma-\rho)_\pm) = T(\rho,\sigma)$, and \eqref{eq:bound-diff-D} follows.

Next, choose $\rho = \diag(1,0,\dotsc,0)$, a pure state which is diagonal in the computational basis, and let $\sigma = \diag( 1 - \eps, \frac{\eps}{d-1},\dotsc, \frac{\eps}{d-1})$. We have that $T(\rho,\sigma) = \eps$ and that $S(\sigma) - S(\rho) = \eps \log d + h(\eps)$. In fact, any two states saturating the Audenaert-Fannes inequality are of this form. Next, let $\omega = \diag(\lambda_-(\omega), \lambda_+(\omega),\dotsc,\lambda_+(\omega))$. Then $(\sigma-\rho)_+ = \diag(0, \frac{\eps}{d-1},\dotsc, \frac{\eps}{d-1})$ while $(\sigma-\rho)_- = \diag(\eps,0,\dotsc,0)$, so that
\[
(\sigma-\rho)_\pm \log \omega = (\sigma-\rho)_\pm \log(\lambda_\pm(\omega)).
\]
Thus, equality holds in \Cref{eq:proof-D-bound} for these states. This completes the proof.
\end{proof}
\begin{remark}
We have not assumed the normalization $\tr(\omega)=1$; while this generalizes the inequality, the tightness proof only goes through in the case $\lambda_- + (d-1) \lambda_+ = 1$. Note also that in the case $\omega=\one$, we recover the Audenaert-Fannes bound.
\end{remark}  
\chapter{Guesswork}\label{sec:guesswork}

In this chapter, we consider a somewhat different entropic quantity: the guesswork. In \Cref{sec:shannon-binary} we review an interpretation of the Shannon entropy as the expected number of guesses required to win a certain binary guessing game. In \Cref{sec:guesswork_no_side_info}, we define the guesswork as the expected number of guesses required to win a  modified version of the guessing game, and establish a tight Lipschitz continuity bound for the quantity. In \Cref{sec:guesswork_with_QSI}, we discuss a variant of the guesswork in which the player has access to quantum side information. In that section, we discuss the equivalence of various guessing strategies, provide several formulations of the quantity as optimization problems, establish its concavity and Lipschitz continuity, and establish some entropic bounds.

\section{The Shannon entropy in a binary guessing game}\label{sec:shannon-binary}

In this section, we review an interpretation of the Shannon entropy in terms of a guessing game\footnote{This interpretation is briefly described in \cite{LHL07} and also appears in \cite[Chapter 4.1]{Mac03}}. Consider a random variable $X\sim p_X$ on some finite alphabet $\cX$. The game has two participants, a referee and a player. The referee, Alice, draws a sample from the random variable and finds that $X=x^*$ for some $x^* \in \cX$. The player, Bob, knows the distribution $p_X$ but not $x^*$, and is allowed to ask the referee binary questions of the form ``Is $x^* \in \tilde\cX$?'' for any subset $\tilde \cX \subseteq \cX$. This repeats until Bob asks such a question with $|\tilde \cX| = 1$ and the answer is `YES', in which case he's learned the true outcome.

We wish to understand: what is the minimal  expected number of guesses required? As we will see, this quantity is characterized in some sense by the Shannon entropy. First, we may assume Bob's strategy is deterministic: Bob starts with the same initial question, and given the answer, always choses his second question according to some preformulated strategy, etc. Otherwise, if Bob's strategy is random, we can see it as drawing a random variable with outcomes in (the finitely many) possible deterministic strategies, and then following that strategy. Since the expectation is more than the minimum, he might as well choose an optimal deterministic strategy to start with. We also assume that if Bob learns $x^* \not \in \tilde\cX$, he will not ask if $x^*$ is in some subset of $\tilde \cX$ again.

We can see Bob's \emph{guessing strategy}, i.e.\@~his sequence of questions, as conducting a binary search algorithm; if he chooses a different sequence of questions (a different guessing strategy), then he conducts this binary search in a different manner. We can see this very explicitly by organizing his questions into a binary tree. The root of the tree corresponds to the whole alphabet $\cX$. Then Bob's first question corresponds to
a partition of $\cX$ into some set $\tilde \cX_1$ and $\cX\setminus \tilde \cX_1$, which are the two nodes on the second level, and so forth. The leaves of the tree are the singleton elements $\{ \{x\} : x \in \cX\}$. Using this binary tree, we may encode a particular sequence of answers as a binary string, where `YES' answers correspond to `0' (and moving to the left child in the tree), while `NO' answer correspond to `1' (and moving to the right child in the tree). Such a sequence of answers terminates when a leaf is reached in the tree; equivalently, when Bob knows $x^*$ lies in a singleton set and can then make the correct singleton guess on the next question.

\begin{figure}
\centering
\includegraphics{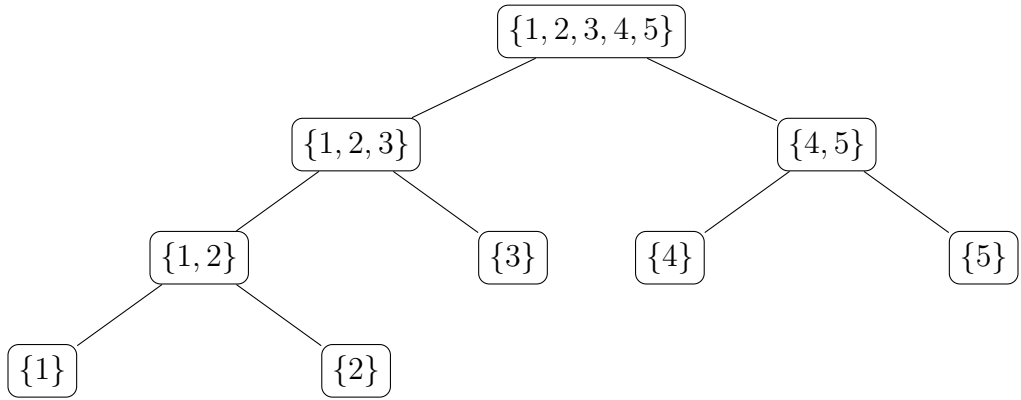}
\caption{An example guessing strategy with $\cX = \{1,2,3,4, 5\}$. With his first guess, Bob asks whether or not $x^* \in \{1,2,3\}$. Let's assume the answer is `YES'; then his second guess asks whether or not $x^* \in \{1,2\}$. Let's assume the answer is `NO'. Then $x^* = 3$, and the sequence of guesses is encoded as $01$.  In this case, it only took two guesses to correctly guess $x^*= 3$; we can see that guessing $4$ or $5$ will similarly be accomplished in two guesses. With this strategy, if $x^* =1$ or $x^* = 2$, however, it will take three guesses to determine them. Hence, this strategy will be better suited for a distribution with more weight on $3$, $4$, and $5$, and less weight on $1$ and $2$.  \label{fig:example_tree}}
\end{figure}

We can interpret this in the language of \emph{source coding}\footnote{See e.g.\@~\cite[Chapter 5]{CT06} for a reference for the source coding results quoted in this section.}. A (binary) code $C$ for $X$ is a map from $\cX \to \{0,1\}^*$, the set of finite binary strings. The map (the \emph{code}) in this case is determined by Bob's sequence of guesses as above. The set of \emph{codewords} is the image $C(\cX)$. The length of a codeword, namely the length of the binary string, corresponds exactly to the number of questions Bob had to ask in order to determine $x^*$. For $x\in \cX$, we denote $l(x)$ as the length of $C(x)$. We are interested, therefore, in the expected length $L =\bE[l(X)]$ of the codewords, with respect to the randomness in drawing $x^*$ from $X$. See \Cref{fig:example_tree} for an example guessing strategy and some of the associated lengths.

The code $C$ described above has the special property that it is a \emph{prefix-free code}, meaning no codeword is a prefix of another codeword. To see this, note that every codeword describes a path from the root of the binary tree to a leaf; a prefix of length $m$ of a codeword corresponds to descending $m$ levels of the tree by choosing to travel to either the left or right child at each step. But each codeword ends in a leaf, and there are no more children to travel to; hence, one codeword cannot be a prefix of another.

The Kraft inequality is a fundamental result about prefix-free codes which states that
\[
\sum_{x\in \cX} 2^{-l(x)} \leq 1.
\]
The converse also holds: for any set of lengths satisfying the above inequality, there is a prefix-free code whose codewords are given by those lengths.

From this inequality, it follows that the expected length of the codewords satisfies
\[
L \geq S(X)
\]
with equality if and only if $2^{-l(x)} = p_X(x)$, and where $S(X) = -\sum_{x\in \cX} p_X(x) \log p_X(x)$ is the Shannon entropy (using base-2 logarithm).

A probability distribution $p_X$ that can be written as $p_X(x) = 2^{-l(x)}$ for some natural numbers $\{l(x) : x \in \cX\}$ is called 2-adic. For such a distribution, we may construct a prefix-free code with lengths given by $l(x)$ (due to the converse of the Kraft inequality), and hence obtain a code with $L = S(X)$. In this sense, the Shannon entropy characterizes the minimum expected number of binary guesses needed to guess a realization of $X$.

This may not be entirely satisfactory, because for distributions that are not 2-adic, however, the inequality can never be saturated. However, by simply choosing $l(x) = \ceil{ \log_2\big(\frac{1}{p_X(X)}\big)}$, the Kraft inequality holds, and hence one can choose a prefix-free code with such lengths. Then its expected length satisfies
\[
S(X) \leq L = \sum_{x\in \cX} p_X(x) l(x) \leq S(X) + 1.
\]
Using this, one can construct a code for $N$ i.i.d.~copies of $X$ in which the expected length of the codebook, $L_N$, satisfies
\[
S(X) \leq \frac{1}{N}L_N \leq S(X) + \frac{1}{N}
\]
and in particular, $\lim_{N\to\infty} \frac{1}{N}L_N = S(X)$.
So for general (non-2-adic) distributions, we at least recover the exact characterization of the Shannon entropy as the minimum expected number of binary guesses in this asymptotic sense.

\begin{remark}
Given a code for the source $X$ with lengths $\{l(x) : x\in \cX\}$ which saturates the Kraft inequality, we can define a probability distribution $q_X$ by $q_X(x) = 2^{-l(x)}$ for $x\in \cX$. Then if this distribution is used to encode symbols from a source with distribution $p_X$, we find
\[
L = \bE_p[l(X)] = S(p) + D(p\|q)
\]
where $S(p) = -\sum_{i} p_i \log p_i$ is again the Shannon entropy, now regarded as a function of the distribution $p$. This provides an interpretation of the relative entropy $D(p\|q) := \sum_{i=1}^d p_i (\log p_i - \log q_i)$ as the number of additional bits needed to encode $X$ using a code whose lengths are associated to distribution $q$ instead of $p$.
\end{remark} 

\section{The guesswork: A different guessing game} \label{sec:guesswork_no_side_info}
The guessing game discussed above provides one sense in which the Shannon entropy captures a notion of uncertainty about the random variable $X$ on some finite alphabet $\cX$. Intuitively, one is more uncertain about the value of $X$ if it takes more guesses to figure out what that value is. 

The binary guessing game described in the previous section, however, isn't the only way to quantify uncertainty through guessing. Let's change the rules. Instead of asking ``Is $x^* \in \tilde \cX$?'', Bob now may only ask questions of the form ``Is $x^* = x$?''. 
The minimal expected number of guesses in this case is called the \emph{guesswork} of $X$, denoted $G(X)$. This clearly provides another measure of the uncertainty in the value of $X$. It has a simple closed-form expression as
\begin{equation}\label{eq:formula-guesswork}
G(X) = \sum_{i=1}^{|\cX|} i \, p_i^\downarrow
\end{equation}
where $X\sim p$. In the following, we will denote $G(X)$ equivalently in terms of the probability mass function $p$ of $X$  as $G(p)$.

Consider the familiar probability vector given by
\[
p_\eps^* = \left(1 - \eps, \frac{\eps}{|\cX|-1}, \dotsc, \frac{\eps}{|\cX|-1}\right)
\]
for some $\eps \in (0, 1-\frac{1}{|\cX|})$, where $p = (1,0,\dotsc,0)$. Then the Shannon entropy satisfies
\[
S(p_\eps^*) = \eps \log (|\cX|-1) + h_2(\eps)
\]
where $h_2(\eps) = -\eps\log\eps - (1-\eps)\log(1-\eps)$ is the binary entropy. The guesswork is given by
\[
G(p_\eps^*) = (1-\eps)\cdot 1 + \sum_{i=2}^{|\cX|} \frac{i \cdot \eps}{|\cX|-1} = 1  + \eps \frac{|\cX|}{2}.
\]
We can see the guesswork scales linearly with $|\cX|$, which makes sense because one essentially needs to perform a linear search over the tail of the distribution, while the entropy scales logarithmically with $|\cX|$.

One might notice that in the case of the guesswork, the $|\cX|$th guess, if it occurs, is unnecessary: after eliminating $|\cX|-1$ outcomes, there can only be one possibility left. This motivates the so-called \emph{modified guesswork} $\tilde G$, given by
\begin{equation}\label{eq:def-modified-guesswork}
\tilde G(p) := \sum_{i=1}^{|\cX|-1} i \cdot p_i^\downarrow + (|\cX|-1) p_{|\cX|}^\downarrow
\end{equation}
\cite{LHL07} showed that the modified guesswork is related to the Shannon entropy by a simple formula:
\begin{equation}\label{eq:modified-guesswork-vs-entropy}
\tilde G(p) = H(p) + D(p^\downarrow \| \{2^{-c_i}\}_{i=1}^{|\cX|})
\end{equation}
where $D(p\|q) := \sum_{i=1}^{|\cX|} p_i (\log p_i - \log q_i)$ is the relative entropy.

We may define a generalized guesswork $G_{\vec c}(X)$ where $\vec c = (c_i)_{i=1}^{|\cX|} \in \bR^{|\cX|}$ is a \emph{cost vector}, satisfying
\[
0 \leq c_1 \leq c_2 \leq \dotsm \leq c_{|\cX|}.
\]
Then we define
\begin{equation}\label{eq:def-generalized-guesswork}
G_{\vec c}(X) := \sum_{i=1}^{|\cX|} c_i p_i^\downarrow.
\end{equation}
Note that this definition unifies \eqref{eq:formula-guesswork} and \eqref{eq:def-modified-guesswork}; choosing $c_i = i$ for each $i$ recovers the guesswork, while choosing
\[
 c_i = \begin{cases}
i & i < d\\
|\cX|-1 & i = |\cX|
\end{cases}
\]
recovers the modified guesswork. The generalized guesswork is a symmetric function of the distribution $p$, as it is only a function of the sorted entries of $p$. Moreover, we can write
\[
G_{\vec c}(X) = \min_{\pi \in S_{|\cX|}} \braket{\pi(p), \vec c}=\min_{\pi \in S_{|\cX|}} \braket{p, \pi\inv(\vec c)}
\]
As a minimum of linear functions, $G_{\vec c}$ is a concave function of $p$. Since $G_{\vec c}$ is symmetric and concave, it is Schur concave. Let us then investigate its continuity with the tools of \Cref{sec:entropic-cty-from-majflow}. Note that $G_{\vec c}$ is a generalization of the case considered in \Cref{ex:non-diff}.

\begin{proposition}\label{prop:guesswork-Lipschitz-nonconditional}
 $G_{\vec c}$ is Lipschitz continuous on $\cP$ with an optimal Lipschitz constant given by $c_{|\cX|} - c_1$.
\end{proposition}
\begin{proof}	
While $G_{\vec c}$ is not differentiable on $\cP_+$, the function $F = \left.G_{\vec c}\right|_{\cP^\downarrow}$ satisfies $F(r) = \sum_{j=1}^{|\cX|} c_j r_j$ and hence has an extension $\tilde F: \cP\to \R$ defined by the same formula. Since $\tilde F$ is differentiable on $\cP_+$, \Cref{cor:non-diff-integral} shows that for $p\in \cP$,
\[
G_{\vec c}(p_\eps^*) - G_{\vec c}(p) = \int_0^\eps \Gamma_{\tilde F}(\mmm_s(p^\downarrow)) \d s.
\]
Hence, as described in \Cref{sec:uniform-bounds-majflow}, bounding $\Gamma_{\tilde F}(\mmm_s(p^\downarrow))$ provides a Lipschitz constant for $G_{\vec c}$. We have that
\[
\Gamma_{\tilde F}(r^\downarrow) = \frac{1}{k_-}\sum_{i: r_i = r_-} c_i - \frac{1}{k_+}\sum_{i: r_i = r_+} c_i.
\]
Since $\frac{1}{k_-}\sum_{i: r_i = r_-} c_i \leq c_{|\cX|}$ and $\frac{1}{k_+}\sum_{i: r_i = r_+} c_i \geq c_1$, we have that $c_{|\cX|} - c_1$ is a Lipschitz constant for $G_{\vec c}$. Moreover, if $p\in \cP^\downarrow$ has that $p_\pm$ are both non-degenerate, and $q = \mmm_\eps(p)$ for $\eps < \delta(p)$ for $\delta$ defined in \Cref{lem:linear-perturbation}, then that Lemma shows that $q = (p_1 - \eps, p_2,\dotsc, p_{|\cX|-1}, p_{|\cX|} + \eps)$, and we immediately have
\[
G_{\vec c}(p_\eps^*) - G_{\vec c}(p) = \eps(c_{|\cX|}- c_1).
\]
Hence, $c_{|\cX|}-c_1$ is the optimal Lipshitz constant.
\end{proof}
\begin{remark}
Note that $\Gamma_{\tilde F}$ defined in the above proof is not symmetric, and hence is not Schur convex on $\cP$. Thus, this choice of $\tilde F$ does not provide a path to establishing a uniform continuity bound by establishing the Schur convexity of $\Delta_\eps^{G_{\vec c}}$ as discussed in \Cref{sec:uniform-bounds-majflow}. However, this does not rule out the Schur convexity of $\Delta_\eps^{H}$. We may calculate for $q = (1,0,\dotsc,0)$ and $0 < \eps < 1 - \frac{1}{|\cX|}$,
\[
\Delta_\eps^{G_{\vec c}}(q) = G_{\vec c}(q_\eps^*) - G_{\vec c}(q) = (1-\eps)c_1 + \frac{\eps}{|\cX|-1}\sum_{i=2}^{|\cX|} c_i  - c_1 = \eps \left( \frac{1}{|\cX|-1}\sum_{i=2}^{|\cX|} c_i  - c_1\right).
\]
On the other hand, by choosing $p$ as in the proof of optimality of the Lipschitz constant for $G_{\vec c}$, we obtain $\Delta_\eps^{G_{\vec c}}(p) = \eps(c_{|\cX|} - c_1)$. As long as for some $i \in \{2,\dotsc, |\cX| -1\}$ one has $c_i \neq c_{|\cX|}$, then $\Delta_\eps^{G_{\vec c}}(p) > \Delta_\eps^{G_{\vec c}}(q)$. Since $p \prec q$, we must have that $\Delta_\eps^{G_{\vec c}}$ is not Schur convex. Thus, in general, one cannot obtain a tight uniform continuity bound for $G_{\vec c}$ via the Schur convexity of $\Delta_\eps^{G_{\vec c}}$. However, since the bound
\begin{equation}\label{eq:Lipschitz-bound-for-Gc}
|G_{\vec c}(p) - G_{\vec c}(q)| \leq \TV(p,q) \cdot (c_{|\cX|}- c_1)
\end{equation}
can in fact be achieved, the Lipschitz bound \eqref{eq:Lipschitz-bound-for-Gc} itself provides a tight uniform continuity bound for $G_{\vec c}$.

Recall that in \Cref{sec:symm-concave-cx} it was shown that there exists a symmetric and concave function $f$ such that its derivative along the path of majorization flow, $\Gamma_f$, is not Schur convex. The Schur convexity of $\Gamma_H$ plays a role in the proof of continuity bounds for a quantity $H$ in \Cref{sec:uniform-bounds-majflow} by establishing the Schur convexity of its integrated form, $\Delta_\eps^H$. Hence, since $\Delta_\eps^{G_{\vec c}}$ is not Schur convex in general, the guesswork provides a counterexample of a similar flavor: namely a quantity $H$ being symmetric and concave is not enough to obtain the Schur convexity of $\Delta_\eps^H$. In this case, however, the quantity of interest is not differentiable on $\cP_+$.
\end{remark}

\section{The guesswork with quantum side information}\label{sec:guesswork_with_QSI}

Let us first briefly discuss the guesswork with classical side information. In this scenario, there are two random variables, $X$ and $Y$. Bob knows the full joint distribution of $XY$. Alice obtains a realization $X = x^*$, and Bob learns the outcome of $Y$, namely $Y = y$. If the two random variables are correlated, Bob has learned something extra about $X$; in fact, knowing $Y = y$, the distribution of $X$ is given by $p_{X|Y}(\cdot| y)$. Then Bob can perform the optimal strategy to guess $x^*$ (namely, sort this conditional distribution in descending order, and guess from most probable to least). The expected number of guesses required to obtain the correct answer is the \emph{guesswork of $X$ with classical side information $Y$}, denoted $G(X|Y)$, and satisfies
\[
G(X|Y) = \sum_{y\in \cY} p_Y(y) G(X|Y=y)
\]
where $\cY$ is the alphabet of $Y$, and $G(X|Y=y) := G(p_{X|Y}(\cdot| y))$ is the guesswork of the conditional distribution. Like with the guesswork itself, this quantity can be generalized to use a cost vector $\vec c$ by simply defining
\[
G_{\vec c}(X|Y) = \sum_{y\in \cY} p_Y(y) G_{\vec c}(X|Y=y).
\]

Now, consider the scenario in which when Alice obtains the realization $X=x^*$, Bob obtains a quantum state $\rho_B^{x^*}$. This occurs, for instance, if a quantum state is measured, and Alice has access to the measurement outcome while Bob has access to the post-measurement state. In this case, the joint distribution $XY$ is replaced by a classical-quantum (c-q) state,
\[
\rho_{XB} = \sum_{x\in \cX}p_X(x) \ket{x}\bra{x}\otimes \rho_B^x.
\]
Given the state $\rho_B^x$, and knowing the joint c-q state $\rho_{XB}$, Bob aims to guess the index $x\in \cX$. The expected number of guesses required to be successful, minimized over all guessing strategies, is the \emph{guesswork with quantum side information}, denoted $G(X|B)_\rho$.

While the case of classical side information admits a very simple optimal strategy (which amounts to simply sorting the conditional probabilities $p_{X|Y}(\cdot|y)$ in non-increasing order and guessing accordingly), the quantum case requires measurement on the quantum system $B$, which potentially disturbs the state of $B$, a priori complicating the analysis of the sequence of guesses in the optimal strategy. We show in \Cref{sec:quantum_strategies}, however, that a general sequential strategy is in fact equivalent to performing a single generalized measurement yielding a classical random variable $Y$ of outcomes and then performing the optimal strategy using this $Y$ as the classical side information. The earlier work~\cite{CCWF15a} instead defined guesswork with quantum side information as the latter quantity, {{i.e.\@}}, a measured version of the guesswork in the presence of classical side information. While these definitions are equivalent, we consider the definition in terms of a sequential protocol to be a more natural one. Moreover, the above-mentioned equivalence is proved via an explicit construction, allowing such a guessing strategy to be implemented sequentially. The single-measurement protocol could in general involve making a measurement with exponentially (in $|\mathcal{X}|$) many outcomes. Hence it may be more efficient to  implement it instead as a sequence of (linearly-many) measurements with linearly-many outcomes, as allowed by the above construction.

This is related to the task of maximizing the ``guessing probability''  $p_{\text{guess}}(X|B)$~\cite{KRS09}. In that task, Bob is only given one attempt to guess the value of $X$ and wishes to maximize the probability of doing so successfully. The guessing probability is related to the so-called \emph{conditional min-entropy} $H_\text{min}(X|B)$ of the c-q state $\rho_{XB}$. In some sense, we can consider the guesswork to be an extension of the guessing probability.  However, the nature of the optimization being done is different: instead of maximizing the probability of success in one attempt, we minimize the total number of guesses required. Therefore, the operations that a guesser performs to minimize the guesswork may be very different from those needed to maximize the guessing probability; for example, one might expect in the case in which only a single guess is being made, one may perform very ``destructive'' measurements to extract maximal information from the state at the cost of having a less informative post-measurement state, whereas in the scenario in which multiple measurements and multiple guesses are being made, the measurements may be chosen to be less destructive. Some of the connections between these two tasks have been investigated in~\cite{CCWF15a}.

We will see that the guesswork $G(X|B)_\rho$ is concave in the state $\rho_{XB}$, but not unitarily invariant, just as for the conditional entropy $H(X|B)_\rho$. This means that the techniques of \Cref{sec:entropic-cty-from-majflow} cannot be applied to obtain a continuity bound for $G(X|B)_\rho$. Moreover, unlike for the guesswork (or the guesswork with classical side-information), we do not have a closed-form expression for $G(X|B)_\rho$, and the quantity is in general not differentiable (since one can see $G(X)$ as a special case, and that quantity is not differentiable on all of $\cP_+$ as discussed in the previous section.) Thus, the techniques of \Cref{sec:general_optimality_conditions} cannot be used to locally maximize $G(X|B)_\rho$, as was done in \Cref{cor:main_thm_for_CE} for the conditional entropy, which otherwise might possibly provide a path towards establishing a (local) continuity bounds.

Instead, we will establish a semidefinite optimization problem (SDP) formulation of the guesswork with quantum side information. By studying the dependence of that formulation on the state $\rho_{XB}$, we quickly obtain a Lipschitz continuity bound for the quantity which has the feature of not depending on the dimension of the $B$-system (like the Alicki-Fannes bound \cite{AlickiFannes04}, or Winter's strengthening thereof \cite{Win16}, for the quantum conditional entropy).

\paragraph{Overview of the remainder of the chapter}
In \Cref{sec:problem_statement}, the guesswork task is described in more detail. In \Cref{sec:strategies} a general framework of guessing strategies is developed which allows formulating a guesswork task with arbitrary kinds of side information; additionally, several types of strategies with quantum side information are described and shown to be equivalent. In \Cref{sec:guesswork_QSI_SDP}, the guesswork with quantum side information is described by a semidefinite optimization problem (SDP) representation. This is used to establish the concavity and Lipschitz continuity of the guesswork with quantum side information in \Cref{sec:concavity_cty_guesswork_QSI}. Lastly, in \Cref{sec:computing_guesswork_QSI} techniques for computing the quantity are discussed, and in \Cref{sec:entropic_bounds} entropic bounds are established.

\subsection{Statement of the problem}\label{sec:problem_statement}

Alice chooses a letter $x\in \cX$ with some probability $p_X(x)$, where $\cX$ is a finite alphabet. This naturally defines a random variable $X\sim p_X(x)$. She then sends a quantum system $B$ to Bob, prepared in the state  $\rho_B^{x}$, which depends on her choice $x$. Bob knows the set of states $\{\rho_B^x : x \in \cX\}$, and the probability distribution $\{p_X(x) : x \in \cX\}$, but he does not know which particular state is sent to him by Alice. Bob's task is to guess $x$ correctly with as few guesses as possible.
From Bob's perspective, he therefore has access to the $B$-part of the c-q state
\begin{equation}\label{eq:cq-state}
\rho_{XB} = \sum_x p_X(x)|x\rangle \langle x|_X \otimes \rho_B^x.
\end{equation}

In the purely classical case, this task reduces to the following scenario: Alice holds the random variable $X\sim p_X(x)$, and Bob holds a correlated random variable $Y$, and knows the joint distribution of $(X,Y)$. In this case $\rho_{XB}$ reduces to the state
\begin{equation} \label{eq:cq-classical}
\rho_{XY} = \sum_x p_X(x)|x\rangle \langle x|_X \otimes \sum_y p_{Y|X}(y|x) |y\rangle \langle y|_Y.
\end{equation}
In this case, if Bob's random variable $Y$ has value $y$, then an optimal guessing strategy would be to sort the conditional distribution $p_{X|Y}(\cdot |y)$ in non-increasing order so that
\begin{equation}
p_{X|Y}(x_1 | y) \geq p_{X|Y}(x_2 | y) \geq \dotsc \geq p_{X|Y}(x_{|\cX|} | y)
\end{equation}
and simply guess first $x_1$, then $x_2$, etc., until he gets it correct~\cite{Ari96}.

In the case in which Bob's system $B$ is quantum, he is allowed to perform any local operations he wishes on $B$, and then make a first guess $x_1$. He is told by Alice whether or not his guess is correct; then he can perform local operations on $B$, and make another guess, and so forth. We are interested in determining the minimal number of guesses needed on average for a given ensemble $\{ p_X(x), \rho_B^x \}_{x\in \cX}$ and the associated optimal strategy.

More generally, we allow Bob to make $K$ guesses, with possibly $K < |\cX|$. Formally, we assume that Bob always makes all $K$ guesses; any guess after the correct guess simply does not factor into the calculation of the minimal number of guesses (see \Cref{sec:strategies} for a more detailed definition of the minimal number of guesses). Thus, Bob makes a sequence of guesses, $g_1,\dotsc,g_K \in \cX^K$ with some probability. 

We could consider the scenario in which Bob makes a guess $x_1$, then learns whether or not the guess was correct, and uses that information to make his second guess $x_2$, and so forth. However, if Bob learns that his $j$th guess $x_j$ is correct, then it does not matter what he guesses subsequently (it has no bearing on the minimal number of guesses). If the guess is incorrect, then his subsequent guesses do matter, and he should make his next guess accordingly. Hence, in such a protocol, the feedback about whether or not the $j$th guess is correct does not help, and Bob might as well assume that each guess is incorrect.

\subsection{Guessing strategies} \label{sec:strategies}

When Alice chooses $x^* \in \cX$, a guessing strategy for Bob outputs a sequence of guesses $\vec g = (g_1,\dotsc,g_K) \in \cX^K$ with some probability $p_{\vec G | X}(\vec g | x^*)$. Hence, formally, a \emph{guessing strategy for $X$ with $K$ guesses} is a random variable $\vec G$ on $\cX^K$ that is correlated with $X$, such that $(X, \vec G)$ has marginal $X \sim p_X$. Note that the definition of a guessing strategy has no reference to the side information (if any) that Bob has access to; instead, the side information dictates the set of guessing strategies Bob has access to. This allows various types of side information to be analyzed within a uniform framework; in particular, the set of strategies available when Bob has access to some classical side information $Y$ is described in \Cref{sec:classical_strategies}, while the case of quantum side information is described in \Cref{sec:quantum_strategies}.

We are interested in the minimal number of guesses required to guess $x^*$ correctly. This is defined as follows:
\begin{equation}\label{eq:def_N}
N(\vec g, x^*) := \begin{cases}
\min \left\{ j :  g_j = x^*\right\} & g_j = x^* \text{ for some }j =1,\dotsc, K\\
\infty & \text{else},
\end{cases}
\end{equation}
where the outcome $\infty$ occurs when none of the $K$ guesses are correct. We can view $N$ as a random variable taking values in $\{1,2,\dotsc,K, \infty\}$. Given a guessing strategy $\vec G$, the quantity of interest is $N(\vec G, X)$, the corresponding random variable. We define
\begin{equation}
\cS_K(X) := \left\{ N(\vec G, X) : X\vec{G} \sim p_{X\vec{G}} \right\}
\end{equation}
to be the set of all possible random variables $N$ associated to all guessing strategies $\vec G$ with $K$ guesses.
We say two guessing strategies $\vec G$ and $\vec G'$ for $X$ with $K$ guesses are \emph{equivalent} if $N(\vec G', X) = N(\vec G, X)$.

Note that if $\vec G$ and $\vec G'$ are two strategies with $K$ guesses for $X$ that  differ only in guesses made after guessing the correct answer, then they are equivalent. This formalizes the notion introduced at the end of the previous section: since guesses made after the correct answer do not change the value of $N(\vec g, x^*)$, feedback of whether or not $g_j = x^*$ can only lead to equivalent strategies.

\subsubsection{Classical strategies} \label{sec:classical_strategies}

Consider a pair of random variables $(X,Y)$ where $X$ has a finite alphabet $\cX$ and $Y$ has a countable alphabet $\cY$. Alice chooses $x^* \in \cX$ (with probability $p_X(x^*)$) and Bob is given $y\in \cY$ (with probability $p_{Y|X}(y|x^*)$). Bob's task is to guess $x^*$. Since Bob's sequence of guesses $(g_1,\dotsc,g_K)$ can only depend on $x^*$ via $y$, a classical guessing strategy $\vec G$ is any random variable $\vec G$ such that the ordered triple $(X,Y,\vec G)$ of random variables  forms a Markov chain, which we denote as $X - Y - \vec G$. Hence, given a joint probability distribution $p_{XY}$, we define the set of random variables $N$ associated to classical guessing strategies as follows:
\begin{equation}
\cS_K^\textnormal{Classical}(p_{XY}) := \left\{ N(\vec G, X) : X - Y - \vec G \right\} \subseteq \cS_K(X).
\end{equation}

\subsubsection{Equivalence of quantum strategies} \label{sec:quantum_strategies}

Let us consider three classes of quantum strategies:
\begin{enumerate}
\item Measured strategy: Bob performs an arbitrary POVM $\{E_y\}_{y \in \cY}$ on the $B$-system. Let $Y$ be the random variable with outcomes in a finite alphabet $\cY$ corresponding to his measurement outcomes, {{i.e.\@}}
\begin{equation}\label{eq:Y-from-X}
  p_{Y|X}(y|x) = \tr[E_{y} \rho_B^x], \quad \forall x\in \cX,\, y \in \cY.
  \end{equation}
Bob then employs a classical guessing strategy on $(X,Y)$. The set of random variables corresponding to the possible number of guesses under such a strategy is given by
\begin{multline}
\cS_K^\textnormal{Measured}(\rho_{XB}) := \Big\{ N(\vec G, X) : X - Y - \vec G, \text{ $Y$ satisfies \eqref{eq:Y-from-X}}\\
\text{for some finite alphabet $\cY$ \& POVM } \{E_y\}_{y \in \cY} \Big\}.
\end{multline}
We then observe that
\begin{equation}
\cS_K^\textnormal{Measured}(\rho_{XB}) \subseteq \cS_K(X).    
\end{equation}

\item Ordered strategy: Bob performs a measurement with outcomes in $\cX^K$, which are identified with guessing orders; {{i.e.\@}}, if the outcome is $(x_1,\dotsc,x_K) \in \cX^K$, Bob first guesses $x_1$, then $x_2$, and so forth. In this case, Bob performs a POVM $\{E_{\vec g}\}_{\vec g \in \cX^K}$ and the guessing strategy $\vec G$ is distributed according to
\begin{equation}\label{eq:G-dist-ordered}
p_{\vec G | X}(\vec g | x) =  \tr[E_{\vec g} \rho_B^{x}].
\end{equation}
As above, we define
\begin{equation}
\cS_K^\textnormal{Ordered}(\rho_{XB}) := \left\{ N(\vec G, X) : (\vec G, X) \text{ satisfy \eqref{eq:G-dist-ordered} for some POVM } \{E_{\vec g}\}_{\vec g \in \cX^K}\right\}
\end{equation}
It is evident that
\begin{equation}
\cS_K^\textnormal{Ordered}(\rho_{XB}) \subseteq \cS_K^\textnormal{Measured}(\rho_{XB})
\end{equation}
because any such ordered strategy is a special type of measured strategy (with $Y = \vec G$). However, any measured strategy can in fact be simulated by an ordered strategy. Suppose we have a measured strategy with alphabet~$\cY$, POVM $\{E_y\}_{y\in \cY}$, and $\vec G$ satisfying $X-Y-\vec G$. Then 
\begin{equation} \label{eq:meas_by_ordered_step}
p_{\vec G|X}( \vec g|x) = \sum_{y \in \cY} p_{\vec G|Y}( \vec g|y) p_{Y|X}(y|x) =  \sum_{y \in \cY} p_{\vec G| Y}(\vec g|y) \tr[E_{y} \rho_B^x],
\end{equation}
where we have used the Markov property for the first equality and \eqref{eq:Y-from-X} for the second equality. 

Let $\tilde E_{\vec g} := \sum_{y\in \cY} p_{\vec G|Y}(\vec g|y) E_{y}$. Note $\{\tilde E_{\vec g}\}_{\vec g \in \cX^K}$ is a POVM: each element is positive semidefinite since $\{E_y\}_{y \in \cY}$ is a POVM, and 
\begin{equation}
\sum_{\vec g \in \cX^K} E_{\vec g} =\sum_{\vec g \in \cX^K}  \sum_{y\in \cY}  p_{\vec G|Y}(\vec g|y) E_{y} =   \sum_{y\in \cY}  \sum_{\vec g \in \cX^K}p_{\vec G|Y}(\vec g|y)E_{y} = \sum_{y \in \cY} E_y = \one_B,
\end{equation}
using again that $\{E_y\}_{y \in \cY}$ is a POVM.
Then substituting the definition of $\tilde E_{\vec g}$ into \eqref{eq:meas_by_ordered_step} yields
\begin{equation}
p_{\vec G|X}(\vec g|x) =  \tr[\tilde E_{\vec g} \rho_B^x]
\end{equation}
and hence \eqref{eq:G-dist-ordered} is satisfied with $E = \tilde E$. Therefore,
\begin{equation}
\cS^\textnormal{Ordered}(\rho_{XB}) = \cS^\textnormal{Measured}(\rho_{XB}).
\label{eq:ordered-=-measured}
\end{equation}

\item \label{it:sequential} Sequential quantum strategy:  
Suppose that Alice chooses $x$ (which occurs with probability $p_X(x)$), and hence Bob has the state $\rho_B^x$. To make his first guess, Bob chooses a set of generalized measurement operators $\{ M_x^{(1)}\}_{x\in \cX}$ and reports the measurement outcome as his guess. He gets outcome $x_1$ with probability
\begin{equation}
p_{G_1|X}(x_1|x)= \tr[ M_{x_1}^{(1)} \rho_B^x  M_{x_1}^{(1)}{}^\dagger]
\end{equation}
and his post-measurement state is
\begin{equation}
\frac{1}{p_{G_1|X}(x_1|x)} M_{x_1}^{(1)} \rho_B^x  M_{x_1}^{(1)}{}^\dagger.
\end{equation}

{\em{Note:}} in general, Bob could perform a unitary operation $U_1$ on his state before measuring it. However, this would simply correspond to measuring with $\{M_{x}^{(1)} U_1\}_{x \in \cX}$ instead. Hence, it suffices to simply consider a generalized measurement $\{ M_x^{(1)}\}_{x\in \cX}$.

Then, after learning the outcome $x_1$, Bob chooses a new set of generalized measurement operators $\{ M_x^{(2 | x_1)}\}_{x\in \cX}$. Note that this set of measurement operators can depend on $x_1$. Without loss of generality, we can keep the same outcome set $\cX$, since Bob could set, {{e.g.\@}}~$M_{x_1}^{(2 | x_1)} = 0$ to avoid guessing the same number twice. Bob measures his state and gets the outcome $x_2$ with probability
\begin{equation}
p_{G_2|G_1 X}(x_2|x_1, x)= \frac{1}{p_{G_1|X}(x_1|x)} \tr[M_{x_2}^{(2| x_1)} M_{x_1}^{(1)} \rho_B^x  M_{x_1}^{(1)}{}^\dagger M_{x_2}^{(2| x_1)}{}^\dagger].
\end{equation}
Multiplying by $p_{G_1|X}(x_1|x)$ we see the joint distribution is given by
\begin{equation}
p_{G_1 G_2|X}(x_1, x_2|x) = \tr[M_{x_2}^{(2| x_1)} M_{x_1}^{(1)} \rho_B^x  M_{x_1}^{(1)}{}^\dagger M_{x_2}^{(2| x_1)}{}^\dagger].
\end{equation}
To make his $j$th guess, we allow Bob to choose a new set of generalized measurement operators $\{M_x^{(j|x_1,\dotsc,x_{j-1})}\}_{x\in \cX}$  which may depend on the previous $j-1$ outcomes. Repeating the previous logic, in the end we find that
\begin{multline} \label{eq:general_prob_G}
p_{G_1 G_2 \dotsm G_K  |X}(x_1, x_2,\dotsc, x_K|x) = \\
\tr[M_{x_K}^{(K| x_1, x_2,\dotsc,x_{K-1})}  \dotsm M_{x_2}^{(2| x_1)} M_{x_1}^{(1)} \rho_B^x  M_{x_1}^{(1)}{}^\dagger M_{x_2}^{(2| x_1)}{}^\dagger\dotsm M_{x_K}^{(K| x_1, x_2,\dotsc,x_{K-1})}{}^\dagger].
\end{multline}
Under such a strategy, the possible random variables giving the number of guesses is given by
\begin{multline}
\cS^{\textnormal{Sequential}}(\rho_{XB}) := \\
\Big\{ N(\vec G, X) : (\vec G, X) \text{ satisfy \eqref{eq:general_prob_G} for some collections of measurement operators }\\
\{M_{x_j}^{(j| x_1, x_2,\dotsc,x_{j-1})}\}_{x_j \in \cX},
\, j =1,\dotsc,K,\,\, x_1,x_2,\dotsc, x_K \in \cX \Big\}.
\end{multline}

\end{enumerate}

\begin{theorem}\label{thm:ordered-from-adaptive}
Let $\rho_{XB}$ be a c-q state as defined in \eqref{eq:cq-state} and $K$ a natural number with $K \leq |\mathcal{X}|$. Then
\begin{equation}\label{eq:all-q-strat-equiv}
 \cS^{\textnormal{Sequential}}_K(\rho_{XB}) = \cS^{\textnormal{Ordered}}_K(\rho_{XB}) = \cS^{\textnormal{Measured}}_K(\rho_{XB}).
\end{equation}
\end{theorem}
Hence, all three sets of random variables of the number of guesses obtained from various classes of strategies all coincide. Hence, we call the single class that of {\em{quantum strategies}}, denoted $\cS^\textnormal{Quantum}_K(\rho_{XB})$.
\begin{proof}
The second equality  was already stated in \eqref{eq:ordered-=-measured} and proven before that, and so it remains to prove the first equality.
Consider a sequential strategy, with the notation of point \ref{it:sequential} above.
Define
\begin{equation} \label{eq:E_from_M}
E_{x_1,\dotsc,x_K} :=  M_{x_1}^{(1)}{}^\dagger M_{x_2}^{(2| x_1)}{}^\dagger\dotsm M_{x_K}^{(K| x_1, x_2,\dotsc,x_{K-1})}{}^\dagger M_{x_K}^{(K| x_1, x_2,\dotsc,x_{K-1})}  \dotsm M_{x_2}^{(2| x_1)} M_{x_1}^{(1)} .
\end{equation}
We see that $E_{x_1,\dotsc,x_K} = A^\dagger A$ for $A =M_{x_K}^{(K| x_1, x_2,\dotsc,x_{K-1})}  \dotsm M_{x_2}^{(2| x_1)} M_{x_1}^{(1)}$, and hence is positive semidefinite. Moreover,
\begin{align}	
\sum_{x_1,\dotsc,x_K \in \cX} E_{x_1,\dotsc,x_K} = I_{B}
\end{align}
as can be seen by first summing \eqref{eq:E_from_M} over $x_K$, using 
\begin{equation}
\sum_{x_K \in \cX} M_{x_K}^{(K| x_1, x_2,\dotsc,x_{K-1})}{}^\dagger M_{x_K}^{(K| x_1, x_2,\dotsc,x_{K-1})}  = I_B
\end{equation}
since $\{M_{x}^{(K| x_1, x_2,\dotsc,x_{K-1})}\}_{x\in \cX}$ is a POVM, and then similarly summing over $x_{K-1}$, $x_{K-2}$,\ldots, and finally $x_1$. Let us write $E_{\vec x}$ where $\vec x=(x_1,\dotsc,x_K)$ for $E_{x_1,
\dotsc, x_K}$. We have shown that $\{ E_{\vec x} \}_{\vec x \in \cX^K}$ is a POVM.
 Moreover,
\begin{equation}
p_{G_1 G_2 \dotsm G_K  |X}(x_1, x_2 ,\dotsc, x_K|x) = \tr[E_{x_1,\dotsc,x_K} \rho_B^x].
\end{equation}
Hence, Bob's strategy is equivalent to simply performing the single POVM $\{ E_{\vec x} \}_{\vec x \in \cX^K}$ once, obtaining an outcome $\vec x = (x_1,\dotsc, x_K)$, and then making $x_1$ his first guess, $x_2$ his second guess, and so forth. That is, any such strategy can be recast as an ordered strategy.

 On the other hand, any such ordered strategy can be reformulated as an adaptive strategy, by the following recursive approach. Suppose that we are given $\left\{ E_{\vec y} \right\}_{\vec y \in \cX^K}$.
 For each $x_1 \in \cX$, define
 \begin{equation}
 M^{(1)}_{x_1} = \sqrt{\sum_{x_2,\dotsc,x_K\in \cX} E_{x_1,\dotsc,x_K}}
 \end{equation}
 where we have chosen the positive semidefinite square root. We have that
 \begin{equation}
 \sum_{x_1\in \cX} M^{(1)}_{x_1}{}^\dagger M^{(1)}_{x_1} =  \sum_{x_1 \in \cX} (M^{(1)}_{x_1})^2 = \sum_{x_1\in \cX} \sum_{x_2,\dotsc,x_K \in \cX} E_{x_1,\dotsc,x_K} = I_B,
 \end{equation}
 so $\left\{  M^{(1)}_{x_1} \right\}_{x_1 \in \cX}$ is indeed a POVM with outcomes in $\cX$. Next, for each $x_1 \in \cX$, corresponding to obtaining outcome $x_1$ on the first measurement, we define a POVM $\{M^{(2| x_1)}_{x_2}\}_{x_2 \in \cX}$ by
 \begin{equation}
 M^{(2| x_1)}_{x_2} = \sqrt{(M^{(1)}_{x_1})\inv\sum_{x_3,\dotsc,x_K \in \cX} E_{x_1,\dotsc,x_K}(M^{(1)}_{x_1})\inv}.
 \end{equation}
 Then
\begin{align}	
 \sum_{x_2\in \cX} (M^{(2| x_1)}_{x_2})^2 &= (M^{(1)}_{x_1})\inv\sum_{x_2\in \cX}\sum_{x_3,\dotsc,x_K\in \cX} E_{x_1,\dotsc,x_K}(M^{(1)}_{x_1})\inv \\
 &= (M^{(1)}_{x_1})\inv (M^{(1)}_{x_1})^2(M^{(1)}_{x_1})\inv = I_B.
\end{align}
Likewise, we define
\begin{equation}
M^{(3| x_1, x_2)}_{x_3} = \sqrt{(M^{(2| x_1)}_{x_2})\inv(M^{(1)}_{x_1})\inv\sum_{x_4,\dotsc,x_K\in \cX} E_{x_1,\dotsc,x_K}(M^{(1)}_{x_1})\inv(M^{(2| x_1)}_{x_2})\inv}.
\end{equation}
Then
\begin{align}	
 \sum_{x_3\in \cX} (M^{(3| x_1, x_2)}_{x_3})^2 &= (M^{(2| x_1)}_{x_2})\inv(M^{(1)}_{x_1})\inv\sum_{x_3\in \cX}\sum_{x_4,\dotsc,x_K\in \cX} E_{x_1,\dotsc,x_K}(M^{(1)}_{x_1})\inv(M^{(2| x_1)}_{x_2})\inv  \\
 &= (M^{(2| x_1)}_{x_2})\inv (M^{(2| x_1)}_{x_2})^2 (M^{(2| x_1)}_{x_2})\inv= I_B.
\end{align}
Repeating this process, we define
\begin{multline}	
  M^{(\ell| x_1, x_2, \dotsc, x_{\ell-1})}_{x_\ell} \\
 =  \sqrt{(M^{(\ell-1| x_1, x_2, \dotsc, x_\ell-2)}_{x_{\ell-1}})\inv\dotsm (M^{(1)}_{x_1})\inv\sum_{x_{\ell+1},\dotsc,x_K\in\cX} E_{x_1,\dotsc,x_K}(M^{(1)}_{x_1})\inv\dotsm M^{(\ell-1| x_1, x_2, \dotsc, x_\ell-2)}_{x_{\ell-1}})\inv}
\end{multline}
to obtain a POVM for step $\ell$ (to use when having obtained outcomes $x_1,\dotsc, x_{\ell-1}$ during the previous steps).
At the last step, $\ell = K$, there is no sum, namely 
\begin{multline}	
  M^{(K| x_1, x_2, \dotsc, x_{K-1})}_{x_K} 
 = \\
 \sqrt{(M^{(K-1| x_1, x_2, \dotsc, x_K-2)}_{x_{K-1}})\inv\dotsm (M^{(1)}_{x_1})\inv E_{x_1,\dotsc,x_K} (M^{(1)}_{x_1})\inv\dotsm M^{(K-1| x_1, x_2, \dotsc, x_K-2)}_{x_{K-1}})\inv}.
\end{multline}
Lastly, we check that by design, \eqref{eq:E_from_M} holds. Thus, we can work backwards from that equation and see that our newly created adaptive strategy yields the same outcomes with the same probabilities as the initial ordered strategy.
\end{proof}

\subsubsection{Success metrics}
Given a random variable $X$ and a maximal number $K$ of allowed guesses, how do we measure the success of a guessing strategy $\vec G$? We will focus on expectations of $N(\vec G, X)$. In particular, we consider the expected number of guesses required to guess correctly:
\begin{equation} \label{eq:expected_guess_inf}
\bE[N(\vec G, X)] = \begin{cases} \sum_{k = 1}^K k\cdot p_{N(\vec G, X)}(k) & \text{if } p_{N(\vec G, X)}(\infty) = 0\\
\infty & \text{if } p_{N(\vec G, X)}(\infty) > 0.
\end{cases}
\end{equation}
As with the generalized guesswork without side information defined in \eqref{eq:def-generalized-guesswork}, we consider a cost vector $\vec c \in (\bR\cup \{\infty\})^{|\cX|}$  with
\begin{equation}\label{eq:QSI_cost_vector}
0 \leq c_1 \leq c_2 \leq \dotsm \leq c_{|\cX|}.
\end{equation}
Then we define the modified expectation
\[
E_{\vec c}(N(\vec G, X)) := \sum_{k = 1}^{|\cX|} c_k\cdot p_{N(\vec G, X)}(k).
\]
Imposing a maximal number $K < |\cX|$ of allowed guesses is equivalent to choosing $c_{K+1} = \dotsm = c_{|\cX|} = \infty$, using the convention $\infty \cdot 0 = 0$. Accordingly, we implicitly associate $K$ with $\vec c$ in all the following via the rule that $K = |\cX|$ if and only if $c_{|\cX|}<\infty$, and otherwise $K=\min \{ i : c_i = \infty \}$. The case $K=|\cX|$ therefore corresponds to $|\cX|$ guesses being allowed, each with finite cost, and the case $K < |\cX|$ corresponds to a limited number of allowed guesses, with a corresponding infinite cost if the correct answer is not obtained in $K$ guesses.

Given a c-q state $\rho_{XB}$ and a cost vector $\vec c$ as in \eqref{eq:QSI_cost_vector}, we define the generalized guesswork with quantum side information as
\begin{equation}\label{eq:def_Ec-quantum}
G_{\vec c}(X|B)_\rho := \inf_{N \in \cS_K^\textnormal{Quantum}(\rho_{XB})} E_{\vec c}(N).
\end{equation}
Likewise, given a joint distribution $p_{XY}$, let
\begin{equation}
G_{\vec c}(X|Y)_p := \inf_{N \in \cS_K^\textnormal{Classical}(p_{XY})} E_{\vec c}(N).
\end{equation}
From the equality
\begin{equation}
\cS_K^\textnormal{Quantum}(\rho_{XB}, K) = \cS_K^\textnormal{Measured}(\rho_{XB}) 
\end{equation}
of \Cref{thm:ordered-from-adaptive} it follows that
\begin{equation} \label{eq:Ec_q_from_c}
 G_{\vec c}(X|B)_\rho = \inf_{\{E_y\}_{y \in \cY}} G_{\vec c}(X|Y)_p
\end{equation}
where the infimum is over all finite alphabets $\cY$ and POVMs  $\{E_y\}_{y \in \cY}$ and $p_{XY}(x, y) = p_X(x) \tr[E_y \rho_B^x]$.

In the standard case in which $\vec c = (1,2,\dotsc,|\cX|)$, we define the \emph{guesswork with quantum side information} as
\begin{equation} \label{eq:def_G}
G(X|B) \equiv G(X|B)_\rho := G_{\vec c}(X|B)_\rho
\end{equation}
and likewise define $G(X|Y)_p = G_{\vec c}(X|Y)_p$ in the case of classical side information $Y$.

\begin{remark}\label{rem:rank_1}
In the work~\cite{CCWF15a}, guesswork with quantum side information was defined by the right-hand side of \eqref{eq:Ec_q_from_c} (with $\vec c = (1,2,\dotsc,|\cX|)$). Moreover, Proposition 1 of that work shows that the infimum in \eqref{eq:Ec_q_from_c} in that case may be restricted to POVMs whose elements are all rank~one.
\end{remark}

\subsection{A semidefinite optimization representation} \label{sec:guesswork_QSI_SDP}

The task of calculating $G_{\vec c}(X|B)_\rho$ as defined in \eqref{eq:def_Ec-quantum} can be written as a semidefinite optimization problem, as was found in~\cite{CCWF15a}. In this section, we present a different derivation of that fact yielding in \eqref{eq:Ec-SDP} a representation dual to the one found in~\cite{CCWF15a}. This representation is used in  \Cref{sec:concavity_cty_guesswork_QSI} to prove that the guesswork $G_{\vec c}(X|B)_\rho$ is a concave and Lipschitz-continuous function of the c-q state $\rho_{XB}$ (if $K = |\cX|$). \Cref{sec:computing_guesswork_QSI} presents other optimization formulations of the guesswork with the aim of computing the quantity.

Consider an ordered strategy $\vec G$ with a set of POVMs $\{ E_{\vec g} \}_{\vec g \in \cX^K}$. Then since $p_{\vec G, X}(\vec g, x) = p_X(x)\tr[ E_{\vec g} \rho_B^x]$, we have
\begin{equation}
    c_k p_{N(\vec G, X)}(k) = c_k \sum_{x \in \cX} p_X(x)\sum_{\substack{\vec g \in \cX^K\\ N(\vec g, x) = k}} \tr[E_{\vec g} \rho_B^x]
\end{equation}
and hence if
\begin{equation}\label{eq:non-zero-prob-of-inf}
 \sum_{x \in \cX} p_X(x)\sum_{\substack{\vec g \in \cX^K\\ N(\vec g, x) = \infty}} \tr[E_{\vec g} \rho_B^x] > 0
 \end{equation}
 then $E_{\vec c}(N(\vec G, X))= \infty$, and otherwise
\begin{align}
E_{\vec c}(N(\vec G, X)) &=  \sum_{k=1}^K c_k \sum_{x \in \cX} p_X(x)\sum_{\substack{\vec g \in \cX^K\\ N(\vec g, x) = k}} \tr[E_{\vec g} \rho_B^x]\\
 &= \sum_{\vec g \in \cX^K}\sum_{x \in \cX}  c_{N(\vec g, x)}  p_X(x)\tr[E_{\vec g} \rho_B^x]\\
 &= \sum_{\vec g \in \cX^K} \tr[R_{\vec g} E_{\vec g}]
\end{align}
where we define $R_{\vec g} := \sum_{x \in \cX}p_X(x) c_{N(\vec g, x)}   \rho_B^x$ for $\vec g\in \cX^K$. Note that if $K=|\cX|$ then \eqref{eq:non-zero-prob-of-inf} does not hold.

Thus,
 \begin{equation}
 \label{eq:Ec-SDP-full}
 \begin{aligned}
 G_{\vec c}(X|B)_\rho\,\,	=\quad	& \text{minimize}	&	& \sum_{\vec g\in \cX^K}\tr[R_{\vec g} E_{\vec g}]\\
 								& \text{subject to}	&	& E_{\vec g} \geq 0 \qquad \forall \vec g \in \cX^K\\
 								&					&	& \sum_{\vec g \in \cX^K} E_{\vec g} = \one_B \\
 								&					&	& \sum_{x \in \cX} p_X(x)\sum_{\substack{\vec g \in \cX^K\\ N(\vec g, x) = \infty}} \tr[E_{\vec g} \rho_B^x] = 0
 \end{aligned}
 \end{equation}
 and where an infeasible problem is associated to the value $+\infty$. Note that in the case $K = |\cX|$ the last constraint is trivially satisfied; in this case, the problem is always feasible.
 The expression in \eqref{eq:Ec-SDP-full} clarifies that $R_{\vec g}$ has an interpretation as a cost operator corresponding to the guessing outcome $\vec g$.
Since $\sum_{\vec g \in \cX^K}\tr[R_{\vec g} E_{\vec g}]$ is linear in each positive semidefinite (matrix) variable $E_{\vec g}$, \eqref{eq:Ec-SDP-full} gives an SDP representation of $G_{\vec c}(X|B)_\rho$. This program has $|\cX|^K$ variables (each $d_B\times d_B$ complex positive semidefinite matrices), subject to two constraints. Note, however, since the cost vector $\vec c$ is increasing, any guess $\vec h \in \cX^K$ with repeated elements is a suboptimal guessing order, in the sense that if $\{E_{\vec g}\}_{\vec g \in \cX^K}$ is a POVM with $E_{\vec h} \neq 0$, and
$\vec h' \in \cX^K$ only differs from $\vec h$ by replacing repeated elements such that $\vec h'$ has no repeated elements, then the POVM defined by
\begin{equation}
    \tilde E_{\vec g} := \begin{cases}
    E_{\vec g} & \vec g \neq \vec h \text{ and }\vec g \neq \vec h' \\
    0 & \vec g = \vec h\\
    E_{\vec h} + E_{\vec h'} & \vec g = \vec h'
    \end{cases}
\end{equation}
has $\sum_{\vec g \in \cX^K} \tr[ R_{\vec g} \tilde E_{\vec g}] \leq \sum_{\vec g \in \cX^K} \tr[ R_{\vec g}  E_{\vec g}]$. Hence, we may restrict to the outcome space
\begin{equation}\label{eq:def_X_K}
    \cX^K_{\neq} := \{ \vec g\in \cX^K : g_i \neq g_j, \forall i \neq j \} \subseteq \cX^K.
\end{equation}
Note $|\cX^K_{\neq}| = \frac{|\cX|!}{(|\cX| - K)!}$, and in the case in which $K = |\cX|$, the set $\cX_{K}$ is just the set of permutations of $\cX$. Hence, \eqref{eq:Ec-SDP-full} can be re-written as the following smaller problem:
 \begin{equation}\label{eq:Ec-SDP}
 \begin{aligned}
 G_{\vec c}(X|B)_\rho\,\,	=\quad	& \text{minimize}	&	& \sum_{\vec g\in \cX^K_{\neq}}\tr[R_{\vec g} E_{\vec g}]\\
 								& \text{subject to}	&	& E_{\vec g} \geq 0 \qquad \forall \vec g \in \cX^K_{\neq}\\
 								&					&	& \sum_{\vec g \in \cX^K_{\neq}} E_{\vec g} = \one_B \\
 								&					&	& \sum_{x \in \cX} p_X(x)\sum_{\substack{\vec g \in \cX^K_{\neq}\\ N(\vec g, x) = \infty}} \tr[E_{\vec g} \rho_B^x] = 0
 \end{aligned}
 \end{equation}
Note that in the case $K < |\cX|$, there exists a feasible point, and hence a solution, if and only if there exists a POVM $\{ E_{\vec g}\}_{\vec g \in \cX^K_{\neq}}$ such that for all $x\in \cX$ and $\vec g \in \cX^K_{\neq}$ with $x \not \in \vec g$, we have $\tr[ E_{\vec g} \rho_B^x] = 0$. Whether or not this holds a priori depends on the particular state $\rho_{XB}$.
However, when $K = |\cX|$, the final constraint can be removed and for any state $\rho_{XB}$, the problem \eqref{eq:Ec-SDP}  trivially has a feasible point (e.g.\@~ $E_{\vec g} = \frac{1}{|\cX^K_{\neq}}\one_X$). In the following, we restrict to the case $K = |\cX|$ for simplicity.
\begin{remark}
This optimization problem has the same form as that of discriminating quantum states in an ensemble, as described in, {{e.g.\@}}, \cite[Section 3.2.1]{Wat18}. 
Note, however, that (1) the $R_{\vec g}$ are positive semidefinite but not normalized, and (2) the case of having two copies of the unknown state, in the guessing framework, does not correspond to $R_{\vec g}^{\otimes 2}$.  Nevertheless,
slight modifications to~\cite[Theorem 3.9]{Wat18} show that a POVM $\{E_{\vec g}\}_{\vec g \in \cX^K_{\neq}}$ is optimal for \eqref{eq:Ec-SDP} if and only if
\begin{equation}
Y = \sum_{\vec g \in \cX^K_{\neq}} R_{\vec g} E_{\vec g}
\end{equation}
satisfies $Y \leq R_{\vec g}$ for all $\vec g \in \cX^K_{\neq}$.
\end{remark}

\begin{remark}
The set of POVMs is convex and since the objective function is linear, any minimizer for \eqref{eq:Ec-SDP} may be decomposed into extremal POVMs which are also minimizers. By~\cite[Corollary 2.2]{Par99}, for any extremal POVM on a Hilbert space of size $d_B$ has at most $d_B^2$ non-zero elements. Hence, there exist minimizers of \eqref{eq:Ec-SDP} with at most $d_B^2$ non-zero elements (even though $|\cX^K_{\neq}|$ could be far larger than $d_B^2$). Let $S \subseteq \cX^K_{\neq}$ be a set of $d_B^2$ points such that there exists $\{\tilde E_{\vec g}\}_{\vec g \in S}$ with $\tilde E_{\vec g}\geq 0$, $\sum_{\vec g \in S} \tilde E_{\vec g} = \one_B$, and
\begin{equation}
G_{\vec c}(X|B)_\rho = \sum_{\vec g \in S} \tr[\tilde E_{\vec g} R_{\vec g}].
\end{equation}
Then \eqref{eq:Ec-SDP} holds with $\cX^K_{\neq}$ replaced by $S$, namely
 \begin{equation}\label{eq:Ec-SDP-small}
 \begin{aligned}
 G_{\vec c}(X|B)_\rho\,\,	=\quad	& \text{minimize}	&	& \sum_{\vec g\in S}\tr[R_{\vec g} E_{\vec g}]\\
 								& \text{subject to}	&	& E_{\vec g} \geq 0 \qquad \forall \vec g \in S\\
 								&					&	& \sum_{\vec g \in S} E_{\vec g} = \one_B.
 \end{aligned}
 \end{equation}
 Note the ``$\leq$'' direction of the equality \eqref{eq:Ec-SDP-small} is trivial, since given a minimizer $\{E_{\vec g}\}_{\vec g \in S}$ for \eqref{eq:Ec-SDP-small}, simply extending it by choosing $E_{\vec g} = 0$ for $\vec g \not \in S$ gives a feasible point for the optimization problem on the right-hand side of \eqref{eq:Ec-SDP}. The ``$\geq$'' direction follows from the existence of the $\{\tilde E_{\vec g}\}_{\vec g \in S}$ described above. While this produces a much smaller problem, the set $S$ depends in general on $\rho$ and the $\{R_{\vec{g}}\}_{\vec g \in \cX^K_{\neq}}$, and the task of finding $S$ is as difficult as solving the original problem.
\end{remark}

\subsection{Concavity and continuity of the guesswork with quantum side information} \label{sec:concavity_cty_guesswork_QSI}

\begin{proposition}\label{prop:guesswork_concave}
For each finite cost vector $\vec c$ satisfying \eqref{eq:QSI_cost_vector}, the function
\begin{equation} \label{eq:Guesswork_fn_rho}
 \rho_{XB} \mapsto G_{\vec c}(X|B)_\rho
\end{equation}
from the set of c-q states of the form \eqref{eq:cq-state} to $\mathbb{R}_{\geq 0}$, is concave. 
\end{proposition}

\begin{proof}
Recall $c_{|\cX|} <\infty$ means $K= |\cX|$.
For $\vec g \in \cX^K_{\neq}$, and $\rho_{XB}$ a c-q state, the quantity $R_{\vec g}^\rho :=  \sum_{x\in \cX}p_X(x)  c_{N(\vec g, x)} \rho_B^x$ can be expressed as
\begin{equation}
R_{\vec g}^\rho = \tr_X\left[\left(\sum_{x\in \cX}c_{N(\vec g, x)}  |x\rangle \langle x|_X \otimes I_B \right)\rho_{XB}\right]
\label{eq:alt-form-R-g}
\end{equation}
and hence is linear in $\rho_{XB}$.
Then for each POVM $(E_{\vec g})_{\vec g \in \cX^K_{\neq}})$, 
\begin{equation}
\rho_{XB} \mapsto \sum_{\vec g \in \cX^K_{\neq}} \tr[ R_{\vec g}^\rho E_{\vec g}]
\end{equation}
is linear in $\rho_{XB}$. The arbitrary infimum of concave functions, and in particular linear functions, is concave, and hence 
\begin{equation}
 G_{\vec c}(X|B)_\rho\equiv \min_{(E_{\vec g})_{\vec g \in \cX^K_{\neq}}} \sum_{\vec g \in \cX^K_{\neq}} \tr[R_{\vec g}^\rho E_{\vec g}],
 \end{equation}
 where the minimum is taken over all POVMs on system $B$ with outcomes in $\cX^K_{\neq}$,
 is concave.
\end{proof}

\begin{proposition}\label{prop:guesswork_Lipschitz}
For each finite cost vector $\vec c$ satisfying \eqref{eq:QSI_cost_vector}, the function
\begin{equation}
 \rho_{XB} \mapsto G_{\vec c}(X|B)_\rho
\end{equation}
from the set of c-q states of the form \eqref{eq:cq-state} to $\mathbb{R}_{\geq 0}$, is Lipschitz continuous, satisfying the bound
\begin{equation}
|G_{\vec c}(X|B)_\rho - G_{\vec c}(X|Y)_\sigma| \leq 2c_{|\cX|} T(\rho_{XB}, \sigma_{XB}).
\end{equation}
Moreover, the optimal Lipschitz constant $k$ satisfies
\begin{equation}\label{eq:guesswork_QSI_bounds_Lip}
c_{|\cX|} - c_1 \leq k \leq 2 c_{|\cX|},
\end{equation}
and the lower bound is tight: there exist $\rho_{XB}$ and $\sigma_{XB}$ such that $|G_{\vec c}(X|B)_\rho - G_{\vec c}(X|Y)_\sigma| = c_{|\cX|} - c_1$.
\end{proposition}
\begin{remark}
In the standard case case $\vec c = (1,\dotsc,|\cX|)$, the optimal Lipschitz constant for $G_{\vec c}(X|B)$ thus satisfies
\[
|\cX| - 1 \leq k \leq 2|\cX|,
\]
and hence the bound of \Cref{prop:guesswork_Lipschitz} is tight up to constant factors in this case. An interesting open question is whether or not equality is achieved in the lower bound of \eqref{eq:guesswork_QSI_bounds_Lip}.
\end{remark}
\begin{proof}
To establish the lower bound of \eqref{eq:guesswork_QSI_bounds_Lip}, note that product states $\rho_{XB} = \omega_X \otimes \sigma_B$ satisfy $G_{\vec c}(X|B)_\rho = G_{\vec c}(X)_\omega$. Hence, \Cref{prop:guesswork-Lipschitz-nonconditional} shows that $k \geq c_{|\cX|} - c_1$, and the tightness of the bound follows from the tightness of \Cref{prop:guesswork-Lipschitz-nonconditional}.

Next, let us establish the upper bound. Define
\begin{equation}
f(\rho_{XB}, \{E_{\vec g}\}_{\vec g \in \cX^K_{\neq}}) :=  \sum_{\vec g \in \cX^K_{\neq}} \tr[ R_{\vec g}^\rho E_{\vec g}].
\end{equation}
Then, by linearity (as discussed in the proof of \Cref{prop:guesswork_concave}),
\begin{align}
f(\rho_{XB}, \{E_{\vec g}\}_{\vec g \in \cX^K_{\neq}})-f(\sigma_{XB}, \{E_{\vec g}\}_{\vec g \in \cX^K_{\neq}})&= \sum_{\vec g \in \cX^K_{\neq}} \tr[ R_{\vec g}^{\rho-\sigma} E_{\vec g}]\\
&= \sum_{\vec g \in \cX^K_{\neq}} \tr[ \tr_X[C^{(\vec g)}_{XB} \Delta_{XB}] E_{\vec g}]
\end{align}
using \eqref{eq:alt-form-R-g}, where $C_{XB}^{(\vec g)} := \sum_{x\in \cX}c_{N(\vec g, x)}  |x\rangle \langle x| \otimes I_B\geq 0$ and $\Delta_{XB} := \rho_{XB}-\sigma_{XB}$. Since $C_{XB}^{(\vec g)}$ and $\Delta_{XB}$ commute, using the c-q structure of each, we have
\begin{align}
    f(\rho_{XB}, \{E_{\vec g}\}_{\vec g \in \cX^K_{\neq}})-f(\sigma_{XB}, \{E_{\vec g}\}_{\vec g \in \cX^K_{\neq}})&=\sum_{\vec g \in \cX^K_{\neq}} \tr[ C^{(\vec g)}_{XB} \Delta_{XB} (I_X \otimes E_{\vec g})] \\
    &= \tr\left[ \Delta_{XB}\sum_{\vec g \in \cX^K_{\neq}}C^{(\vec g)}_{XB} (I_X\otimes E_{\vec g})\right].
\end{align}
Set
\begin{equation}
    F_{XB} := \sum_{\vec g \in \cX^K_{\neq}}C^{(\vec g)}_{XB} (I_X\otimes E_{\vec g}) =\sum_{x\in \cX}\sum_{\vec g \in \cX^K_{\neq}}  c_{N(\vec g, x)}  |x\rangle \langle x| \otimes E_{\vec g}.
\end{equation}
Since $c_{N(\vec g, x)}\leq c_{|\cX|}$ for each $x \in \cX$ and $\vec g \in \cX^K_{\neq}$, we have that $F_{XB} \leq c_{|\cX|} \sum_{x\in \cX}\sum_{\vec g \in \cX^K_{\neq}}  |x\rangle \langle x| \otimes E_{\vec g} $ in semidefinite order. Performing the sums, we have $F_{XB} \leq c_{|\cX|} \, I_X \otimes I_B$ and hence $\|F_{XB}\|_\infty \leq c_{|\cX|}$. Thus,
\begin{align}
    f(\rho_{XB}, \{E_{\vec g}\}_{\vec g \in \cX^K_{\neq}})-f(\sigma_{XB}, \{E_{\vec g}\}_{\vec g \in \cX^K_{\neq}}) 
    &= \tr\left[ \Delta_{XB} F_{XB}\right]\\
    &\leq \|  \Delta_{XB} F_{XB}\|_1\\
    &\leq \|\Delta_{XB}\|_1 \,\|F_{XB}\|_\infty\\
    &\leq c_{|\cX|} \|\rho_{XB}-\sigma_{XB}\|_1\\
    &= 2 c_{|\cX|} T(\rho_{XB}, \sigma_{XB}).
\end{align}
using H\"older's inequality in the third to last inequality.
Swapping the role of $\rho_{XB}$ and $\sigma_{XB}$ completes the proof.
\end{proof}

This completes the main goal of this chapter: establishing the Lipschitz continuity of the guesswork with quantum side information. In the remainder of the chapter, we will look at some further aspects of the guesswork with quantum side information in order to establish methods for computing it (or computing bounds of it), and to obtain entropic bounds on the quantity.

\subsection{Computing the guesswork} \label{sec:computing_guesswork_QSI}

The SDP formulation of the guesswork described in \eqref{eq:Ec-SDP} provides the first method of computing the quantity: by solving the semidefinite optimization problem. The difficulty in this approach is that the number of variables in the problem scales as $|\cX_{\neq}^K| = \frac{|\cX|!}{(|\cX|-K)!}$. For $K$ scaling with $|\cX|$ (e.g.\@, $K = |\cX|$), the number of variables grows extremely quickly. In this section, we present two alternatives in the case $K = |\cX|$: an algorithm using an active-set method to establish an upper bound, and a reformulation of the problem as a \emph{mixed-integer SDP}  with polynomial dependence on $|\cX|$.

\subsubsection{The dual problem}\label{sec:dual}
First, we compute the dual problem to \eqref{eq:Ec-SDP}, in the case $K= |\cX|$. Consider the Lagrangian
\begin{align}	
 \cL((E_{\vec g})_{\vec g \in \cX^K_{\neq}}, (\lambda_{\vec g})_{\vec g \in \cX^K_{\neq}}, \nu) &= \sum_{\vec g\in \cX^K_{\neq}}\braket{ R_{\vec g}, E_{\vec g}}  - \sum_{\vec g\in \cX^K_{\neq}}\braket{\lambda_{\vec g}, E_{\vec g}} + \Braket{\nu, \sum_{\vec g \in \cX^K_{\neq}} E_{\vec g} - \one_B}\\
 &= \sum_{\vec g\in \cX^K_{\neq}}\braket{R_{\vec g}-\lambda_{\vec g} + \nu, E_{\vec g}}  - \tr[\nu]
\end{align}
where we have introduced the Hilbert--Schmidt inner product $\braket{A,B} = \tr[A^\dagger B]$, and where $\lambda_{\vec g}\geq 0$ is the dual variable to the inequality constraint $E_{\vec g}\geq 0$, and $\nu = \nu^\dagger$ is the dual variable to the equality constraint $\sum_{\vec g \in \cX^K_{\neq}} E_{\vec g} = \one_B$. As shown in, {{e.g.\@}}, \cite{BV04}, the primal problem \eqref{eq:Ec-SDP} can be expressed as
\begin{equation}\label{eq:primal-formula}
 \min_{(E_{\vec g})_{\vec g \in \cX^K_{\neq}}}\max_{\lambda_{\vec g} \geq 0, \nu}\cL((E_{\vec g})_{\vec g \in \cX^K_{\neq}}, (\lambda_{\vec g})_{\vec g \in \cX^K_{\neq}}, \nu)
\end{equation}
while the dual problem is given by
\begin{equation}\label{eq:dual-formula}
\max_{\lambda_{\vec g} \geq 0, \nu}\min_{(E_{\vec g})_{\vec g \in \cX^K_{\neq}}}\cL((E_{\vec g})_{\vec g \in \cX^K_{\neq}}, (\lambda_{\vec g})_{\vec g \in \cX^K_{\neq}}, \nu).
\end{equation}
If $R_{\vec g} - \lambda_{\vec g} + \nu \neq 0$ for any $\vec g\in \cX^K_{\neq}$, then the inner minimization in \eqref{eq:dual-formula} yields $-\infty$. Hence, 
\begin{equation}
\min_{(E_{\vec g})_{\vec g \in \cX^K_{\neq}}}\cL((E_{\vec g})_{\vec g \in \cX^K_{\neq}}, (\lambda_{\vec g})_{\vec g \in \cX^K_{\neq}}, \nu) = \begin{cases}
- \infty & R_{\vec g} - \lambda_{\vec g} + \nu \neq 0 \quad \exists \vec g \in \cX^K_{\neq}\\
- \tr[\nu] & \text{else.}
\end{cases}
\end{equation}
The constraint $\lambda_{\vec g} \geq 0$ and $R_{\vec g} - \lambda_{\vec g} + \nu = 0$ imply the semidefinite inequality $-\nu \leq R_{\vec g}$. Writing $Y = - \nu$ and maximizing over $\lambda_{\vec g} \geq 0$, \eqref{eq:dual-formula} becomes
\begin{equation} \label{eq:dual_SDP}
\begin{aligned}
\text{maximize} \quad & \tr[Y]\\
\text{subject to} \quad & Y = Y^\dagger \\
& Y \leq R_{\vec g} \qquad \forall \vec g \in \cX^K_{\neq}
\end{aligned}
\end{equation}
Since \eqref{eq:Ec-SDP} is strictly feasible ({{e.g.\@}}, $E_{\vec g} = \one_B \frac{1}{|\cX^K_{\neq}|}$ is a strictly feasible point), Slater's condition proves that strong duality holds. Hence, \eqref{eq:dual_SDP} obtains the same optimal value as \eqref{eq:Ec-SDP}. The formulation of the problem as given in  \eqref{eq:dual_SDP} was previously found in the work~\cite[Proposition 3]{CCWF15a}.

\subsubsection{A simple algorithm to compute upper bounds}\label{sec:ub_algo}

The dual form of the SDP can be used to generate upper bounds on $G_{\vec c}(X|B)_\rho$ simply by removing constraints. This provides an algorithm to find an upper bound on the objective function: Decide on some number of constraints $\kappa$ to impose in total. Then,
\begin{enumerate}
	\item  Initialize an empty list $L=\{\}$ corresponding to constraints to impose.
	\item Set $Y$ to be the identity matrix, as a first guess at the optimal dual variable.
	\item If $Y$ satisfies $Y \leq R_{\vec g}$ for all $\vec g \in \cX^K_{\neq}$, then $Y$ is the maximizer of the dual problem \eqref{eq:dual_SDP}, and the optimization is solved. Otherwise, find $\vec g\in \cX^K_{\neq}$ such that $Y \not \leq R_{\vec g}$, and add $\vec g$ to the list $L$.
	\item  Solve the problem
	\begin{equation} \label{eq:dual_SDP-L}
	\begin{aligned}
	\text{maximize} \quad & \tr[Y]\\
	\text{subject to} \quad & Y = Y^\dagger \\
	& Y \leq R_{\vec g} \qquad \forall \vec g \in L
	\end{aligned}
	\end{equation}
	and set $Y$ to be its maximizer.
	\item Repeat steps 2 and 3 until the list $L$ has length $\kappa$.
	\item Solve the problem one last time, and return the output.
\end{enumerate}
In order to find a constraint that $Y$ violates, a heuristic technique such as simulated annealing can be used. Moreover, in the case that there are too many constraints to fit into memory or check exhaustively, using an iterative technique (such as simulated annealing) is essential. If this algorithm was continued without imposing a limit on the total number of constraints $\kappa$ to impose, it would eventually yield the true value $G_{\vec c}(X|B)_\rho$. When a total number of constraints is limited, it provides an upper bound (since it is a relaxation of \eqref{eq:dual_SDP}). Note that since it takes exponential time to check if $Y \leq R_{\vec g}$ for every $\vec g \in \cX_\neq^K$, in practice $\kappa$ must be chosen low enough so that violated constraints can be found quickly enough (or $\kappa$ can be chosen adaptively by adding constraints until finding a new violated constraint violates some predetermined time limit).

However, even with a limit $\kappa$ on the total number of constraints, this algorithm can in theory yield the true value $G_{\vec c}(X|B)_\rho$. Note that the dual problem to \eqref{eq:Ec-SDP-small} is 
\begin{equation} \label{eq:dual_SDP-S}
\begin{aligned}
\text{maximize} \quad & \tr[Y]\\
\text{subject to} \quad & Y = Y^\dagger \\
& Y \leq R_{\vec g} \qquad \forall \vec g \in S
\end{aligned}
\end{equation}
where $S \subseteq \cX^K_{\neq}$ has $|S| = d_B^2$ and is described in the remark above. Hence, if $L$ in \eqref{eq:dual_SDP-L} equals $S$, then the algorithm finds the true value $G_{\vec c}(X|B)_\rho$, not just an upper bound. Thus, $\kappa = d_B^2$ suffices if the constraints $\vec g$ can be chosen precisely to obtain $L=S$. In general, finding $S$ is as difficult as solving the original problem. Nonetheless, this motivates why choosing a relatively small value of $\kappa$ (such as $d_B^2$) can still yield good upper bounds.

\subsubsection{A mixed-integer reformulation}\label{sec:misdp}

The problem can be formulated another way as a mixed-integer SDP, {i.e.\@}~an SDP that has additional integer or binary constraints. Consider a POVM $\{F_j\}_{j=1}^M$ with $M$ outcomes. When outcome $j$ is obtained, Bob guesses in some order $\vec g^{(j)} \in \cX^K_{\neq}$. Then consider the problem
\begin{equation} \label{eq:misdp-1}
\begin{aligned}
	& \text{minimize}	&	& \sum_{x\in \cX, j = 1,\dotsc, M}p_X(x)  c_{N(\vec g^{(j)}, x)}   \tr[ F_j \rho_B^x]\\
 								& \text{subject to}	&	& F_j \geq 0 \qquad j = 1,\dotsc, M,\\
 								&					&	& \vec g^{(j)} \in \cX^K_{\neq}, j = 1,\dotsc, M,\\
 								&					&	& \sum_{j=1}^M F_j = \one_B.
\end{aligned}
\end{equation}
This optimization is not an SDP, since the dependence on the optimization variables $\{\vec g^{(j)}\}_{j=1}^M$ and $\{F_j\}$ is not linear, and $\vec g^{(j)} \in \cX^K_{\neq}$ is a discrete constraint. In the case that $K = |\cX|$, we will be able to remove the nonlinearity, although not the discrete variables. This yields a \emph{mixed-integer} SDP: an optimization problem such that if all integral constraints were removed, the result would be an SDP. We proceed as follows.

Under the condition $K = |\cX|$, we may restrict to considering guessing orders that are permutations without loss of generality;  other guessing orders have repeated guesses, which can only increase the objective function. In this case, the outcome $\infty$ never occurs, and for each $\vec g \in S_{|\cX|}$, the quantity $( c_{N(\vec g, x)} )_{x \in \cX}$ satisfies
\begin{equation}
( c_{N(\vec g, x)} )_{x \in \cX} = \vec g\inv(c),
\end{equation}
where $\vec g \inv$ is the inverse permutation to $\vec g$, and $c$ is the cost vector. Here, $S_n$ is the set of permutations on $\{1,\dotsc,n\}$. Let $P^{(j)}$ be an $|\cX|\times |\cX|$ matrix representation of the permutation $\vec g^{(j)}{}\inv$. Then $(P^{(j)} c)_x = \sum_{y\in \cX} P^{(j)}_{xy} c_y = c_{N(\vec g, x)}$. Hence, the optimization \eqref{eq:misdp-1} can be reformulated as
\begin{equation}\label{eq:misdp-2}
\begin{aligned}
	& \text{minimize}	&	&  \sum_{x,y\in \cX, j = 1,\dotsc, M}p_X(x)  P^{(j)}_{xy} c_y   \tr[ F_j \rho_B^x]\\
 								& \text{subject to}	&	& F_j \in \cM_{d_B} \qquad \forall\,j \in [M],\\
 								&					&	& P^{(j)}_{xy} \in \{0,1\}, \quad \forall\,j \in [M], x,y \in \cX\\
 								& 	&	& F_j \geq 0 \qquad \forall\,j \in [M],\\
 								&					&	& \sum_{j=1}^M F_j = \one_B,\\
 								&					&	& \sum_{x\in \cX}P^{(j)}_{xy} = 1, \quad \forall\,j \in [M], y \in \cX\\
 								&					&	& \sum_{y\in \cX}P^{(j)}_{xy} = 1, \quad \forall\,j \in [M], x \in \cX\\
 					\end{aligned}
\end{equation}
Note that all the constraints are semidefinite or linear, except that each element $P^{(j)}_{xy}$ is a binary variable: $P^{(j)}_{xy} \in \{0,1\}$, which is a particularly simple type of discrete constraint. The non-linearity in the objective function, however, persists. To remove this, we take advantage of the  fact that the  $P^{(j)}_{xy}$ are binary. In particular,~\cite[Equations (22)--(24)]{BDNK19} provide  a clever trick to turn objective functions with terms of the form $z x$ where $z$ is a binary variable and $x$ a continuous variable into objective functions of a continuous variable $y$ subject to four affine constraints (in terms of $x$ and $z$), as long as $x$ is bounded by known constants. We reproduce this argument in the following.

 We first write the objective function entirely in terms of scalar quantities:
\begin{equation}
\sum_{x,y\in \cX, j \in [M]}p_X(x)  P^{(j)}_{xy} c_y   \tr[ F_j \rho_B^x] = \sum_{k,\ell \in [d_B]}\sum_{x,y\in \cX, j \in [M]}p_X(x)(\rho_B^x)_{k\ell}  c_y\,  P^{(j)}_{xy}  (F_j)_{\ell k} 
\end{equation}
Let $x = (F_j)_{\ell k} $ and $z = P^{(j)}_{xy} \in \{0,1\}$. Then $|x| \leq \tr[F_j]/2 \leq d_B/2$. Then $x_L := - d_B/2$ and $x_U := d_B/2$ constitute lower and upper bounds to $x$, respectively. Hence, the following four inequalities hold trivially:
\begin{equation}
\begin{aligned}\label{eq:ref1}
z (x - x_L) \geq 0,\\
(z-1)(x - x_U) \geq 0,\\
z(x - x_U) \leq 0,\\
(z-1)(x - x_L) \leq 0.
\end{aligned}
\end{equation}
Now, let $y = xz$. Then we have
\begin{equation}\label{eq:ref2}
\begin{aligned}
y - zx_L \geq 0,\\
y - z x_U \geq x - x_U,\\
y - z x_U \leq 0,\\
y - zx_L \leq x - x_L.
\end{aligned}
\end{equation}
On the other hand, let us remove the constraint $y= xz$, and consider $y$ as another variable. Then if $z = 0$, the first equation of \eqref{eq:ref2} implies that $y \geq 0$, while the third implies $y\leq 0$, so $y= 0$. On the other hand, if $z=1$, then the second equation of \eqref{eq:ref2} implies that $y \geq x$ while the fourth implies that $y \leq x$. Hence, either way, $y = xz$. Thus, \eqref{eq:ref2} is equivalent to $y = xz$.

With this transformation, \eqref{eq:misdp-2} can be reformulated as the following. 
\begin{equation} \label{eq:misdp-final}
\begin{aligned}
	& \text{minimize}	&	&  \sum_{k,\ell \in [d_B]}\sum_{x,y\in \cX, j \in [M]}p_X(x)(\rho_B^x)_{k\ell}  c_y\, y_{xy \ell k j}\\
 								& \text{subject to}	&	& F_j \in \cM_{d_B} \qquad \forall\,j \in [M],\\
 								&					&	& y_{xy \ell k j} \in \mathbb{R}, \quad  \forall \,x,y \in \cX, \ell,k \in [d_B], j \in [M],\\
 								&					&	& P^{(j)}_{xy} \in \{0,1\}, \quad \forall\,j \in [M], x,y \in \cX\\
 								& 	&	& F_j \geq 0 \qquad \forall\,j \in [M],\\
 								&					&	& \sum_{j=1}^M F_j = \one_B,\\
 								&					&	& \sum_{x\in \cX}P^{(j)}_{xy} = 1, \quad \forall\,j \in [M], y \in \cX\\
 								&					&	& \sum_{y\in \cX}P^{(j)}_{xy} = 1, \quad \forall\,j \in [M], x \in \cX\\
&					&	& y_{xy \ell k j} + P^{(j)}_{xy} \frac{d_B}{2} \geq 0\quad  \forall \,x,y \in \cX, \ell,k \in [d_B], j \in [M],\\
&					&	& y_{xy \ell k j} - P^{(j)}_{xy}  \frac{d_B}{2} \geq (F_j)_{\ell k}  - \frac{d_B}{2}\quad  \forall \,x,y \in \cX, \ell,k \in [d_B], j \in [M],\\
&					&	& y_{xy \ell k j} - P^{(j)}_{xy}  \frac{d_B}{2} \leq 0\quad  \forall \,x,y \in \cX, \ell,k \in [d_B], j \in [M],\\
&					&	& y_{xy \ell k j} + P^{(j)}_{xy} \frac{d_B}{2} \leq (F_j)_{\ell k}  + \frac{d_B}{2}\quad  \forall \,x,y \in \cX, \ell,k \in [d_B], j \in [M].
\end{aligned}
\end{equation}
This is a mixed-integer SDP, with a number of constraints and variables that is polynomial in $M, d_B, |\cX|$. Moreover, if $M \geq d_B^2$, then as follows from the remark below \eqref{eq:Ec-SDP}, the mixed-integer SDP \eqref{eq:misdp-final} obtains the same optimal value as \eqref{eq:Ec-SDP}, namely $G_{\vec c}(X|B)_\rho$. Note, however, that mixed-integer SDPs are not in general efficiently solvable; they encompass mixed integer linear programs, which are NP-hard. However, in practice they can sometimes be quickly solved. Since the original SDP formulation \eqref{eq:Ec-SDP} involves an exponential (in $|\cX|$) number of variables (or an exponential number of constraints in its dual formulation \eqref{eq:dual_SDP}), \eqref{eq:misdp-final} which instead has a polynomial (in $|\cX|$) number of variables may provide a more practical approach in some cases. Mixed-integer SDPs can be solved in various ways; in the code \cite{guesswork_code}, the problem \eqref{eq:misdp-final} is solved using the library Pajarito.jl \cite{CLV18}, which proceeds by solving an alternating sequence of mixed-integer linear problems and SDPs.

\subsubsection{The ellipsoid algorithm}

The ellipsoid algorithm (see e.g.\@~\cite{GLS93}) provides a theoretical proof that under a strict feasibility assumption, semi-definite programs can be solved in time that scales as a polynomial in: the number of scalar variables and constraints, the log of a 2-norm bound on the feasible points, $\log(1/\eps)$ where $\eps$ is the solution tolerance, and the maximum bit length of the scalar entries of the objective and constraints (see e.g.\@ \cite[Theorem 4]{Wat09}). 

In fact, the ellipsoid algorithm applies quite generally to the optimization of a linear objective function over a convex \emph{feasible region} (which could be described by a domain and constraint functions). The ellipsoid algorithm only requires a \emph{separation oracle} for the feasible region, a subroutine which either asserts that a given point lies within the feasible region, or provides a separating hyperplane between the given point and the feasible region. When the separation oracle can be evaluated in polynomial time, the overall ellipsoid algorithm runs in polynomial time as well (see \cite[Corollary 4.2.7]{GLS93}).

In the case of a single positive semi-definite constraint, e.g.\@ $Y \geq 0$, a simple separation oracle is given by computing the eigendecomposition of $Y$ and checking if all of its eigenvalues are non-negative. If so, it returns that $Y$ is indeed feasible, and otherwise returns the matrix $C := U \text{diag}(f(\lambda_1),\dotsc,f(\lambda_d)) U^\dagger$ where $Y = U\text{diag}(\lambda_1,\dotsc,\lambda_d)U^\dagger$ is the eigendecomposition of $Y$, $U$ is unitary, $\lambda_1,\dotsc,\lambda_d$ are the eigenvalues, and $f(x) = 1$ if $x < 0$ and $f(x) = 0$ otherwise. The matrix $C$ has the properties that $C\geq 0$, $\|C\|_\infty = 1$, and $\tr[C^\dagger Y] = \sum_{i=1}^d f(\lambda_i)\lambda_i = \sum_{i: \lambda_i < 0}\lambda_i <0$.

In the case of the dual problem \eqref{eq:dual_SDP} with $K=|\cX|$, we have $|\cX|!$ positive semidefinite constraints. Thus, we cannot check all of them together in polynomial time.

The feasibility problem $Y\leq R_{\pi}$ for each $\pi \in S_{|\cX|}$ can be written as the following mixed-integer non-linear problem,
\begin{equation*}	
\begin{aligned}
\eta := \text{minimize} &\quad \braket{\psi, \left(\sum_{x\in \cX}(P c)_{x} p_X(x) \rho_B^x - Y\right) \psi}\\
\text{subject to} &\quad \sum_{i\in \cX} P_{ij} = \sum_{j\in \cX} P_{ij} = 1\\
&\quad P_{ij}\in \{0\}, \,\,i,j \in \cX \\
&\quad \psi \in \bC^{d_B}
\end{aligned}
\end{equation*}
where $\eta \geq 0$ if and only if $Y \leq R_{\pi}$ for all $\pi \in S_{|\cX|}$, using that a matrix $M$ satisfies $M\geq 0$ if and only if $\braket{\psi, M \psi}\geq 0$ for all $\psi \in \bC^{d_B}$. Since the convex hull of the permutation matrices is given by the doubly stochastic matrices, the discrete constraints can be relaxed, yielding the following reformulation
\begin{equation}\label{eq:nl_feasibility}
\begin{aligned}
\eta = \text{minimize} &\quad \braket{\psi, \left(\sum_{x\in \cX}(Dc)_{x} p_X(x) \rho_B^x - Y\right) \psi}\\
\text{subject to} &\quad \sum_{i\in \cX} D_{ij} = \sum_{j\in \cX} D_{ij} = 1\\
&\quad D_{ij} \geq 0, \,\,i,j \in \cX \\
&\quad \psi \in \bC^{d_B}.
\end{aligned}
\end{equation}
If $\eta\geq 0$, then $Y$ is feasible. Otherwise, the optimal value $D^*$ can be decomposed as a convex combination of permutations, $D^* = \sum_i \alpha_i P_i$, and we must have $\sum_{x} (P_i c)_x p_X(x) \rho_B^x \not \geq Y$ for some $P_i$, using that the objective is an affine function of $D$. The problem \eqref{eq:nl_feasibility} can be solved by global non-linear optimization solvers such as EAGO.jl \cite{WS17} or SCIP \cite{GamrathEtal2020OO}, but not in general in polynomial time.

At each iteration of the ellipsoid algorithm, one must evaluate the separation oracle for some Hermitian matrix $Y$. In order to avoid solving \eqref{eq:nl_feasibility}, one may attempt to prove the feasibility or infeasibility of a point $Y$ by other means. For example, one may search over permutations $\pi$ heuristically, in order to find $R_\pi$ such that $Y\not\leq R_\pi$. If such a permutation can be identified, then $Y$ is not feasible, and the problem \eqref{eq:nl_feasibility} does not need to be solved. Likewise, if one can show that for some $k \in (1,\dotsc,|\cX|)$,
\begin{equation}
    Y \leq \sum_{i=1}^k c_{|\cX|-i} p_{x_{i}} \rho_B^{x_{i}} + c_1 \sum_{x \in \cX \setminus \{x_{i_1},\dotsc,x_{i_k}\}}p_x \rho_B^x \forall (x_1,\dotsc,x_k)\in \cX^k_{\neq}
\end{equation}
then $Y$ must be feasible, and again \eqref{eq:nl_feasibility} does not need to be solved. The number of comparisons required scales as $|\cX|^k$; for small choices of $k$, this provides an efficient check for feasibility (which may, however, be inconclusive).

\subsubsection{Numerical experiments} \label{sec:guesswork_numerical_experiments}

We compare numeric implementations of several of the above algorithms on a set of 12 test problems. The code for these experiments can be found at \cite{guesswork_code}. Each problem has $p \equiv u$, the uniform distribution $u := (1/|\cX|, \dotsc, 1/|\cX|)$, for simplicity. The states are chosen as
\begin{enumerate}
	\item Two random qubit density matrices
	\item Two random qutrit density matrices
	\item Three pure qubits chosen equidistant within one  plane of the Bloch sphere (the ``Y-states''), i.e.\@
	\[
	\cos\left(j \tfrac{2\pi}{3}\right) \ket{0} + \sin\left(j \tfrac{2\pi}{3}\right) \ket{1}, \qquad j = 1,2,3
	\]
	\item Three random qubit density matrices
	\item Three random qutrit density matrices
	\item The four BB84 states, $\ket{0}, \ket{1}$, and $\ket{\pm} = \frac{1}{\sqrt{2}}(\ket{0} \pm \ket{1})$
\end{enumerate}
as well as the ``tensor-2'' case of
\begin{equation}\label{eq:tensor-2}
\{ \rho \otimes \sigma : \rho, \sigma \in S\}
\end{equation}for each of the six sets $S$ listed above, corresponding to the guesswork problem with quantum side information associated to $\rho_{XB}^{\otimes 2}$, where $\rho_{XB}$ is the state associated to the original guesswork problem with quantum side information. The random states were chosen uniformly at random (i.e.\@~according to the Haar measure).

The exponentially-large SDP formulation (and its dual), the mixed-integer SDP algorithm, and the active set method were compared, with several choices of parameters and underlying solvers. The mixed-integer SDP formulation was evaluated with $M=d_B$ (yielding an upper bound), $M=d_B^2$ (yielding the optimal value), with the Pajarito mixed-integer SDP solver \cite{CLV18}, using Convex.jl (version 0.12.7) \cite{Convex.jl-2014} to formulate the problem. Pajarito proceeds by solving mixed-integer linear problems (MILP) and SDPs as subproblems, and thus uses both a MILP solver and an SDP solver as subcomponents. Pajarito provides two algorithms: an iterative algorithm, which alternates between solving MILP and SDP subproblems, and solving a single branch-and-cut problem in which SDP subproblems are solved via so-called lazy callbacks to add cuts to the mixed-integer problem. The latter is called ``mixed-solver drives'' (MSD) in the Pajarito documentation. We tested three configurations of Pajarito (version 0.7.0):

\begin{description}
\item[(c1)] Gurobi (version 9.0.3) as the MILP solver and MOSEK
(version 8.1.0.82) as the SDP solver, with Pajarito's MSD algorithm
\item[(c2)] Gurobi as the MILP solver and MOSEK as the SDP solver, with Pajarito's iterative algorithm, with a relative optimality gap tolerance of $0$,
\item[(o)] Cbc \cite{Cbc} (version 2.10.3) as the MILP solver, and SCS \cite{SCS} (version 2.1.1) as the SDP solver, with Pajarito's iterative algorithm
\end{description}

Here, `c' stands for commercial, and `o' for open-source. In the configuration (c1), Gurobi was set to have with a relative optimality gap tolerance of $10^{-5}$ and in (c2), a relative optimality gap tolerance of $0$. In both configurations, Gurobi was given an absolute linear-constraint-wise feasibility tolerance of $10^{-8}$, and an integrality tolerance of $10^{-9}$. These choices of parameters match those made in \cite{CLV18}. Cbc was given an integrality tolerance of $10^{-8}$, and SCS's (normalized) primal, dual residual and relative gap were set to $10^{-6}$ for each problem. The default parameters were used otherwise. Note the MSD option was not used with Cbc, since the solver does not support lazy callbacks.

For the (exponentially large) SDP primal and dual formulations, the problems were solved with both MOSEK and SCS, and likewise with the active-set upper bound.

The active set method uses simulated annealing to iteratively add violated constraints to the problem to find an upper bound, as described in \Cref{sec:upper-bound}, and uses a maximum-time parameter $t_\text{max}$ to stop iterating when the estimated time of finding another constraint to add would cause the running time of to exceed the maximum-time\footnote{The maximum time can still be exceeded, since at least one iteration must be performed and the estimate can be wrong.}. This provides a way to compare the improvement (or lack thereof) of running the algorithm for more iterations. The algorithm also terminates when a violated constraint cannot be found after 50 runs of simulated annealing (started each time with different random initial conditions). Here, the problems were solved with three choices of $t_\text{max}$, $20$\,s, $60$\,s, and $240$\,s.

The exact answer was not known analytically for most of these problems, so the average relative error was calculated by comparing to the mean of the solutions (excluding the active-set method and the MISDP with $M=d_B$, which only give an upper bound in general). In the BB84 case, in which the solution is known exactly (see \cite{HKDW20}), the solutions obtained here match the analytic value to a relative tolerance of at least $10^{-7}$.

The problems were run sequentially on a 4-core desktop computer (Intel i7-6700K 4.00GHz CPU, with 16 GB of RAM, on Ubuntu-20.04 via Windows Subsystem for Linux, version 2), via the programming language Julia \cite{BEKS17} (version 1.5.1), with a 5 minute time limit. The results are summarized in \Cref{tab:summary}, and presented in more detail in \Cref{tab:first_problems} and \Cref{tab:second_problems}.

One can see that the MISDP problems were harder solve than the corresponding SDPs for these relatively small problem instances. The MISDPs have the advantage of finding extremal solutions, however, in the case $M=d_B$, and may scale better to large instances. Additionally, the active-set upper bound performed fairly well, finding feasible points within 20\,\% of the optimum in all cases, with only $t_\text{max}=20$\,s, and often finding near-optimal solutions. It was also the only method able to scale to the largest instances tested, such as two copies of the BB84 states (which involves 16 quantum states in dimension $4$, and which the SDP formulation has 16!\@ variables.). In general, the commercial solvers performed better than the open source solvers, with the notable exception of the active-set upper bound with MOSEK, in which 2 more problems timed out than with SCS. This could be due to SCS being a first-order solver which can therefore possibly scale to larger problem instances than MOSEK, which is a second-order solver.

\bigskip

\begin{table}
\begin{adjustbox}{center}
\begin{tabular}{p{2.75cm}p{2.25cm}p{1.75cm}p{1.75cm}p{1.75cm}p{1.75cm}p{1.75cm}}
Algorithm & Parameters & average rel. error & average time & number solved & number timed out & number errored out\\ \toprule
MISDP ($d_B$) & Pajarito (c1) & 0\,\% & 23.74\,s & 6 & 6 & 0\\
 & Pajarito (c2) & 0\,\% & 24.03\,s & 6 & 6 & 0\\
 & Pajarito (o) & 0\,\% & 45.05\,s & 6 & 6 & 0\\\addlinespace[0.75em]
MISDP ($d_B^2$) & Pajarito (c1) & 0\,\% & 35.27\,s & 5 & 7 & 0\\
 & Pajarito (c2) & 0\,\% & 27.97\,s & 4 & 8 & 0\\
 & Pajarito (o) & 0\,\% & 131.35\,s & 4 & 8 & 0\\\addlinespace[0.75em]
SDP & MOSEK & 0\,\% & 8.99\,s & 8 & 3 & 1\\
 & SCS & 0\,\% & 9.08\,s & 8 & 3 & 1\\\addlinespace[0.75em]
SDP (dual) & MOSEK & 0\,\% & 8.74\,s & 8 & 3 & 1\\
 & SCS & 0\,\% & 8.59\,s & 8 & 3 & 1\\\addlinespace[0.75em]
\multirow[t]{3}{2.75cm}[0pt]{Active set upper bound (MOSEK)} & $t_\text{max}=20$\,s & 6.80\,\% & 16.08\,s & 10 & 2 & 0\\
 & $t_\text{max}=60$\,s & 6.79\,\% & 19.03\,s & 10 & 2 & 0\\
 & $t_\text{max}=240$\,s & 6.80\,\% & 26.17\,s & 10 & 2 & 0\\\addlinespace[0.75em]
\multirow[t]{3}{2.75cm}[0pt]{Active set upper bound (SCS)} & $t_\text{max}=20$\,s & 6.09\,\% & 33.24\,s & 12 & 0 & 0\\
 & $t_\text{max}=60$\,s & 6.09\,\% & 35.78\,s & 12 & 0 & 0\\
 & $t_\text{max}=240$\,s & 6.09\,\% & 34.30\,s & 11 & 1 & 0
\end{tabular}
\end{adjustbox}

\bigskip

\caption{\label{tab:summary} Comparison of average relative error and average solve time for the 12 problems discussed above. A problem is considered ``timed out'' if an answer is not obtained in 5 minutes, and ``errored out'' if the solution was not obtained due to errors (such as running out of RAM). The average relative error, which was rounded to two decimal digits, and the time taken are calculated only over the problems which were solved by the given algorithm and choice of parameters. ``MISDP ($d_B$)'' refers to the choice $M=d_B$, and likewise ``MISDP ($d_B^2$)'' refers to the choice $M=d_B^2$.}
\end{table}
\begin{table}
\begin{adjustbox}{center}
\begin{tabular}{p{2.75cm}p{2.25cm}p{3cm}@{\hspace{1cm}}p{3cm}@{\hspace{1cm}}p{3cm}}
Algorithm & Parameters & Two\mbox{ }random qubits & Two\mbox{ }random qutrits & Y-states\\ \toprule
MISDP ($d_B$) & Pajarito (c1) & 23.63\,s, timeout & 23.60\,s, timeout & 23.56\,s, timeout\\
& Pajarito (c2) & 22.99\,s, timeout & 23.31\,s, timeout & 23.21\,s, timeout\\
& Pajarito (o) & 23.47\,s, timeout & 24.77\,s, timeout & 26.15\,s, timeout\\\addlinespace[0.75em]
MISDP ($d_B^2$) & Pajarito (c1) & 24.49\,s (0.00\,\%), timeout & 31.40\,s (0.00\,\%), timeout & 26.27\,s (0.00\,\%), timeout\\
 & Pajarito (c2) & 25.02\,s (0.00\,\%), timeout & 31.39\,s (0.00\,\%), timeout & 29.08\,s (0.00\,\%), timeout\\
 & Pajarito (o) & 26.54\,s (0.00\,\%), timeout & 212.79\,s (0.00\,\%), timeout & 141.14\,s (0.00\,\%), timeout\\\addlinespace[0.75em]
SDP & MOSEK & 8.69\,s, 8.84\,s & 8.78\,s, 9.23\,s & 9.33\,s, timeout\\
& SCS & 9.00\,s, 8.90\,s & 8.44\,s, 11.22\,s & 8.98\,s, timeout\\\addlinespace[0.75em]
SDP (dual) & MOSEK & 8.46\,s, 8.63\,s & 8.54\,s, 8.83\,s & 9.11\,s, timeout\\
 & SCS & 8.76\,s, 8.32\,s & 8.33\,s, 9.20\,s & 8.74\,s, timeout\\\addlinespace[0.75em]
\multirow[t]{3}{2.75cm}[0pt]{Active set upper bound (MOSEK)} & $t_\text{max}=20$\,s & 8.76\,s\mbox{ }(19.5\,\%), 10.41\,s\mbox{ }(1.5\,\%) & 8.89\,s\mbox{ }(19.5\,\%), timeout & 9.72\,s\mbox{ }(0\,\%), 34.25\,s\mbox{ }(?\@\,\%)\\
 & $t_\text{max}=60$\,s & 10.91\,s\mbox{ }(19.5\,\%), 10.41\,s\mbox{ }(1.9\,\%) & 8.87\,s\mbox{ }(19.5\,\%), timeout & 9.66\,s\mbox{ }(0\,\%), 31.00\,s\mbox{ }(?\@\,\%)\\
 & $t_\text{max}=240$\,s & 9.47\,s\mbox{ }(19.5\,\%), 10.40\,s\mbox{ }(1.5\,\%) & 8.90\,s\mbox{ }(19.5\,\%), timeout & 9.81\,s\mbox{ }(0\,\%), 30.26\,s\mbox{ }(?\@\,\%)\\\addlinespace[0.75em]
\multirow[t]{3}{2.75cm}[0pt]{Active set upper bound (SCS)} & $t_\text{max}=20$\,s & 9.04\,s\mbox{ }(19.5\,\%), 10.92\,s\mbox{ }(1.5\,\%) & 8.70\,s\mbox{ }(19.5\,\%), 101.06\,s\mbox{ }(1.1\,\%) & 9.23\,s\mbox{ }(0\,\%), 82.84\,s\mbox{ }(?\@\,\%)\\
 & $t_\text{max}=60$\,s & 9.07\,s\mbox{ }(19.5\,\%), 10.22\,s\mbox{ }(1.9\,\%) & 8.66\,s\mbox{ }(19.5\,\%), 32.94\,s\mbox{ }(1.2\,\%) & 9.18\,s\mbox{ }(0\,\%), 50.79\,s\mbox{ }(?\@\,\%) \\
 & $t_\text{max}=240$\,s & 9.04\,s\mbox{ }(19.5\,\%), 10.02\,s\mbox{ }(1.9\,\%) & 8.79\,s\mbox{ }(19.5\,\%), 22.69\,s\mbox{ }(1.2\,\%) & 9.31\,s\mbox{ }(0\,\%), 37.36\,s\mbox{ }(?\@\,\%)
\end{tabular}
\end{adjustbox}

\bigskip

\caption{\label{tab:first_problems} The individual timings for each algorithm and choice of settings on problems (1)--(3), and the corresponding ``tensor-2'' problems discussed at \eqref{eq:tensor-2}. For each algorithm, the running time of the original problem is given followed by the running time on the ``tensor-2'' problem, e.g.\@ the SDP formulation with MOSEK on the two random qubits problem was solved in 8.69 seconds, and in 8.84 seconds for the corresponding tensor-2 problem. ``timeout'' is written whenever the problem was not solved within 5 minutes. For the active set algorithms, the relative error is also given for each problem in parenthesis. Note that the MISDP formulation with $M=d_B$ is also only known to be an upper bound, but a relative error of less than $10^{-5}$ in each instance, so the relative errors are omitted. Lastly, the relative error is written as {?\@\,\%} in the case that only an upper bound was obtained.}
\end{table}

\begin{table}
\begin{adjustbox}{center}
\begin{tabular}{p{2.75cm}p{2.25cm}p{3cm}@{\hspace{1cm}}p{3cm}@{\hspace{1cm}}p{3cm}}
Algorithm & Parameters & Three\mbox{ }random qubits & Three\mbox{ }random qutrits & BB84\mbox{ }states\\ \toprule
MISDP ($d_B$) & Pajarito (c1) & 23.63\,s, timeout & 24.49\,s, timeout & 23.51\,s, timeout\\
 & Pajarito (c2) & 23.17\,s, timeout & 26.50\,s, timeout & 25.01\,s, timeout\\
 & Pajarito (o) & 27.11\,s, timeout & 95.01\,s, timeout & 73.80\,s, timeout\\\addlinespace[0.75em]
MISDP ($d_B^2$) & Pajarito (c1) & 25.82\,s (0.00\,\%), timeout & timeout, timeout & 68.35\,s (0.00\,\%), timeout\\
& Pajarito (c2) & 26.38\,s (0.00\,\%), timeout & timeout, timeout & timeout, timeout\\
& Pajarito (o) & 144.93\,s (0.00\,\%), timeout & timeout, timeout & timeout, timeout\\\addlinespace[0.75em]
SDP & MOSEK & 9.35\,s, timeout & 8.82\,s, timeout & 8.87\,s, error\\
 & SCS & 9.09\,s, timeout & 8.46\,s, timeout & 8.54\,s, error\\ \addlinespace[0.75em]
SDP (dual) & MOSEK & 9.14\,s, timeout & 8.55\,s, timeout & 8.62\,s, error\\
& SCS & 8.86\,s, timeout & 8.25\,s, timeout & 8.23\,s, error\\\addlinespace[0.75em]
\multirow[t]{3}{2.75cm}[0pt]{Active set upper bound (MOSEK)}  & $t_\text{max}=20$\,s & 9.73\,s\mbox{ }(1.3\,\%), 32.29\,s\mbox{ }(?\@\,\%) & 9.50\,s\mbox{ }(5.8\,\%), timeout & 9.51\,s\mbox{ }(0\,\%), 27.72\,s\mbox{ }(?\@\,\%)\\
& $t_\text{max}=60$\,s & 9.69\,s\mbox{ }(1.3\,\%), 24.12\,s\mbox{ }(?\@\,\%) & 9.45\,s\mbox{ }(5.8\,\%), timeout & 9.77\,s\mbox{ }(0\,\%), 66.47\,s\mbox{ }(?\@\,\%)\\
& $t_\text{max}=240$\,s & 9.69\,s\mbox{ }(1.3\,\%), 30.42\,s\mbox{ }(?\@\,\%) & 9.45\,s\mbox{ }(5.8\,\%), timeout & 9.49\,s\mbox{ }(0\,\%), 133.81\,s\mbox{ }(?\@\,\%)\\\addlinespace[0.75em]
\multirow[t]{3}{2.75cm}[0pt]{Active set upper bound (SCS)}  & $t_\text{max}=20$\,s & 9.44\,s\mbox{ }(1.3\,\%), 35.66\,s\mbox{ }(?\@\,\%) & 9.67\,s\mbox{ }(5.8\,\%), 84.43\,s\mbox{ }(?\@\,\%) & 9.09\,s\mbox{ }(0\,\%), 28.76\,s\mbox{ }(?\@\,\%)\\
& $t_\text{max}=60$\,s & 9.44\,s\mbox{ }(1.3\,\%), 54.60\,s\mbox{ }(?\@\,\%) & 9.11\,s\mbox{ }(5.8\,\%), 155.51\,s\mbox{ }(?\@\,\%) & 9.29\,s\mbox{ }(0\,\%), 70.57\,s\mbox{ }(?\@\,\%)\\
& $t_\text{max}=240$\,s & 8.89\,s\mbox{ }(1.3\,\%), 75.43\,s\mbox{ }(?\@\,\%) & 9.01\,s\mbox{ }(5.8\,\%), timeout & 9.15\,s\mbox{ }(0\,\%), 177.64\,s\mbox{ }(?\@\,\%)
\end{tabular}
\end{adjustbox}

\bigskip

\caption{\label{tab:second_problems} The individual timings for each algorithm and choice of settings on problems (4)--(6). See \Cref{tab:first_problems} for a description of the quantities shown. Here, ``error'' means the solution was not obtained due to an error (such as running out of memory).}
\end{table}

\subsection{Entropic bounds} \label{sec:entropic_bounds}

In this section, we use the results of \Cref{sec:strategies} to obtain one-shot and asymptotic entropic bounds on the guesswork with side information in terms of measured versions of bounds known in the classical case. 

\subsubsection{One-shot bounds}

Arikan~\cite{Ari96} showed that 
\begin{equation} \label{eq:Arikan-1shot-bound}
\frac{1}{1 + \ln|\cX|} \exp(H_{\frac{1}{2}}^\uparrow(X|Y)_p) \leq G(X|Y)_p \leq \exp(H_{\frac{1}{2}}^\uparrow(X|Y)_p)
\end{equation}
where $H_{\alpha}^\uparrow(X|Y)_p$ for $\alpha\in (0,1)\cup(1,\infty)$ denotes the following $\alpha$-conditional entropy of a joint distribution $p_{XY}$ given by
\begin{equation}  \label{eq:def_CEalpha_classical}
H_{\alpha}^\uparrow(X|Y) =  \frac{\alpha}{1-\alpha}\ln \left( \sum_{y \in \cY} \left( \sum_{x\in \cX} p_{XY}(x,y)^\alpha \right)^{1/\alpha} \right) = \sup_{q_Y} \left[- D_\alpha( p_{XY} \| \one_X \otimes q_Y )\right]
\end{equation}
where the supremum is over probability distributions $q_Y$ on $\cY$, and $D_\alpha$ is the $\alpha$-R\'enyi relative entropy,
\begin{equation}
D_\alpha(p_X \| q_X) = \frac{1}{\alpha-1} \ln \left( \sum_x p_X(x)^{\alpha} q_X(x)^{1-\alpha} \right).
\end{equation}
The second equality of \eqref{eq:def_CEalpha_classical} follows from  of~\cite[Theorem~4]{FB14}.

Arikan's bound \eqref{eq:Arikan-1shot-bound} applies to each $G(X|Y)_p$ in \eqref{eq:Ec_q_from_c}, and hence by minimizing over the POVMs $\{E_y\}_{y\in \cY}$, we obtain
\begin{equation} \label{eq:1-shot-bounds}
\frac{1}{1 + \ln|\cX|} \exp(H_{\frac{1}{2}}^{\uparrow, M}(X|B)_{\rho}) \leq  G(X|B)_\rho \leq \exp(H_{\frac{1}{2}}^{\uparrow, M}(X|B)_\rho),
\end{equation}
where for $\alpha \in (0,1)\cup(1,\infty)$, $H_{\alpha}^{\uparrow, M}(X|B)_\rho$ is the $B$-measured conditional $\alpha$-R\'enyi entropy, defined by
\begin{equation}
H_{\alpha}^{\uparrow, M}(X|B)_\rho := \inf_{\{E_y\}_{y \in \cY}} H_{\alpha}^\uparrow(X|Y)_p,
\end{equation}
where the infimum is taken over POVMs $\{E_y\}_{y \in \cY}$ and $p_{XY}(x,y) = p_X(x) \tr[E_y \rho_B^x]$ is the joint probability distribution obtained by measuring the $B$ part of $\rho_{XB}$ via $\{E_y\}_{y \in \cY}$.

\begin{remark}
We may expand this quantity as
\begin{equation}
H_{\alpha}^{\uparrow, M}(X|B)_\rho = \inf_{\{E_y\}_{y \in \cY}}\sup_{q_Y} \left[ -D_{\alpha}(p_{XY} \| \one_X \otimes q_Y)\right],
\end{equation}
where $p_{XY}$ is induced by the measurement of $\rho_{XB}$.
This quantity seems to be different from the \emph{conditional entropy induced by the measured R\'enyi divergence}, namely
\begin{equation}
 H_{D_\alpha^M}^\uparrow(X|B)_\rho := \sup_{\sigma_B}-D^M_\alpha( \rho_{XB} \| \one_X \otimes \sigma_B),
\end{equation}
where the supremum is over states on the $B$ system, and for any pair of states $(\rho, \sigma)$,
\begin{equation}
D^M_\alpha(\rho\| \sigma) := \sup_{\{E_z\}_{z}} D_\alpha ( \{\tr[E_z \rho]\}_{z} \| \{\tr[E_z \sigma]\}_{z} )
\end{equation}
is the \emph{measured $\alpha$-R\'enyi divergence}. Indeed, the latter quantity may be expanded to obtain
\begin{equation}
 H_{D_\alpha^M}^\uparrow(X|B)_\rho = \sup_{\sigma_B} \inf_{\{E_z\}_{z}} \left[ -D_\alpha ( \{\tr[E_z \rho_{XB}]\}_{z} \| \{\tr[E_z \one_X \otimes \sigma_B]\}_{z} )\right].
\end{equation}
From the min-max inequality, and the fact that collective measurements on $XB$ can simulate measurements on $B$ alone, we have
\begin{equation}
 H_{D_\alpha^M}^\uparrow(X|B)_\rho  \leq H_{\alpha}^{\uparrow, M}(X|B)_\rho.
\end{equation}
\end{remark}
\subsubsection{Asymptotic analysis}
We can consider the asymptotic setting in which a Bob receives a sequence of product states $\rho_B^{\vec x} := \rho_B^{x_1} \otimes \dotsm \otimes \rho_B^{x_n}$, with probability $p_X(x_1)\dotsm p_X(x_n)$ and aims to guess the full sequence $\vec x = (x_1,\dotsc, x_n)$. In this case, the problem is characterized by the c-q state $\rho_{XB}^{\otimes n}$. The 1-shot bounds \eqref{eq:1-shot-bounds} give us
\begin{equation}
 -\frac{1}{n}\ln\left(1 + n\ln(|\cX|)\right) + \frac{1}{n}H_{\frac{1}{2}}^{\uparrow, M}(X^n | B^n)_{\rho^{\otimes n}} \leq   \frac{1}{n}\ln G(X^n|B^n)_{\rho^{\otimes n}} \leq  \frac{1}{n}H_{\frac{1}{2}}^{\uparrow, M}(X^n | B^n)_{\rho^{\otimes n}}
\end{equation}
where $H_{\frac{1}{2}}^{\uparrow, M}(X^n | B^n)_{\rho^{\otimes n}}$ can involve collective measurements on the system $B^n$. Taking $n\to \infty$, we obtain
\begin{equation} \label{eq:asymptotic-equality}
     \lim_{n\to\infty} \frac{1}{n}\ln G(X^n|B^n)_{\rho^{\otimes n}} = \lim_{n\to\infty} \frac{1}{n} H_{\frac{1}{2}}^{\uparrow, M}(X^n | B^n)_{\rho^{\otimes n}},
\end{equation}
assuming that the limit on the right-hand side exists.

Note that we can bound
\begin{align}
\frac{1}{n} H_{\frac{1}{2}}^{\uparrow, M}(X^n | B^n)_{\rho^{\otimes n}}
&\leq \frac{1}{n} \inf_{\{ E_{y} \}_{y\in\cY}} H_{\frac{1}{2}}^\uparrow(X^n|Y^n)_{p^{\otimes n}}\\
&=  \inf_{\{ E_{y} \}_{y\in\cY}} H_{\frac{1}{2}}^\uparrow(X|Y)_{p} \\
&= H_{\frac{1}{2}}^{\uparrow, M}(X | B)_{\rho}
\end{align}
where the first inequality follows from the fact that product measurements are a special case of collective measurements, and the first equality follows from the additivity of the classical R\'enyi entropy (Proposition 1 of~\cite{Ari96}), and the third by the definition of $H_{\frac{1}{2}}^{\uparrow, M}(X | B)_{\rho}$. Moreover, by the data-processing inequality \cite{FL13},
\begin{equation}
\label{eq:relate-ce-to-asymptotic-measured-ce}
\frac{1}{n} H_{\frac{1}{2}}^{\uparrow, M}(X^n | B^n)_{\rho^{\otimes n}} \geq \frac{1}{n} \widetilde H_{\frac{1}{2}}^{\uparrow}(X^n | B^n)_{\rho^{\otimes n}} = \widetilde H_{\frac{1}{2}}^{\uparrow}(X | B)_{\rho},
\end{equation}
where the conditional R\'enyi entropy
$\widetilde H_{\alpha}^{\uparrow}(C | D)_{\sigma}$ of a bipartite state $\sigma_{CD}$ is defined as
\begin{equation}
\widetilde H_{\alpha}^{\uparrow}(C | D)_{\sigma}
= \sup_{\omega_D} \left[-\widetilde D_{\alpha} (\sigma_{CD} \Vert \one_C \otimes \omega_D)\right],
\end{equation}
with the optimization with respect to states $\omega_D$
and the sandwiched R\'enyi relative entropy defined as \cite{MDS+13,WWY14}:
\begin{equation}
    \widetilde D_{\alpha}(X \Vert Y) = \frac{1}{\alpha-1}
    \ln \tr[ (Y^{(1-2\alpha)/\alpha} X Y^{(1-2\alpha)/\alpha})^\alpha].
\end{equation}
The equality in \eqref{eq:relate-ce-to-asymptotic-measured-ce} follows from the additivity of $\widetilde H_{\frac{1}{2}}^{\uparrow}$ under tensor products (see, {{e.g.\@}}, Corollary~5.2 of~\cite{Tom16}). Hence, we obtain
\begin{equation} \label{eq:asymptotic-bounds}
 \widetilde H_{\frac{1}{2}}^{\uparrow}(X | B)_{\rho} \leq \lim_{n\to\infty} \frac{1}{n}\ln G(X^n|B^n)_{\rho^{\otimes n}}  \leq  H_{\frac{1}{2}}^{\uparrow, M}(X | B)_{\rho}.
\end{equation}
In the classical case, \eqref{eq:cq-classical}, both the left and right-hand sides reduce to
\begin{equation}
H_{\frac{1}{2}}^{\uparrow}(X | Y)_{p}
\end{equation}
where $p$ is the underlying classical distribution of \eqref{eq:cq-classical}. Hence, these bounds recover Proposition 5 of~\cite{Ari96}. 

\chapter{General concave functions maximized over the \texorpdfstring{$\eps$}{epsilon}-ball} \label{sec:general_optimality_conditions}

The main results of sections \Cref{sec:geometry_trace_ball,sec:applications}, are essentially fully classical in that the first step of the proofs in the case of quantum states simply consists of a reduction to the case of probability vectors in $\R^d$. In this section, we prove \Cref{thm:short_optimality_conditions} which was used to motivate the construction of $\rho\majmin$ by finding necessary and sufficient conditions for maximizing concave functions over the $\eps$-ball. Moreover, we can treat---to some extent---functions which are not unitarily invariant and depend on the eigenprojections as well as the eigenvalues of a state. In particular, in \Cref{cor:main_thm_for_CE} we obtain necessary and sufficient conditions for a state $\xi$ to maximize the conditional entropy over the $\eps$-ball. Unfortunately, we will not be able to determine a formula for a specific maximizer, and we do not expect such a maximizer to have the universality of $\rho\majmin$ (which maximizes any Schur concave function over the $\eps$-ball $\Be(\rho)$). 
A similar approach of investigating the critical points of entropic quantities is undertaken in \cite{Teh20}.

\section{Tools from convex optimization}\label{tools-convex}
Let $\cX$ be a  real Hilbert space with inner product $\braket{\cdot,\cdot}$ and  $f: \cX\to \R \cup \{+\infty\}$ be a  function. In our applications, we  take $\cX = \Bsa(\H)$ equipped with the Hilbert-Schmidt inner product $\braket{\cdot,\cdot}_\text{HS}$. Let $\dom f = \{ x\in \cX: f(x) < \infty\}$ and assume $\dom f \neq \emptyset$. Let $\cX^* = \cB(\cX, \R)$ the set of bounded linear maps from $\cX$ to $\R$, equipped with the dual norm $\|\ell\|_* = \sup_{\|x\|=1} |\ell(x)|$ for $\ell \in \cX^*$. Since $\cX$ is a Hilbert space, by the Riesz-Fr\'echet representation, for each $\ell \in \cX^*$ there exists a unique $u_\ell \in \cX$ such that $\ell(x) = \braket{u_\ell, x}$ for all $x\in \cX$. We call $u_\ell$ the dual vector for $\ell$ (in particular the Hilbert-Schmidt dual in the case of $\cX = (\Bsa(\H), \braket{\cdot,\cdot}_\text{HS})$). We say $f$ is \emph{lower semicontinuous} if $\liminf_{x\to x_0} f(x) \geq f(x_0)$ for each $x_0\in \cX$. The \emph{directional derivative} of $f$ at $x\in \dom f$ in the direction $h\in \cX$ is given by
\begin{equation} \label{eq:def_dir_deriv}
  f'(x;h) = \lim_{t \downarrow 0} \frac{1}{t}[f(x+th) - f(x)].
\end{equation}
If $f$ is convex, this limit exists in $\R\cup\{\pm \infty\}$. If the map $\phi_x(h):= \cX\ni h \mapsto f'(x;h)$ is linear and continuous, then $f$ is called \emph{G\^ateaux-differentiable} at $x\in \cX$. Moreover, if $f$ is G\^ateaux differentiable at every $x\in A\subset \cX$, then $f$ is said to be G\^ateaux-differentiable on $A$. If $f$ is convex and continuous at $x$, it can be shown that the map $\phi_x$ is finite and continuous. However, a continuous and convex function may not be G\^ateaux-differentiable. For example, $f:\R\to \R$, $f(x) = |x|$ has $\phi_0(h) = \lim_{t\downarrow 0} \frac{1}{t}[ f(x+th) - f(x)] = \lim_{t\downarrow 0} \frac{1}{t}|th| = |h|$ which is nonlinear. If $f$ is G\^ateaux-differentiable at $x$, we call the dual to $\phi_x$ as the \emph{G\^ateaux gradient} of $f$ at $x$, written $\nabla f (x)\in \cX$. That is, $\braket{\nabla f(x), h} = \phi_x(h)$ for each $h\in \cX$.

The function $f$ is Fr\'echet-differentiable at $x$ if there is $y\in \cX$ such that
\begin{equation}
  \lim_{\|r\|\to 0} \frac{1}{\|r\|} | f(x+r) - f(x) - \braket{y,r}| = 0 \label{eq:def_Frechet_deriv}
\end{equation}
and in this case, one writes $D f(x)\in \cX^*$ for the map $\cX\ni z\mapsto D f(x) z = \braket{y,z}$.
If $f$ is Fr\'echet-differentiable at $x$, then by taking $r = t h$ in \eqref{eq:def_Frechet_deriv} we find $\braket{y,h} = f'(x,h)$ and therefore $f$ is G\^ateaux-differentiable at $x$ with $\braket{\nabla f(x),h}  = D f(x)h$.

By regarding differentiability as the existence of a linear approximation at a point, we can generalize it by defining a notion of a linear subapproximation at a point.
\begin{definition}
  The \emph{subgradient} of a function $f:\cX\to \R$ at $x$ is the set
  \begin{equation}\label{def:subgradient}
    \partial f (x) = \{ u \in \cX: f(y) - f(x) \geq \braket{u,y-x} \,\,\forall y\in \cX \} \subset \cX.
  \end{equation}
\end{definition}
For a convex function $f$, the following properties hold:
\begin{itemize}
  \item if $f$ is continuous at $x$, then $\partial f(x)$ is bounded and nonempty.
  \item if $f$ is G\^ateaux-differentiable at $x$, then $\partial f(x) = \{ \nabla f(x)\}$.
\end{itemize}
We briefly prove the second point here. Assume $f$ is G\^ateax-differentiable at $x$. Then
\[
  \braket{\nabla f (x), y-x} = f'(x,y-x) = \lim_{t \downarrow 0} \frac{1}{t}[f((1-t)x+ty) - f(x)].
\]
By convexity, $f((1-t)x + ty) \leq (1-t) f(x) + t f(y)$. Therefore, $\braket{\nabla f (x), y-x} \leq f(y) - f(x)$; hence, $\nabla f(x)\in \partial f(x)$.

On the other hand, given $u\in \partial f$, $h\in \cX$, and $t>0$, we can set $ y = th+x$. Then
\[
  \braket{u,th} = \braket{u,y-x} \leq f(y) -f(x)  = f(x+th) - f(x).
\]
Dividing by $t$ and taking the limit $t\downarrow 0$ yields $u = \nabla f(x)$.

Fermat's Rule of convex optimization theory, stated below, provides a simple characterization of the minimum of a function in terms of the zeroes of its subgradient. Moreover, since
the subgradient of a G\^ateaux differentiable function consists of a single element, namely, its G\^ateaux gradient, finding its minimizer amounts to showing that its G\^ateaux gradient
is equal to zero.
\begin{theorem}[Fermat's Rule] \label{thm:Global_minimizer_convex_fn}
  Consider a function $f: \cX\to \R\cup\{+\infty\}$ with $\dom f \neq \emptyset$. Then $\hat x\in \cX$ is a global minimizer of $f$ if and only if $0_{\cX} \in \partial f (\hat x)$, where $0_{\cX}$ is the zero vector of $\cX$.
\end{theorem}
\begin{proof}
  $0_{\cX}\in \partial f(\hat x)$ if and only if
  \[
    f(y) - f(\hat x) \geq \braket{0,y-\hat x} = 0
  \]
  for every $y \in \cX$,
  i.e.\@ if and only if $\hat x$ is a global minimizer of $f$.
\end{proof}

The following result proves useful in computing the subgradient of a sum of convex functions (see e.g.\@ \cite[Theorem 3.30]{Peypouquet2015} for the proof).
\begin{theorem}[Moreau-Rockafellar] \label{thm:Moreau-Rockafellar}
  Let $f,g: \cX\to \R\cup \{+\infty\}$ be convex and lower semicontinuous, with non-empty domains. For each $A\in \cX$, we have
  \begin{equation}
    \partial f(A) + \partial g(A) := \{a+b: a\in \partial f(A), b\in \partial g(A) \} \subset \partial(f+g)(A).
  \end{equation}
  Equality holds for every $A\in \cX$ if $f$ is continuous at some $A_0\in \dom(g)$.
\end{theorem}

\section{G\^ateaux-differentiable functions \label{sec:classF}}
Let $\cF$ denote the class of functions $\varphi:\cD(\cH) \to \RR$ which are concave and continuous, and G\^ateaux-differentiable on $\cD_+(\cH)$. Note that these functions need \emph{not} be Schur concave, and in particular, need not be unitarily invariant. These include the following:
\begin{itemize}
  \item The von Neumann entropy $\xi \mapsto S(\xi):= -\tr(\xi \log \xi)$. In \Cref{lem:Lrho_VN}, we show for $\xi>0$,
        \[
          \nabla S (\xi) = - \log \xi - \tfrac{1}{\log_{\mathrm{e}}(2)}\one.
        \]
  \item The conditional entropy $\xi_{AB} \mapsto S(A|B)_\xi := S(\xi_{AB}) - S(\xi_B)$, for which
        \[
          \nabla S(A|B) (\xi_{AB})=  - (\log \xi_{AB} - \one_A\otimes \log \xi_B)
        \]
        as shown in \Cref{cor:Lrho_CE}. Note that for $\varphi(\xi) := S(A|B)_\xi$, the conditional entropy satisfies the interesting property that
        \begin{equation}
          \varphi(\xi) = \braket{\nabla \varphi (\xi),\xi}_\text{HS}.
        \end{equation}
  \item The $\alpha$-R\'enyi entropy for $\alpha \in (0,1)$. The map $\xi\mapsto S_\alpha(\xi)$ is concave for $\alpha\in (0,1)$, continuous, and has G\^ateaux gradient
  \[
    \nabla S_\alpha (\xi) = \frac{\alpha}{1-\alpha} \frac{1}{\ln 2} \frac{1}{\tr [\xi^\alpha]} \xi^{\alpha-1}
  \]
  by \Cref{lem:grad_Renyi_ent}.
  Note that the $\alpha$-R\'enyi entropy for $\alpha>1$ is not concave.
  \item The function $-T_\alpha$, for $\xi \mapsto T_\alpha(\xi):= \tr [\xi^\alpha]$ and $\alpha > 1$. This map is concave for $\alpha>1$, continuous, and by \Cref{lem:grad_Renyi_ent},  has G\^ateaux gradient
  \[
    \nabla ({-T_\alpha})(\xi) = -\alpha \xi^{\alpha-1}.
  \]
  A state $\xi^* \in \Be(\rho)$ maximizes $-T_\alpha$ over $\Be(\rho)$ if and only if $\xi^*$ minimizes $T_\alpha$ over $\Be(\rho)$. For $\alpha>1$, minimizing $T_\alpha$ is equivalent to maximizing $S_\alpha(\xi)$, using that $x\mapsto \frac{1}{1-\alpha}\log x $ is decreasing. Therefore, $\xi^*$ maximizes $-T_\alpha$ over $\Be(\rho)$ if and only if it maximizes $S_\alpha$ over $\Be(\rho)$. Thus, by considering the function $f(\xi)  =-T_\alpha(\xi)$ for $\alpha>1$ in \Cref{thm:short_optimality_conditions}, one can establish conditions for $\xi^*$ to maximize the $\alpha$-R\'enyi entropy for $\alpha>1$.

\end{itemize}

\section{Optimality conditions for maximizing concave functions over the trace-ball \label{sec:optimality_conditions}}

\Cref{thm:short_optimality_conditions} is a condensed form of the following theorem, whose proof we include below.
\begin{theorem} \label{thm:optimality_conditions}
  Let $f:\cD(\cH) \to \RR$ be a concave, continuous function which is G\^ateaux-differentiable on $\cD_+(\cH)$.  Given a state $\rho\in \cD(\cH)$ and $\eps> 0$, for any $\xi\in \Bep(\rho)$, we have
  \begin{equation} \label{eq:optimality_condition_ineq}
    \tr [\nabla f(\xi) (\xi-\rho)] \leq \eps [\lambda_+(\nabla f(\xi)) - \lambda_-(\nabla f(\xi))].
  \end{equation}
  Furthermore, the following are equivalent:

  \begin{enumerate}
    \item Equality in \eqref{eq:optimality_condition_ineq},
    \item $\xi \in \argmax_{\Be(\rho)} f$,
    \item
          \begin{enumerate}
            \item Either $\frac{1}{2}\|\xi-\rho\|_1= \eps$ or $\nabla f(\xi) = \lambda \one$ for some $\lambda \in \R$, and
            \item  we have
                  \[
                    \pi_\pm \nabla f(\xi) \pi_\pm = \lambda_\pm(\nabla f(\xi)) \pi_\pm
                  \] where $\pi_\pm$ is the projection onto the support of $\Delta_\pm$, and where $\Delta= \Delta_+ - \Delta_-$ is the Jordan decomposition of $\Delta:= \xi-\rho$.
          \end{enumerate}
  \end{enumerate}

\end{theorem}
A corollary of this theorem concerns the conditional entropy
\[
 S(A|B)_\xi := S(AB)_\xi - S(B)_\xi \equiv S(\xi_{AB}) - S(\xi_B), \qquad \xi_B := \tr_A \xi_{AB}
 \] of a bipartite state $\xi_{AB}$. It corresponds to the choice $\varphi(\xi_{AB}) = S(A|B)_\xi$ whose G\^ateaux gradient $L_\xi := \nabla \varphi(\xi_{AB})$ is given by $L_{\xi_{AB}} = - (\log \xi_{AB} -\one_A \otimes \log \xi_B) $ (see \Cref{cor:Lrho_CE}).

    \begin{corollary} \label{cor:main_thm_for_CE}
    Given a state $\rho_{AB}\in \cD_+(\cH_A\otimes\cH_B)$ and $\eps> 0$, a state $\xi_{AB}\in \Bep(\rho_{AB})$ has maximum conditional entropy if and only if
    \[
    S(A|B)_\xi - S(A|B)_\rho + D(\rho_{AB}\|\xi_{AB}) - D(\rho_B\|\xi_B) = \eps (\lambda_+(L_\xi) - \lambda_-(L_\xi))
    \]
    where $L_\xi = \one_A \otimes \log \xi_B-\log \xi_{AB}  $.

    \end{corollary}

We first prove the inequality \eqref{eq:optimality_condition_ineq} by considering the Jordan decomposition of $\Delta:=\xi-\rho$. Next, we convert the constrained optimization problem $\max_{\xi\in\Be(\rho)} f(\xi)$ to an unconstrained optimization problem $\min h$ for a non-G\^ateaux differentiable function $h$ defined on all of $\Bsa(\H)$ by adding appropriate indicator functions for the sets $\cD(\cH)$ and $\{A\in \Bsa(\H): \frac{1}{2}\|A-\rho\|_1 \leq \epsilon \}$.

The Moreau-Rockafellar Theorem (\Cref{thm:Moreau-Rockafellar}) allows us to compute $\partial h(\xi)$ in terms of $\partial f(\xi) = \{\nabla f(\xi)\}$ and the subgradients of the indicator functions. We then show $0\in \partial h(\xi)$ if and only if equality is achieved in \eqref{eq:optimality_condition_ineq}.

\paragraph{Proof of \Cref{thm:optimality_conditions}}
We start with the following general result, which does not use convex optimization.

\begin{lemma}
  Let $\eps>0$ and $\Delta\in\Bsa$ with $\tr (\Delta) = 0$, $\frac{1}{2}\|\Delta\|_1 \leq \eps$. Let $L\in \Bsa$. Then
  \begin{equation} \label{eq:UB_LDelta}
    \tr( L \Delta ) \leq \eps (  \lambda_+(L)  -  \lambda_-(L) )
  \end{equation}
  with equality if and only if
  \begin{enumerate}
    \item Either $\frac{1}{2}\|\Delta\|_1= \eps$ or else $L = \lambda \one$ so that $\lambda:= \lambda_+(L) = \lambda_-(L)$, and
    \item  we have
          \[
            \pi_\pm L \pi_\pm = \lambda_\pm(L) \pi_\pm
          \] where $\pi_\pm$ is the projection onto the support of $\Delta_\pm$, and where $\Delta= \Delta_+ - \Delta_-$ is the Jordan decomposition of $\Delta$.
  \end{enumerate}
\end{lemma}
\begin{proof}
  Note $|\Delta| = \Delta_+ + \Delta_-$, and since $\Delta$ has trace zero, $\tr(\Delta_+) = \tr(\Delta_-)$.
  We expand
  \[
    \tr(L \Delta) = \tr(L \Delta_+)- \tr(L \Delta_-).
  \]
  Now, we use that $L$ is self-adjoint so that
  \[
    \lambda_-(L)\one \leq L \leq  \lambda_+(L)\one.
  \]
  Since $\Delta_\pm\geq 0$, we have
  \begin{align}
    \tr(L \Delta_+) - \tr(L \Delta_-)
     & \leq \tr(L \Delta_+) - \tr( \lambda_-(L)\Delta_-) \label{eq:lambda_min_ieq}            \\
     & \leq \tr(\lambda_+(L) \Delta_+) - \tr( \lambda_-(L)\Delta_-) \label{eq:lambda_max_ieq} \\
     & = (\lambda_+(L) - \lambda_-(L) )\tr(\Delta_+), \nonumber
  \end{align}
  where the last line follows from the fact that $\tr(\Delta_+) =\tr(\Delta_-)$. However, $\tr(\Delta_+)\leq \eps$, so we have
  \begin{equation} \label{eq:epsilon_ieq}
    (\lambda_+(L) - \lambda_-(L) )\tr(\Delta_+)  \leq \eps (  \lambda_+(L)  -  \lambda_-(L) ).
  \end{equation}
  Thus, \eqref{eq:UB_LDelta} follows.
  To obtain equality in (\ref{eq:UB_LDelta}), we require equality in \eqref{eq:lambda_min_ieq}, \eqref{eq:lambda_max_ieq} and \eqref{eq:epsilon_ieq}. Equality in \eqref{eq:epsilon_ieq} is equivalent to condition $1$ in the statement of the lemma. We now show that equality in \eqref{eq:lambda_min_ieq} and \eqref{eq:lambda_max_ieq} is equivalent to condition $2$.

  Set $\lambda_+ = \lambda_+(L)$. Then since $\pi_+$ is the projection onto the support of $\Delta_+$, we have $\Delta_+ = \pi_+\Delta_+\pi_+$. Using cyclicity of the trace, we obtain
  \[
    \tr(L \Delta_+) = \tr(L \pi_+ \Delta_+\pi_+ ) = \tr(\pi_+ L \pi_+ \Delta_+).
  \]
  Equality in \eqref{eq:lambda_max_ieq} implies $\tr(L \Delta_+) = \lambda_+ \tr(\Delta_+)$, so
  \[
    \tr((\pi_+ L \pi_+)\Delta_+) = \tr(\lambda_+ \Delta_+)
  \]
  and thus
  \begin{equation}
    \tr(\Delta_+ (\lambda_+ \pi_+ - \pi_+ L \pi_+ )) = 0. \label{eq:tr_product_zero_proof}
  \end{equation}
  Since $\lambda_+$ is the largest eigenvalue of $L$ which is self-adjoint, we have $L \leq \lambda_+ \one$. Since conjugating by any operator preserves the semidefinite order, we have $\lambda_+ \pi_+ - \pi_+ L \pi_+ \geq 0$. Then since $\Delta_+$ restricted to $\pi_+$ is positive definite, \eqref{eq:tr_product_zero_proof} implies $\pi_+ L \pi_+ = \lambda_+\pi_+$.  Similarly, equality in \eqref{eq:lambda_min_ieq} implies $\pi_- L \pi_- = \lambda_-(L) \pi_-$.

  Conversely, if $\pi_\pm L\pi_\pm = \lambda_\pm \pi_\pm$,  we immediately obtain equality in \eqref{eq:lambda_min_ieq} and \eqref{eq:lambda_max_ieq}.
\end{proof}

This lemma with the choices $\Delta = \xi-\rho$ and $L \equiv L_\xi := \nabla f(\xi)$ gives the inequality \eqref{eq:optimality_condition_ineq} of \Cref{thm:optimality_conditions}. It also gives the equivalence between the conditions $1$ and $3$ of the theorem.

We now turn to the theory of convex optimization to establish the equivalence between conditions $1$ and $2$.
Recall $f: \cD \to \R$ is a continuous, concave function which is G\^ateaux-differentiable on $\cD_+$. Let us write $\tilde{f} = -f$ which is convex, and note $L_\xi =- \nabla \tilde{f} (\xi)$.
With this choice, it remains to be shown that  $\xi^* \in \argmin_{\xi\in \Be(\rho)} \tilde{f}(\xi)$ if and only if
\begin{equation}
  \tr (L_{\xi^*} (\xi^* -\rho)) \geq  \eps (\lambda_+(L_{\xi^*}) - \lambda_-(L_{\xi^*})). \label{eq:remains_to_show}
\end{equation}

The Tietze extension theorem (e.g.\@~\cite[Theroem 2.2.5]{simon2015a}) allows one to extend a bounded continuous function defined on a closed set of a normal topological space (such as a normed vector space) to the entire space, while preserving continuity and boundedness.  We use this to extend $\tilde{f}$ (which is bounded as it is a continuous function on the compact set $\cD$) to the whole space $\Bsa$, using that $\cD$ is a closed set in $\Bsa$. We consider the closed and convex sets $\cD$ and
\begin{equation}
  T := \left\{ A \in \Bsa: \|A - \rho_{AB}\|_1 \leq 2 \eps \right\},\\
\end{equation}
and note $\Be = \cD\cap T$.
We define $h: \Bsa\to \R\cup \{\infty\}$ as
\[
  h =   \tilde{f} + \delta_{\cD} + \delta_T
\]
where for $S\subset \Bsa$, the indicator function $\delta_S(A) = \begin{cases}
    0       & A\in S            \\
    +\infty & \text{otherwise}.
  \end{cases}$

We have the important fact that, by construction,
\begin{equation}
  \argmin_{\omega\in\Be(\rho)} \tilde{f}(\omega) = \argmin_{A\in \Bsa} h(A).
\end{equation}

By Theorem \ref{thm:Global_minimizer_convex_fn} $\hat A\in \Bsa$ is a global minimizer of $h$ if and only if $0\in \partial h(\hat A)$. Note each of $\tilde{f}, \delta_{\cD}, \delta_T$ is lower semicontinuous, convex, and has non-empty domain. Moreover, both $\tilde{f}$ and $\delta_T$ are continuous at any faithful state  $\omega \in \Bep(\rho) \subset \dom \delta_{\cD}$. Hence, by \Cref{thm:Moreau-Rockafellar},
\[
  \partial h = \partial \tilde{f} + \partial \delta_{\cD} + \partial \delta_T := \{\ell_f + \ell_{\cD} + \ell_T: \ell_f \in \partial \tilde{f},\, \ell_{\cD} \in \partial \delta_{\cD}, \,\ell_T \in \partial\delta_T\}.
\]
Hence, to obtain a complete characterization of
\[
  \argmin h = \{A\in \Bsa: 0\in \partial h(A)\}
\]
we need to evaluate the subgradients of $\tilde{f}$, $\delta_T$, and $\delta_{\cD}$.

Since $\tilde{f}$ is G\^ateaux-differentiable on $\cD_+$, for any $\omega\in \cD_+$, we have $\partial \tilde{f}(\omega) = \{-L_\omega\}$ for $L_\omega := -\nabla \tilde{f}(\omega)$.  The following two results (proven in \Cref{sec:proof_of_lemmas_convex}) describe the other two relevant subgradients.

\begin{lemma} \label{lem:subgradT}
  For $A\in \Bsa$ with $0 < \frac{1}{2}\|A - \rho_{AB}\|_1 \leq \eps$,
  \[
    \partial \delta_{T}(A)  =  \left\{ u \in \Bsa:  2 \eps \|u\|_\infty = \braket{u,A - \rho_{AB}}  \right\}
  \]
  where
  \[
    \|u\|_\infty :=\lambda_+(|u|)= \sup_{A\in \Bsa,\, \|A\|_1=1} |\!\braket{u,A}\!|.
  \]
\end{lemma}

\begin{lemma} \label{lem:subgradD}
  Let $\omega \in \cD_+$. Then $\partial \delta_{\cD}(\omega) = \left\{ x \one: x\in \R \right\}$.
\end{lemma}

Putting together these results, we have, for $\xi^* \in \Bep(\rho)$,
\[
  0 \in \partial h(\xi^*) \iff 0 = -L_{\xi^*} + G + x \one
\]
for some $G\in \partial \delta_T({\xi^*})$ and $x\in\R$, where $G$ satisfies
\[
  2 \eps \|G\|_\infty=  \tr(G (\xi^* - \rho)).
\]
Then $G =  L_{\xi^*} - x \one$, which implies that
\[
  \tr( L_{\xi^*} (\xi^* - \rho)) = 2 \eps \|L_{\xi^*} - x \one\|_\infty .
\]
Set $\alpha = \frac{\tr( L_{\xi^*} (\xi^* - \rho))}{2 \eps}.$
Note $\alpha$ and $L_{\xi^*}$ depend on $\xi^*$. Then we have
\[
  0\in \partial h(\xi^*) \iff \exists\,  x\in \R: \alpha = \|L_{\xi^*}-x \one\|_\infty.
\]
Since $\|L_{\xi^*}-x \one\|_\infty = \max_{\lambda \in \spec L_{\xi^*}} |\lambda- x|$, we have that $0\in \partial h({\xi^*})$ if and only if
\begin{equation}
  \exists\, x\in \R: \alpha =  \max_{\lambda \in \spec L_{\xi^*}} |\lambda- x|. \label{eq:cond_x}
\end{equation}
Let
\begin{equation}
  q(x) = \max_{\lambda \in \spec L_{\xi^*}} |\lambda - x| \label{eq:def_qx}
\end{equation}
We immediately see that $q$ is continuous, $q(0) = \lambda_+(|L_{\xi^*}|)$, and $\lim_{z\to\pm\infty} q(z) = +\infty$. In fact, we can write a simple form for $q(x)$ as the following lemma, which is proven in \Cref{sec:proof_of_lemmas_convex}, shows. Set $\lambda_+ \equiv \lambda_+(L_{\xi^*})$ and $\lambda_- \equiv \lambda_-(L_{\xi^*})$ in the following.
\begin{lemma} \label{lem:q(x)}
  The quantity $q(x)$ defined by \eqref{eq:def_qx} can be expressed as follows:
  \[
    q(x) = \frac{\lambda_+ - \lambda_-}{2} + \left| \frac{\lambda_+ + \lambda_-}{2} - x \right|.
  \]
\end{lemma}

Thus, Lemma \ref{lem:q(x)} implies that the range of the function $q$ is $[\frac{\lambda_+ - \lambda_-}{2}, \infty)$. Hence, \eqref{eq:cond_x} holds if and only if
\[
  \frac{\lambda_+ - \lambda_- }{2} \leq \alpha.
\]
Substituting  $\alpha = \frac{\tr( L_{\xi^*} (\xi^* - \rho))}{2 \eps}$ in the above expression yields \eqref{eq:remains_to_show} and therefore concludes the proof of Theorem \ref{thm:optimality_conditions}.\qed

\bigskip

Below, we collect some results (proven in \Cref{sec:proof_of_lemmas_convex}) relating to the G\^ateaux gradients of relevant entropic functions which are candidates for $f$ in \Cref{thm:optimality_conditions}.
\begin{lemma}\label{lem:Lrho_VN}
  The von Neumann entropy $\xi\mapsto S(\xi):=-\tr (\xi \log \xi)$ is G\^ateaux-differentiable at each $\xi >0$ and $\nabla S(\xi) = - \log \xi - \tfrac{1}{\log_{\mathrm{e}}(2)}\one$.
\end{lemma}

\begin{corollary}\label{cor:Lrho_CE}
  The conditional entropy $\xi_{AB}\mapsto S(A|B)_\xi:=S(\xi_{AB}) - S(\xi_B)$ is G\^ateaux-differentiable at each $\xi_{AB} >0$ and $\nabla S(A|B)(\xi_{AB}) =-(\log \xi_{AB} - \one_A\otimes \log\xi_B)$.
\end{corollary}

\begin{lemma}\label{lem:grad_Renyi_ent}

  For $\alpha\in (0,1)\cup (1,\infty)$, the map $T_\alpha$ and the R\'enyi entropy $S_\alpha$ are G\^ateaux-differentiable at each $\xi \in \cD_+$, with G\^ateaux gradients
  \[
    \nabla T_\alpha(\xi) = \alpha \xi^{\alpha-1}, \qquad
    \nabla S_\alpha (\xi) = \frac{\alpha}{1-\alpha}\frac{1}{\ln 2}\frac{1}{\tr [\xi^\alpha]} \xi^{\alpha-1}.
  \]
\end{lemma}
\section{Proofs of the remaining lemmas and corollaries \label{sec:proof_of_lemmas_convex}}

\paragraph{Proof of \Cref{cor:main_thm_for_CE}}
We write the conditional entropy as the map
\begin{equation}
  \begin{aligned}
    S(A|B)_\cdot\quad: \quad & \cD(\H_{AB}) \to \R     \\
                             & \xi \mapsto S(A|B)_\xi.
  \end{aligned} \label{eq:CE_as_map}
\end{equation}
By \Cref{cor:Lrho_CE}, for any $\xi_{AB}>0$ we have
$$\nabla S(A|B) (\xi) =  - (\log \xi_{AB} - \one_A\otimes \log \xi_B) = - L_\xi.$$
Substituting this in the left-hand side of \eqref{eq:optimality_condition_ineq}, we obtain
\begin{align*}
  \tr (L_\xi (\rho_{AB} - \xi_{AB})) & =\tr[\rho_{AB} \log \xi_{AB}] -\tr[\xi_{AB}\log \xi_{AB}] - \tr[\rho_B\log \xi_B] + \tr[\xi_B\log \xi_B] \\
                                     & = S(A|B)_\xi + \tr[\rho_{AB} \log \rho_{AB}] - \tr[\rho_{AB} (\log \rho_{AB} - \log \xi_{AB})]           \\
                                     & \qquad\qquad\quad - \tr[\rho_B\log\rho_B] + \tr[\rho_B(\log \rho_B - \log \xi_B)]                        \\
                                     & = S(A|B)_\xi - S(A|B)_\rho - D(\rho_{AB}\|\xi_{AB}) + D(\rho_B\|\xi_B).
\end{align*}
The statement of \Cref{cor:main_thm_for_CE} then follows from \Cref{thm:optimality_conditions}.\qed

\paragraph{Proof of Lemma \ref{lem:subgradT}}

Let $\cA := \left\{ u \in \Bsa:  2 \eps \|u\|_\infty = \braket{u, A - \rho_{AB}})  \right\}$ and set
\[
  C :=\{ B\in \Bsa: \| B -\rho_{AB} \|_1 \leq \eps\}\subset \Bsa.
\]
Let $u \in \cA$. Then for $y\in C$,
\begin{align*}
  \braket{u,y - A} & = \braket{u,y - \rho_{AB}} + \braket{u,\rho_{AB} - A} = \braket{u,y - \rho_{AB}} - 2 \eps \|u\|_\infty \\
                   & \leq \|u\|_\infty \|y-\rho_{AB}\|_1 - 2 \eps \|u\|_\infty                                              \\
                   & = \|u\|_\infty (\|y-\rho_{AB}\|_1 - 2 \eps) \leq 0
\end{align*}
and thus $u \in \partial \delta_T(A)$.
On the other hand, take $u \in \partial \delta_T(A)$. Then
\[
  \|u\|_\infty = \sup_{\|x\|_1 = 1}|\!\braket{u,x}\!|
\]
is achieved at some $x^* \in \Bsa$ since the set $\{x\in \Bsa: \|x\|_1 = 1\}$ is compact, using that $\Bsa$ is finite-dimensional. Then $y = \rho_{AB} + 2\eps x \in C$. Hence,
\[
  0 \geq \braket{u,y-A} = \braket{u, \rho_{AB} - A} + 2 \eps \braket{u,x} = \braket{u,\rho_{AB} - A} + 2 \eps \|u\|_\infty
\]
and thus $\braket{u,A- \rho_{AB}}  \geq   2 \eps \|u\|_\infty$.
Then by the bound
\[
  \braket{u,A-\rho_{AB}} \leq \|u\|_\infty \|A-\rho_{AB}\|_1 \leq 2 \eps \|u\|_\infty
\] we have $u \in \cA$.\qed

\paragraph{Proof of Lemma \ref{lem:subgradD}}

By definition,
\begin{align*}
  \partial \delta_{\cD}(\omega) & = \left\{ u\in \Bsa: \braket{u, y-\omega} \leq 0 \, \text{ for all }y\in \cD  \right\}.
\end{align*}
Since $u$ is self-adjoint, we can write its eigen-decomposition as $u = \sum_{k=1}^d \alpha_k \ketbra{k}{k}$. If $\alpha_k = \alpha_j$ for each $j,k$, then $u \propto \one$. Conversely, $\alpha_k \one \in \partial \delta_{\cD}(\omega)$ since $\tr[\alpha_k \one (y-\omega)] = \alpha_k (\tr(y)  - \tr (\omega)) = 0$.

Otherwise, assume $\alpha_k > \alpha_j$ for some $k,j\in \{1,\dotsc, d\}$.
Let
\[
  y := \sum_{i, \, i\neq k,\, i \neq j} \braket{i| \omega| i} \ketbra{i}{i} + ( \braket{k| \omega| k} + \braket{j| \omega| j} )\ketbra{k}{k} \in \Bsa.
\]
Then $\tr (y) = \sum_i \braket{i| \omega| i}  = \tr (\omega) = 1$, and $y\geq 0$ since all of its eigenvalues are non-negative. Note $\ket{j}$ is an eigenvector of $y$ with eigenvalue zero. Next,
\begin{align*}
  \tr(u ( y -\omega)) & = \sum_i \braket{i| (y-\omega) |i}                                                                                              \\
                      & = \sum_{i, \, i\neq k,\, i \neq j} (\braket{i| \omega| i} - \braket{i|\omega| i})                                               \\
                      & \qquad +\alpha_k ( \braket{k| \omega| k} + \braket{j| \omega| j}  - \braket{k|\omega| k}) + \alpha_j ( 0 -\braket{j|\omega| j}) \\
                      & =(\alpha_k - \alpha_j)\braket{j|\omega| j} > 0.
\end{align*}
Thus, $u \not \in \partial \delta_{\cD} (\omega)$.\qed

\paragraph{Proof of Lemma \ref{lem:q(x)}}
First, we establish
\begin{equation}
  q(x) = \max(|\lambda_+- x|,|\lambda_--x|). \label{eq:formq1}
\end{equation}
Given $\lambda\in \spec L$ and $x\in \R$, if $x < \lambda$, then
\[
  |x- \lambda| = \lambda - x \leq \lambda_+ - x = |\lambda_+-x|
\]
and otherwise
\[
  |x- \lambda| = x-\lambda  \leq x -\lambda_- = |\lambda_--x|,
\]
yielding \eqref{eq:formq1}.
Next, set $r := \frac{\lambda_+ - \lambda_-}{2}$ and $m := \frac{\lambda_+ + \lambda_-}{2}$. Then $\lambda_\pm =  m \pm r$. Therefore, for $x\in \R$,
\[
  |\lambda_\pm - x| = |m \pm r -x| \leq r + |m - x|,
\]
yielding $q(x) = r + |m-x|$ as claimed. \qed

\paragraph{Proof of Lemma \ref{lem:Lrho_VN}}

Let us introduce the Cauchy integral representation of an analytic function.
If $g$ is analytic on a domain $G\subset \bC$ and $A$ is a matrix with $\spec A \subset G$, then we can write
\[
  g(A) = \frac{1}{2\i \pi} \int_\Gamma g(z) (z \one - A)\inv \d z,
\]
and
\begin{equation}
  g'(A)= \frac{1}{2\i \pi} \int_\Gamma g(z) (z \one - A)^{-2} \d z, \label{eq:gprimeA}
\end{equation}
where $\Gamma\subset G$ is a simple closed curve with $\spec A\subset \Gamma$, and $g': G\to \bC$ is the derivative of $g$. \cite{STICKEL1987} shows that with these definitions, the Fr\'echet derivative of $g$ at $A$ exists, and when applied to a matrix $X$ yields
\begin{equation*}
  D(g)(A)X = \frac{1}{2\i \pi}\int_\Gamma g(z) (z \one - A)\inv X (z\one - A)\inv \d z. \label{eq:DgA_integral_rep}
\end{equation*}
Therefore, by cyclicity of the trace,
\begin{equation} \label{eq:trace-Frechet-deriv-formula}
  \tr(D(g)(A)X) = \tr \left[\frac{1}{2\i \pi}\int_\Gamma g(z) (z \one - A)^{-2} \d z\, X\right] = \tr(g'(A) X)
\end{equation}
using \eqref{eq:gprimeA} for the second equality.

We may write the von Neumann entropy as the map
\[
  S: \cD \to \R, \qquad \xi \mapsto S(\xi) = - \tr[ \xi \log \xi].
\]
Then we may write $S = \tr \circ \eta$ for $\eta(x) = - x \log x$ which is analytic on $\{\zeta\in \bC: \re \zeta > 0 \}$, with derivative $\eta'(x) = - \log x - \frac{1}{\log_{\mathrm{e}}(2)}$. Then for $\xi > 0$,
\[
  D(S)(\xi) = D(\tr \circ \eta)(\xi) = D(\tr)(\eta(\xi)) \circ D(\eta)(\xi).
\]
by the chain rule for Fr\'echet derivatives. By the linearity of the trace, $D(\tr)B(X) = \tr(X)$ for any $X,B\in \Bsa$. So for $X\in \Bsa$,
\[
  D(S)(\xi) X  = \tr[D(\eta)(\xi)X].
\]
By \eqref{eq:trace-Frechet-deriv-formula} for $g = \eta$, we have
\begin{equation*}
  D(S)(\xi)X = \tr[\eta'(\xi) X] = \tr [ (- \log \xi - \tfrac{1}{\log_{\mathrm{e}}(2)}\one)X].\qed \end{equation*}

\paragraph{Proof of Corollary \ref{cor:Lrho_CE}}

We decompose the conditional entropy map (\Cref{eq:CE_as_map}) by writing
\[
  S(A|B)_\cdot = S_{AB}(\cdot) - S_B \circ \tr_A (\cdot): \quad \D(H_{AB})\to \R
\]
where we have indicated explicitly the domain of each function in the subscript. That is, $S_{AB}: \D(\H_{AB})\to \R$ is the von Neumann entropy on system $AB$, and $S_B:\D(\H_B)\to \R$ is the von Neumann entropy on $B$.

Since $\tr_A: \Bsa(\H_{AB})\to \Bsa(\H_B)$ is a bounded linear map and $S_B$ concave and continuous, the chain rule for the composition with a linear map (see Prop. Prop.~3.28 of \cite{Peypouquet2015}) gives
\[
  \nabla (S_B\circ \tr_A)(\omega_{AB}) = \tr_A^* \circ \nabla S_B  (\omega_{B})= \one_A \otimes \nabla S_B(\omega_B)
\]
for any $\omega_{AB}\in \D(\H_{AB})$, where $\tr_A^*$ is the dual to the map $\tr_A$.
As $\nabla S_B(\omega_B) = - (\log \omega_B+\tfrac{1}{\log_{\mathrm{e}}(2)} \one_B)$, we have
\begin{align*}
  \nabla S(A|B) (\omega_{AB}) & = \nabla S_{AB} (\omega_{AB}) - \nabla (S_{B}\circ \tr_A)(\omega_{AB})                                                                   \\
                                    & = - (\log \omega_{AB} + \tfrac{1}{\log_{\mathrm{e}}(2)}\one_{AB})+ \one_A\otimes (\log \omega_B + \tfrac{1}{\log_{\mathrm{e}}(2)}\one_B) \\
                                    & = -(\log \omega_{AB} -\one_A\otimes \log \omega_B). \qed                                                                                 \end{align*}

\paragraph{Proof of Lemma~\ref{lem:grad_Renyi_ent}}

The $\alpha$-R\'enyi entropy $\alpha\in(0,1)\cup(1,\infty)$ can be described by the map
\[
  S_\alpha: \cD \mapsto \R, \qquad \xi \mapsto S_\alpha(\xi) :=  \frac{1}{1-\alpha} \log \tr (\xi^\alpha).
\]
Let us use the notation $\pow_\alpha : \Bsa\to \Bsa$, $A\mapsto \pow_\alpha(A) := A^\alpha$. Then
\[
  T_\alpha= \tr \circ \pow_\alpha
\]
and $(1-\alpha) S_\alpha = \log \circ T_\alpha$.
For $\xi \in \cD_+$, the function $\pow_\alpha$ is analytic on $\{\zeta\in \bC: \re \zeta > 0\}$. As in the proof of \Cref{lem:Lrho_VN} then, using the chain rule and linearity of the trace we find
\begin{equation*}
  D(T_\alpha)(\xi) = \tr [D(\pow_\alpha)(\xi) X].
\end{equation*}
Invoking \eqref{eq:trace-Frechet-deriv-formula} for $g(z) = z^\alpha$, with $\Gamma \subset \{\zeta\in \bC: \re \zeta > 0\}$  a simple closed curve enclosing the spectrum of $\xi$, we have
\begin{equation}
  D(T_\alpha)(\xi)X = \alpha \tr (\xi^{\alpha-1} X).
\end{equation}
Moreover,
\begin{align*}
  D(S_\alpha)(\xi)X & = \frac{1}{1-\alpha}D(\log \circ T_\alpha) (\xi)                                 \\
                    & = \frac{1}{1-\alpha} D(\log)(\tr (\xi^\alpha)) \circ D(T_\alpha)(\xi)            \\
                    & =  \frac{1}{1-\alpha}\frac{1}{\ln 2}\frac{\alpha}{\tr [\xi^\alpha]} \tr(\xi^{\alpha-1 }X). \qed \end{align*}

\chapter{Open questions from Part I}
\begin{enumerate}
	\item \Cref{sec:other-p-norms} shows that balls in $p$-norm with $1<p<\infty$ do not admit majorization-minimizers, while \Cref{sec:geometry_trace_ball} is dedicated to proving that balls in $1$-norm do admit majorization-minimizers (and majorization-maximizers). Do balls in Hellinger distance, Bures distance, the (square-root) Shannon-Jensen divergence, Wasserstein distances, or other metrics on $\cP$ admit majorization-minimizers?
	\item In \Cref{chap:majflow_ctybounds}, the notion of majorization flow is defined in terms of the majorization-minimizer. Can a similar notion be defined in terms of the majorization-maximizer? Would it be useful?
	\item In \Cref{sec:applications}, the path of majorization flow (defined in \Cref{chap:majflow_ctybounds}) is used for understanding the continuity properties of various Schur concave functions. Does it have other applications?
	\item \Cref{thm:Delta-eps-Schur-convex} shows that \typeone{} $(h,\phi)$-entropies $H$, which are symmetric and concave, are such that their derivative along the path of majorization flow, $\Gamma_H$, is strictly Schur convex. \Cref{prop:symm-concave-cx} shows that there exist symmetric and concave functions $f$ such that $\Gamma_f$ is not Schur convex. Is there a simple characterization of functions $f: \cP \to \bR$ such that $\Gamma_f$ is Schur convex?
	\item \Cref{cor:Renyi-lipschitz} provides upper and lower bounds on the optimal Lipschitz constant for the $\alpha$-R\'enyi entropy on $\cP_d$ for $\alpha > 1$, and an exact expression for $\alpha\in \{2,\infty\}$. The lower bound scales with $d$ as $(d-2)^{1-1/\alpha}$ while the upper bound scales as $d$. What is the correct scaling for $\alpha \in (1,2)\cup(2,\infty)$?
	\item \Cref{prop:LOCC} shows that the majorization-minimizer and -maximizer can be used to provide an approximate source or target state for LOCC conversions of bipartite pure states. Are these states optimal with respect to some criterion?
	\item \Cref{prop:guesswork_Lipschitz} provides upper and lower bounds on the optimal Lipschitz constant for the guesswork with quantum side information which match up to constant factors. What is the exact optimal Lipschitz constant?
	\item \Cref{sec:entropic_bounds} provides entropic bounds on the guesswork with quantum side information. Is there a single-letter expression for the asymptotic quantity \\{$\lim_{n\to\infty}\frac{1}{n}G(X^n|B^n)_{\rho^{\otimes n}}$?}
	\item \Cref{cor:main_thm_for_CE} presents necessary and sufficient conditions for a state to maximize the conditional entropy over a ball in trace distance around a given state. Can these conditions be ``inverted'' to obtain an expression for the maximizer, or the maximum value of the conditional entropy? See also \cite{AS19,Wil19} for recent progress on uniform continuity bounds for the conditional entropy.
\end{enumerate}

\part{Eventually Entanglement-Breaking Channels}
\chapter{Introduction}

In this part of the thesis, we study the behavior of the composition of quantum channels $\Phi_n\circ \Phi_{n-1}\circ\dotsm\circ\Phi_1$ in two settings.

In the first setting, discussed in \Cref{chap:char_EEB}, the channels are all identical, and we investigate for which channels $\Phi$ there exists $n\in \mathbb{N}$ such that $\Phi^n := \underbrace{\Phi\circ\dotsm\circ\Phi}_{n \text{ times}}$ is \emph{entanglement-breaking}, meaning that when applied to a state, any initially present entanglement with a reference system is destroyed. Such channels are called \emph{eventually entanglement-breaking} (EEB), and provide the title of this part of the thesis. Under the assumption that $\Phi$ admits a full-rank invariant state (i.e.\@~that $\Phi$ is \emph{faithful}), we characterize which channels $\Phi$ are eventually entanglement-breaking in \Cref{prop:charEEB-discrete}. 

Let us briefly discuss the relevance of this result to quantum information theory.
First, many quantum information-theoretic protocols require entangled qubits  (ebits) shared between separated parties Alice and Bob as a resource. However, if  Alice or Bob's laboratory is subject to local noise, initially entangled states may soon become unentangled, and Alice and Bob may not be able to carry out the protocol. As quantum channels can be seen as a model of noise, characterizing which channels are eventually-entanglement breaking provides a means to understand which types of noise lead to complete loss of entanglement in finite time.

The noise model presented by an $n$-fold composition $\Phi^n$ is not generic, however, but instead possesses a discrete-time Markovianity, represented by the semi-group property $\Phi^{n+m} = \Phi^n \circ \Phi^m$. This can be given a physical interpretation as follows. The Stinespring dilation theorem states that any quantum channel $\Phi$, the model for general time evolution of a quantum system $\sys$, can be written as
\[
\Phi(\rho_\sys) = \tr_{\env}[ U_{\sys \env} (\rho_\sys \otimes \psi_\env) U_{\sys\env}^*]
\]
for some environment system $\env$, joint unitary $U_{\sys\env}$, and pure state $\psi_\env$. The time evolution of $\rho_\sys$, therefore, can be represented by a noiseless interaction with an environment system $\env$, followed by the loss of $\env$ itself, inducing noise on the system. The composition $\Phi^n$, therefore, models an evolution of $\rho_\sys$ which consists of $n$ consecutive identical lossless interactions with separate environment systems, followed by the loss of each. Thus, between subsequent interactions the environment does not ``remember'' the details of the previous interaction; instead, a fresh environment is procured.

This Markovianity aids in the mathematical analysis of the evolution, and provides a good representation of some physical systems, such as the Haroche experiments \cite{GKG+07}, but is not a faithful model of all types of time evolution. This model is, however, more general than the oft-considered model provided by a continuous time quantum Markov semigroup, in which time evolution from $0$ to $t\in \mathbb{R}_{>0}$ is given by $\Phi_t = \exp(t \cL)$ for a Lindblad generator $\cL$. The reduction follows from the fact that $\Phi_1$ is a quantum channel such that $\Phi_{n+m} = \Phi_1^{n+m}$. Note that while the continuous-time model can represent non-integer times as well, for the purposes of determining whether or not a channel is eventually entanglement-breaking, one can consider the decomposition $\Phi_t = \Phi_{t-\floor{t}} \circ \Phi_{\floor{t}}$ to see that as soon as $\Phi_{\floor{t}}$ is entanglement-breaking, $\Phi_t$ is too. On the other hand, not all quantum channels $\Phi$ can be obtained in such a manner from a continuous-time quantum Markov semigroup; this is the topic of the divisibility of quantum channels, discussed in e.g.\@~\cite{WC08}.

The second information-theoretic application of this result relates to quantum key repeaters introduced in \cite{BCHW15}. Consider the task of distilling private states between Alice and Bob, who each individually share a state with an intermediary, Charlie: Alice shares $n$ copies of state $\rho$ with Charlie, and Bob shares  $n$ copies of a state $\rho'$ with Charlie. Charlie acts as a ``repeater'', facilitating communication between Alice and Bob. The three parties use a tripartite LOCC protocol to approximately recover $nR$ private states shared between Alice and Bob; here, $R$ is the \emph{rate} of the task.

This task is related to the so-called PPT$^2$ conjecture \cite{PPT2}, which conjectures that the composition of two PPT quantum channels (i.e.\@ those whose Choi-Jamiolkowski state has a positive partial transpose) is entanglement-breaking. Translated to the language of bipartite states via the Choi-Jamiolkowski isomorphism (discussed in \Cref{sec:CJ}), this conjecture states that the quantum state obtained by entanglement swapping of PPT states (those with positive partial transpose) is separable. If true, the PPT$^2$ conjecture shows if $\rho$ and $\rho'$ are PPT states in the quantum key repeater task described above, the asymptotic rate is zero \cite{BCHW15,CF17}, in contrast to the task of distilling private states from PPT states (without an intermediary), which can have a strictly positive rate \cite{HHHO05a}.

\Cref{thm:CP_PPT_EB_discrete}, which can be seen as a corollary of the characterization of eventually entanglement-breaking channels discussed above, makes a step towards establishing the PPT$^2$ conjecture by showing that any faithful PPT completely positive map is eventually entanglement-breaking. This extends the recent results of \cite{RJP18}, which establish this statement for unital quantum channels (i.e.\@~quantum channels such that the completely mixed state is invariant under the action of the channel).

The second setting, discussed in \Cref{chap:RIS}, is a slight generalization of the first, in which the maps $\Phi_1,\Phi_2,\dotsc,\Phi_n$ are not necessarily identical, but have a particular extended form
\[
\Phi_k(\rho_\sys) = \tr_{\env_k}[ U_{\sys \env_k} (\rho_\sys \otimes \xi_{\env_k}) U_{\sys\env_k}^*]
\]
in which the state of the environment, $\xi_{\env_k}$, is a thermal state instead of a pure state.
This setup yields a \emph{repeated interaction system} (or RIS), which is a system where the environment consists of a sequence of ``probes'' $\env_k$, $k=1,\ldots,T$, initially in a thermal state at inverse temperature $\beta_k$, and $\sys$ interacts with $\env_k$ (and only $\env_k$) during the time interval $\big[ k\tau, (k+1)\tau\big)$ for some interaction time $\tau > 0$. Here, $T\in\mathbb{N}$ is the total number of probes.

\Cref{chap:RIS} discusses the results of \cite{HJPR18}, which studies a refinement of Landauer's Principle in terms of a two-time measurement protocol (better known as ``full counting statistics'') for repeated interaction systems, in an adiabatic regime. We briefly describe each of these elements below.

Landauer's Principle is a universal principle for which the most common formulation is as a lower bound on the energetic cost of erasing a bit of information in a fixed system $\sys$ by interaction with an environment $\env$ which is initially at thermal equilibrium. It was first stated by Landauer in \cite{La}. A recent, mathematically sound derivation (in \cite{RW13}, later extended to the case of infinitely extended systems in  \cite{JP14}) is based on the entropy balance equation, given by $\Delta S_{\sys}+ \sigma = \beta \Delta Q_{\env}$ where $\Delta S_{\sys}$ is the average decrease in entropy of $\sys$ during the process, $\Delta Q_{\env}$ the average increase in energy of $\env$, and $\beta$ is the inverse temperature of the environment\footnote{we will always set the Boltzmann constant to $1$, so that $\beta=1/\Theta$, $\Theta$ the temperature.}. The term $\sigma$ is called the \emph{entropy production} of the process.  As it can be written as a relative entropy, the entropy production is non-negative  which yields the inequality $\Delta S_{\sys} \leq  \beta \Delta Q_{\env}$. One of the questions of interest regarding Landauer's Principle concerns the saturation of that identity, i.e.\@ the vanishing of $\sigma$. It is a general physical principle that when the system--environment coupling is described by a time-dependent Hamiltonian, the entropy production $\sigma$ vanishes in the adiabatic limit, that is, when the coupling between $\sys$ and $\env$ is a slowly varying time-dependent function. More precisely, if the typical time scale of the coupling is $T$, one considers the regime $T\to\infty$.

For a repeated interaction system,  the entropy balance equation becomes $\sum_{k=1}^T \Delta S_{\sys} + \sum_{k=1}^T \sigma_k = \sum_{k=1}^T \beta_k \Delta Q_{\env,k}$, where each term with index $k$ corresponds to the interaction between $\sys$ and $\env_k$. We describe the repeated interaction system as an ``adiabatic RIS'' when the various parameters of the probes are sampled from sufficiently smooth functions on $[0,1]$ as the values at times $k/T$, $k=1,\ldots,T$. This is the setup that was studied in \cite{HJPR17}; there it was shown that the total entropy production $\lim_{T\to\infty} \sum_{k=1}^T \sigma_k$ is finite only under the condition $X(s)=0$ for all $s\in[0,1]$, where $X(s)$ is a quantity depending on the probe parameters at time $s\in[0,1]$ discussed below (see~\eqref{eq:def-X}). The proof of this result relied mostly on a new discrete, non-unitary adiabatic theorem  to estimate the behavior of $\Phi_T\circ\dotsm\circ\Phi_2\circ\Phi_1$, where $\Phi_1,\Phi_2\dotsc,\Phi_T$ are slowly varying quantum channels that represent the reduced dynamics acting on $\sys$.

A refinement of the above formulation of Landauer's Principle is however possible using the so-called full counting statistics. Full  counting statistics were first introduced in the study of charge transport, and have met with success in the study of fluctuation relations and work in quantum mechanics (see Kurchan \cite{Ku00} and Tasaki \cite{Ta00}). An example of their use in improving Landauer's Principle was given in \cite{BFJP,GCGPVP}. In the present situation, the formulation of Landauer's Principle in terms of full counting statistics can be stated by defining random variables $\Delta s_{\sys}$ and $\Delta q_{\env_k}$ which are outcomes of simple physical experiments, which we now describe. In such an experiment, one initially measures the quantity $-\log \rho_\sys$ ($\rho_\sys$ is the state of the system) and the energies $h_{\env_k}$ for each $k$ ($h_{\env_k}$ is the free Hamiltonian of $\env_k$), then lets the system interact with the chain of probes, then measures again the same quantities. Note that it makes no difference if one measures each probe immediately after it interacts with the system, or measures all the probes at the end of the protocol.
With the right sign conventions, the changes in these quantities are random variables which we denote $\Delta s_{\sys}$ and $\Delta q_{\env_k}$. Our refinement discusses the connections between the probability distributions of $\Delta s_{\sys}$ and $\sum_k \beta_k \Delta q_{\env_k}$. One can show that the expectations of these distributions are $\Delta S_{\sys}$ and $\sum_k \beta_k \Delta Q_{\env_k}$ respectively; there is, therefore, more information in these distributions than in the previously considered scalar quantities.

We consider an adiabatic repeated interaction system and study the limiting distributions of the above random variables as $T\to\infty$. Again, we show that in the case $X(s)\equiv 0$ we have the expected refinement of Landauer's Principle, which is essentially that when $T\to\infty$, one has $\Delta s_{\sys}=\sum_k \beta_k \Delta q_{\env_k}$ almost-surely.
In the case $X(s)\not\equiv 0$, we show that $\sum_k \beta_k \Delta q_{\env_k}$ satisfies a law of large numbers, a central limit theorem, and a large deviation principle, all of these for the time scale $T$, and with explicit parameters. In particular, $\sum_k \beta_k \Delta q_{\env_k}$ is of order $T$, whereas $\Delta s_\sys$ is a bounded quantity. All results in the case $X(s)\not\equiv 0$ can actually be extended to the case where the probe observables measured at each step $k$ are not simply $\beta_k h_{\env_k}$ but a more general observable, or when the system observables are not $-\log\rho_\sys$.

We show in addition that the random variable $\varsigma_T= \sum_k \beta_k \Delta q_{\env_k}- \Delta s_{\sys}$ can be expressed as a relative information random variable between the probability measure describing the experiment outcomes, and the probability measure corresponding to a backwards experiment. Since we obtain a full large deviation principle for this random variable as $T\to \infty$, this connects these results with the appearance of the arrow of time (see \cite{ABL,BJPP18}). We discuss in particular the appearance of symmetries in the moment generating functions, and their implications in terms of Gallavotti--Cohen type symmetries.

This approach gives an improvement over~\cite{HJPR17} in various aspects. First of all, \Cref{prop:mgf_X_zero} (in the case $X(s)\equiv 0$) and \Cref{theo_ldp} together with \Cref{coro_LLN} and \Cref{theo_clt} (in the general case) characterize the limiting distributions of relevant random variables, whereas in~\cite{HJPR17}  only  information about the behaviour of their expectations was derived. We recover those former results (and more) about these expectations, as \Cref{prop:mgf_X_zero} implies in particular the convergence of $\lim_{T\to\infty} \sum_{k=1}^T \sigma_k$ to an explicit quantity when $X(s)\equiv0$, and \Cref{theo_clt} gives the divergence of the same quantity under generic assumptions when $X(s)$ does not vanish identically. In addition, \Cref{cor:adiab_for_alpha=0} gives an expression for the adiabatic evolution of any initial state.

\chapter{Linear Algebra and entanglement theory}
In this chapter, we review fundamental areas of linear algebra and entanglement-theory used in the following chapters. Much of this chapter is written with an eye towards precision, at the cost of notational complexity, which is relaxed in the later chapters.

Let $\cX$ and $\cY$ be a finite-dimensional vector spaces. We write
\begin{itemize}
 	\item $\cB(\cX \to \cY)$ for the set of linear maps from $\cX$ to $\cY$
 	\item $\cB(\cX) \equiv \cB(\cX\to \cX)$ for the set of linear operators on $\cX$
 	\item $\Bsa(\cX)$ for the set of self-adjoint operators on $\cX$
    \item $\Bp(\cX) \subset \Bsa(\cX)$ for the set of positive semidefinite
    operators on $\cX$,
    \item $\Bpp(\cX) \subset  \Bp(\cX)$ for the set of full-rank positive semidefinite operators on $\cX$
    \item $\cD(\cX)$ for the set of unit trace
    positive semidefinite operators, i.e.\@~\emph{density matrices} or \emph{states}, on $\cX$, and
   	\item $\cD_+(\cX)\subset \cD(\cX)$ for the set of full-rank density
    matrices. We say also call a state $\sigma \in \cD_+(A)$ a \emph{faithful state}.
 \end{itemize}

In the following, $\cH$, $A$, $A'$, $B$, and $B'$ represent finite-dimensional Hilbert
spaces, with dimensions $d_\cH, d_A, d_{A'}, d_B, $ and $d_{B'}$ respectively, and
inner products $\ip_{\cH}$, $\ip_A$, $\ip_{A'}$, $\ip_B$, and $\ip_{B'}$. We often use $\cH$ in the case when there is only one Hilbert space under consideration.

We may equip $\cB(A\to B)$ with the Hilbert-Schmidt inner product $\ip\HS$ by
\begin{equation}\label{eq:def-HS}
\braket{X,Y}\HS = \tr[X^* Y]
\end{equation}
where $X^* \in \cB(B \to A)$ is the adjoint of $X$, defined by
\begin{equation}\label{eq:def_adjoint}
\braket{\psi_B, X \phi_A}_B = \braket{X^* \psi_B, \phi_A}_A \qquad \forall
\phi_A \in A, \, \psi_B \in B.
\end{equation}
Then $X^* Y \in \cB(A\to A)$, and $\tr[X^*Y]$ is defined as
\[
\tr[X^*Y] := \sum_{i=1}^{d_A} \braket{\psi_i, X^*Y \psi_i}_A
\]
for any orthonormal basis $\{\ket{\psi_i}\}_{i=1}^{d_A}$ of $A$.

In fact, $\cX := (\cB(A\to B), \ip\HS)$ is itself a Hilbert space. Therefore, given $\cX := (\cB(A\to B), \ip\HS)$ and $\cY := (\cB(C\to D), \ip\HS)$, the space $\cB( \cX\to \cY)$ can be equipped with its own Hilbert-Schmidt inner product, and so forth. Therefore, for $X,Y \in \cB(A\to B)$, we sometimes write $\braket{X,Y}_{\cB(A\to B), \, \text{HS}} \equiv \braket{X,Y}\HS$ for clarity.

\section{Vectors, operators, and superoperators}
A superoperator is a linear map $\Phi \in \cB( \cB(A) \to \cB(B))$, where $A$
and $B$ are finite-dimensional Hilbert spaces. The space $\cB( \cB(A) \to \cB(B))$ has a lot of structure which we will explore in this section.
It is helpful to start with the least structure and then see what each
additional piece of structure adds to the theory. As a high-level overview, we will see that:
\begin{enumerate}
    \item Vector spaces $\cX$ have scalar multiplication and vector
    addition, and admit direct sum decompositions. They can also be normed.
    \item The set of linear operators on $\cX$, $\cB(\cX)$, admits another
    operation, composition (i.e.\@ vector multiplication), turning $\cB(\cX)$
    into an algebra.  Additionally, an operator in $\cB(\cX)$ can induce direct
    sum decompositions on $\cX$, such as the spectral decomposition. As a
    vector space, $\cB(\cX)$ can be normed, and norms on $\cX$ induce
    operator norms on $\cB(\cX)$.
    \item Setting $\cX = A$, a Hilbert space, adds an inner product to $A$
    and hence an adjoint to $\cB(\cX)$, as well as a norm on $A$ (the norm induced by the inner product). This gives rise to the adjoint and
    hence the notion of self-adjointness, as well as positive
    semidefiniteness. Together, $\cB(A)$ with the operator norm $\|\cdot\|_\infty$
    induced by the Hilbert space norm on $A$ and the adjoint $^*$ induced by the
    inner product on $A$ yield a $C^*$-algebra (i.e.\@~an algebra in which the norm, adjoint, and the composition operation satisfy key relations, such as the submultiplicativity of the norm, and the C$^*$-identity $\|X^*X\|=\|X\|^2$ for $X\in A$).
    \item Raising one more level to the level of superoperators, i.e.\@ the set
    $\cB(\cB(A))$, allows the consideration of which maps commute with
    the adjoint and which maps preserve positive semidefiniteness.
    Moreover, such maps act on an algebra and not only a vector
    space, and vector multiplication can play a role. Norms on $\cB(A)$
    induce operator norms on $\cB(\cB(A))$, and $\cB(\cB(A))$ can be
    equipped with its own norms as well.
\end{enumerate}
We see that at each level properties are added but generally not removed. For example,
$\cB(\cB(A))$ can be seen as a vector space, as the set of operators on the
vector space $\cB(A)$, as the set of operators on the Hilbert space $(\cB(A),
\ip\HS)$, as the set of operators on the $C^*$-algebra $(\cB(A),
\|\cdot\|_\infty)$, as superoperators on a Hilbert space $A$, or even as a
Hilbert space itself. Additionally, norms at lower levels induce operator norms
at higher levels.

\subsection{Vector spaces}
The starting point is a vector space $\cX$ over $\bR$ or $\bC$. Some examples:
\begin{itemize}
	\item $A$ is a complex vector space of dimension $d_A$
    \item $\cB(A)$ is a complex vector space of (complex) dimension
    $d_A^2$, while  $\Bsa(A)$ is a real vector space of (real) dimension
    $d_A^2$.
    \item $\cB(\cB(A) \to \cB(B))$ is a complex vector space of dimension
    $d_A^2 d_B^2$.
\end{itemize}

Vector spaces are closed under scalar multiplication and vector addition. Given
two subsets $S_1, S_2 \subset \cX$, we define
\[
S_1 + S_2 = \left\{ u + v : u \in S_1, v \in S_2 \right\}.
\]
A \emph{subspace} $V \subset \cX$ is a subset which is closed under scalar
multiplication and vector addition. If $V_1, V_2 \subset \cX$ are subspaces,
then so is $V_1 + V_2$ and $V_1 \cap V_2$. We say $\cX$ is the \emph{direct sum}
of subspaces $V_1,\dotsm, V_n$ if
\[
\cX = V_1 + \dotsm + V_n
\]
and $V_i \cap V_j = \{0\}$ for each $i\neq j$. In that case, we write $\cX = V_1 \oplus V_2
\oplus \dotsm \oplus V_n$. Moreover, for any $u \in \cX$, there exists a unique
decomposition $u = \sum_{i=1}^n v_i$ where $v_i \in V_i$, $i=1,\dotsc, n$. We
can equip $\cX$ with a \emph{norm} which is a function $\|\cdot\| : \cX\to
\R_{\geq 0}$ which is subadditive (i.e.\@~satisfies triangle inequality), positive definite
($\|x\|=0 \Rightarrow x =0_\cX$), and absolutely homogeneous (i.e.\@ $\|\alpha
x\| = |\alpha|\|x\|$).

\subsection{Operators on a vector space}
Let $\cX$ be a vector space over $\bC$ with finite dimension $d$. Now, we
consider $\cB(\cX)$. Given a norm $\|\cdot\|_\cX$ on $\cX$, the \emph{operator
norm} is defined by
\[
\|R\|_{\cX \to \cX} := \sup_{\|x\|_{\cX} \leq 1} \| R x\|_\cX
 \]
 for $R\in \cB(\cX)$ is a norm on $\cB(\cX)$. We denote the identity operator on $\cX$ as $\one_\cX\in \cB(\cX)$. Next, we say
\begin{enumerate}
    \item An operator $P \in \cB(\cX)$ is a \emph{projection} if $P^2 = P$.
    For any projection $P$, we have $\cX = P\cX \oplus P^\perp \cX$ where
	\[
    P \cX := \left\{ Pu : u \in \cX \right\}, \qquad P^\perp :=
    \one_{\cB(\cX)} - P.
	\]
    On the other hand, if $\cX = V_1\oplus \dotsm \oplus V_n$, then $P_i$
    defined as $P_i u = v_i$  where $v_i$ is the unique element of $V_i$
    such that $u = \sum_{i=1}^n v_i$ for $v_i \in V_i$, is a projection.
    Moreover, $\sum_{i=1}^n P_i = \one_{\cX} \in \cB(\cX)$ is the identity
    operator, and $P_i P_j = \delta_{ij}P_i$.

    Restating this, if $\cX = V_1 \oplus \dotsm \oplus V_n$, then any $\Phi
    \in \cB(\cX)$ can be decomposed as
	\begin{equation} \label{eq:decomp-Phi}
	\Phi = \sum_{ij} \Phi_{ij}
	\end{equation}
	where $\Phi_{ij} = P_j\Phi P_i$, and the $\Phi_{ij}$ commute with each other.
    \item An operator $N\in \cB(\cX)$ is \emph{nilpotent} if $N^n = 0$ for
    some $n \in \mathbb{N}$. The \emph{order} of a nilpotent operator $N$ is
    the minimal $n$ such that $N^n = 0$. The nilpotent order is always at
    most $d$ and is denoted $\ord N$.
    \item An operator $R \in \cB(\cX)$ is a \emph{contraction} on $(\cX, \|\cdot\|_\cX)$ if $\|R\|_{\cX\to\cX} \leq 1$.
\end{enumerate}
 Let $\Phi \in \cB(\cX)$ be a linear operator on $\cX$. Then the \emph{image}
 \[
 \Phi(\cX) := \left\{ \Phi(x): x \in \cX \right\}
 \]
 and \emph{kernel}
 \[
 \ker \Phi := \left\{ x\in \cX : \Phi(x) = 0 \right\}
 \]
 are subspaces of $\cX$. The dimension of the image is called the rank of
 $\Phi$, denoted $\rank(\Phi)$, and the Rank-Nullity Theorem states that $\rank(\Phi) + \dim \ker \Phi =
 d$.

Now, let $\cX$ be a complex vector space of finite dimension $d$, and let $\Phi \in \cB(\cX)$. Then
$\Phi$ has $d$ (possibly repeated) \emph{eigenvalues} $\lambda \in \bC$ and
associated \emph{eigenvectors} $X \in \cX\setminus \{0\}$ which are defined by
the relationship
\[
\Phi(X) = \lambda X.
\]
The set $\left\{ X \in \cX: \Phi(X) = \lambda X \right\}$ is a subspace of $\cX$
called the \emph{eigenspace} of $\Phi$ associated to the eigenvalue $\lambda$,
and its dimension is the \emph{geometric multiplicity} of $\lambda$.

Moreover, $\Phi$ has a Jordan decomposition, defined as follows. Let $n \leq d$
denote the number of distinct eigenvalues of $\Phi$, and let
$\mu_1,\dotsc,\mu_n$ denote the distinct eigenvalues of $\Phi$. Then there
exists $V_1,\dotsc, V_n$ such that
\begin{equation}\label{eq:eig_directsum}
\cX = V_1 \oplus \dotsm \oplus V_n
\end{equation}
and each $V_i$ is invariant under $\Phi$, namely $\Phi(V_i) \subset V_i$ for
each $i=1,\dotsc,n$. The dimension of $V_i$ is called the \emph{algebraic
multiplicity} of $\mu_i$, and is at least the geometric multiplicity.

Considering the decomposition of $\Phi$ given in \eqref{eq:decomp-Phi} induced by the direct sum decomposition of \eqref{eq:eig_directsum}, we have $\Phi_{ij} = 0$ for
$i\neq j$, since $\Phi P_i \subset P_i \cX$ and hence $P_j \Phi P_i \subset P_j
P_i \cX = \{0\}$, where $P_i$ is the projection onto $V_i$. Thus, it remains to
understand the $\Phi_{ii}$. In fact, we have that $\Phi_{ii} = \mu_i \, P_i + D_i$
where $D_i$ is nilpotent. The projection $P_i$ is called \emph{spectral
projection} of $\Phi$ for eigenvalue $\mu_i$, and $D_i$ the
\emph{eigen-nilpotent}. If $\dim V_i = 1$, then $\mu_i$ is called a \emph{simple}
eigenvalue, and $D_i$ is necessarily zero. If $\dim V_i \geq 1$ and $D_i = 0$,
then $\mu_i$ is called a \emph{semi-simple} eigenvalue. The resulting decomposition
\begin{equation}\label{eq:abstract-Jordan-decomp}
\Phi = \sum_i \mu_i P_i + D_i
\end{equation}
is known as the \emph{Jordan decomposition} of $\Phi$.
The \emph{spectral radius} of a map $\Phi \in \cB(\cX)$ is defined by
\begin{equation}\label{eq:def_spr}
\spr(\Phi) := \max_i |\mu_i|
\end{equation}
as the maximum modulus of its eigenvalues.

So far, we have investigated how the vector space structure of $\cX$ and the linearity of $\Phi$ determines the structure of $\Phi$ via its Jordan canonical form. Next, we show one way in which norms on $\cX$ can be used to obtain information about the Jordan decomposition of $\Phi$. \begin{lemma} \label{lem:periph_evals_contraction_semisimple}
Let $(\cX, \|\cdot\|_{\cX})$ be a finite-dimensional Banach space (i.e.\@~a complete normed vector space). The peripheral eigenvalues of a contraction $\Phi \in \cB(\cX)$ with spectral radius 1 on $(\cX,\|\cdot\|_{\cX})$ are semi-simple.
\end{lemma}
\begin{remark}
The following proof is reproduced from my master's thesis \cite[Lemma 3.1]{Han16} for the sake of self-containedness; see also Prop. 6.2 of
\cite{Wol12}.
\end{remark}
\begin{proof}  
Write $\Phi$ in its
Jordan canonical form, $\Phi = \sum_i \mu_i P_i + D_i$, where each $\mu_i$ is an
eigenvalue of $\Phi$, $P_i$ is the associated eigenprojection, and $D_i$ the
associated nilpotent, summed from $i=1$ to $\dim \cX$. Let $m_i = \ord D_i$, and
$m =  \max \{ m_i: 1\leq i\leq \dim \cX\}$.  Assume for some eigenvalue $\mu_i$ with
$|\mu_i|=1$ that $m_i \geq 2$, i.e.\@, $\mu_i$ has an eigen-nilpotent. We wish to
derive a contradiction, implying $D_i\equiv 0$.

Note that for $n>m$, using $D_i P_j = P_j D_i = \delta_{ij} D_i$, and $P_i P_j =
P_j P_i = \delta_{ij} P_i$, a binomial expansion yields
\begin{align*}  
\Phi^n =\sum_j  \sum_{k=0}^m {n \choose k} \mu_j^{n-k} P_j(D_j)^k.
\end{align*}
Our approach to deriving a contradiction will be to use that the binomial coefficient ${n\choose k}$
becomes large with $n$, while $\mu_i^n$ stays on the unit circle, and $\|\Phi^n\| \leq \|\Phi\|^n
=1$ remains bounded.

Since $ D_i^{m_i-1} \neq 0 $, for some vector $v$ we have $D_i^{m_i-1} v
\neq 0$. Let $n\in \bN$ large enough so that $ n \|D_i^{m_i-1} v\|_{\cX} >
\|D_i^{m_i-2} v \|_{\cX}$ and $n>m$. Then, since $\Phi$ is a contraction,
\begin{align*}  
\|D_i^{m_i-2} v\|_{\cX} &\geq \|\Phi^n D_i^{m_i-2} v\|_{\cX} = \left\|  \sum_j  \sum_{k=0}^m {n
\choose k} \mu_j^{n-k} P_j D_j^k D_i^{m_i-2} v  \right\|_{\cX}.
\end{align*}
Note $P_j P_i = 0$ implies $P_j D_i^{m_i-2} v = P_j P_i D_i^{m_i-2} v = 0$ for
$i\neq j$, so we have
\begin{align*}  
 \|D_i^{m_i-2} v\|_{\cX} &\geq \left\| \sum_{k=0}^m {n \choose k} \mu_i^{n-k} P_i(D_i)^k
 D_i^{m_i-2} v  \right\|_{\cX}.
\end{align*}
But, by choice of $m_i$, only two terms in the sum survive: $k=0$ and $k=1$:
\begin{align*}  
\|D_i^{m_i-2}v \|_{\cX} &\geq  \left\|   \mu_i^{n}  D_i^{m_i-2} v + n \mu_i^{n-1}
D_i^{m_i-1} v \right\|_{\cX}.
\end{align*}
Then, by reverse triangle inequality and using $|\mu_i|=1$,
\begin{align*}  
\|D_i^{m_i-2}v \|_{\cX} &\geq  \left|  \left\|  D_i^{m_i-2} v \right\|_{\cX} -n  \left\|
D_i^{m_i-1} v \right\| \right|\geq  n  \left\|  D_i^{m_i-1} v \right\|_{\cX}.
\end{align*}
By our choice of $v$, we have $\left\|  D_i^{m_i-1} v \right\|_{\cX}\neq 0$, hence
\begin{align*}  
\frac{\| D_i^{m_i-2} v\|_{\cX}}{\left\|  D_i^{m_i-1} v \right\|_{\cX}} \geq n
\end{align*}
for all $n$ large enough. This is a contradiction to our choice of $n$;
moreover, we could take $n\to\infty$, while the LHS remains bounded.
\end{proof}

\subsection{Operators on a Hilbert space}

Let $\cX = A$, a Hilbert space with an inner product
$\ip_A$. For example:
\begin{itemize}
    \item $\cH = \mathbb{C}^{d_\cH}$ for some $d_\cH\in \NN$, equipped with the inner product $\braket{u,v}_\cH := \sum_{i=1}^{d_\cH} \bar u_i v_i$.
    \item $\cB(\cH)$, for $\cH$ as in the previous example, equipped with Hilbert-Schmidt inner product \eqref{eq:def-HS}.
\end{itemize}

The Hilbert space $A$ has a privileged norm $\|\psi\|_A := \sqrt{\braket{\psi, \psi}_A}$ for $\psi \in A$; this norm satisfies the parallelogram law
\[
2 \|\psi\|^2_A + 2 \|\phi\|^2_A = \|\psi + \phi\|^2_A + \|\psi - \phi\|^2_A \qquad \forall \psi,\phi \in A.
\]
The set of operators on $\cB(A)$ inherit the operator norm induced by $\|\cdot\|_A$, namely
\begin{equation}\label{eq:2-norm-as-operator-norm}
\|X\|_{\infty} \equiv \|X\|_{\|\cdot\|_A \to \|\cdot\|_A} := \sup_{\substack{\psi \in A \\ \|\psi\|_A \leq 1}} \|A\psi\|_A.
\end{equation}
The inner product on $A$ also induces the adjoint $* : \cB(A) \to \cB(A)$, an anti-linear map defined via \eqref{eq:def_adjoint}. The adjoint provides for a notion of self-adjointness and positive semidefiniteness. We say $X \in \cB(A)$ is \emph{self-adjoint} if $X = X^*$, and is \emph{positive semidefinite} if $X = Y^* Y$ for some $Y \in \cB(A)$, denoted $X \geq 0$. We see that if $X\geq 0$ then
\begin{equation}\label{eq:X_psd_ip_geq_0}
 \braket{\psi, X \psi}_A = \braket{\psi, Y^*Y \psi}_A = \braket{Y \psi, Y \psi}_A = \| Y\psi\|_A \geq 0 \qquad \forall \psi \in A.
\end{equation}
As mentioned at the start, the set of positive semidefinite operators on $A$ are denoted $\cB_+(A)$, and the set of full-rank positive semidefinite operators are denoted $\cB_{++}(A)$. Note that if $X\in \cB_{++}(A)$, then the inequality in \eqref{eq:X_psd_ip_geq_0} becomes strict unless $\psi = 0_A$.

The adjoint and the operator norm induced by the Hilbert space norm satisfy the so-called C$^*$ identity,
\begin{equation}\label{eq:C*-identity}
\|X\|_\infty^2 = \|X^*\|_\infty^2 = \|X^*X\|_\infty \qquad \forall X \in \cB(A).
\end{equation}

The operator norm is one member of an interpolating family of norms on $\cB(A)$, the Schatten $p$-norms, defined as
\[
\|X\|_p := (\tr[|X|^p])^{1/p}
\]
for $p\geq 1$,  $X \in \cB(A)$, and $|X| := \sqrt{X^*X}$, where the function $\sqrt{\phantom{n}}: \cB_+(A) \to \cB_+(A)$ is defined as follows. Let $X = \sum_i \mu_i P_i$ have a Jordan form \eqref{eq:abstract-Jordan-decomp} with no eigen-nilpotents (i.e.\@~$D_i=0$ for all $i$); this holds, for example, if $X$ is self-adjoint. Then if $f: D\subseteq\mathbb{C} \to \mathbb{C}$ is defined on a set $D$ which includes the eigenvalues $\{\mu_i\}_i$, we define
\begin{equation}\label{eq:functional-calculus}
f(X) := \sum_{i} f(\mu_i) P_i.
\end{equation}
Since $X^*X\geq 0$ for any $X\in \cB(A)$, we have that the eigenvalues of $X^*X$ are all non-negative, and hence the square-root $\sqrt{X^*X}$ indeed can be defined in this manner.

As it turns out, the operator norm induced by the Hilbert space norm, \eqref{eq:2-norm-as-operator-norm}, is exactly the limit $p\to\infty$ of the Schatten $p$-norms, motivating the notation $\|\cdot\|_\infty$. Note also that the Schatten $1$-norm is used to define the trace distance used extensively in Part I, and the Schatten $2$-norm is the norm induced by the Hilbert-Schmidt inner product.

\subsection{Superoperators}
A \emph{superoperator} is a map $\Phi \in \cB( \cB(A) \to \cB(A'))$ for some Hilbert spaces $A$ and $A'$. We denote the identity superoperator from $\cB(A) \to \cB(A)$ as $\id_A$. Note that $\id_A = \one_{\cB(A)}$. Given $\Phi \in \cB( \cB(A) \to \cB(A'))$, we say
\begin{enumerate}
    \item $\Phi$ is \emph{positive} (P) if $\Phi( \cB_+(A)) \subseteq \cB_+(A')$;
    \item $\Phi$ is \emph{hermitian-preserving} if $\Phi( \Bsa(A)) \subseteq \Bsa(A')$;
    \item $\Phi$ is \emph{completely positive} (CP) if $\Phi \otimes \id_{E} : \cB(A \otimes E) \to \cB(A' \otimes E)$ is positive, for any finite-dimensional system $E$;
    \item $\Phi$ is \emph{trace-preserving} (TP) if $\tr[\Phi(X)] =\tr[X]$ for all $X\in \cB(A)$;
    \item $\Phi$ is \emph{trace non-increasing} (TNI) if $\tr(\Phi(X)) \leq \tr(X)$  for all $X \in \cB_+(A)$;
    \item $\Phi$ is \emph{unital} (U) if $\Phi(\one_A) = \one_{A'}$;
    \item $\Phi$ is a \emph{quantum channel} if it is completely positive and trace-preserving (CPTP)
    \item A quantum channel $\Phi$ is \emph{faithful} if it has a faithful invariant state (i.e.\@ $\sigma \in \cD_+(A)$ such that $\Phi(\sigma) =\sigma$).
\end{enumerate}
Additionally, $\Phi^*$ will denote the adjoint of $\Phi$ with respect to the Hilbert-Schmidt inner product.
\begin{remark}
$\Phi$ being hermitian-preserving is equivalent to the property that $\Phi(X^*) = \Phi(X)^*$ for all $X\in A$, and is implied by positivity. The first statement follows from the fact that there is a unique decomposition $X = \Re(X) + i\cdot\Im(X)$ such that $\Re(X), \Im(X) \in \Bsa(A)$, which is satisfied by $\Re(X) := \frac{X + X^*}{2}$ and $\Im(X) := \frac{X - X^*}{2i}$. If $\Phi$ is hermitian-preserving, then
\[
\Phi(X)^* = [\Phi( \Re(X)) + i \Phi(\Im(X))]^* = \Phi(\Re(X)) - i\Phi(\Im(X)) = \Phi(X^*).
\]
Conversely, if $\Phi(X^*) = \Phi(X)^*$ for all $X \in \cB(A)$, then for $Y\in \Bsa(A)$, we have
\[
2i\Im(\Phi(Y)) = 2i[ \Phi(Y) - \Phi(Y)^*] = 2i [ \Phi(Y - Y^*)] = 2i\Phi(0)=0.
\]
The second statement follows from the decomposition $X = X_+ - X_-$ with $X_\pm \in \cB_+(A)$ which exists for $X\in \cB(A)$ if and only if $X\in \Bsa(A)$. Then if $X\in \Bsa(A)$, we have $\Phi(X) = \Phi(X_+) - \Phi(X_-) \in \Bsa(A)$ as the difference of positive operators.

Note also that $\Phi$ being positive implies that if $X,Y\in \Bsa(A)$ satisfy $X\leq Y$, then $\Phi(X) \leq \Phi(Y)$. This follows immediately from the fact that $Y-X\geq 0$, so $\Phi(Y-X)\geq 0$.
\end{remark}

One particular superoperator of relevance is that of conjugation. If $R \in \cB(A)$, define $\Ad_R \in \cB( \cB(A) \to \cB(A))$  by $\Ad_R(X) := RXR^*$ for $X\in \cB(A)$. The notation for this operation is borrowed from \cite{SS12}.

The first result of interest concerns induced norms on $\Phi$. In the following, $(\cX,\|\cdot\|_\cX)$ represents a normed vector space.
\begin{theorem}[Russo Dye \cite{RD66}]
Let $\Phi: (\cB(A), \|\cdot\|_\infty) \to (\cX, \|\cdot\|_\cX)$ be a linear map.
Then
\[
\|\Phi \|_{\|\cdot\|_\infty \to \|\cdot\|_\cX} = \sup_{\substack{U \in \cB(A)\\
U \text{unitary}}} \|\Phi(U)\|_\cX.
\]
If $(\cX, \|\cdot\|_\cX) = (\cB(B), \|\cdot\|_\infty)$ and $\Phi$ is unital, then $\Phi$ is positive if and only if $\|\Phi
\|_{\|\cdot\|_\infty \to \|\cdot\|_\infty} = 1$.
\end{theorem}

The following is a well-known consequence of the Russo-Dye theorem: PTP linear maps are contractions on $(\cB(A), \|\cdot\|_1)$.
\begin{corollary}\label{cor:PTP_contraction}
Let $\Phi: (\cB(A),\|\cdot\|_1) \to (\cB(A),\|\cdot\|_1)$ be positive and
trace-preserving. Then
\[
\|\Phi \|_{\|\cdot\|_1 \to \|\cdot\|_1} = 1.
\]
\begin{proof}	
Let
\[
\Phi' : (\cB(A),\|\cdot\|_1)' \to (\cB(A),\|\cdot\|_1)'
\]
denote the \emph{Banach space adjoint} of $\Phi$ (also known as the topological
dual), where $(\cB(A),\|\cdot\|_1)'$ is the set of continuous linear functionals
on $\cB(A)$ equipped with the norm
\[
\| \ell \|_1' = \sup_{x \in \cB(A)} | \ell(x) |
\]
and where $\Phi'$ is defined by $\Phi'(\ell)(x) = \ell(\Phi(x))$. In particular,
$\Phi'(\tr)(x) = \tr(\Phi(x)) = \tr(x)$, so $\Phi'(\tr) = \tr$. 

In fact, the Banach dual of $(\cB(A),\|\cdot\|_1)$ is isomorphic to $(\cB(A),\|\cdot\|_\infty)$ (see \cite[Theorem VI.26]{RSI}) and each $\ell \in (\cB(A),\|\cdot\|_1)'$ is of the form
\[
\ell(x) = \tr[ y_\ell^* x] = \braket{y_\ell, x}\HS
\]
for some $y_\ell \in \cB(A)$, and of course each $y_\ell\in \cB(A)$ gives such a
linear functional. Under the map $\iota : \ell \mapsto y_\ell$, we see that
$\tr\mapsto \one_A$. Additionally,
\[
\| \ell \|_1' = \sup_{x \in \cB(A)} | \ell(x) | = \sup_{x\in \cB(A)} |
\tr[y_\ell x] | = \|y_\ell\|_\infty.
\]
Thus, $\iota$ is an anti-linear isometry from  $(\cB(A), \|\cdot\|_1)'$ to
$(\cB(A), \|\cdot\|_\infty)$. Moreover, $\tilde\Phi = \iota \circ \Phi' \circ
\iota\inv$ satisfies
\[
 \tilde \Phi(y_\ell) = \iota \circ \Phi' \circ \iota\inv \circ \iota (\ell) =
 \iota \circ \Phi' (\ell) = y_{\Phi' (\ell)}.
\]
Hence,
\[
\braket{ \tilde \Phi(y_\ell), x}\HS = \braket{y_{\Phi' (\ell)}, x}\HS.
\]
Then expanding the right-hand side,
\[
\braket{y_{\Phi' (\ell)}, x}\HS = \Phi' (\ell)(x) = \ell(\Phi(x)) =
\braket{y_\ell, \Phi(x)}\HS.
\]
Thus, we've found that $\tilde \Phi$ is in fact $\Phi^*$, the Hilbert-Schmidt
adjoint. Since Hilbert-Schmidt adjoints preserve positivity, we see that $\tilde
\Phi$ is positive. Thus, by the Russo-Dye theorem,
\[
\|\tilde \Phi\|_{\infty\to\infty} = \| \Phi^*\|_{\infty\to\infty} = 1.
\]
Moreover,
\begin{align*}
\|\tilde \Phi\|_{\infty\to\infty} &= \sup_{\|x\|_\infty = 1} \|\tilde
\Phi(x)\|_\infty = \sup_{\|\ell\|_1' = 1} \|\tilde \Phi(\iota(\ell))\|_\infty
\\&=\sup_{\|\ell\|_1' = 1} \| \iota \circ \Phi' (\ell))\|_\infty  =
\sup_{\|\ell\|_1' = 1} \| \Phi' (\ell))\|_1' = \|\Phi'\|_{\|\cdot\|_1' \to
\|\cdot\|_1'}.
\end{align*}
But since the Banach space adjoint is an isometry, we have $
\|\Phi'\|_{\|\cdot\|_1' \to \|\cdot\|_1'} = \|\Phi\|_{\|\cdot\|_1 \to
\|\cdot\|_1}$, concluding the proof.
\end{proof}

\end{corollary}

We also find that the spectral radius of a PTP linear map $\Phi$ is $1$. Clearly, the spectral radius cannot be greater than one; if $\Phi(X) = \lambda X$ for $|\lambda| > 1$, then $\|\Phi(X)\|_{1} = |\lambda| \|X\|_1$ and hence $\|\Phi\|_{\|\cdot\|_1 \to \|\cdot\|_1} \geq |\lambda| > 1$, contradicting \Cref{cor:PTP_contraction}. On the other hand, since $\Phi^*$ is unital we have $1\in\spec (\Phi^*) = \overline{\spec(\Phi)}$ and hence $1\in \spec(\Phi)$ as well. Thus, $\spr(\Phi) = 1$, and in fact, $1$ itself is an eigenvalue of $\Phi$. These properties and \Cref{lem:periph_evals_contraction_semisimple} show that the peripheral eigenvalues of a PTP linear map are semi-simple, meaning their associated eigen-nilpotents vanish. We can also check that the spectrum is closed under complex conjugation, as follows. If $\Phi$ is positive, then it is Hermitian preserving. Hence $\Phi(X) = \lambda X$ implies $\Phi(X^*) = \Phi(X)^* = (\lambda X)^* = \bar \lambda X^*$.

We have started to build an understanding of the spectral properties of PTP linear maps: they are contractions in 1-norm, $1$ is the spectral radius and an eigenvalue, all peripheral eigenvalues are semi-simple, and the spectrum is closed under complex conjugation. We will return to this study of the spectral properties of these maps, in the completely positive case, in \Cref{sec:periph_superoperators}. First, we will review the notions of separability and entanglement.

\section{Separability and entanglement}
In this section, we define separable operators and correspondingly, entanglement-breaking superoperators, and investigate some of their basic properties and relationships.

\subsection{Separability at the level of operators}

\begin{definition}[$\SEP(A:B)$]
We define the cone of separable operators in $\cB(A\otimes B)$ with respect to
the partition $A:B$ as
\[
\SEP(A:B) := \conv \left\{ X_A \otimes Y_B : X_A \in \Bp(A), Y_B \in \Bp(B)
\right\} \subset \Bp(A\otimes B).
\]
\end{definition}

\begin{definition}[Entanglement witness] \label{lem:entanglement-witness}
Let $X_{AB} \in \Bp(A\otimes B)$. An \emph{entanglement witness} $W_{AB} \in
\Bsa(A\otimes B)$ for $X_{AB}$ is any operator such that
\begin{enumerate}
	\item $\braket{W_{AB}, X_{AB}}\HS < 0$
	\item $\braket{W_{AB}, Z_{AB}}\HS  \geq 0$ for all $Z_{AB} \in \SEP(A:B)$.
\end{enumerate}
\end{definition}

As the name suggests, entanglement witnesses provide a ``witness'' (or ``certificate'') that a given operator is entangled. We can see directly from the definition that an entanglement witness for $X_{AB}$ cannot exist if $X_{AB}\in \SEP(A:B)$. The following proposition shows that for any entangled $X_{AB}$, an entanglement witness always exists. This is a consequence of the fact that $\SEP(A:B)$ is a closed convex set, and if $X\not \in \SEP(A:B)$, then there exists a hyperplane separating $X_{AB}$ from $\SEP(A:B)$.
\begin{proposition}[\cite{HHH96}]
$X_{AB} \in \Bp(A\otimes B)$ has an entanglement witness if and only if
$X_{AB}\not\in \SEP(A:B)$.
\end{proposition}

The following is a simple but very useful lemma regarding the structure of SEP.
Roughly, it states that if the sum of two positive operators is separable, and
locally they live in orthogonal subspaces, then each is separable.
\begin{lemma} \label{lem:SEP_orthog}
If  $X_{AB}, Y_{AB} \in \Bp(A\otimes B)$ satisfy
\[
X_{AB} + Y_{AB} \in \SEP(A:B)
\]
and are such that  for some orthogonal projection $P \in \cB(A)$ we have $X_{AB}
\in \cB_+(PA \otimes B)$ and $Y_{AB} \in \cB_+(P^\perp A \otimes B)$, then $X_{AB}
\in \SEP(A:B)$ and $Y_{AB} \in \SEP(A:B)$. Here, $P^\perp = \one_A - P$, and
e.g.\@
\[
PA \otimes B := \left\{ (P\otimes \one_B) \,\ket{\psi_{AB}} : \ket{\psi_{AB}} \in
A \otimes B\right\}\subseteq A\otimes B.
\]
\end{lemma}
\begin{proof}
Assume $X_{AB} \not \in \SEP(A:B)$. Then there exists an entanglement witness
$W_{AB}$ for $X_{AB}$ by \Cref{lem:entanglement-witness}. Let $\tilde W_{AB} =
\Ad_P\otimes \id (W_{AB})$; recall $\Ad_P$ denotes the superoperator $X\mapsto P X P^*$. Then $\tilde W_{AB}$ is an entanglement witness for
$X_{AB} + Y_{AB}$:
\[
\tr[ \tilde W_{AB} (X_{AB} + Y_{AB})] = \tr[ W_{AB} (X_{AB} + \Ad_P \otimes \id
(Y_{AB}))] = \tr[W_{AB} X_{AB}] < 0
\]
since $\Ad_P^* = \Ad_{P^*} = \Ad_P$ and $P(I-P)=0$, and similarly for $Z_{AB}
\in \SEP(A:B)$,
\[
\tr[\tilde W_{AB} Z_{AB}] = \tr[W_{AB} \Ad_P \otimes \id (Z_{AB})] \geq 0
\]
since $\Ad_P \otimes \id (Z_{AB}) \in \SEP(A:B)$ by
\Cref{lem:local_preserve_sep}.
\end{proof}

\subsection{Choi-Jamiolkowski (CJ) isomorphism: translate between operators and superoperators}\label{sec:CJ}
\begin{definition}
A linear map $\Phi: \cB(A) \to \cB(A')$ is  \emph{entanglement-breaking} (EB) if $\Phi\otimes \id_B (\Bp(A\otimes B))\subseteq \SEP(A:B)$ for all $B$.
\end{definition}
In fact, it suffices to take the dimension of the $B$ system equal to $d_A$. Note also that (EB) implies (CP).

\begin{definition}[CJ]
$J_b(\Phi) := (\Phi \otimes \id_{\tilde A}) (\Omega_{A:\tilde A}^b) \in \cB(A'
\otimes \tilde A)$ where $\tilde A$ is a copy of $A$, and
	\[
    \ket{\Omega_{A:B}^b} := \sum_{i=1}^{\min(d_A, d_B)} \ket{i_A}\otimes
    \ket{i_B}, \qquad \Omega_{A:B}^b := \ket{\Omega_{A:B}}\bra{\Omega_{A:B}}
	\]
    and $\{\ket{i_A}\}_{i=1}^{d_A}$ and $\{\ket{i_B}\}_{i=1}^{d_B}$ are
    orthonormal bases (ONB) of $A$ and $B$ respectively, and $b
    =(\{\ket{i_A}\}_{i=1}^{d_A}, \{\ket{i_B}\}_{i=1}^{d_B})$ denotes the
    pair of ONBs. We usually regard $J_b(\Phi) \in \cB(A'\otimes A)$ for
    notational simplicity, and write $J \equiv J_b$ when the choice of basis
    does not matter (which is most of the time), and likewise $\Omega_{A:B}
    \equiv \Omega_{A:B}^b$.
\end{definition}

\begin{proposition}
$J_b : (\cB( \cB(A) \to \cB(A') ), \ip\HS) \to (\cB(A'\otimes A) ), \ip\HS)$ is
a linear bijective isometry, i.e.\@ a Hilbert space isomorphism, and the value
$\|J_b(\Phi)\|\HS$ is independent of the choice of bases $b$.
\end{proposition}
\begin{proof}	
Again letting $\tilde A$ be a copy of $A$ for explicitness, we have
\begin{align*}	
\|J_b(\Phi)\|_{\cB(A'\otimes \tilde A),\,\text{HS}}^2 &=
\braket{(\Phi\otimes\id)(\Omega_{A:\tilde A}^b), (\Phi\otimes\id)(\Omega_{A:\tilde
A}^b)}_{\cB(A'\otimes \tilde A),\,\text{HS}}\\ &=\braket{\Omega_{A:\tilde A}^b,(
\Phi^* \Phi\otimes\id)(\Omega_{A:\tilde A}^b)}_{\cB(A\otimes \tilde A),\,
\text{HS}}\\ &= \sum_{i,j, i', j' = 1}^{d_A}
\braket{\ket{i_A}\bra{j_A} \otimes \ket{i_{\tilde A}}\bra{j_{\tilde A}},
(\Phi^*\Phi\otimes\id)(\ket{i'_A}\bra{j'_A} \otimes \ket{i'_{\tilde
A}}\bra{j'_{\tilde A}})}_{\cB(A\otimes \tilde A),\, \text{HS}}\\ &= \sum_{i,j,
i', j' = 1}^{d_A} \braket{\ket{i_A}\bra{j_A}, \Phi^*
\Phi(\ket{i'_A}\bra{j'_A})}_{\cB(A),\, \text{HS}} \underbrace{\braket{
\ket{i_{\tilde A}}\bra{j_{\tilde A}},  \ket{i'_{\tilde A}}\bra{j'_{\tilde
A}}}_{\cB(\tilde A),\, \text{HS}}}_{\delta_{ii'}\delta_{jj'}}\\ &= \sum_{i,j =
1}^{d_A} \braket{\ket{i_A}\bra{j_A}, \Phi^* \Phi(\ket{i_A}\bra{j_A})}_{\cB(A),\,
\text{HS}}
\end{align*}
Since $\{\ket{i_A}\bra{j_A}\}_{i,j=1}^{d_A}$ is an ONB of $\cB(A)$, then
\[
\sum_{i,j = 1}^{d_A} \braket{\ket{i_A}\bra{j_A}, \Phi^*
\Phi(\ket{i_A}\bra{j_A})}_{\cB(A),\, \text{HS}} = \tr[\Phi^*\Phi] =
\braket{\Phi, \Phi}_{\cB(\cB(A)\to \cB(A')),\,\text{HS}}
\]
 completing the proof that $J_b$ is an isometry. Note also that the value
 $\|J_b(\Phi)\|\HS$ does not depend on the choice of basis $b$ since the trace
 is independent of basis.
\end{proof}

\begin{definition}
If $\Phi : \cB(A) \to \cB(A')$ is a linear map, we say $\{R_1,\dotsc, R_n\}$
with $R_j \in \cB(A \to A')$ is a Kraus decomposition for $\Phi$ if $\Phi =
\sum_{i=1}^n \Ad_{R_i}$.
\end{definition}

The following is a standard result which provides useful characterizations of complete positivity. The proof is omitted for the sake of brevity.
\begin{proposition}
$\Phi \in \cB(\cB(A) \to \cB(A'))$ has a Kraus decomposition if and only if it is CP if and only if $J(\Phi) \geq 0$. 
\end{proposition}

The maximally-entangled operator $\Omega_{A:B}$ satisfies some properties that will be helpful in the following: the ``reflection trick'' (also known as the ``ricochet property''), and that any pure state $\ket{\psi}_{AB}$ can be obtained from $\ket{\Omega}_{A:B}$ via a local operation.

\begin{lemma}[Reflection trick] \label{lem:reflection-trick}
For any $A$ and $B$ and $R \in \cB(B)$
\[
(\one_A \otimes R) \ket{\Omega_{A:B}} = (R^{T_b} \otimes \one_B) \ket{\Omega_{A:B}}
\]
where $R^{T_b} \in \cB(A)$ is a ``pseudo-transpose'' (in the sense of the Moore-Penrose
pseudo-inverse) of $R$ induced by the pair of bases $b$: if $R = \sum_{i,j=1}^{d_B}
R_{ij} \ket{i}_B\bra{j}_B$, then
\[
 R^{T_b} =  \sum_{i,j=1}^{\min(d_A,d_B)} R_{ji} \ket{i}_A\bra{j}_A \in \cB(A).
 \]
 Likewise, if $S \in \cB(A)$, then
 \[
 (S \otimes \one_B) \ket{\Omega_{A:B}} = (\one_A \otimes S^{T_b}) \ket{\Omega_{A:B}}
 \]
 where for $S = \sum_{i,j=1}^{d_A} S_{ij} \ket{i}_A\bra{j}_A$
 \[
 S^{T_b} =  \sum_{i,j=1}^{\min(d_A,d_B)} S_{ji} \ket{i}_B\bra{j}_B \in \cB(B)
 \]
\end{lemma}

\begin{lemma} \label{lem:pure-to-omega} For any $A$ and $B$ and pure state
$\ket{\phi_{AB}} \in A\otimes B$, there exist $R_1 \in \cB(A)$ and $R_2\in
\cB(B)$ such that
\[
\ket{\phi_{AB}} = (R_1 \otimes \one_B) \ket{\Omega_{A:B}} = (\one_A \otimes R_2)\ket{\Omega_{A:B}}
\]
where $\tr[R_1^\dagger R_1] = \tr[R_1^\dagger R_1] = 1$.
\end{lemma}
\begin{proof}	
If $\ket{\phi_{AB}} \in A\otimes B$, then the Schmidt decomposition yields
\[
\ket{\phi_{AB}} = \sum_{i=1}^{\min(d_A,d_B)} \sqrt{\lambda_i} \ket{e_i}\otimes
\ket{f_i}
\]
where $\sum_{i=1}^{\min(d_A,d_B)} \lambda_i = \tr \phi_{AB} = \braket{\phi_{AB}
| \phi_{AB}}$, and $\ket{e_i}_{i=1}^{d_A}$ is an ONB of $A$, and
$\ket{f_i}_{i=1}^{d_B}$ is an ONB of $B$. Then
\[
\ket{\phi_{AB}} = (R_A \otimes U_B) \ket{\Omega_{A:B}}
\]
where we define $U_B \ket{i} = \ket{f_i}$ for $i=1,\dotsc,d_B$, and $R_A \ket{i}
= \sqrt{\lambda_i} \ket{e_i}$ for $i=1,\dotsc,d_A$, where $\lambda_i = 0$ for
$i>\min(d_A,d_B)$. Then by \Cref{lem:reflection-trick}, $R_1 := R_A U_B^{T_b}$
suffices. We see also $\tr[R_1^\dagger R_1] = \tr[R_A^\dagger R_A] = \sum_{i=1}^{d_A}\lambda_i = 1$. $R_2$ can be constructed similarly.
\end{proof}

\subsection{EB maps: separability at the level of superoperators}

\begin{lemma}[Local maps cannot create entanglement]
\label{lem:local_preserve_sep} For any positive maps $\Phi_1: \cB(A) \to
\cB(A')$ and $\Phi_2: \cB(B) \to \cB(B')$,
\[
(\Phi_1 \otimes \Phi_2) (\SEP(A: B)) \subset \SEP(A':B').
\]
\end{lemma}
\begin{proof}	
If $X_{AB} \in \SEP(A:B)$, then $X_{AB} = \sum_{i=1}^n \lambda_i\, x_i \otimes
y_i$ for $n\in \bN$, $x_i \in \Bp(A)$, $y_i \in \Bp(B)$ and $\lambda_i \geq 0$.
Then
\[
(\Phi_1\otimes \Phi_2)(X_{AB}) = \sum_{i=1}^n \lambda_i\, \Phi_1(x_i) \otimes
\Phi_2(y_i) \in \SEP(A':B')
\]
since $\Phi_1(x_i) \in \Bp(A')$ and $\Phi_2(y_i) \in \Bp(B')$.
\end{proof}
\begin{remark}
In particular, this means that if $\Phi$ is entanglement-breaking and $\Psi$ is positive, then $\Psi\circ\Phi$ is entanglement-breaking, since $J(\Psi\circ\Phi) = (\Psi\otimes \id)(J(\Phi))$, and hence $J(\Psi\circ\Phi)$ is the result of local positive operations applied to a separable state. Additionally, if $\Psi$ is CP, then $\Phi\circ \Psi$ is entanglement-breaking too, since $J(\Phi\circ \Psi) = (\Phi\otimes \id) J(\Psi) \in \SEP$ as $J(\Psi)\geq 0$.
\end{remark}

\begin{corollary}If $\Phi: \cB(A)\to \cB(A')$ is a linear map such that
$J_b(\Phi) \in \SEP(A':A)$ for some pair of bases $b$, then $J_{b'}(\Phi) \in
\SEP(A':B)$ for any other choice of pairs of bases $b'$.
\end{corollary}
\begin{proof}	
$J_{b'}(\Phi) = \Ad_{U_A} \otimes \Ad_{U_B} ( J_{b}(\Phi))$ for some unitaries
$U_A \in \cB(A)$ and $U_B \in \cB(B)$.
\end{proof}

We may use CJ to lift \Cref{lem:SEP_orthog} to the superoperator level.
\begin{corollary}[Corollary to \Cref{lem:SEP_orthog}] \label{cor:orth_image_sep}
If $\Phi_1$ and $\Phi_2$ are CP maps from $\cB(A)$ to $\cB(A')$ such that their
images are orthogonal, then $J(\Phi_1 + \Phi_2) \in \SEP(A':A)$ if and only if
both $J(\Phi_1)\in \SEP(A':A)$ and $J(\Phi_2)\in \SEP(A':A)$.

In particular, $J(\Phi_1\oplus \Phi_2) \in \SEP(A':A)$ if and only if
$J(\Phi_1), J(\Phi_2) \in \SEP(A':A)$.
\end{corollary}
\begin{proof}	
Let $P \in \cB(A')$ be the projection onto the image of $\Phi_1$. Then
$J(\Phi_1) = \Ad_P \otimes \id (J(\Phi_1))$ and $J(\Phi_2) = \Ad_{P^\perp}
\otimes \id (J(\Phi_2))$ for $P^\perp = I-P$. The result follows from
\Cref{lem:SEP_orthog}.
\end{proof}

The following proposition shows that the CJ isomorphism preserves notions of separability between states and maps.

\begin{proposition} \label{prop:EB_equiv_J}
Let $\Phi: \cB(A)\to \cB(A')$ be a CP map. Then $\Phi$ is EB if and only if $J(\Phi) \in
\SEP(A':A)$.
\end{proposition}
\begin{remark}
\cite{HSR03} proves this in the TP case. The fact is known in the general case, and I include the proof for completeness.
\end{remark}
\begin{proof}The implication
$\Rightarrow$ is immediate. For the reverse implication, let $\ket{\phi_{AB}} \in A\otimes
B$. Then $\phi_{AB} = \id_A \otimes \Ad_R (\Omega_{A:B})$ for some $R\in \cB(B)$
by \Cref{lem:pure-to-omega}. Then
\[
\Phi \otimes \id_B (\phi_{AB}) = \Phi \otimes \Ad_R(\Omega_{A:B}) = \id \otimes
\Ad_R (\tilde J)
\]
for
\[
 \tilde J := \Phi \otimes \id_B (\Omega_{A:B}) = \sum_{i,j=1}^{\min(d_{A},d_B)}
 \Phi(\ket{i}\bra{j}) \otimes \ket{i}\bra{j}.
 \]
By \Cref{lem:local_preserve_sep}, it suffices to prove that $\tilde J \in
\SEP(A:B)$. Let us compare to
 \[ J(\Phi)= \sum_{i,j=1}^{d_A} \Phi(\ket{i}\bra{j}) \otimes \ket{i}\bra{j} \in
 \SEP(A':A)
 \]
 \begin{itemize}
    \item  If $d_A > d_B$, then we can regard $B = \linspan \{
    \ket{i}\}_{i=1}^{d_B}$ as a subspace of $A =\linspan \{
    \ket{i}\}_{i=1}^{d_A}$. Letting $P$ denote the orthogonal projection
    onto this subspace (i.e.\@ $P = \sum_{i=1}^{d_B} \ket{i}\bra{i} \in
    \cB(A)$), then $\tilde J = \id_{A'} \otimes \Ad_P ( J(\Phi))$ and hence
    is separable by \Cref{lem:local_preserve_sep}.
    \item  If $d_A \leq d_B$ then in the same way, we can regard $A$ as a
    subspace of $B$, and note that $\tilde J$ is in fact an element of
    $\cB(A'\otimes A)\subset \cB(A' \otimes B)$. Moreover, $\tilde J =
    J(\Phi) \in \SEP(A':A)$, concluding the proof.
 \end{itemize}
\end{proof}

\begin{remark} \Cref{prop:EB_equiv_J} implies that \Cref{cor:orth_image_sep}
directly yields statements about entanglement-breakingness.
\end{remark}

\section{Peripheral spectral properties of superoperators}\label{sec:periph_superoperators}
In this section we consider the peripheral spectral properties of superoperators.  Note that in this section, we mostly restrict to the case of endomorphisms in order to study spectral properties, which require the input and output space to be the same. The peripheral spectrum, i.e.\@~the set of eigenvalues of maximal modulus, of a map $\Phi \in \cB(\cB(A))$ plays a fundamental role in understanding the action of $\Phi^n$ for large $n$.
This arises in the notion of an \emph{eventually} entanglement-breaking map. A map $\Phi\in \cB(\cB(A))$ is \emph{eventually-entanglement breaking} (EEB) if the $n$-fold composition $\Phi^n$ is entanglement-breaking for some $n\in \mathbb{N}$. If $\Phi,\Psi \in \cB(\cB(A))$, and either is entanglement-breaking, then $\Phi\circ \Psi$ is entanglement-breaking too. Thus, if $\Phi^n$ is EB, so is $\Phi^{n'}$ for all $n' \geq n$. Therefore, to know whether or not $\Phi$ is EEB, it suffices to check for large $n$.

It is then natural to introduce the \textit{phase subspace} $\tilde{\cN}(\Phi)$, which is the union of the eigenspaces associated to peripheral eigenvalues. It is defined as
\begin{equation*}
    \tilde{\cN}(\Phi)=\text{span}\{X\in \cB(\cH):\exists\, \phi\in \R \text{ s.t. }\Phi(X)=e^{i\phi}X\},
\end{equation*}
i.e.\@ the linear span of the peripheral points and denote by $P$ the projection onto it. First, it's important to note that the phase subspace of a quantum channel always contains a density matrix. The following is proven in \cite[Theorem 6.11]{Wol12} as an immediate corollary to Brouwer’s fixed point theorem.

\begin{proposition}[Existence of an invariant state]\label{prop:exist-invar}
Let $\Phi \in \cB(\cB(A))$ be positive and trace-preserving. Then $\Phi$ admits an invariant state $\sigma\in \cD(A)$, i.e.\@~such that $\Phi(\sigma)=\sigma$.
\end{proposition}
\begin{remark}
By a very different proof technique, this can be generalized to remove the trace-preserving assumption; see \cite[Theorem 2.5]{EH78}. In this case, the spectral radius of $\Phi$ may not be 1, and one obtains $\sigma \geq 0$ such that $\Phi(\sigma) = \spr(\Phi)\sigma$.
\end{remark}
The invariant state $\sigma$ guaranteed to exist by \Cref{prop:exist-invar} may not be full-rank. Recall that if a quantum channel $\Phi$ admits a full-rank invariant state, it is called \emph{faithful}.

Moreover, for any quantum channel $\Phi:\cB(\cH)\to \cB(\cH)$, the phase subspace is known to possess the following structure (Theorem 6.16 of \cite{Wol12}, Theorem 8 of \cite{WP10}): there exists a decomposition of $\cH$ as $\cH=\bigoplus_{j=1}^K\cH_j\otimes \cK_j\oplus\cK_0$ such that
\begin{equation} \label{eq:decohere-decomp}
    \tilde{\cN}(\Phi):=\bigoplus_{i=1}^K\,\cB(\cH_i)\otimes \tau_i\oplus 0_{\cK_0}\,,~~~P(\rho\oplus 0_{\cK_0})=\sum_{i=1}^K\tr_{\cK_i}(p_i\rho p_i)\otimes \tau_i\,,
\end{equation}
where $p_i$ is the orthogonal projector onto the $i$-th subspace, for some fixed states $\tau_i\in\cD_+(\cK_i)$. Moreover, there exist unitaries $U_i \in \cH_i$, and a permutation $\pi \in S_K$ such that for any element $X\in \tilde \cN(\Phi)$, we have
\begin{equation} \label{eq:Phi-on-decohere-decomp}
    \Phi(X) = \bigoplus_{i=1}^K\, U_i X_{\pi(i)} U_i^\dagger \otimes \tau_i\oplus 0_{\cK_0}
\end{equation}
where $X$ is decomposed as $X=\bigoplus_{i=1}^K\,X_i \otimes \tau_i\oplus 0_{\cK_0}$ according to \eqref{eq:decohere-decomp}.
The permutation $\pi$ permutes within subsets of $\{1,\dotsc,K\}$ for which the corresponding $\cH_i$'s have equal dimension.
Note that the space $\cK_0 =\{0\}$ if and only if the quantum channel is faithful.

To further explore the spectral properties of $\Phi \in \cB(\cB(A))$, we need to introduce the notions of primitivity and irreducibility.

\begin{definition}[Primitive map]\label{def:prim_map}
Let $\Phi \in \cB(\cB(A))$ be positive and trace-preserving. Then we say $\Phi$ is \emph{primitive} if there exists a full-rank state $\sigma\in \cD(A)$ such that we have the long-time convergence
\begin{equation}\label{eq:def_prim_eeb}
    \Phi^n(\rho) \xrightarrow{n\to\infty} \sigma, \qquad \forall \rho \in \cD(A).
\end{equation}
\end{definition}
\begin{remark}
The limit \eqref{eq:def_prim_eeb} implies
\begin{equation}\label{eq:def_prim_eeb_X}
\Phi^n(X) \xrightarrow{n\to\infty}  \tr[X] \sigma, \qquad \forall X\in \cB(A).
\end{equation}
To see this, let $X\in \cB(A)$ and consider the decomposition $X = X_+ - X_-$ where $X_\pm \geq 0$. Then $\rho_\pm = \frac{X_\pm}{\tr[X_\pm]} \in \cD(A)$, so $\lim_{n\to\infty}\Phi^n(\rho_\pm) = \sigma$. Hence, by linearity of $\Phi$, $X = \tr[X_+] \rho_+ - \tr[X_-]\rho_-$ satisfies
\[
\lim_{n\to\infty}\Phi^n(X) = \tr[X_+] \sigma - \tr[X_-] \sigma = \tr[X]\sigma
\]
as desired.
\end{remark}

\begin{proposition}[Spectral characterization of primitivity] \label{prop:spec_char_prim}
A positive trace-preserving map $\Phi \in \cB(\cB(A))$ is primitive if and only if $1$ is a non-degenerate eigenvalue of $\Phi$, $1$ is the only eigenvalue of $\Phi$ on the unit circle, and the  eigenvector associated to $1$ can be chosen to be strictly positive definite. Moreover, if $\Phi$ is primitive, then the state $\sigma$ in \eqref{eq:def_prim_eeb} is the strictly positive-definite eigenvector of $\Phi$ associated to the eigenvalue $1$.
\end{proposition}
\begin{remark}
The definition of primitivity given in \Cref{def:prim_map} is a priori difficult to check, while the characterization in \Cref{prop:spec_char_prim} can be checked in time polynomial in $d_A$ by computing the peripheral eigenvalues of $\Phi$ and possibly the eigenvalues of the eigenvector associated to the eigenvalue $1$ (if it is indeed an eigenvalue).
\end{remark}
\begin{proof}    
Assume $\Phi$ is primitive. Then if $\Phi$ has an eigenvalue $\lambda \in \bC$  and corresponding eigenvector $X\in \cB(A)$ with $|\lambda|=1$, we have $\Phi^n(X) = \lambda^n X$ and thus
\[
\tr[X] \sigma = \lim_{n\to\infty} \Phi^n(X) = \lim_{n\to\infty} \lambda^n X = X \lim_{n\to\infty} \lambda^n.
\]
Since $|\lambda|=1$ and $X$ is an eigenvector, the right-hand side must not be zero. Moreover, for the limit to exist, we  must have $\lambda=1$; this follows from the fact that
\[
\lim_{n\to\infty}\lambda^n = \lim_{n\to\infty}\lambda^{n+1} = \lambda \lim_{n\to\infty}\lambda^n
\] and that $\lambda\neq 0$. Thus, the only eigenvalue on the unit circle must be $1$. Next, we have $X = \tr[X] \sigma$ and hence $X$ is proportional to $\sigma$. Thus, the geometric multiplicity of $1$ must be $1$: there is a one-dimensional eigenspace. Lastly, we may choose $\sigma$ as the eigenvector to $1$, which is strictly positive definite.

On the other hand, assume $\Phi$ satisfies the spectral conditions of the proposition, and call $\tilde \sigma\in \cD(A)$ the normalized positive definite eigenvector associated to the eigenvalue 1. Then the Jordan decomposition \eqref{eq:abstract-Jordan-decomp} of $\Phi$ gives
\[
\Phi = P_1 + \sum_{i} \mu_i Q_i + D_i
\]
where $P_1$ is the eigenprojection onto $\tilde \sigma$, and each $|\mu_i|<1$. Then, using the orthogonality relations $P_1 (Q_i + D_i) = (Q_i + D_i)(Q_j + D_j) = 0$ for $i\neq j$, we have
\[
\Phi^n = P_1 + \sum_{i} \mu_i^n (Q_i + D_i)^n.
\]
The limit $n\to\infty$ yields $\Phi^n \to P_1$. Since $P_1$ is rank-1, as the projection onto a one-dimensional subspace spanned by $\tilde\sigma$, we have $P_1(X) = \ell(X) \tilde\sigma$ for some linear functional $\ell$. Since $\Phi$ is trace-preserving, we must have $\ell = \tr$. Thus, we obtain \eqref{eq:def_prim_eeb_X} with $\sigma=\tilde\sigma$ and in particular \eqref{eq:def_prim_eeb}.
\end{proof}

We are also interested in a generalization of primitivity called \emph{irreducibility}.

\begin{definition}[Irreducible map] \label{def:irred_map}
Let $\Phi \in \cB(\cB(A))$ be positive. Then we say $\Phi$ is \emph{irreducible} if any orthogonal projection $P\in \cB(A)$ satisfies
\begin{equation}\label{eq:irred-preserves-no-diag-blocks}
\Phi(P \cB(A) P) \subseteq P \cB(A) P \,\,\implies\,\, P \in \{0,\one_A\}.
\end{equation}
\end{definition}
\begin{remark}
The condition given in \eqref{eq:irred-preserves-no-diag-blocks} is equivalent to
\begin{equation}\label{eq:irred_proj_lt}
\exists \,c > 0 \,\text{ s.t. }\,\Phi(P) \leq c P \, \,\implies\,\, P \in \{0,\one_A\}.
\end{equation}
To see this equivalence, note that if $\Phi(P)\leq c P$, then for any $X\in \cB(A)$,
\[
PXP \leq \|X\|_\infty P
\]
so $\Phi(PXP) \leq \|X\|_\infty \Phi(P) \leq c \|X\|_\infty P \in P\cB(A)P$, so $P\cB(A)P$ is invariant under $\Phi$. On the other hand, if $P\cB(A)P$ is invariant under $\Phi$, then $\Phi(P) \in P\cB(A)P$, so $\Phi(P)\leq \|\Phi(P)\|_\infty P$.
\end{remark}

The following characterization in the case of TP irreducible maps parallels the definition of primitivity in \Cref{def:prim_map}. The following can be found as Corollary 6.3 of \cite{Wol12}.
\begin{proposition}[Ergodic sum characterization of irreducibility] \label{prop:irred-ergodic}
Let $\Phi \in \cB(\cB(A))$ be positive and trace-preserving. Then $\Phi$ is irreducible if and only if there is a full-rank state $\sigma \in \cD(A)$ such that
\[
\lim_{N\to\infty} \frac{1}{N}\sum_{n=1}^N \Phi^n(\rho) = \sigma, \qquad \forall \,\rho \in \cD(A).
\]
\end{proposition}
Note that \Cref{prop:irred-ergodic} implies that if $\Phi \in \cB(\cB(A))$ is positive, trace-preserving, and primitive, then it is irreducible.

The following spectral characterization can be found in \cite[Theorem 6.4]{Wol12}, and provides an analog to \Cref{prop:spec_char_prim} in the case of irreducibility.
\begin{proposition}[Spectral characterization of irreducibility] \label{prop:spec_char_irred}
Let $\Phi\in \cB(\cB(A))$ be a positive map with spectral radius $r$. Then $\Phi$ is irreducible if and only if $r$ is a non-degenerate eigenvalue with strictly positive-definite left- and right- eigenvectors, i.e.\@~there exists $X,Y\in \cB(A)$ such that $\Phi(X) = rX$ and $\Phi^*(Y) = rY$.
\end{proposition}
\begin{remark}
Note that if $\Phi$ is a quantum channel, then $\spr(\Phi)=1$ is indeed an eigenvalue with corresponding left-eigenvector $Y=\one_A>0$ (since $\Phi$ is TP), and there always exists a corresponding right-eigenvector $\sigma \in \cD(A)$ due to \Cref{prop:exist-invar}. However, while $\sigma \geq 0$, in general the state may not be full-rank, and the eigenvalue $1$ could be degenerate. An irreducible quantum channel, therefore, is one in which $1$ is indeed non-degenerate, and the associated eigenvector $\sigma$ satisfies $\sigma > 0$.
\end{remark}

The following property of irreducible maps will prove useful later on.
\begin{proposition}[Irreducible maps preserve faithfulness]\label{prop:irred-preserve-faithful}
If $\Phi\in \cB(\cB(A))$ is positive and irreducible, then $\Phi(\cB_{++}(A)) \subseteq \cB_{++}(A)$.
\end{proposition}
\begin{proof}  
Assume for the sake of contradiction that for some $X \in \cB_{++}(A)$ we have that $\Phi(X)$ is not full-rank. Let $\pi\in\cB(A)$ be the projection onto the support of $\Phi(X)$. Then $\pi \leq c X$ for some constant $c>0$. Hence, by positivity, $\Phi(\pi) \leq c\Phi(X) \leq c' \pi$ for some $c' >0$. Since $\pi\not\in \{0,\one_A\}$, this contradicts the irreducibility of $\Phi$ via \eqref{eq:irred_proj_lt}.
\end{proof}

The study of the peripheral spectral properties of irreducible maps is the subject of the non-commutative Perron-Frobenius theory for irreducible completely positive maps; see \cite{EH78}, \cite{FP09}, or \cite[Section 6.2]{Wol12}. See also \cite[Appendix A]{HJPR18} for a summary of this theory and extensions to deformations of irreducible CPTP maps. Together with the Jordan decomposition \eqref{eq:abstract-Jordan-decomp}, this theory provides a useful decomposition of irreducible quantum channels. In the next proposition, we recall this decomposition and provide a minimal set of quantities needed to construct such a map. This will prove useful in \Cref{chap:char_EEB} for constructing examples of irreducible quantum channels which are not eventually entanglement-breaking.

\begin{proposition}[Irreducibility via a decomposition] \label{prop:decomp-irred}
    Consider
    \begin{enumerate}
        \item An integer $z\in \{1,\dotsc,d_\cH\}$,
        \item An orthogonal resolution of the identity $\{ p_n\}_{n=0}^{z-1}$, i.e.\@, $\sum_{n=0}^{z-1}p_n = \one$ and $p_n^\dagger = p_n^2 = p_n$ for each $n$,
        \item A faithful state $\sigma$ such that $[\sigma,p_n]=0$ and $\tr[\sigma p_n] = \frac{1}{z}$, for each $n=0,\dotsc,z-1$,
        \item A linear map $ \Phi_Q$ such that:
              \begin{enumerate}
                  \item \label{it:Phi_Q-spr} $\spr ( \Phi_Q) < 1$
                  \item \label{it:Phi_Q-J} $J( \Phi_Q) \geq - z\left(\sigma \otimes \one\right)L_{1}$, where for $k=0,\dotsc,z-1$, we define $L_k := \sum_{n=0}^{z-1}  p_{n-k} \otimes  p_{n}$ where the subscripts are taken modulo $z$.
                  \item \label{it:Phi_Q-Pj} We have
                        \begin{equation} \label{eq:Phi_Q_kills_pn}
                            \Phi_Q( \sigma p_n) = \Phi_Q^*(p_n) = 0, \qquad \forall n = 0,1,\dotsc,z-1.
                        \end{equation}
              \end{enumerate}
    \end{enumerate}
    Let
    \begin{equation} \label{eq:Phi-irred}
        \Phi:= \sum_{n=0}^{z-1}  \theta^n P_n + \Phi_Q
    \end{equation}
    where $P_n(\cdot) = \tr[ u^{-n} \, \cdot\, ] u^n \sigma$ for $u := \sum_{k=0}^{z-1}  \theta^k  p_k$ and $ \theta := \exp(2\iu \pi / z)$.
    Then  $\Phi$ is an irreducible quantum channel. On the other hand, any irreducible quantum channel $\Phi$ can be decomposed as \eqref{eq:Phi-irred} for some choices of $z, \{p_n\}_{n=0}^{z-1}$, $\sigma$, and $\Phi_Q$ as in (1)--(4).
    Moreover, in either case, $\sigma$ is the unique fixed point\footnote{up to a multiplicative constant} of $\Phi$; $P_n(\cdot)$ are its peripheral eigenprojections, associated to eigenvalues $\theta^n$ and eigenvectors $u^n \sigma$; and, for any $j,k =0,\dotsc,z-1$, we have the intertwining relations
    \begin{equation} \label{eq:irred_on_p_blocks}
        \Phi(p_j X p_k) = p_{j-1}\Phi(X) p_{k-1}, \quad \text{and}\quad \Phi^*(p_j X p_k) = p_{j+1}\Phi^*(X) p_{k+1}, \quad \forall X\in \cB(\cH)
    \end{equation}
    where the subscripts are interpreted modulo $z$. Additionally, for $\Phi_P:= \sum_{n=0}^{z-1}  \theta^n P_n$,
    \begin{equation} \label{eq:irred-choi-P}
        J(\Phi_P^k) = \hat J_k :=  z\left(\sigma \otimes \one\right) L_{k}= \sum_{m=0}^{z-1} \tr[p_m] \frac{p_{m-k}\sigma p_{m-k}}{\tr[ p_{m-k} \sigma]} \otimes \frac{p_m}{\tr[p_m]}.
    \end{equation}

\end{proposition}
\begin{proof}
Let us note that \eqref{it:Phi_Q-Pj} is equivalent to the property that
\begin{equation} \label{eq:Phi_Q-kills-Pj}
    \Phi_Q\circ P_j =  P_j \circ \Phi_Q = 0, \qquad \forall\, j=0,\dotsc,z-1.
\end{equation}
To see this, note that the generalized discrete Fourier transform
\[
    \cF : \bigoplus_{j=0}^{z-1} \cB(\cH) \to \bigoplus_{j=0}^{z-1} \cB(\cH)
\]
given by $\cF( (X_0,\dotsc, X_{z-1} ) ) = (Y_0,\dotsc, Y_{z-1} )$ for $Y_n = \sum_{j=0}^{z-1} \theta^{nj}X_j$ is an invertible linear transformation, with inverse $\cF\inv( (Y_n)_{n=0}^{z-1} ) = \frac{1}{z}\cF( (Y_{z-\ell})_{\ell=0}^{z-1} )$. All indices are taken mod $z$. Next, using the definition of the $P_n$, \eqref{eq:Phi_Q-kills-Pj} is equivalent to
\[
    \Phi_Q( u^{n} \sigma) = 0, \qquad  \Phi_Q^*(u^n)=0, \qquad \forall\, j=0,\dotsc,z-1.
\]
Since $\vec 0 = (\Phi_Q(u^n \sigma))_{n=0}^{z-1} = \cF( (\Phi_Q(p_j \sigma) )_{j=0}^{z-1} )$ and $\vec 0 = (\Phi_Q(u^{n}))_{n=0}^{z-1} = \cF( (\Phi_Q(p_j))_{j=0}^{z-1} )$, and $\cF$ has trivial kernel, \eqref{eq:Phi_Q-kills-Pj} implies \eqref{it:Phi_Q-Pj}. The converse follows similarly.
\Cref{eq:irred_on_p_blocks} follows from a simple computation. The fact that an irreducible map can be decomposed as (\ref{eq:Phi-irred}) for some choices of $z$, $\{p_n\}_{n=0}^{z-1}$, $\sigma$ and $\Phi_Q$ as in (1)-(4b) and (\ref{eq:Phi_Q-kills-Pj}) is not new, and we refer to \cite[Section 6.2]{Wol12} for more details. We believe however that the forward implication is novel, however, and include the proof below.

Let us show that given $z, \{p_n\}_{n=0}^{z-1}$, $\sigma$, and $\Phi_Q$, the decomposition \eqref{eq:Phi-irred} gives an irreducible quantum channel.
Note, by the definition of $P_n$, for any $X\in\mathcal{B}(\cH)$,
\[
    P_j \circ P_k (X) = \tr[ u^{-k} X] \tr[ u^{k-j}\sigma] u^j\sigma  = \delta_{jk} P_j(X)
\]
using $\tr[ \sigma p_n ] = \frac{1}{z}$ and the formula
\begin{equation}\label{eq:roots-of-unity-sum}
    \sum_{n=0}^{z-1} \theta^{n m} = \begin{cases}
        z & m = zk \text{ for some } k\in \mathbb{Z} \\
        0 & \text{otherwise.}
    \end{cases}
\end{equation}
Since $P_0(X) = \tr[X]\sigma$, we have for $j\neq 0$, $0 = P_0\circ P_j (X) = \tr[P_j(X)] \sigma$, yielding that $P_j$ is trace-annihilating: $\tr[P_j(X)] =0$ for all $X\in \cB(\cH)$. In the same way, using assumption~\labelcref{it:Phi_Q-Pj},  $\Phi_Q$ is trace-annihilating. Thus, $\Phi = P_0 + \sum_{n=1}^{z-1} \theta^n P_n + \Phi_Q$ is trace-preserving.

Next, we prove \eqref{eq:irred-choi-P}, which will prove $\Phi$ is CP via assumption~\labelcref{it:Phi_Q-J}. For $\Phi_P := \sum_{m=0}^{z-1} \theta^m P_m$, we have $\Phi_P^k = \sum_{m=0}^{z-1} \theta^{km} P_m$. Then, for any $X\in \cB(\cH)$, we have the discrete Fourier-type computation,
\begin{align}
    \Phi_P^k(X) & = \sum_{m=0}^{z-1}\theta^{km} \tr[ u^{-m} X]u^m \sigma = \sum_{m,n,\ell=0}^{z-1} \theta^{km} \tr[\theta^{-m n}  p_n X] \theta^{\ell m} p_\ell \sigma\nonumber \\
                & =\sum_{m,n,\ell=0}^{z-1} \theta^{m(k-n+\ell)} \tr[ p_n X]  p_\ell \sigma =\sum_{n,\ell=0}^{z-1} z\delta_{\ell = n-k} \tr[ p_n X]  p_\ell \sigma\nonumber      \\
                & =z\sum_{n=0}^{z-1} \tr[ p_n X]  p_{n-k} \sigma
    \label{eq:PhikP_formula}
\end{align}
using \eqref{eq:roots-of-unity-sum}. Next, let $\{\ket{i}\}_{i=0}^{d_{\cH}-1}$ be an orthonormal basis of $\cH$ such that the first $\rank(p_0)$ elements are a basis for $p_0 \cH$, the next $\rank(p_1)$ elements are a basis for $p_2\cH$, and so on. We have $p_0 = \sum_{i=0}^{\rank(p_0)-1} \ket{i}\bra{i}$, and so forth. Thus,
\begin{align}
    J(\Phi_P^k) & =\sum_{i,j = 0}^{d-1} P^k(\ket{i}\bra{j}) \otimes \ket{i}\bra{j} = z\sum_{i,j}\sum_{n=0}^{z-1} \tr[p_n \ket{i}\bra{j}] p_{n-k} \sigma \otimes \ket{i}\bra{j}\nonumber \\
                & = z \sum_{n=0}^{z-1} \sum_{i = 0}^{d-1}\braket{i | p_n | i} p_{n-k} \sigma \otimes \ket{i}\bra{i} = z\sum_{n=0}^{z-1} \sigma p_{n-k}  \otimes p_n\label{eqJPhiP}      \\
                & =z \big(\sigma \otimes\one \big) L_{k}\nonumber.
\end{align}
In particular, $J(\Phi_P) =z \big(\sigma \otimes \one\big) L_{1}$. Thus, by assumption \labelcref{it:Phi_Q-J},
\[
    J(\Phi) = J(\Phi_P) + J(\Phi_Q) \geq J(\Phi_P) - z \big(\sigma \otimes \one\big) L_{1} =0
\]
and hence $\Phi$ is CP. Since $\Phi$ is CPTP, we can use \Cref{prop:irred-ergodic} to prove $\Phi$ is irreducible.  We have
\begin{align*}
    \frac{1}{M}\sum_{n=0}^{M-1} \Phi^n = P_0 + \frac{1}{M} \sum_{m=1}^{z-1} \frac{1 - \theta^{Mm}}{ 1 - \theta^m} P_m + \frac{1}{M}\sum_{n=0}^{M-1} \Phi_Q^n
\end{align*}
using the geometric series $\sum_{n=0}^{M-1} \theta^{mM} = \frac{1 - \theta^{mM}}{1 - \theta^m}$ for $m\neq 0$, which is valid as $\theta^m \neq 1$. Since $P_0[X] = \tr[X]\sigma$, it remains to show that the latter two terms vanish in the limit $M\to \infty$. In fact, since $ \sum_{m=1}^{z-1} \frac{1 - \theta^{Mm}}{ 1 - \theta^m} P_m$ is bounded in norm uniformly in $M$, the second term vanishes asymptotically. Next, since $\ell := \spr(\Phi_Q) < 1$ by assumption~\labelcref{it:Phi_Q-spr}, for $\eps = \frac{1 - \ell}{2} > 0$,  Gelfand's theorem gives that there is $n_0 > 0$ such that (in any matrix norm $\|\cdot\|$), for all $n\geq n_0$,
\[
    \| \Phi_Q^n \|\leq (\ell+\eps)^n < 1.
\]
We may write
\[
    \frac{1}{M}\sum_{n=0}^{M-1} \Phi_Q^n = \frac{1}{M}\sum_{n=0}^{n_0} \Phi_Q^n + \frac{1}{M}\sum_{n=n_0 + 1}^{M-1} \Phi_Q^n.
\]
Since $\sum_{n=0}^{n_0} \Phi_Q^n $ is bounded in norm independently of $M$, the first term vanishes asymptotically; the second term is bounded in norm by the triangle inequality and the geometric series $\sum_{n=0}^\infty (\ell+\eps)^n = \frac{1}{1 - (\ell + \eps)}$. Thus, the limit
\[
    \lim_{M\to \infty} \frac{1}{M}\sum_{n=0}^{M-1} \Phi^n = P_0
\]
holds in any norm. In particular, by \Cref{prop:irred-ergodic} $\Phi$ is irreducible. 
\end{proof}
\begin{remarks}
~\begin{itemize}
\item The proof of \Cref{prop:irred-ergodic} given in \cite[Corollary 6.3]{Wol12} relies on the reverse direction of \Cref{prop:decomp-irred}, while the proof of the forward direction of \Cref{prop:decomp-irred} relies on \Cref{prop:irred-ergodic}. Thus, circular logic is avoided by proving first the reverse direction of  \Cref{prop:decomp-irred}, then \Cref{prop:irred-ergodic}, then the forward direction of  \Cref{prop:decomp-irred}.
\item Under the assumptions of the proposition, $z=1$ if and only if $\Phi$ is primitive. This follows from the spectral characterization of primitivity given in \Cref{prop:spec_char_prim}.
 \item The intertwining property \eqref{eq:irred_on_p_blocks} holds for $\Phi$ and $\Phi_P$ (which itself is an irreducible channel), and therefore for $\Phi_Q$. This implies $J(\Phi_Q) = L_1 J(\Phi_Q) L_1$, i.e.\@, the Choi matrix of $\Phi_Q$ is supported on the same subspace as that of $\Phi_P$. Additionally, \cite[Theorem 5.4]{FP09} shows that for any Kraus decomposition $\{V_i\}_i\subset \cB(A)$, i.e.\@~such that $\Phi(X) = \sum_{i} V_i X V_i^*$ holds for all $X\in \cB(A)$, the intertwining relations
 \begin{equation}\label{eq:irred-kraus-intertwine}
       p_mV_i = V_{i}p_{m+1} \quad \forall m,i
 \end{equation}
 also hold.
        \item Given a map $\Phi_Q$ which intertwines with $\{p_n\}_{n=0}^{z-1}$, a sufficient condition for $\Phi_Q\ge - z\left(\sigma \otimes \one\right)L_{1}\equiv -J(\Phi_P)$ is given by
              \begin{equation}\label{eq:Phi_Q-suff-cond-for-CPB}
                  \|\Phi_Q\|_2 =\|J(\Phi_Q)\|_2 \leq z \lambda_\text{min}(\sigma)\,,
              \end{equation}
              since in that case
              \begin{multline}
              J(\Phi_Q) \geq - \spr(\Phi_Q) L_1 \geq -\|J(\Phi_Q)\|_2 L_1 \\
              \geq  -z \lambda_\text{min}(\sigma) L_1 \geq -z L_1(\sigma\otimes\id)L_1 =- J(\Phi_P).
              \end{multline}
                \item The matrix $\hat J_k$ is separable, and thus $\Phi_P^k$ is entanglement-breaking, for any $k\geq 1$.
        \item If $z >1$, then $\frac{1}{d_\cH}\hat J_k$ does not have full support. Thus, $\frac{1}{d_\cH}\hat J_k$ is on the boundary of the set of density matrices, and thus on the boundary of $\SEP$ and $\PPT$ as well. In fact, we can say something stronger than this: whenever $z>1$, there exist entangled density matrices arbitrarily close to each $\frac{1}{d_\cH} \hat J_k$, $k=0,1,\dotsc,z-1$. To see this, note that $\frac{1}{d_\cH}\hat J_k \in L_k \cD(\cH \otimes \cH ) L_k$. However, we can construct entangled states in $L_j \cD(\cH \otimes \cH ) L_j$ for any $j\neq k$. For instance, let $\ket{0} \in p_0 \cH$, $\ket{1} \in p_1 \cH$, $\ket{ j_0 } \in p_{-j} \cH$, and $\ket{j_1} \in p_{1-j} \cH$ be normalized vectors. Then
              \[
                  \ket{\Omega_j} :=\frac{1}{\sqrt{2}}\big( \ket{j_0}\otimes \ket{0} + \ket{j_1}\otimes \ket{1}  \big)
              \]
              is (local-unitarily equivalent to) a Bell state, and has
              \[
                  L_j \ket{\Omega_j} = \sum_{n=0}^{z-1} (p_{n-j} \otimes p_n )\ket{\Omega_j} = \ket{\Omega_j},
              \]
              and thus $L_j \ket{\Omega_j}\bra{\Omega_j} L_j \in L_j \cD(\cH \otimes \cH ) L_j$. Thus, for any $t\geq 0$,
              \[
                  (1-t)\frac{1}{d_\cH} \hat J_k + t \ket{\Omega_j}\bra{\Omega_j} = (1-t) \frac{1}{d_\cH} \hat J_k \Big|_{L_k} \,\oplus \,t\ket{\Omega_j}\bra{\Omega_j}\Big|_{L_j}
              \]
              is an entangled density matrix, and can be made arbitrarily close to  $\frac{1}{d_\cH} \hat J_k $ by sending $t\to 0$.

\end{itemize}
\end{remarks}

One can relate the decompositions given in \Cref{prop:decomp-irred} and \eqref{eq:Phi-on-decohere-decomp}, as follows. The following result is straightforward, but apparently novel (first appearing in \cite{HRS20}).
\begin{proposition} \label{prop:char-irred-asymp}
    The following are equivalent:
    \begin{enumerate}
        \item There exists a decomposition \eqref{eq:decohere-decomp} such that $\Phi$ satisfies \eqref{eq:Phi-on-decohere-decomp} with $\cK_0=\{0\}$ and $d_i =1$ for each $i$, and $\pi$ is a $K$-cycle
        \item $\Phi$ is irreducible.
    \end{enumerate}
\end{proposition}
\begin{proof}
    First, assume $\Phi$ is irreducible, and adopt the notation of \Cref{prop:decomp-irred}. Then
    \[
        \tilde N(\Phi) =  \linspan \{ u^n \sigma : n = 0, \dotsc,z-1\}\,.\,
    \]
    Let $X\in \tilde \cN(\Phi)$ be given by $X =\sum_{j=0}^{z-1} \lambda_j u^j \sigma$ for some $\lambda_j \in \bC$. Since $\sum_{n=0}^{z-1}p_n = \one$, we have
    \begin{equation} \label{eq:proof-X-irred}
        X = \sum_{n,j=0}^{z-1} \lambda_j  p_n u^j \sigma = \sum_{n,j=0}^{z-1} \lambda_j \theta^{nj} p_n  \sigma = \bigoplus_{n=0}^{z-1} \left(\sum_{j=0}^{z-1} \lambda_j \theta^{nj} \right) \sigma|_{p_n}
    \end{equation}
    where $\sigma|_{p_n}$ is $\sigma$ restricted to the subspace $p_n \cH$, and the direct sum decomposition is with respect to the decomposition $\cH = \bigoplus_{n=0}^{z-1} p_n\cH$.  Note moreover, $\sum_{j=0}^{z-1} \lambda_j \theta^{nj}$ is the $n$th coefficient of the discrete Fourier transform $F_z:\bC^z\to \bC^z$ of the vector $\lambda:=(\lambda_j)_{j=0}^{z-1}$. Since the Fourier transform is invertible, as $\lambda$ ranges over $\bC^z$, the vector of Fourier coefficients range over $\bC^z$ as well. We therefore find
    \[
        \tilde \cN(\Phi) = \bigoplus_{n=0}^{z-1} \bC \,  \sigma|_{p_n}
    \]
    which is a decomposition of the form given by \eqref{eq:decohere-decomp} with $\cK_0=\{0\}$ and $d_n =1$ for each $n$. Moreover, for $X\in \tilde N(\Phi)$ decomposed as $X = \bigoplus_{n=0}^{z-1}  \gamma_n   \sigma|_{p_n} = \sum_{n=0}^{z-1} \gamma_n p_n \sigma p_n$, we have
    \[
        \Phi(X) = \sum_{n=0}^{z-1} \gamma_n \Phi(p_n \sigma p_n) = \sum_{n=0}^{z-1} \gamma_n p_{n-1} \Phi( \sigma)p_{n-1}= \sum_{n=0}^{z-1} \gamma_n p_{n-1} \sigma p_{n-1}
    \]
    by \eqref{eq:irred_on_p_blocks}. Thus, \eqref{eq:Phi-on-decohere-decomp} holds, where $\pi$ is the cyclic permutation $k\to k+1$.

    On the other hand, assume we are given such a decomposition,
    \[
        \tilde \cN(\Phi) = \bigoplus_{i=0}^{K-1} \bC \tau_i
    \]
    for some states $\tau_i \in \cD_+(\cK_i)$ where $\cH = \bigoplus_{i=0}^{K-1} \cK_i$, such that\footnote{If $\pi$ is a cycle, we may reorder the index of the direct sum so that $\pi$ maps $i$ to $i+1$.}
    \[
        \Phi(X) = \bigoplus_{i=0}^{K-1} \gamma_{i+1} \tau_i,\quad \text{ for } \quad X =  \bigoplus_{i=0}^{K-1} \gamma_{i} \tau_i.
    \]
    Define $\sigma := \bigoplus_{i=0}^{K-1} \frac{1}{z}\tau_i$. Then by \eqref{eq:Phi-on-decohere-decomp},
    \[
        \Phi(\sigma) = \bigoplus_{i=0}^{K-1}\frac{1}{z}\tau_i = \sigma
    \]
    as suggested by the name. On the other hand, assume $X\in \cB(\cH)$ is also invariant under $\Phi$. Then $X \in \tilde N(\Phi)$, so $X = \bigoplus_{i=0}^K\lambda_i \tau_i$ for some $\lambda_i \in \bC$. But then
    \[
        \Phi(X) = \bigoplus_{i=0}^K\lambda_{i+1} \tau_i = X \implies \lambda_i = \lambda_{i+1} \, \forall i.
    \]
    Thus, $X$ is proportional to $\sigma$, and $\Phi$ must have a unique invariant state. Since $\sigma$ is additionally full-rank (by construction), we conclude that $\Phi$ is irreducible by \Cref{prop:spec_char_irred}.
\end{proof}

\chapter{Characterizing EEB channels}\label{chap:char_EEB}

\section{Introduction through (counter)examples}\label{sec:ris-intro-examples}
In this chapter, we are interested in characterizing which quantum channels are eventually entanglement-breaking (EEB). 
Here, we will explore this question via a sequence of motivating examples and counterexamples.
Recall that a linear map $\Phi$ is called \emph{eventually entanglement-breaking} (EEB) if $\Phi^n$ is entanglement-breaking for some $n \in \mathbb{N}$.

Let's first see an example of an EEB map: a primitive quantum channel. As defined in \Cref{def:prim_map},
a quantum channel $\Phi$ on $\cB(\cH)$ to be \emph{primitive} if there exists a full-rank state $\sigma\in \cD(\cH)$ such that we have the asymptotic convergence
\begin{equation}\label{def:prim_eeb_intro}
    \Phi^n(\rho) \xrightarrow{n\to\infty} \sigma, \qquad \forall \rho \in \cD(\cH).
\end{equation}

Let us see that if $\Phi \in \cB( \cB(A) \to \cB(A))$ is primitive, it is indeed EEB. Let $R\cong A$ be a reference system, and set $d := d_A = d_R$. Then to check if $\Phi^n$ is EB, we compute its Choi state,
\begin{align*}
\Phi^n \otimes \id_R (\Omega_{A:R}) & = \frac{1}{d}\sum_{ij} \Phi^n(\ket{i}\bra{j}_A)  \otimes \ket{i}\bra{j}_{R}          \\
& \to\frac{1}{d} \sum_{ij} \tr[\ket{i}\bra{j}_A] \,\sigma_A  \otimes \ket{i}\bra{j}_{R} \\
& = \sigma_A  \otimes \frac{1}{d}\sum_{ij} \delta_{ij} \ket{i}\bra{j}_{R}                  \\
& =  \sigma_A  \otimes \frac{1}{d}\sum_{i} \ket{i}\bra{i}_{R}= \sigma_A \otimes \tau_R
\end{align*}
where $\tau_R := \frac{1}{d_{R}} \one_{R}$. We see this is indeed a product state. However, this does not yet complete the proof, since we need to show that $\Phi^n$ is entanglement-breaking for finite $n$, not only in the limit $n\to\infty$. To complete the proof, we use the following corollary to \cite[Theorem 1]{GB02}.
\begin{lemma}\label{lemma1}
    Let $\omega_A,\,\sigma_B > 0$. Then  $\omega_A\otimes \sigma_B + \Delta_{AB}$ is separable for any Hermitian operator $\Delta_{AB}$ such that $\|\Delta_{AB}\|_2\le \lambda_{\min}(\sigma_A)\,\lambda_{\min}(\omega_B)$, where $\lambda_{\min}(\omega_A)$, resp. $\lambda_{\min}(\sigma_B)$, stands for the smallest eigenvalue of $\omega_A$, resp. $\sigma_B$.
\end{lemma}
\begin{proof}
    Theorem 1 of \cite{GB02} shows that if $\|\tilde \Delta_{AB}\|_2 \leq 1$ then $\one_A\otimes \one_B + \tilde \Delta_{AB} \in \SEP(A:B)$. Since
    \[
    \omega_A\otimes \sigma_B + \Delta_{AB} \in \SEP \iff \one_A\otimes \one_B + \omega^{-1/2}_A\otimes \sigma^{-1/2}_B \Delta_{AB} \omega^{-1/2}_A\otimes \sigma^{-1/2}_B \in \SEP(A:B)
    \]
    the proof is concluded by the fact that
    \[
    \|\omega^{-1/2}_A\otimes \sigma^{-1/2}_B \,\Delta_{AB}\, \omega^{-1/2}_A\otimes \sigma^{-1/2}_B\|_2 \leq \|\omega\inv_A\|_\infty \|\sigma\inv_B\|_\infty\|\Delta_{AB}\|_2.
    \]
\end{proof}
\begin{remark}
This short and simple proof is due to an anonymous referee of the paper \cite{HRS20}.
\end{remark}

Returning to our primitive map $\Phi$, we saw that 
\[
\Phi^n \otimes \id_R (\Omega_{A:R}) \xrightarrow{n \to \infty} \sigma_A \otimes \tau_R.
\]
Hence, for $n$ large enough, it holds that
\begin{equation*}
  \|\Phi^n(\Omega_{A:R}) - \sigma_A \otimes \tau_R \|_2\leq  \lambda_\text{min}(\sigma_A)\lambda_\text{min}(\tau_R) = \frac{1}{d}\lambda_\text{min}(\sigma_A).
\end{equation*}
Thus, \Cref{lemma1} proves that $\Phi^n(\Omega_{A:R}) \in \SEP(A:R)$ for $n$ sufficiently large; hence, $\Phi$ is EEB.

\bigskip

Now that we have found an example of a class of EEB quantum channels, let us speculate this is the only such class\footnote{This turns out to be correct in the simpler case of continuous-time Markov quantum semigroups \cite{HRS20}}.

\medskip
\noindent\textit{Wrong conjecture 1:} $\Phi$ is EEB if and only if it is primitive.

\medskip
\noindent\textit{Counterexample:} Consider the simple quantum channel
\[
    \Phi(\rho) = \tr(\rho)\,\ket{0}\bra{0}.
\]
Then $\Phi$ has a rank-1 invariant state (and thus is not primitive), but is also entanglement-breaking and hence eventually entanglement-breaking.

\bigskip

What if we restrict our considerations to channels with a full-rank invariant state? 

\medskip
\noindent\textit{Wrong conjecture 2:} If $\Phi$ has a full-rank invariant state, then it is EEB if and only if it is primitive.

\medskip
\noindent\textit{Counterexample:}
Let $\cH = \cH^1 \oplus \cH^2$, and $\Phi = \Phi_1 \oplus \Phi_2$. That is, for a state $\rho\in \cD(\cH)$,
\[
    \Phi\left( \begin{pmatrix}
            \rho_{11} & \rho_{12} \\ \rho_{21} & \rho_{22}
        \end{pmatrix}\right) := \begin{pmatrix}
        \Phi_1(\rho_{11}) & 0 \\ 0 & \Phi_2(\rho_{22})
    \end{pmatrix}
\]
where each $\rho_{ij}$ is a block-matrix, namely the part of $\rho$ mapping $\cH^j \to \cH^i$. If $\Phi_1$ satisfies \eqref{def:prim_eeb_intro} with some $\sigma_1$ and $\Phi_2$ satisfies \eqref{def:prim_eeb_intro} with some state $\sigma_2$, then similarly as before, $\Phi$ is EEB.
However,
\[
  \Phi^n(\rho) \xrightarrow{n\to\infty} \tr[\rho_{11}] \sigma_1 \oplus \tr[\rho_{22}]\sigma_2,
\]
so $\Phi$ does not satisfy \eqref{def:prim_eeb_intro}: the final state still depends on the input.

\bigskip

In this example, however, $\Phi$ is the direct sum of primitive channels. Let us allow that too.

\medskip
\noindent\textit{Wrong conjecture 3:} 
If $\Phi$ has a full-rank invariant state, then it EEB if and only if it is primitive or the direct sum of primitive channels.

\medskip
\noindent\textit{Counterexample:} Consider if for some $k$, $\Phi^k$ is the sum of primitive channels, but $\Phi$ itself is not. For example,
\[
    \Phi\left( \begin{pmatrix}
            \rho_{11} & \rho_{12} \\ \rho_{21} & \rho_{22}
        \end{pmatrix}\right) := \begin{pmatrix}
        \tr(\rho_{22})\sigma_1 & 0 \\ 0 & \tr(\rho_{11})\sigma_2
    \end{pmatrix}
\]
for some full-rank states $\sigma_1, \sigma_2$. Then
\[
    \Phi^2\left( \begin{pmatrix}
            \rho_{11} & \rho_{12} \\ \rho_{21} & \rho_{22}
        \end{pmatrix}\right) = \begin{pmatrix}
        \tr(\rho_{11})\sigma_1 & 0 \\ 0 & \tr(\rho_{22})\sigma_2
    \end{pmatrix}
\]
is the direct sum of primitive channels. Thus, $\Phi^2$ is eventually entanglement-breaking, as in the previous example, and so $\Phi$ is EEB too.

\bigskip

It turns out the next logical conjecture, that if $\Phi$ has a full-rank invariant state, then it is EEB if and only if it is \emph{eventually} primitive or the direct sum of primitive channels, is correct, as shown by \Cref{prop:charEEB-discrete}. 
 
\section{Entanglement classes of quantum channels}

A map $\Phi:\cB(\cH) \to \cB(\cH)$ is of \textit{positive partial transpose} (PPT) if $(\cT\circ \Phi)\otimes\id_{\cH}$ is a positive operator, where $\cT$ is the partial transpose w.r.t. to some basis. The class of PPT channels on $\cH$ is called $\operatorname{PPT}(\cH)$, and the class of entanglement-breaking channels on $\cH$ is denoted $\EB(\cH)$. As the set of separable bipartite states is a subset of the set of bipartite states with a positive partial transpose, it follows that $\operatorname{EB}(\cH)\subset\operatorname{PPT}(\cH)$.

Given a quantum channel $\Phi:\cB(\cH)\to \cB(\cH)$, the sequence $\{\Phi^n\}_{n\in\NN}$ is called a \textit{discrete time quantum Markov semigroup} (discrete time QMS). Here, semigroup refers simply to the property that $\Phi^{n+m}=\Phi^n\circ \Phi^m$. We will often discuss properties of a discrete time QMS $\{\Phi^n\}_{n\in\NN}$ by simply referring to the associated quantum channel $\Phi$.

In this chapter, we study the entanglement properties of quantum Markovian evolutions in discrete time. A discrete time QMS $\{\Phi^n\}_{n\in\NN}$ is said to be \textit{eventually entanglement breaking} (EEB) if there exists $n_0\in\NN$ such that for any $n\ge n_0$, $\Phi^n$ is entanglement breaking. The class of eventually entanglement breaking Markovian evolutions is denoted by $\operatorname{EEB}(\cH)$. We also say a quantum channel $\Phi$ is eventually entanglement breaking if the discrete QMS $\{\Phi^n\}_{n=1}^\infty$ is eventually entanglement breaking. On the other hand, Markovian evolutions which are not entanglement breaking at any finite time are called \emph{entanglement saving}, using language introduced by Lami and Giovannetti \cite{LG16}; the class of entanglement saving Markovian evolutions is denoted by $\operatorname{ES}(\cH)$. Thus, the set of all discrete-time Markovian evolutions  decomposes into two disjoint classes,
\begin{equation} \label{eq:EEB-ES-decomp}
    \EEB(\cH) \sqcup \ES(\cH).
\end{equation}
Lami and Giovannetti also introduce the notion of \emph{asymptotically entanglement saving} evolutions in the discrete-time case. They showed that every discrete time QMS has at least one limit point, and either all of the limit points of a discrete time QMS $\{\Phi^n\}_{n=1}^\infty$ are entanglement breaking, or none of them are. They term the latter case as asymptotically entanglement saving, and we denote the set of asymptotically entanglement saving evolutions on $\cH$ as $\AES(\cH)$. In analogy, we call the former case by \emph{asymptotically entanglement breaking}, denoted $\AEB(\cH)$. Thus, the set of discrete time QMS on $\cH$ decomposes into the disjoint classes
\begin{equation}\label{eq:AES-AEB-decomp}
    \AES(\cH)\sqcup \AEB(\cH).
\end{equation}
It is interesting to compare \eqref{eq:EEB-ES-decomp} and \eqref{eq:AES-AEB-decomp}. A discrete time QMS $\{\Phi^n\}_{n=1}^\infty$ is AES if the limit points of the sequence $\{J(\Phi^n)\}_{n=1}^\infty$ are all entangled. Since $J(\Phi^{n+1}) = \Phi\otimes \id (J(\Phi^n))$, if $J(\Phi^{n+1})$ is entangled, $J(\Phi^n)$ must be as well. In particular, if $\{\Phi^n\}_{n=1}^\infty\in \AES(\cH)$, then $J(\Phi^n)$ is entangled for every $n$, and the discrete time QMS is entanglement saving. So we see $\AES(\cH)\subset \ES(\cH)$. However, a priori, an entanglement saving QMS could be asymptotically entanglement breaking: at any finite $n$, $J(\Phi^n)$ could be entangled, but in the limit, $J(\Phi^n)$ could be in the set of separable states (though necessarily on the boundary). We therefore define $\EB_\infty(\cH) = \AEB(\cH) \cap \ES(\cH)$, the set of discrete time QMS which are asymptotically entanglement breaking, but not entanglement breaking for any finite $n$. With this notion, we may relate \eqref{eq:EEB-ES-decomp} and \eqref{eq:AES-AEB-decomp}. We have the disjoint decomposition of the set of all discrete time QMS, denoted $\operatorname{dQMS}(\cH)$, satisfies
\begin{equation} \label{eq:QMC-full-decomp}
    \operatorname{dQMS}(\cH) = \UOLoverbrace{\EEB(\cH) \operatorname{\sqcup}}[\EB_\infty(\cH)]^{\AEB(\cH)}  \UOLunderbrace{ \sqcup \AES(\cH)}_{\ES(\cH)}.
\end{equation}
\begin{remark}
    The isomorphism between bipartite states and CP maps given by the Choi matrix and the equivalence between EB maps and separable states discussed in \Cref{sec:CJ} allows us to directly translate results on the entanglement loss of CP maps to statements about the separability of a bipartite state. We will mostly state our results in the picture of CPTP maps (see also \Cref{sec:CPnonTP} for a method to remove the TP assumption), but it should be straightforward to derive the corresponding statements for bipartite states.
\end{remark}

\begin{figure}
    \centering
    \begin{tabular}{cc}
        Discrete-time   & \begin{tikzcd}
            \text{Primitive} \rar \arrow[drr, bend right] & \text{Irreducible} \rar & \text{Faithful} \arrow[d, dashed, "\text{ \Cref{cor:pptfulleeb}}"'] & \lar \text{Unital} \arrow[ld, bend left, dashed, "\text{ \cite{RJP18}}"]\\
            & & \text{EEB}
        \end{tikzcd}
\end{tabular}
    \caption{Relations between classes of quantum Markov semigroups, in which arrows represent subsets; e.g.\@, primitive discrete-time quantum Markov semigroups are a subset of irreducible discrete-time quantum Markov semigroups. The dashed arrows indicate relations which only hold for quantum Markov semigroups $\{\Phi^n\}_{n=1}^\infty$ associated to a PPT channel $\Phi$. The arrows without annotations are standard and are discussed in the text.}
    \label{fig:diagram}
\end{figure}
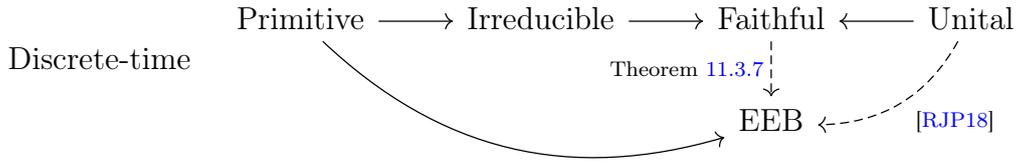

Some of the definitions introduced in this chapter, along with a preview of some of the results, are depicted diagrammatically in \Cref{fig:diagram}.

Let us now consider an example to illustrate these definitions and show that all sets arise naturally in physical systems.

\begin{example} \label{ex:RWA}
    Consider a discrete time quantum Markov semigroup $\{\Phi^n\}_{n\in\NN}$ associated to a \emph{repeated interaction system}, in which a system $\sys$ interacts with a chain of identical \emph{probes} $\env_k$, one at a time, for a duration $\tau$. During the interaction, the dynamics of the system are modeled by a Hamiltonian evolution, and at the times $(k \tau)_{k\geq 1}$, the evolution forms a semigroup. In this example, the system and each probe are 2-level systems, with associated Hilbert spaces $\cH_\sys = \cH_\env = \bC^2$. We define Hamiltonians $h_\sys = E a^*a$ and $h_\env = E_0 b^*b$, where $a/a^*$, resp. $b/b^*$, are the annihilation/creation operators for $\sys$, resp. $\env$, and $E\in \R_{>0}$ (resp. $E_0\in \R_{>0}$) corresponds to the energy of the excited state of $\sys$ (resp. $\env$). We can express these operators in the (ground state, excited state) basis of each system by
    \[
        a = b= \begin{pmatrix}
            0 & 1 \\ 0 & 0
        \end{pmatrix}, \qquad a^* = b^* = \begin{pmatrix}
            0 & 0 \\ 1 & 0
        \end{pmatrix}, \qquad a^*a = b^*b = \begin{pmatrix}
            0 & 0 \\ 0 & 1
        \end{pmatrix}.
    \]
    We consider the initial state of each probe to be a \emph{thermal state},
    \[
        \xi_\beta  = \frac{\exp(-\beta h_\env)}{\tr[\exp(-\beta h_\env)]} = \begin{pmatrix}
            \frac{1}{1 + \e^{-\beta E_0}} & 0 \\ 0 & \frac{\e^{-\beta E_0}}{1+\e^{-\beta E_0}}
        \end{pmatrix}
    \]
    where $\beta \in [0,\infty]$ represents the inverse temperature (setting Boltzmann's constant to one). In the case of zero-temperature ($\beta=\infty$), we take
    \[
        \xi_\infty = \lim_{\beta\to\infty} \xi_\beta = \begin{pmatrix}
            1 & 0 \\ 0 & 0
        \end{pmatrix}.
    \]
    We consider an interaction modeling the two systems coupling through their dipoles, in the \emph{rotating wave approximation}. In this setting, the system and each probe interact via the potential $\lambda v_\text{RW} \in \cB(\cH_\sys \otimes \cH_\env)$, where $\lambda \geq 0$ is a coupling constant, and
    \[
        v_\text{RW} = \frac{u_1}{2}(a^* \otimes b + a \otimes b^*)
    \]
    where $u_1$ is a constant, which we take to be equal to $1$ with units of energy.
    This is a common approximation in the regime $|E - E_0 | \ll \min\{E,E_0\}$ and $\lambda \ll |E_0|$.

    The system begins in some state $\rho\init$, couples with the first probe (in thermal state $\xi_\beta$), and evolves for a time $\tau >0$ according to the unitary operator
    \[
        U := \exp(-i\tau (h_\sys \otimes \id + \id \otimes h_\env + \lambda v_\text{RW})).
    \]
    That is, $\rho\init\otimes \xi_\beta$ evolves to $U\,( \rho\init \otimes \xi_\beta )\,U^*$. Then we trace out the probe to obtain
    \[
        \rho_1 := \tr_\env( U\,( \rho\init \otimes \xi_\beta )\,U^*).
    \]
    This process is repeated, and at the end of the $k$th step, the system is in the state
    \[
        \rho_k =  \tr_\env( U \,(\rho_{k-1} \otimes \xi_\beta )\,U^*) = \Phi^k(\rho\init)
    \]
    where $\Phi$ is the quantum channel given by
    \[
        \Phi(\eta) := \tr_\env( U(\, \eta \otimes \xi_\beta\,) U^*) = \Phi^k(\rho\init).
    \]
    What class in the decomposition \eqref{eq:QMC-full-decomp} does the discrete time QMS $\{\Phi^n\}_{n\in\NN}$ lie in? To answer that question, we first compute the eigenvectors and eigenvalues of $\Phi$, yielding
    \[
        \Phi(\rho_{\beta^*}) = \rho_{\beta^*}, \qquad \Phi(a) = \gamma a, \qquad \Phi(a^*) = \bar \gamma a^*, \qquad \Phi(\sigma_z) = |\gamma|^2 \sigma_z
    \]
    where $\sigma_z = \begin{pmatrix}
            1 & 0 \\ 0 & -1
        \end{pmatrix}$ is the Pauli-$z$ matrix,
    \[
        \gamma := \e^{-\frac{1}{2}i \tau (E_0 + E)} \left(\cos \frac{\tau \nu}{2} + i \frac{(E_0 - E)}{\nu} \sin \frac{\tau\nu}{2}\right)
    \]
    for $\nu := \sqrt{ (E_0 -E)^2 + \lambda^2}$,
    and, defining the rescaled inverse temperature $\beta^* := \frac{E_0}{E}\beta$,
    \[
        \rho_{\beta^*} := \begin{pmatrix}
            \frac{1}{1 + \e^{-\beta^* E}} & 0 \\ 0 & \frac{\e^{-\beta^* E}}{1+\e^{-\beta^* E}}
        \end{pmatrix}=  \xi_\beta
    \]
    is a thermal state on $\cH_\sys$ represented by the same matrix as $\xi_\beta$. In the case $\beta=\infty$, we set $\rho_{\beta^*} := \begin{pmatrix}
            1 & 0 \\ 0 & 0
        \end{pmatrix}= \xi_\infty$. Note
    \[
        |\gamma| = \sqrt{ 1 - \frac{\lambda^2}{\nu^2}\sin^2 \left( \frac{\nu \tau}{2} \right) } \leq 1
    \]
    as required, since $\Phi$ is a CPTP map.
    In \cite[Prop. 23]{LG16} the authors show that a qubit channel is AES if and only if it is a unitary channel, which in turn is equivalent to it having determinant $1$.
    Thus, we have $\{\Phi^n\}_{n\in\NN} \in \AES(\cH)$ if and only if $|\gamma|=1$.

    To analyze whether or not $\Phi^n$ is entanglement-breaking, it suffices to check if $J(\Phi^n)$ is PPT, as $\Phi$ is a qubit channel. To that end, define the Gibbs factor $g = \exp(-\beta E_0)$ and  partition function $Z = 1 + g$. Then we have
    \[
        \rho_{\beta^*} +g Z\inv \sigma_z= \begin{pmatrix}
            1 & 0 \\ 0 & 0
        \end{pmatrix}, \qquad \rho_{\beta^*}  -  Z\inv \sigma_z = \begin{pmatrix}
            0 & 0 \\ 0 & 1
        \end{pmatrix}_.
    \]
    Thus, taking $(\ket{0},\ket{1})$ to be the (ground state, excited state) basis for each system,
    \begin{align*}
        (\id \otimes \mathcal{T}) J(\Phi^n) & = \sum_{i,j=0}^1 \Phi^n(\ket{i}\bra{j})\otimes \ket{j}\bra{i} \\
                                            & = \begin{pmatrix}
            \Phi^n(\rho_{\beta^*} + Z\inv \sigma_z) & \Phi^n(a^*)                                \\
            \Phi^n(a)                               & \Phi^n(\rho_{\beta^*}  - g Z\inv \sigma_z)
        \end{pmatrix}                                  \\
                                            & = \begin{pmatrix}
            \rho_{\beta^*}+|\gamma|^{2n} Z\inv\sigma_z) & \bar \gamma^n a^*                              \\
            \gamma^n a                                  & \rho_{\beta^*}- |\gamma|^{2n}g Z\inv \sigma_z)
        \end{pmatrix}                                  \\
                                            & = \begin{pmatrix}
            Z\inv( g+ |\gamma|^{2n}) & 0                          & 0                          & 0                        \\
            0                        & Z\inv ( 1 - |\gamma|^{2n}) & \bar \gamma^n              & 0                        \\
            0                        & \gamma^n                   & Z\inv g(1 - |\gamma|^{2n}) & 0                        \\
            0                        & 0                          & 0                          & Z\inv(1+g |\gamma|^{2n})
        \end{pmatrix}
    \end{align*}
    which has eigenvalues
    \[
        Z\inv( g+ |\gamma|^{2n}), \quad Z\inv(1+g |\gamma|^{2n}), \quad \frac{1 - |\gamma|^{2n}}{2} \pm \frac{1}{2}\sqrt{ (1 - |\gamma|^{2n})^2 \left( \frac{g - g\inv}{g + g\inv} \right)^2 + 4 |\gamma|^{2n} }
    \]
    for $g\in (0,1]$, and
    \[
        1, \quad 1, \quad \pm |\gamma|^{2n}
    \]
    for $g=0$.
    In either case, the eigenvalues only depend on the independent parameters $|\gamma| \in [0,1]$ and $g \in [0,1]$. Since $(\id \otimes \mathcal{T}) J(\Phi^n) \geq 0$ is equivalent to $\Phi^n \in \EB(\cH)$, we find
    \begin{itemize}
        \item If $|\gamma| = 1$, $\{\Phi^n\}_{n\in\NN} \in \AES(\cH)$. This occurs when $\nu\tau\in 2\pi\bZ$; in this case, $\Phi$ is a unitary channel.
        \item If $|\gamma| = 0$, then $\Phi\in \EB(\cH)$. This occurs when $E=E_0$, and $\nu\tau\in\pi + 2\pi \bZ$. In this case, $\Phi =\rho_{\beta^*}\tr[\cdot]$.
        \item If $|\gamma| \in (0,1)$ and $g=0$, $\{\Phi^n\}_{n\in\NN} \in \EB_\infty(\cH)$.  In this case, $\beta=\infty$ and $\Phi$ has a unique peripheral eigenvalue, namely $1$, with non-full-rank invariant state, $\ket{1}\bra{1}$.
        \item If $|\gamma|\in (0,1)$ and $g\in (0,1]$,  then $\{\Phi^n\}_{n=1}^\infty \in \EEB(\cH)$, and in particular, the minimal $n$ such that $\Phi^n \in \EB(\cH)$ is given by $n =\max(1, \ceil{ \frac{1}{2}\frac{\log B(g)}{\log|\gamma|}})$ where
              \[
                  B(g) := \frac{1 + 4 g^2 + g^4 - (1 + g^2) \sqrt{1 + 6 g^2 + g^4}}{2 g^2} \in (0, 3-2 \sqrt{2}] \quad \text{for}\quad g\in (0,1].
              \]
              In this case, $\Phi$ is primitive, with faithful invariant state $\rho_{\beta^*}$.
    \end{itemize}
     Repeated interaction systems are returned to in \Cref{chap:RIS} and this example is revisited in \Cref{sec:RIS_examples}.
\end{example}

\section{Which quantum channels are eventually\texorpdfstring{\\}{ }entanglement-breaking?}

In order to better characterize discrete-time evolutions for which asking the titular question makes sense, we first need to leave aside those evolutions for which the phenomenon does not occur, that is, evolutions that either destroy entanglement after an infinite amount of time (EB$_\infty$), or even those of never-vanishing output entanglement (AES).

A large part of this question was already answered in the discrete time case by \cite{LG16}. In that work (Theorem 21), the authors showed that, given a quantum channel $\Phi:\cB(\cH)\to\cB(\cH)$ with $\dim(\ker\Phi)<2(d_\cH-1)$, $\{\Phi^n\}_{n\in\NN}\in \operatorname{ES}(\cH)$ if and only if either it has a non-full rank positive fixed point, or the number of peripheral eigenvalues is strictly greater than $1$, which itself is equivalent to the existence of $1\le n\le d_\cH$ such that $\Phi^n$ has a non-full rank positive fixed point. In the same paper, the authors showed that if $\Phi$ has more than $d_\cH$ peripheral eigenvalues, then $\{\Phi^n\}_{n\in\NN}$ is asymptotically entanglement saving. These interesting results clearly show the link between the spectral properties of the quantum channel $\Phi$ and the entanglement properties of the corresponding discrete time quantum Markov semigroup $\{\Phi^n\}_{n\in\NN}$. In the next subsections, we further develop this intuition. First, we prove the following simple consequence of a result from \cite{LG16}.

\begin{lemma} \label{d-periph-ES}
    Let $\Phi$ be a quantum channel on $\cH$ with $d_\cH$ peripheral eigenvalues counted with multiplicity and at least one non-zero non-peripheral eigenvalue. Then $\{\Phi^n\}_{n\in\NN}\in\operatorname{ES}(\cH)$.
\end{lemma}

\begin{proof}
    For any $N$, $\Phi^N$ has $d_\cH$ peripheral eigenvalues and at least one non-peripheral eigenvalue non-zero. Thus, if $\{\lambda_k\}_{k=1}^{d^2}$ are the eigenvalues of $\Phi$ counted with multiplicity, we have
    \[
        \|\Phi^N\|_1 \geq \sum_{k=1}^{d^2} |\lambda_k^N| = d_\cH + \sum_{\lambda_k : |\lambda_k|<1} |\lambda_k^N| > d_\cH\,.
    \]
    The result follows from the fact that a quantum channel $\Psi:\cB(\cH)\to \cB(\cH)$,
     \begin{equation} \label{eq:1norm_not_EB}
        \|\Psi\|_1 > d_{\cH}\implies \Psi\not\in \EB(\cH)
    \end{equation}
    which is due to the reshuffling criterion \cite{CW03} (see, e.g.\@ \cite[eq.~(47)]{LG16}).
\end{proof}

\subsection{Irreducible evolutions} \label{sec:irred}

The previous result suggest looking at channels with less than $d_\cH$ peripheral eigenvalues counted with multiplicity in order to characterize EEB. Since primitive channels are eventually entanglement-breaking, a natural next step is to consider the wider class of irreducible channels introduced in \Cref{sec:periph_superoperators}.

As a first step, \Cref{prop:decomp-irred} shows that the peripheral eigenvectors of irreducible channels $\Phi$ commute. In Theorem 32 of \cite{LG16}, the authors show that asymptotically entanglement saving channels are characterized by the fact that they possess at least two noncommuting phase points. This implies $\{\Phi^n\}_{n\in\NN} \in \AEB(\cH)$, which generalizes Corollary 6.1 of \cite{RJP18} to the non-unital case. There, the authors show that a unital irreducible quantum channel is AEB if and only if its phase space is commutative.

Next, the intertwining property \eqref{eq:irred_on_p_blocks} of irreducible maps is very useful for understanding their entanglement breaking properties. In fact, a slight generalization of this property will prove useful. Recall that an orthogonal resolution of the identity is a collection of orthogonal projections which sum to the identity.

\begin{definition}[$\big(\{p_i\}_{i=0}^{z-1}, \{\tilde p_i\}_{i=0}^{z-1}\big)$-block preserving]
    Given two orthogonal resolutions of the identity, $\{p_i\}_{i=0}^{z-1}$ and $\{\tilde p_i\}_{i=0}^{z-1}$, we say that a quantum channel $\Phi$ is $\big(\{p_i\}_{i=0}^{z-1}, \{\tilde p_i\}_{i=0}^{z-1}\big)$-block preserving if for all $i,j\in \{0,\dotsc,z-1\}$,
    \[
        \Phi( p_i \cB(\cH) p_j) \subset \tilde p_i \cB(\cH)\tilde p_j.
    \]
\end{definition}
From \eqref{eq:irred_on_p_blocks}, if $\Phi$ is irreducible, then $\Phi^k$ is $\big(\{p_i\}_{i=0}^{z-1}, \{ p_{i-k}\}_{i=0}^{z-1}\big)$-block preserving. Using this notion, the following result shows that PPT channels which are block preserving in the above sense must annihilate off-diagonal blocks. In this work, this is the key property that links PPT channels to eventually entanglement-breaking channels. We note that results of a similar flavor were shown in~\cite{,Cariello2016,Cariello2018}.
\begin{lemma} \label{lem:block preserving-PPT}
    If $\Phi$ is a $\operatorname{PPT}$ quantum channel and  is $\big(\{p_i\}_{i=0}^{z-1}, \{ \tilde p_{i}\}_{i=0}^{z-1}\big)$-block preserving, then
    \begin{equation}\label{eq:PPT-kills-offdiag}
        \Phi( p_i \cB(\cH) p_j) = \{0\}
    \end{equation}
    for all $i\neq j$.
\end{lemma}
\begin{proof}
    Let us prove the contrapositive. Assume for some $i\neq j$, $\Phi(p_i \cB(\cH) p_j) \neq \{0\}$; without loss of generality, take $i=0$ and $j=1$. Then let $\ket{0}\bra{1}\in p_0 \cB(\cH) p_1$ such that $\Phi(\ket{0}\bra{1})\neq 0$. Then also $\Phi(\ket{1}\bra{0})\neq 0$.
    Consider $\Omega_{01} = \ket{\Omega_{01}}\bra{\Omega_{01}}$, the unnormalized Bell state associated to $\ket{\Omega_{01}} := \ket{00}+\ket{11}$. The block-preserving assumption yields
    \[
        \Phi(\ket{i}\bra{j}) =\Phi(p_i\ket{i}\bra{j}p_j) = \tilde p_i \,\Phi(\ket{i}\bra{j}) \tilde p_j
    \]
    for each $i,j\in \{0,1\}$.
    Then, neglecting rows and columns of all zeros, $(\Phi\otimes\id)(\Omega_{01})$ can be written as
    \[
        (\Phi\otimes \id)(\Omega_{01}) = \sum_{i,j=0}^1 \Phi(\ket{i}\bra{j})\otimes\ket{i}\bra{j}= \left(\begin{array}{@{}c|c@{}}
                \begin{matrix}
                    \Phi(\ket{0}\bra{0}) & 0 \\
                    0                    & 0
                \end{matrix}
                                           & \begin{matrix}
                    0 & \Phi(\ket{1}\bra{0}) \\
                    0 & 0
                \end{matrix} \\
                \hline
                \begin{matrix}
                    0                    & 0 \\
                    \Phi(\ket{0}\bra{1}) & 0
                \end{matrix} &
                \begin{matrix}
                    0 & 0                    \\
                    0 & \Phi(\ket{1}\bra{1})
                \end{matrix}
            \end{array}\right)
    \]
    as $\{\tilde p_0,\tilde p_1\}$ blocks in the  $\{\ket{0},\ket{1}\}$ basis.
    Now, taking the partial transpose on the first system,
    \[
        (\mathcal{T}\otimes \id )\circ (\Phi\otimes \id)(\Omega_{01}) = \left(\begin{array}{@{}c|c@{}}
                \begin{matrix}
                    \Phi(\ket{0}\bra{0}) & 0 \\
                    0                    & 0
                \end{matrix}
                                           & \begin{matrix}
                    0                    & 0 \\
                    \Phi(\ket{1}\bra{0}) & 0
                \end{matrix} \\
                \hline
                \begin{matrix}
                    0 & \Phi(\ket{0}\bra{1}) \\
                    0 & 0
                \end{matrix} &
                \begin{matrix}
                    0 & 0                    \\
                    0 & \Phi(\ket{1}\bra{1})
                \end{matrix}
            \end{array}\right).
    \]
    The eigenvalues of this matrix are the eigenvalues of $  \Phi(\ket{0}\bra{0})$, together with the eigenvalues of $\Phi(\ket{1}\bra{1})$, and the eigenvalues of the block matrix
    \[
        X =\begin{pmatrix}
            0                    & \Phi(\ket{1}\bra{0}) \\
            \Phi(\ket{0}\bra{1}) & 0
        \end{pmatrix}_.
    \]
    Since $X$ is non-zero, self-adjoint, and traceless, it must have both strictly positive and strictly negative eigenvalues. Thus, $(\mathcal{T}\otimes \id )\circ(\Phi\otimes \id)(\Omega_{01}) $ has negative eigenvalues, so $(\Phi\otimes \id)(\Omega_{01})$ is not PPT.
\end{proof}

\begin{proposition}\label{theoremEB}
    Let $\Phi$ be an irreducible $\operatorname{CPTP}$ map, $k\geq 1$, and let us adopt the notation of \Cref{prop:decomp-irred}. Assume $\Phi$ is not primitive (i.e $z\geq 2$). Then
    \begin{equation} \label{eq:irred-nec_for_EB}
        \Phi^k\in \PPT(\cH) \implies \Phi^k(p_i \cB(\cH) p_j) = \{0\} \quad \forall\, i\neq j.
    \end{equation}
    On the other hand, if
    \begin{equation}\label{eq:assume_PhiQ_kills_offdiag}
        \Phi^k(p_i \cB(\cH) p_j) = \{0\} \quad \forall\, i\neq j
    \end{equation}
    and additionally, for each $j$ such that $\rank p_j \geq 2$,
    \begin{equation} \label{eq:irred-JQ-block-EB}
        \left\|J(\Phi_Q^k|_{p_j \cB(\cH)p_j})\right\|_2\leq z\lambda_{\min}(\sigma|_{p_j\cH})
    \end{equation}
    then $\Phi^k\in \EB(\cH)$.

    In the case $z=d_\cH$, we may write $p_j=\ket{j}\bra{j}$ for $j=0,\dotsc,z-1$. In this case,  $\Phi_Q^k(\ket{i}\bra{j})=0$ for all $i\neq j$, if and only if $\Phi^k\in \EB(\cH)$.
\end{proposition}

\begin{remark}
    Under the assumption \eqref{eq:assume_PhiQ_kills_offdiag},
    \begin{align}\label{EB}
        \|J(\Phi_Q^k)\|_2  \leq z\lambda_\text{min}(\sigma)
    \end{align}
    implies \eqref{eq:irred-JQ-block-EB}, as
    \[
        \|J(\Phi_Q^k|_{p_j \cB(\cH)p_j})\|_2 \leq \sum_{n=0}^{z-1}\|J(\Phi_Q^k|_{p_n \cB(\cH)p_n})\|_2 = \|J(\Phi_Q^k)\|_2  \leq z\lambda_\text{min}(\sigma) \leq  z\lambda_\text{min}(\sigma|_{p_j}).
    \]
\end{remark}

\begin{proof}
    \eqref{eq:irred-nec_for_EB} follows immediately from the fact that if $\Phi$ is irreducible, then $\Phi^k$ is $\big(\{p_i\}_{i=0}^{z-1}, \{ p_{i-k}\}_{i=0}^{z-1}\big)$-block preserving, and \Cref{lem:block preserving-PPT}.

    Next, assume \eqref{eq:assume_PhiQ_kills_offdiag} holds. Let $\{\ket{i}\}_{i=0}^{d_\cH-1}$ be an orthonormal basis of $\cH$ such that there are disjoint index sets $\{I_n\}_{n=0}^{z-1}$ such that for each $n$, $\ket{i}\in p_n \cH$ for $i\in I_n$. Taking the Choi matrix in the $\ket{i}$ basis,
    \begin{align*}
        J(\Phi^k) & = \sum_{i,j}\Phi^k(\ket{i}\bra{j})\otimes \ket{i}\bra{j}                        \\
                  & = \sum_{n=0}^{z-1}\sum_{i,j\in I_n}\Phi^k(\ket{i}\bra{j})\otimes \ket{i}\bra{j}
    \end{align*}
    using that $\Phi^k(\ket{i}\bra{j})=0$ whenever $i$ and $j$ do not share an index set $I_n$, which follows from \eqref{eq:assume_PhiQ_kills_offdiag}. Then (see e.g.\@ \eqref{eqJPhiP}):
    \begin{align*}
        J(\Phi^k) & = \sum_{n=0}^{z-1}\sum_{i,j\in I_n}\left(\delta_{i,j}\,z\, p_{n-k}\,\sigma + \Phi^k_Q(\ket{i}\bra{j})\right)\otimes \ket{i}\bra{j}  \\
                  & = \sum_{n=0}^{z-1}\left(z \,p_{n-k}\,\sigma \otimes p_n + \sum_{i,j\in I_n}\Phi^k_Q(\ket{i}\bra{j})\otimes \ket{i}\bra{j}\right)\,.
    \end{align*}
    We note that $\sum_{i,j\in I_n}\Phi^k_Q(\ket{i}\bra{j})\otimes \ket{i}\bra{j}\in p_{n-k}\otimes p_{n}\cB(\cH\otimes \cH)p_{n-k}\otimes p_{n}$, since both $\Phi$ and $\Phi_P$ map $p_n\cB(\cH)p_n$ to $p_{n-1}\cB(\cH)p_{n-1}$ (see \cref{eq:irred_on_p_blocks,eq:PhikP_formula}).
    Since by assumption
    \[
        \|J(\Phi^k_Q|_{p_n\cB(\cH)p_n})\|_2\leq z\lambda_\text{min}(\sigma|_{p_n\cH})
    \]
    then \Cref{lemma1} applied to the Hilbert space $p_{n-k} \cH\,\otimes\, p_n\cH$ gives that $z \sigma \otimes \one |_{p_{n-k}\cH\,\otimes\, p_n\cH} + J(\Phi^k_Q|_{p_n\cB(\cH)p_n})$ is separable on that space. 
    We may embed this state in $\cB(\cH\otimes \cH)$ (without changing the tensor product structure) yielding that
    \[
        p_{n-k}\otimes p_n \left(z \sigma \otimes \one  + J(\Phi^k_Q)\right)p_{n-k}\otimes p_n
    \]
    is a non-full-rank separable state on $\cB(\cH\otimes \cH)$. By \Cref{lem:SEP_orthog},
    summing over $n$ yields that $J(\Phi^k)$ is separable, so $\Phi^k\in \EB(\cH)$.

    For the case $z=d$, we simply note that $\rank p_j = 1$ for all $j$, and hence the statement follows from the above two results.
\end{proof}

Since the limit points of $\{J(\Phi^n)\}_{n=1}^\infty$ are separable but arbitrarily close to entangled states (as shown in the remarks following \Cref{prop:decomp-irred}) the question arises of whether or not there are quantum channels that are both irreducible and in $\operatorname{EB}_\infty(\cH)$. In the case that $\Phi$ has maximal period $z=d_\cH$, \Cref{d-periph-ES} resolves this affirmatively as long as $\Phi_Q$ is not nilpotent. In the case when the period is much less than the dimension; say $z=2 <d_\cH$, then the underlying argument (relying on the reshuffling criterion via \eqref{eq:1norm_not_EB}) provides little help: $\|\Phi^n\|_1 = z + o(n) < d_\cH$ for large $n$. However, using \Cref{theoremEB}, we can design $\EB_\infty(\cH)$ irreducible channels rather easily, as shown in the following example.
\begin{example}
    Let us construct an irreducible quantum channel $\Phi$ via \Cref{prop:decomp-irred}. with period $z=2$. We choose any full rank state $\sigma$ as the invariant state, and any pair of orthogonal projections $\{p_1,p_2\}$ which commute with $\sigma$ as the Perron-Frobenius projections. Let $d_j=\rank p_j$ for $j=0,1$, and let $\{\ket{e_i}\}_{i=0}^{d_0-1}$ be an eigenbasis of $\sigma p_0$ and likewise $\{\ket{f_i}\}_{i=0}^{d_1-1}$ be an eigenbasis of $\sigma p_1$. For some $\lambda\in \bC$ with $|\lambda|<1$, define $\Phi_Q$ by
    \begin{gather*}
        \Phi_Q(\ket{e_i}\bra{e_j}) = \Phi_Q(\ket{f_i}\bra{f_j}) = 0, \\
        \Phi_Q(\ket{e_i}\bra{f_j})^* =
        \Phi_Q(\ket{f_i}\bra{e_j}) = \delta_{i,0}\delta_{j,0}\bar\lambda \ket{e_0}\bra{f_0}
    \end{gather*}
    for each $i,j$. Then $\Phi_Q(\sigma p_j)=\Phi_Q^*(p_j)=0$, $\spr(\Phi_Q)=|\lambda|<1$. Since
    \[
        J(\Phi_Q) = \lambda \ket{f_0}\bra{e_0} \otimes \ket{f_0}\bra{e_0}  +\bar\lambda \ket{e_0}\bra{f_0}\otimes  \ket{f_0}\bra{e_0} \,,
    \]
    we have $\|J(\Phi_Q)\|_2 = \sqrt{2}\,|\lambda|$. Thus, choosing $\lambda$ to satisfy $0<|\lambda| < \min(1,  \sqrt{2} \,\lambda_\text{min}(\sigma))$, we have  $J(\Phi_Q)\geq -J(\Phi_P)$ by \eqref{eq:Phi_Q-suff-cond-for-CPB}, and  $\Phi$ is an irreducible CPTP map of period 2. Moreover,
    \[
        \Phi_Q^n(\ket{e_0}\bra{f_0}) = \begin{cases}
            \lambda^n \ket{f_0}\bra{e_0} & n\text{ odd}  \\
            \lambda^n \ket{e_0}\bra{f_0} & n\text{ even}
        \end{cases}
    \]
    which is non-zero for any $n$. Thus, $\Phi\in \EB_\infty(\cH)$ by \Cref{theoremEB}.

    \begin{remark}
    It was proven in Theorem 21 of \cite{LG16} that if the number of zero eigenvalues of $\Phi$ is strictly less than $2(d_\cH-1)$, the fact that $\Phi$ has at least two peripheral eigenvalues implies that it is entanglement saving. However, in the present example,
        $\Phi$ has four non-zero eigenvalues (two peripheral, and $\pm |\lambda|$), and therefore $d_\cH^2-4$ zero eigenvalues. Thus, for $d_\cH\geq 3$, Theorem~11 of \cite{LG16} does not apply to $\Phi$.
    \end{remark}
\end{example}
\medskip
\Cref{theoremEB} implies the following corollary, which is extended to the non-irreducible case in the next section. \begin{corollary} \label{cor:irred-PPT-EEB}
    Any irreducible $\operatorname{PPT}$ channel $\Phi:\cB(\cH)\to \cB(\cH)$ is eventually entanglement breaking.
\end{corollary}
\begin{proof}
    Since the channel is PPT, \eqref{eq:assume_PhiQ_kills_offdiag} holds. Then setting $\ell:=\spr(\Phi_Q)<1$, by Gelfand's formula we have that for any $\eps\in(0,1-\ell)$, there exists $n_0>0$ such that for all $k\geq n_0$,
    \[
        \|J(\Phi_Q^k)\|_2=\|\Phi_Q^k\|_2 \leq (\ell +\eps)^k.
    \]
    Thus, for $k$ large enough, $\|J(\Phi_Q^k)\|_2  \leq z\lambda_\text{min}(\sigma)$ and \eqref{eq:irred-JQ-block-EB} holds. Hence, $\Phi^k \in \EB(\cH)$.
\end{proof}
\begin{remark}
\cite{HRS20} provides many quantitative estimates of entanglement-breaking times which are omitted here.
\end{remark}
\subsection{Beyond irreducibility}
A non-irreducible channel can be decomposed into \emph{irreducible components}, on which it acts irreducibly. More specifically, we may decompose the identity $\one$ of $\cH$ into maximal subspaces with corresponding orthogonal projections $D, P_1,\dots,P_n$ such that $\Phi$ restricted to $P_i\cB(\cH)P_i$ is irreducible, and $D\cH$ is orthogonal to the support of every invariant state of $\Phi$ \cite{CP16a}. In particular, $D=0$ is equivalent to $\Phi$ being a faithful quantum channel (that is, possessing a full-rank invariant state). In general, $\Phi$ may act non-trivially on $P_i \cB(\cH)P_j$ for $i\neq j$. The following proposition shows that this is not the case for PPT maps $\Phi$.

\begin{theorem}\label{cor:pptfulleeb}
    Any faithful $\operatorname{PPT}$ channel $\Phi:\cB(\cH)\to \cB(\cH)$ is the direct sum of irreducible $\operatorname{PPT}$ quantum channels, and therefore is eventually entanglement breaking.
\end{theorem}

\begin{remark}
    This result extends that of \cite{KMP17} where it was shown that PPT maps are AEB. It also completes Theorem 4.4 of \cite{RJP18} where it was shown that any unital PPT channel is EEB.
\end{remark}

\begin{proof}
    As $\Phi$ has an invariant state of full rank, it follows from~\cite{CP16a} that we may decompose the identity $\one$ of $\cH$ into maximal subspaces with corresponding orthogonal projections $P_1,\dots,P_n$ s.t. $\Phi$ restricted to $P_i\cB(\cH)P_i$ is irreducible. We will call $P_i\cB(\cH)P_i$ a maximal irreducible component. From \cite[Proposition 5.4]{CP16a}, we have that
    \[
        \Phi(P_i \cD(\cH)P_j) \subset P_i \cB(\cH)P_j
    \]
    and hence linearity yields
    \[
        \Phi(P_i \cB(\cH)P_j) \subset P_i \cB(\cH)P_j.
    \]
    Thus, $\Phi$ is $\big(\{P_i\}_{i=1}^n,\{P_i\}_{i=1}^n\big)$-block preserving, and by \Cref{lem:block preserving-PPT},
    \[
        \Phi(P_i \cB(\cH)P_j) = \{0\}
    \]
    for each $i\neq j$. Hence, $\Phi=\bigoplus_{i=1}^n \Phi_i$  for $\Phi_i =\Phi|_{P_i \cB(\cH)P_i}$. Each $\Phi_i$ is irreducible and PPT and hence EEB by \Cref{cor:irred-PPT-EEB}. Thus, $\Phi$ is EEB as well.
\end{proof}

Combining \Cref{cor:pptfulleeb} with that observation that for an irreducible channel $\Phi$ of period $z$, the channel $\Phi^z$ is the direct sum of primitive channels leads to the following structural result.
\begin{theorem} \label{prop:charEEB-discrete}
    Let $\Phi:\cB(\cH)\to \cB(\cH)$ be a faithful quantum channel. The following are equivalent.
    \begin{enumerate}
        \item  $\Phi\in \EEB(\cH)$,
        \item $\Phi$ is eventually PPT, meaning $\Phi^n \in \PPT(\cH)$ for some $n\in \mathbb{N},$
        \item $\Phi^{d_\cH^2 Z}$ is the direct sum of primitive channels,  where $Z$ is the least common multiple of the periods of $\Phi$ restricted to each of its irreducible components,
        \item $\Phi^{n}$ is the direct sum of primitive channels for some $n\leq d_\cH^2 \exp(1.04\,d_\cH)$.
    \end{enumerate}
\end{theorem}
\begin{remark}
    This provides a characterization of $\EEB(\cH)$  for faithful channels.
\end{remark}
\begin{proof}
    We will show the chain of implications
    \[
    (4) \implies (1) \implies (2) \implies (3) \implies (4),
    \]
    establishing the stated equivalence.

    Since primitive channels are eventually entanglement breaking\footnote{As shown in \Cref{sec:ris-intro-examples}}, the direct sum of primitive channels $\Phi=\bigoplus_i \Phi_i$ is also eventually entanglement breaking, as
    \[
        J(\Phi^n) = \bigoplus_i J(\Phi_i^n)
    \]
    is separable once each $J(\Phi_i^n)$ is separable. Thus, (4) implies (1).

    Next, that (1) implies (2) is immediate. Let us see that (2) implies (3).
    Let $\Phi \in \EEB(\cH)$ be a faithful quantum channel which is eventually PPT. Without loss of generality, assume $\Phi$ is not irreducible. Recall from the proof of \Cref{cor:pptfulleeb} that there exist $P_1, \dotsc, P_k\in \cB(\cH)$ distinct orthogonal projections such that $\Phi(P_i\cB(\cH)P_i)\subset\Phi(P_i\cB(\cH)P_i)$ with $\sum_i P_i = \one$, which moreover satisfy
    \[
        \Phi(P_i \cB(\cH)P_j)\subset \Phi(P_i\cB(\cH)P_j)
    \]
    for all $i$ and $j$, and $\Phi_i = \Phi|_{P_i\cB(\cH)P_i}$ is irreducible, with some period $z_i$, and Perron-Frobenius projections $\{p_k^{(i)}\}_{k=0}^{z_i-1}$. Let $Z$ be the least common multiple of the periods $z_i$. Then
    \begin{equation} \label{eq:proof-cov}
        \Phi^Z ( p^{(j)}_i P_j \cB(\cH) P_k p^{(k)}_\ell) \subset  p^{(j)}_i P_j \cB(\cH) P_k p^{(k)}_\ell
    \end{equation}
    for each $i,j,k,\ell$. Moreover, since $\Phi$ is eventually PPT, $\Phi^{nZ}$ is PPT for some $n\in \mathbb{N}$. Hence, by \Cref{lem:block preserving-PPT} and \eqref{eq:proof-cov},
    \[
        \Phi^{nZ} ( p^{(j)}_i P_j \cB(\cH) P_k p^{(k)}_\ell) = \{0\}
    \]
    unless $j=k$ and $i=\ell$. For $j\neq k$ and $i\neq \ell$, $\Phi^Z|_{ p^{(j)}_i P_j \cB(\cH) P_k p^{(k)}_\ell}$ is therefore a nilpotent linear endomorphism (on the vector space $ p^{(j)}_i P_j \cB(\cH) P_k p^{(k)}_\ell$). Since $\dim( p^{(j)}_i P_j \cB(\cH) P_k p^{(k)}_\ell) \leq d_\cH^2$, we have
    \[
        \Phi^{d^2 Z} ( p^{(j)}_i P_j \cB(\cH) P_k p^{(k)}_\ell) = \{0\}
    \]
    Hence, $\Phi^{d^2 Z} = \bigoplus_{i=1}^n \bigoplus_{j=0}^{z_i-1} \Phi_{ij}$   is the direct sum of primitive quantum channels  $\Phi_{ij} := \left.\Phi^{d^2Z}\right|_{ \cB( p_j P_i \cH)}$.

    To prove (3) implies (4), we upper bound $Z$. Since each period is at most $d_\cH$, $Z$ is at most the least common multiple of the natural numbers $1,\dotsc,d_\cH$. By \cite[Theorem 12]{RS62}, we thus have
    \[
        Z \leq \exp(1.03883\,d_\cH).
    \]
\end{proof}

\subsection{Reduction to the trace-preserving case} \label{sec:CPnonTP}

When considering quantum channels as a form of time evolution, trace-preservation is a natural assumption as an analog of conserving total probability. When considering maps resulting from the inverse Choi isomorphism $J\inv$ applied to bipartite operators, however, the map $J\inv(\rho_{A \tilde A})$ is trace-preserving if and only if the first marginal is completely mixed: $\rho_{A} =\one_A$. Thus, a priori, results about quantum channels only provide results about a restricted class of bipartite operators. However, as shown by \Cref{prop:similarity_transform} below, a large class of CP maps are (up to normalization) similar to CPTP maps. Moreover, this similarity transformation preserves many properties.

Recall that $\spr(\Phi)$ denotes the \emph{spectral radius} of a map $\Phi$, defined as $\spr(\Phi) := \max_{\lambda} |\lambda|$, where $\lambda$ ranges over the eigenvalues of $\Phi$, and that the spectral radius of a quantum channel is 1.

\begin{proposition} \label{prop:similarity_transform}
    Let $\Phi$ be a $\operatorname{CP}$ map such that for some $X>0$, $\Phi^*(X) = \spr(\Phi)X$ and for some $\sigma > 0$, $\Phi(\sigma) = \spr(\Phi)\sigma$. Then
    \begin{equation} \label{eq:def_Petz}
        \cP_X(\Phi) := \frac{1}{\spr(\Phi)} \Gamma_X \circ \Phi \circ \Gamma_X\inv, \qquad \text{where} \quad \Gamma_X(Y) := X^{1/2}Y X^{1/2}
    \end{equation}
    has the following properties:
    \begin{enumerate}
        \item \label{it:faithfulCPTP} $\cP_X(\Phi)$ is a faithful quantum channel
        \item \label{it:powers} $\cP_X (\Phi^n) = \cP_X(\Phi)^n$, where powers denote repeated composition
        \item \label{it:EB} $\cP_X(\Phi)$ is $\operatorname{EB}$ if and only if $\Phi$ is $\operatorname{EB}$
        \item \label{it:PPT} $\cP_X(\Phi)$ is $\operatorname{PPT}$ if and only if $\Phi$ is $\operatorname{PPT}$
    \end{enumerate}
\end{proposition}
\begin{remarks}
    ~\begin{itemize}
        \item This transformation is not novel; the map $\Gamma_{\sigma} \circ \Phi^* \circ \Gamma\inv_{\Phi(\sigma)}$ is known as the Petz recovery map \cite{Pet03}, and is usually considered in the case of CPTP maps $\Phi$. Replacing $\Phi$ with $\Phi^*$ and $\sigma$ with $X$ yields \eqref{eq:def_Petz} in the case that $\Phi^*(X) = \spr(\Phi)X$. This transformation has also been considered in \cite[Appendix A]{HJPR18} in order to study the spectral properties of deformed CP  maps as a function of the deformation. See \cite[Theorem 3.2]{Wol12} for a similar but slightly different approach (using $\Phi^*(\one)$ instead of $X$).
        \item By the Perron-Frobenius theory (see \cite[Theorem 2.5]{EH78}), if $\Phi$ is a positive map, then the spectral radius $\spr(\Phi)$ is an eigenvalue of $\Phi$, and $\Phi$ admits a positive semidefinite eigenvector $\sigma$, which can be normalized to have unit trace. Since $\spr(\Phi) = \spr(\Phi^*)$, the same logic applied to $\Phi^*$ yields $X\geq 0$ such that $\Phi^*(X) = \spr(\Phi)X$. Hence, the assumption of \Cref{prop:similarity_transform} is that both of these eigenvectors are full rank.
        \item This transformation cannot be applied in general to a CPTP map $\Phi$ in order to obtain a CPTP unital map $\tilde \Phi$, since trace-preservation will be lost. An intuitive way to see this is that a similarity transformation corresponds to a choice of (non-orthogonal) basis, and by fixing $\Phi^*(\one)=\one$, we choose a particular basis for $\Phi$ in which the dual eigenvector for $\spr(\Phi)$ is represented by the identity matrix. Thus, in general, we cannot simultaneously choose a basis to fix a representation for the eigenvector of $\spr(\Phi)$.
    \end{itemize}
\end{remarks}
\begin{proof}
    \begin{enumerate}
        \item $\cP_X(\Phi)$ is the positive multiple of the composition of CP maps and hence is CP. Since
              \[
                  \cP_X(\Phi)^* = \frac{1}{\spr(\Phi)} \Gamma_X\inv \circ \Phi^* \circ \Gamma_X
              \]
              we have that $\cP_X(\Phi)^*$ is unital:
              \[
                  \cP_X(\Phi)^*(\one) = \frac{1}{\spr(\Phi)}\Gamma_X\inv \circ \Phi^* (X) = \Gamma_X\inv(X) = \one
              \]
              and hence $\cP_X(\Phi)$ is TP. Lastly, $\rho := \frac{\Gamma_X(\sigma)}{\tr[\ldots]}$ is a full-rank invariant state for $\cP_X(\Phi)$: up to normalization,
              \[
                  \cP_X(\Phi)(\rho) =  \frac{1}{\spr(\Phi)} \Gamma_X \circ \Phi \circ \Gamma_X\inv \circ \Gamma_X(\sigma) = \frac{1}{\spr(\Phi)} \Gamma_X \circ \Phi (\sigma) = \Gamma_X(\sigma) = \rho
              \]
              and hence $\cP_X(\Phi)$ is faithful.

        \item We have
              \begin{align*}
                  \cP_X(\Phi^n) & = \frac{1}{\spr(\Phi^n)} \Gamma_X \circ \Phi^n \circ \Gamma_X\inv                                                                                                     \\
                                & = \frac{1}{\spr(\Phi)^n} \Gamma_X \circ \Phi^{n-1} \circ \Gamma_X\inv \circ \Gamma_X \circ \Phi \circ \Gamma_X\inv                                                    \\
                                & = \ldots                                                                                                                                                              \\
                                & = \frac{1}{\spr(\Phi)^n} \Gamma_X \circ \Phi \circ \Gamma_X\inv \circ \Gamma_X \circ \Phi \circ \Gamma_X\inv \circ \dotsm\circ \Gamma_X \circ \Phi \circ \Gamma_X\inv \\
                                & = \cP_X(\Phi)^n.
              \end{align*}
        \item If $\Phi$ is EB, then $\Phi\otimes \id (\cB(\cH_A \otimes \cH_B)_+ ) \subseteq \SEP(A:B)$. Since $X$ is invertible, $\Gamma_X\inv \otimes \id (\cB(\cH_A \otimes \cH_B)_+) = \cB(\cH_A \otimes \cH_B)_+$. Hence $\Phi\circ \Gamma_X\inv \otimes \id$ is EB. Since $\SEP$ is invariant under local positive operations, $\cP_X(\Phi)$ is EB too. The converse follows the same way.
        \item The same proof holds with $\SEP(A:B)$ replaced by $\PPT(A:B)$ and EB replaced by PPT.
    \end{enumerate}
\end{proof}

This transformation allows us to establish our most general result on the entanglement-breakability of PPT maps.

\begin{theorem} \label{thm:CP_PPT_EB_discrete}
    Let $\Phi$ be a $\operatorname{CP}$ map such that for some $X>0$, $\Phi^*(X) = \spr(\Phi)X$ and for some $\sigma > 0$, $\Phi(\sigma) = \spr(\Phi)\sigma$. Then if $\Phi$ is $\operatorname{PPT}$, it is $\operatorname{EEB}$.
\end{theorem}
\begin{proof}
    By (\ref{it:faithfulCPTP}), $\cP_X(\Phi)$ is faithful, CPTP, and by (\ref{it:PPT}) PPT. Hence, for some $n$, $\cP_X(\Phi)^n = \cP_X(\Phi^n)$ is EB, using (\ref{it:powers}) and \Cref{cor:pptfulleeb}. Then $\Phi^n$ is EB by (\ref{it:EB}).
\end{proof}

  \chapter{Repeated interaction systems}\label{chap:RIS}

In the previous chapter, we investigated the finite-time entanglement breaking properties of discrete quantum Markov semigroups (QMS) $\{\Phi^n\}_{n=1}^\infty$. 
In this chapter, we will discuss repeated interaction systems, which provide a physical model from which such QMS, and others, arise. We will see how delving into more of the details of the physical model permits us to make a more refined analysis of the evolution of an initial state. In particular, we will use a two-time measurement protocol in which \emph{indirect} measurements are performed on the system. The state of the system and other quantities of interest can then be seen as random variables with respect to the distribution of measurement outcomes. We will chiefly investigate the so-called \emph{entropy production} of the evolution, however, rather than entanglement-breaking properties of the maps in question.

We will also consider a generalization of the setting of the previous chapter. In the systems considered in that chapter, the evolution of a state after $n$ timesteps is given by $\underbrace{\Phi\circ \Phi\circ\dotsm\Phi}_{n \text{ times}}$. In this chapter, we consider systems in which the evolution over $n$ timesteps is given by a sequence $\Phi_n \circ \Phi_{n-1}\circ\dotsm\circ \Phi_1$ where one may have $\Phi_1 \neq \Phi_{2}$ and so forth. However, the maps $\{\Phi_k\}_k$ will be obtained in such a way as to guarantee that consecutive maps differ only by a small amount $O(1/T)$ when the total number of timesteps is given by $T$. This is the so-called adiabatic setting discussed in \Cref{sec:adiabatic_limit}.

This chapter discusses the results of \cite{HJPR18} with the aim of illustrating how the use of the two-time measurement protocol (defined in \Cref{sec:FS_two_time_measurements}) yields more detailed information about the evolution of the system and its statistics. For the sake of focus and brevity, many of the proofs of the technical results have been suppressed (but may be found in the full text of \cite{HJPR18}).

\section{Repeated interaction systems}
A repeated interaction system (RIS) consists of the system of interest $\sys$, along with a structured environment $\env$ which consists of a chain of ``probes'' $\env_k$, $k=1,2\ldots$. The system $\sys$ interacts with $\env_k$ (and only $\env_k$) during the time interval $\big[ k\tau, (k+1)\tau\big)$, where $\tau > 0$ is the interaction time. 

More specifically, we describe the quantum system~$\sys$ by a finite-dimensional Hilbert space~$\cH_\sys$, a (time-independent) Hamiltonian~$h_\sys=h_\sys^*\in \cB(\cH_\sys)$, and an initial state~$\sysstate\init \in \D(\cH_\sys)$. Likewise, the $k$th quantum probe~$\env_k$ is described by a finite dimensional Hilbert space~$\cH_{\env,k}$, Hamiltonian~$\henv[k]=\henv[k]^* \in \cB(\cH_{\env,k})$, and initial state $\envstate[k]\init \in \D(\cH_{\env,k})$. We assume the probe Hilbert spaces $\cH_{\env,k}$ are all identical, $\cH_{\env,k} \equiv \cH_\env$, and that the initial state of each probe is a Gibbs state at inverse temperature $\beta_k>0$:
$$
\envstate[k]\init = \frac{\Exp{-\beta_k \henv[k]}}{\tr(\Exp{-\beta_k \henv[k]})}.
$$
We often use $Z_{\beta,k}$ to denote the partition function $\tr(\Exp{-\beta_k \henv[k]})$. We will consider a finite number $T\in \mathbb{N}$ of probes as well as the limit $T\to\infty$.

The state of the system~$\sys$ evolves by interacting with each probe, one at a time, as follows. Assume that after interacting with the first $k-1$ probes the state of the system is~$\sysstate[k-1]$. Then the system and the $k$th probe, with joint initial state $\sysstate[k-1] \otimes \envstate[k]\init$, evolve for a time $\tau$ via the free Hamiltonian plus interaction $v_k$ according to the unitary operator
$$
U_k := \exp\big({-\i}\tau (\hsys \otimes \id + \id \otimes \henv[k] +  v_k)\big),
$$
yielding a joint final state $U_k(\sysstate[k-1] \otimes \envstate[k]\init)U_k^*$. The probe~$\env_k$ is traced out, resulting in the system state
$$
\sysstate[k] := \tr_\env\big(U_k(\sysstate[k-1] \otimes \envstate[k]\init)U_k^*\big),
$$
where $\tr_\env$ is the partial trace over $\cH_\env$; likewise, $\tr_\sys$ denotes the partial trace over $\cH_\sys$. We also introduce $\envstate[k]\fin := \tr_\sys\big(U_k(\sysstate[k-1] \otimes \envstate[k]\init)U_k^*\big)$.
A diagram representing a repeated interaction system is shown in \Cref{fig:RIS}.
\begin{figure}
\centering

\begin{tikzpicture}[scale=0.5]
\draw  (0,0) ellipse (1 and 1);
\node at (0,0) {$\rho_{k-1}$};
\node at (0,-1.5) {$\mathcal{S},h_\sys$};

\draw (0,4) -- (0,1);
\node at (1,2.5) {$v_k$};

\draw  (0,5.5) ellipse (1 and 1);
\node at (0,7) {$\mathcal{E}_{k}, h_{\env_k}$};
\node at (0,5.5) {$\envstate[k]\init$};

\draw[gray,pattern=north west lines, pattern color=gray]  (-3.5,6) ellipse (1 and 1);
\node at (-3.5,7.5) {$\color{gray}\mathcal{E}_{k-1}, h_{\env_{k-1}}$};

\draw[gray,pattern=north west lines, pattern color=gray]  (-7,7) ellipse (1 and 1);
\node at (-7,8.5) {$\color{gray}\mathcal{E}_{k-2}, h_{\env_{k-2}}$};

\draw  (7,7) ellipse (1 and 1);
\node at (7,8.5) {$\mathcal{E}_{k+2}, h_{\env_{k+2}}$};
\node at (7,7) {${\envstate[k+2]\init}$};

\draw  (3.5,6) ellipse (1 and 1);
\node at (3.5,7.5) {$\mathcal{E}_{k+1}, h_{\env_{k+1}}$};
\node at (3.5,6) {${\envstate[k+1]\init}$};

\node at (-10,8) {$\color{gray}\cdots$};
\node at (10,8) {$\cdots$};

\draw (-1.1491,4.5358) arc (-140.0003:-40:1.5);
\end{tikzpicture}
 \caption{A schematic representation of a repeated interaction system, in which the system $\sys$ interacts, one a time, with a chain of probes $\env_1,\env_2,\ldots$, at step $k$ of the interaction. The $k$th probe has a Hamiltonian $\henv[k]$ and initial state $\xi\init_k$, and interacts with the system via a coupling $v_k$. Before the $k$th interaction, the system, which has its own Hamiltonian $\hsys$, starts in state $\sysstate[k-1]$.}\label{fig:RIS}
\end{figure}

 The evolution of the system~$\sys$ during the $k$th step is given by the \emph{reduced dynamics}
\begin{align}
\cL_k : \cB(\cH_\sys) &\to \cB(\cH_\sys) \notag \\
\eta &\mapsto \tr_\env\big(U_k(\eta \otimes \envstate[k]\init)U_k^*\big), \label{eq_defLk}
\end{align}
that is $\sysstate[k] = \cL_k(\sysstate[k-1])$. The reduced dynamics $\cL_k$ is completely positive and trace-preserving, and hence maps $\D(\cH_\sys)$ to $\D(\H_\sys)$. By iterating this evolution, we find that the state of the system~$\sys$ after $k$ steps is given by the composition
\begin{equation}\label{eq:final-state}
\sysstate[k] = (\cL_k \circ \dotsb \circ \cL_1)(\sysstate\init).
\end{equation}
We often omit the parentheses and composition symbols. In the case that the probes have identical Hamiltonians, $\henv\equiv \henv[k]$, and identical temperatures $\beta\equiv\beta_k$, and the system and each probe interacts with the system via identical couplings $v \equiv v_k$, one obtains identical reduced dynamics $\cL\equiv \cL_k$. In this case, $\rho_k = \cL^k(\rho\init)$, where $\cL^k$ denotes the $k$-fold composition of $\cL$ with itself. Hence, in this case the evolution of the system forms a discrete-time QMS as discussed in the previous chapter. On the other hand, the Stinespring dilation shows that any discrete-time QMS can be formulated as a repeated interaction system with reduced dynamics $\cL\equiv \cL_k$ (albeit one in which the probe has zero temperature, i.e.\@~$\beta\equiv \beta_k = +\infty$).

\section{Two-time measurement protocols} \label{sec:FS_two_time_measurements}

We now describe a two-time measurement protocol for repeated interaction systems with~$T$ probes. By studying quantities of interest as random variables on the distribution of measurement outcomes, we obtain the so-called \emph{full statistics}. We will see that in many cases this provides a generalization of the setting in which measurements are not performed, as the analogous quantities in the non-measured setting can be obtained as expectation values of the random variables obtained in the measured setting.

Note that a similar protocol was considered in~\cite{HorPar} (see also~\cite{BJPP18}).
For the purpose of defining the full statistics measure for an RIS, we will consider observables to be measured on both the system~$\sys$ and the probes~$\env_k$, $k\in\nn$.

First, we assume that we are given two observables $\Ai$ and $\Af$ in $\cB(\H_\sys)$ with spectral decompositions
\[
\Ai = \sum_{\ai} \ai \, \pi\init_{\ai}, \qquad
\Af = \sum_{\af} \af \, \pi\fin_{\af}
\]
where~$\ai$, $\af$ run over the distinct eigenvalues of~$\Ai$, $\Af$ respectively, and~$\pi\init_{\ai}$, $\pi\fin_{\af}$ denote the corresponding spectral projectors. When we consider increasing the number of probes $T$, we assume that the observable $\Ai$ is independent of $T$ (as we measure it on $\sys$ before the system interacts with any number of probes), but allow $\Af$ to depend on $T$, as long as the family $(\Af)_{T=1}^\infty$ is uniformly bounded in $T$.

On the chain, we consider probe observables $Y_k \in \cB(\H_\env)$ to be measured on the probe $\env_k$. We require that each observable commutes with the corresponding probe Hamiltonian: $$[Y_k, \henv[k]] = 0.$$ We write the spectral decomposition of each $Y_k$ as
\[Y_k=\sum_{i_k} y_{i_k} \Pi^{(k)}_{i_k}.\]
If the $k$th probe is initially in the state $\xi$, a measurement of~$Y_k$ before the time evolution will  yield~$y_{i_k}$ with probability $\tr(\xi \Pi_{i_k}^{(k)})$.

Associated to the observables~$\Ai$,~$\Af$ and~$(Y_k)_{k=1}^T$ and the state~$\sysstate\init$, we define two processes: the \emph{forward process}, and the \emph{backward process}.

\paragraph*{The forward process} The system $\sys$ starts in some initial state~$\sysstate\init \in \D(\H_\sys)$ and the probe~$\env_k$ starts in the initial Gibbs state $\envstate[k]\in\D(\H_{\env_{k}})$; we write the state of the chain of~$T$ probes $\Xi = \bigotimes_{k=1}^T \envstate[k]$. We measure~$\Ai$ on~$\sys$ and measure~$Y_k$ on~$\env_k$ for each $k = 1, \dotsc, T$. We obtain results~$\ai$ and~$\vec{\imath} = (i_k)_{k=1}^T$ with probability
$$
\tr\big((\sysstate\init \otimes \Xi)(\pi\init_{\ai} \otimes \Pi_{\vec{\imath}})\big),
$$
where $\Pi_{\vec{\imath}} := \bigotimes_{k=1}^T \Pi_{i_k}^{(k)}$. Then the system interacts with each probe, one at a time, starting at $k=1$ until $k=T$, via the time evolution
\[
U_{k} := \exp\big(-\i\tau (\hsys + \henv[k] + v_{k})\big).
\]
Next, we measure~$\Af$ on the system and measure~$Y_k$ on~$\env_k$ for each $k = 1, \dotsc, T$, yielding outcomes~$\af$ and~$\vec \jmath =(j_k)_{k=1}^T$. The probability of measuring the sequence~$(\ai,\af,\vec{\imath},\vec{\jmath})$ of outcomes is given by
$$
\tr\big( U_{T}\dotsm U_1   ( \pi\init_{\ai} \otimes \Pi_{\vec \imath}) (\rho\init \otimes \Xi)( \pi\init_{\ai} \otimes \Pi_{\vec \imath}) U_1^* \dotsm U_T^*  (\pi\fin_{\af}\otimes \Pi_{\vec \jmath})\big).
$$
We emphasize that the outcomes are labelled by $(\ai,\af,\vec{\imath},\vec{\jmath})$ which refers to the eigenprojectors of the operators involved, but not to the corresponding eigenvalues which only need to be distinct. Also, we may write the second measurement projector $\pi\fin_{\af}\otimes \Pi_{\vec \jmath}$ only once by cyclicity of the trace.

\paragraph{The backward process} The system starts in the state
\[
\rho\fin_T := \tr_\env\big( U_{T}\dotsm U_1    (\sysstate\init \otimes \Xi) U_1^* \dotsm U_T^* \big),
\]
and the probe $\env_k$ starts in the state $\xi_k$. We measure observable $\Af$ on~$\sys$ and $Y_k$ on~$\env_k$ for each $k = 1, \dotsc, T$, yielding outcomes $\af$ and $(j_k)_{k=1}^T$. Then the system interacts with each probe, one at a time, starting with $k=T$ until $k=1$, via the time evolution
\[
U_k^* = \exp\big(\i\tau(\hsys + \henv[k] + v_k)\big).
\]
Then we measure~$\Ai$ on~$\sys$ and $Y_k$ on~$\env_k$ for each $k = 1, \dotsc, T$, yielding outcomes $\ai$ and $(i_k)_{k=1}^T$. The probability of these outcomes is given by
$$
\tr \big( U_1^*\dotsm U_T^*(\pi\fin_{\af} \otimes \Pi_{\vec \jmath})(\rho\fin_T \otimes \Xi) (\pi\fin_{\af} \otimes \Pi_{\vec \jmath}) U_T \dotsm U_1 (\pi\init_{\ai} \otimes \Pi_{\vec \imath})\big).
$$

\paragraph{The full statistics associated to the two-step measurement process}

For notational simplicity, we assume that the cardinality of~$\spec Y_k$
does not depend on $k$. We can therefore use the same index set~$\mathfrak{I}$ for all eigenvalue sets: $\spec Y_k=(y_{i_k})_{i_k\in \mathfrak{I}}$ for all $k=1,\ldots,T$.
We define the space
\[
\Omega_T := \spec \Ai \times \spec \Af \times \mathfrak{I}^T \times \mathfrak{I}^{T}
\]
and equip it with the maximal $\sigma$-algebra $\sP(\Omega_T)$. We will refer to elements $(\ai,\af,\vec{\imath},\vec{\jmath})$ of~$\Omega_T$ as \emph{trajectories}, and denote them by the letter~$\omega$.

\begin{definition}
	On $\Omega_T$, we call the law of the outcomes for the forward process,
	\begin{equation}\label{eq:def_P_F^T}
	\bP^F_{T}(\ai,\af,\vec \imath,\vec \jmath) := \tr\big( U_{T}\dotsm U_1   ( \pi\init_{\ai} \otimes \Pi_{\vec \imath}) (\rho\init \otimes \Xi)( \pi\init_{\ai} \otimes \Pi_{\vec \imath}) U_1^* \dotsm U_T^*  (\pi\fin_{\af}\otimes \Pi_{\vec \jmath})\big),
	\end{equation}
	the forward \emph{full statistics measure}. We denote by~$\E_T$ the expectation with respect to~$\bP^F_{T}$. We also consider the backward full statistics measure
	\begin{equation}\label{eq:def_P_B^T}
	\bP^B_{T}(\ai,\af,\vec \imath,\vec \jmath) := \tr \big( U_1^*\dotsm U_T^*(\pi\fin_{\af} \otimes \Pi_{\vec \jmath})(\rho\fin_T \otimes \Xi) (\pi\fin_{\af} \otimes \Pi_{\vec \jmath}) U_T \dotsm U_1 (\pi\init_{\ai} \otimes \Pi_{\vec \imath})\big)
	\end{equation}
	which is the law of the outcomes for the backward process. Let us emphasize here that $\pp^F_{T}$ and~$\pp^B_{T}$ depend on the spectral projectors $(\pi\init_{\ai})_{\ai}$ of $\Ai$, $(\pi\fin_{\af})_{\af}$ of $\Af$, and $(\Pi_{\vec \imath})$ of the $(Y_k)_k$, and not on the spectral values of these operators. In particular, the probabilities $\pp^F_{T}$ and $\pp^B_{T}$ associated with two families of observables $(Y(s))_{s\in [0,1]}$, $(\tilde Y(s))_{s\in [0,1]}$ that have the same spectral projectors (as e.g.\@\ $Y(s) = \beta(s) h_\env(s)$ and $\tilde Y(s) = h_\env(s)$) will be the same.

	To $(Y_k)_{k=1}^T$, $\Ai$, and $\Af$, we associate two generic classical random variables on $(\Omega_T, \sP(\Omega_T))$:
	\begin{gather}
		\rvW(\ai, \af, \vec \imath, \vec \jmath) := {\ai-\af}, \label{eq_defW} \\
		\rvY(\ai, \af, \vec \imath, \vec \jmath) := \sum_{k=1}^T (y^{(k)}_{j_k} - y^{(k)}_{i_k}) \label{eq_defY}.
	\end{gather}
\end{definition}
Note that the assumption that $(\Af)_{T=1}^\infty$ has uniformly bounded norm yields that the random variable $\rvW$ has $L^\infty$ norm  uniformly bounded in $T$.

\section{Landauer's Principle and the balance equation}
Next, we define the quantities of interest whose full statistics we will subsequently study. At this stage, we are studying the repeated interaction system without performing any measurements. For each step~$k$ of the RIS process, we define the quantities \begin{gather*}
\Delta S_k := S(\sysstate[k-1]) - S(\sysstate[k])
,\\
\Delta Q_k := \tr_\env(\henv[k]\envstate[k]\fin) - \tr_\env( \henv[k] \envstate[k]\init),
\end{gather*}
that represent the decrease in entropy of the small system, and the increase in energy of probe~$k$, respectively, and
\begin{equation} \label{eq:def_sigmak}
\sigma_k := D\big(U_k (\sysstate[k-1]\otimes \envstate[k]\init) U_k^* \|  \sysstate[k]\otimes \envstate[k]\init\big),
\end{equation}
the \emph{entropy production} of step~$k$. Recall that $D(\eta\|\zeta) := \tr\big(\eta(\log\eta -\log\zeta)\big)$ is the relative entropy, which satisfies $D(\eta\|\zeta)\geq 0$ for $\eta,\zeta\in\cD(\cH)$, and that $S(\eta) := -\tr[\eta\log\eta]\geq 0$ for $\eta\in\cD(\cH)$ is the von Neumann entropy. For notational simplicity, we at times omit the ``i'' superscript in $\envstate[k]\init$. Also, we omit tensored identities for operators acting trivially on the environment or on the system, whenever the context is clear.

These quantities are related through the \emph{entropy balance equation}
\begin{equation} \label{eq:step-balance}
\Delta S_k + \sigma_k = \beta_k \Delta Q_k,
\end{equation}
(see {e.g.\@} \cite{RW13,Han16} for its brief derivation). This equation, together with $\sigma_k\geq 0$, {i.e.\@} the nonnegativity of the entropy production term, encapsulates the more general \emph{Landauer principle}: when a system undergoes a state transformation by interacting with a thermal bath, the average increase in energy of the bath is bounded below by $\beta^{-1}$ times the average decrease in entropy of the system. This principle was first presented in 1961 by Landauer \cite{La} and its saturation in quantum systems has more recently been investigated by Reeb and Wolf \cite{RW13} and Jak\v{s}i\'{c} and Pillet \cite{JP14}, the latter providing a treatment of the case of infinitely extended quantum systems.

If we consider a RIS with $T$ steps, then summing \eqref{eq:step-balance} over $k = 1, \dotsc, T$ yields the \emph{total entropy balance equation}
\begin{equation} \label{eq:balance}
\Delta S_{\sys,T}+\oldsigmaT  =  \sum_{k=1}^T \beta_k \Delta Q_k,
\end{equation}
where $\Delta S_{\sys,T}=S(\sysstate\init) - S(\sysstate\fin)$ and $\sysstate\fin = \sysstate_T$ is the state of~$\sys$ after the final step of the RIS process (see~\eqref{eq:final-state}) and
\begin{equation}\label{eq:def_sigma_tot}
\oldsigmaT := \sum_{k=1}^T \sigma_{k},
\end{equation}
is the expected \emph{total entropy production}.

\section{The balance equation at the level of trajectories} \label{subsec:bal-eq}

We turn to obtaining an analogue of~\eqref{eq:balance} for random variables on the probability space~$(\Omega_T, \sP(\Omega_T), \bP^F_{T})$.
Remark first that $\bP^F_{T}(\ai,\af,\vec{\imath},\vec{\jmath})$ and $\bP^B_{T}(\ai,\af,\vec{\imath},\vec{\jmath})$ are of the form
$\bP^F_{T}(\ai,\af,\vec{\imath},\vec{\jmath})=\tr\big((\rho\init\otimes \Xi) \, S^* S\big)$ and $ \bP^B_{T}(\ai,\af,\vec{\imath},\vec{\jmath})=\tr\big((\rho_T\fin\otimes \Xi) \, S S^*\big)$.
Under the assumption that  $\rho\init$ and $\rho_T\fin$ are faithful (full-rank) we therefore have
\[\bP^F_{T}(\ai,\af,\vec{\imath},\vec{\jmath})=0 \mbox{ if and only if }\bP^B_{T}(\ai,\af,\vec{\imath},\vec{\jmath})=0.\]
Since  \Cref{prop:irred-preserve-faithful} shows that the image of a faithful state by an irreducible CPTP map is faithful,  $\rho\init$ and $\rho_T\fin$ will be faithful as long as $\rho\init$ is faithful and each map $\cL_k$, for $k=1,\dotsc,T$, is irreducible.

This allows us to give the following definition:
\begin{definition} \label{def:class-rv}
	If $\rho\init$ and $\rho_T\fin$ are faithful, we define the classical random variable
	\begin{gather*}
	\varsigma_{T}(\ai,\af,\vec{\imath},\vec{\jmath}) :=\log \frac{\bP^F_{T}(\ai,\af,\vec{\imath},\vec{\jmath})}{\bP^B_{T}(\ai,\af,\vec{\imath},\vec{\jmath})},
	\end{gather*}
	on $(\Omega_T, \sP(\Omega_T),\pp^F_{T})$, which we call the entropy production of the repeated interaction system associated to the trajectory~$\omega = (\ai,\af,\vec{\imath},\vec{\jmath})$.
\end{definition}

Note that the random variable $\varsigma_{T}$ is the logarithm of the ratio of likelihoods, also known as the relative information random variable between $\pp^F_{T}$ and $\pp^B_{T}$ (see e.g.\@ \cite{CT06}). It is well-known that the distribution of such a random variable is related to the distinguishability of the two distributions (here $\pp^F_{T}$ and $\pp^B_{T}$): see e.g.\@\ \cite{BicDok}. Distinguishing between $\pp^F_{T}$ and $\pp^B_{T}$ amounts to testing the arrow of time; we refer the reader to \cite{JOPS,BJPP18} for a further discussion of this idea.

We have the following result, essentially present in~\cite{HorPar}.
\begin{lemma}\label{lem:balance_eq_traj}
	Assume $\sysstate\init$ and $\sysstate\fin_T$ are faithful. If
	\begin{enumerate}[label=\roman*.]
		\item $\pi\init_{\ai}\sysstate\init\pi\init_{\ai} = \frac{\tr(\sysstate\init\pi\init_{\ai})}{\dim \pi\init_{\ai}}\,\pi\init_{\ai}$ for each $\ai$,
		\item $\pi\fin_{\af}\sysstate\fin_T\pi\fin_{\af} = \frac{\tr(\sysstate\fin_T\pi\fin_{\af})}{\dim \pi\fin_{\af}}\,\pi\fin_{\af}$ for each $\af$,
		\item for each $k = 1, \dotsc, T$, the state  $\envstate[k]$ (or equivalently $h_{\env_k}$) is a function of $Y_k$,
	\end{enumerate}
	then
	\begin{equation}\label{eq:balance_eq_for_trajectories}
	\varsigma_{T}(\ai,\af,\vec{\imath},\vec{\jmath}) = \log \Big( \frac{\tr(\pi\init_{\ai} \sysstate\init)}{ \tr(\pi\fin_{\af} \sysstate\fin_T)} \,\frac{\dim \pi\fin_{\af}}{\dim \pi\init_{\ai}}\Big) + \sum_{k=1}^T \beta_k (E_{j_k}^{(k)} - E_{i_k}^{(k)}),
	\end{equation}
	where $E_{i_k}^{(k)} = \frac{\tr(\henv[k]\Pi_{i_k}^{(k)})}{\dim \Pi_{i_k}^{(k)}}$ are the energy levels of the $k$th probe.
\end{lemma}

\begin{remark}
	The first two hypotheses are automatically satisfied if, for example, $\Ai$ and $\Af$ are non-degenerate (all their spectral projectors are rank-one). All three hypotheses are automatically satisfied if, for example, $\sysstate\init$, $\sysstate\fin_T$ and $\envstate[k]$ can be written as functions of $\Ai$, $\Af$ and $Y_k$ (for each $k =1, \dotsc, T$) respectively.
\end{remark}

\begin{proof}[Proof of Lemma~\ref{lem:balance_eq_traj}]
	By definition
	\begin{align*}
	&\varsigma_{T}(\ai,\af,\vec{\imath},\vec{\jmath}) \\
   &\quad = \log \frac{\tr\big( U_{T}\dotsm U_1   ( \pi\init_{\ai} \otimes \Pi_{\vec \imath}) (\rho\init \otimes \Xi)( \pi\init_{\ai} \otimes \Pi_{\vec \imath}) U_1^* \dotsm U_T^*  (\pi\fin_{\af}\otimes \Pi_{\vec \jmath})\big)}{\tr\big( U_1^*\dotsm U_T^*(\pi\fin_{\af} \otimes \Pi_{\vec \jmath})(\rho\fin_T \otimes \Xi) (\pi\fin_{\af} \otimes \Pi_{\vec \jmath}) U_T \dotsm U_1 (\pi\init_{\ai} \otimes \Pi_{\vec \imath})\big)} \\
	\intertext{using assumptions {i.}, {ii.} and {iii.},}
	&\quad =  \log \frac{\frac{\tr (\sysstate\init \pi\init_{\ai})}{\dim \pi\init_{\ai}} \prod_{k=1}^T \frac{\tr(\envstate[k] \Pi_{i_k}^{(k)})}{\dim \Pi_{i_k}^{(k)}} \tr\big( U_{T}\dotsm U_1   ( \pi\init_{\ai} \otimes \Pi_{\vec \imath})  U_1^* \dotsm U_T^*  (\pi\fin_{\af}\otimes \Pi_{\vec \jmath})\big)}
	{\frac{\tr (\sysstate\fin \pi\fin_{\af})}{\dim \pi\fin_{\af}} \prod_{k=1}^T \frac{\tr(\envstate[k] \Pi_{j_k}^{(k)})}{\dim \Pi_{j_k}^{(k)}} \tr\big( U_1^*\dotsm U_T^*(\pi\fin_{\af} \otimes \Pi_{\vec \jmath})U_T \dotsm U_1 (\pi\init_{\ai} \otimes \Pi_{\vec \imath})\big)} \\
	&\quad =  \log\frac{\tr (\sysstate\init \pi\init_{\ai}) \dim \pi\fin_{\af}}{\tr (\sysstate\fin \pi\fin_{\af}) \dim \pi\init_{\ai}} + \log \frac{  \tr\big( U_{T}\dotsm U_1   ( \pi\init_{\ai} \otimes \Pi_{\vec \imath})  U_1^* \dotsm U_T^*  (\pi\fin_{\af}\otimes \Pi_{\vec \jmath})\big)}
	{ \tr\big( U_1^*\dotsm U_T^*(\pi\fin_{\af} \otimes \Pi_{\vec \jmath})U_T \dotsm U_1 (\pi\init_{\ai} \otimes \Pi_{\vec \imath})\big)} \\
	&\quad \qquad {} + \log \prod_{k=1}^T \frac{ \tr(\Exp{-\beta_k \henv[k]} \Pi_{i_k}^{(k)})}{Z_k \dim \Pi_{i_k}^{(k)}} - \log \prod_{k=1}^T \frac{ \tr(\Exp{-\beta_k \henv[k]} \Pi_{j_k}^{(k)})}{Z_k \dim \Pi_{j_k}^{(k)}}  \\
	\intertext{
      and because $\frac{\tr(\e^{-\beta_k h_{\env_k}} \Pi_{i_k}^{(k)})}{\dim \Pi_{i_k}^{(k)}}= \exp\big({-\beta_k \frac{\tr(\e^{-\beta_k h_{\env_k}\Pi_{i_k}^{(k)}})}{\dim \Pi_{i_k}^{(k)}}}\big)$ by assumption {iii.} again,
      }
	&\quad =  \log\frac{\tr (\sysstate\init \pi\init_{\ai}) \dim \pi\fin_{\af}}{\tr (\sysstate\fin \pi\fin_{\af}) \dim \pi\init_{\ai}} + \log \frac{  \tr\big( U_{T}\dotsm U_1   ( \pi\init_{\ai} \otimes \Pi_{\vec \imath})  U_1^* \dotsm U_T^*  (\pi\fin_{\af}\otimes \Pi_{\vec \jmath})\big)}
	{ \tr\big( U_1^*\dotsm U_T^*(\pi\fin_{\af} \otimes \Pi_{\vec \jmath})U_T \dotsm U_1 (\pi\init_{\ai} \otimes \Pi_{\vec \imath})\big)}  \\
	&\quad \qquad {} + \sum_{k=1}^T \frac{ \tr({-\beta_k \henv[k]} \Pi_{i_k}^{(k)})}{\dim \Pi_{i_k}^{(k)}} - \sum_{k=1}^T \frac{ \tr({-\beta_k \henv[k]} \Pi_{j_k}^{(k)})}{\dim \Pi_{j_k}^{(k)}}
	\end{align*}
	and the last term vanishes by cyclicity of the trace.
\end{proof}

Again, $\varsigma_{T}$ depends on the spectral projectors of the observables~$\Ai$,~$\Af$ and~$(Y_k)_{k=1}^T$, but not on their eigenvalues. However, with the choices~$\Ai = - \log \sysstate\init$, $\Af = - \log \sysstate\fin_T$ and~$Y(s) = \beta(s) \henv(s)$, and writing the spectral decompositions $\rho\init=\sum r_a\init \pi\init_a$, $\rho\fin_T=\sum r_a\fin \pi\fin_a$, the relation~\eqref{eq:balance_eq_for_trajectories} takes the simpler form of a sum of differences of the obtained eigenvalues (measurement results):
\begin{align*}
\varsigma_{T}(\omega) &= (-\log r_{a\fin}\fin)-(-\log r_{a\init}\init)  + \sum_{k=1}^T \beta_k (E_{j_k}^{(k)} - E_{i_k}^{(k)}),
\end{align*}
which is the random variable introduced earlier as $\rvZ(\omega)$ (again, in the case $Y=\beta h_\env$). In this case, $\rvW = (-\log r_{a\init}\init)-(-\log r_{a\fin}\fin)$ is a classical random variable that is the difference of outcomes of measurements of entropy observables on the system~$\sys$, which we call $\Delta s_{\sys,T}(\omega)$. On the other hand, $\sum_{k=1}^T \beta_k (E_{j_k}^{(k)} - E_{i_k}^{(k)})$ is a classical random variable that encapsulates Clausius' notion of the entropy increase of the chain $(\env_k)_{k=1}^T$ at the level of trajectories, which we call $\Delta s_{\env,T}(\omega)$.  Then,
\begin{equation}\label{eq:decomp-entropy-prod-traj}
	\Delta s_{\sys,T}(\omega) +\varsigma_{T}(\omega) = \Delta s_{\env,T}(\omega),
\end{equation}
and $\varsigma_{T}(\omega)$ measures the difference between these two entropy variations, on the trajectory~$\omega$.

Moreover, Proposition~\ref{prop:averaged_balance_equation} {below}, whose proof is also left for the Appendix, links expression~\eqref{eq:balance_eq_for_trajectories} to the entropy balance equation \eqref{eq:balance}. Indeed, by showing that under suitable hypotheses the two terms on the right hand side of~\eqref{eq:balance_eq_for_trajectories} average to the corresponding terms in~\eqref{eq:balance}, we show that $\E_T( \varsigma_{T}) = \oldsigmaT$. In other words, $\oldsigmaT$ coincides with the (classical) relative entropy or Kullback--Leibler divergence $D(\bP^F_{T}\|\, \bP^B_{T})$ between the classical distributions $\bP^F_{T}$ and $\bP^B_{T}$. Recall that $D(\bP^F_{T}\|\, \bP^B_{T})=0$ if and only if $\bP^F_{T}=\bP^B_{T}$. Hence, we will refer to~\eqref{eq:balance_eq_for_trajectories} as the \emph{entropy balance equation at the level of trajectories}.

\begin{proposition} \label{prop:averaged_balance_equation}
	Assume that $\rho\init$ is faithful and a function of $\Ai$, that $\rho\fin_T$ is faithful and a function of $\Af$, and the state~$\envstate[k]$ (or equivalently~$h_{\env_k}$) is a function of~$Y_k$ for each~$k = 1, \dotsc, T$, then
	\begin{gather}
	\E_T\Big[ \log \Big( \frac{\tr(\pi\init_{\ai} \sysstate\init)}{ \tr(\pi\fin_{\af} \sysstate\fin)} \frac{\dim \pi\fin_{\af}}{\dim \pi\init_{\ai}}\Big) \Big] = -\E_T(\Delta s_{\sys,T}) = S(\sysstate\fin) - S(\sysstate\init) \label{eq:ssys-avg}
	\intertext{and}
	\E_T\Big( \sum_{k=1}^T \beta_k (E_{j_k}^{(k)} - E_{i_k}^{(k)}) \Big) = \E_T(\Delta s_{\env,T}) = \sum_{k=1}^T \beta_k \Delta Q_k. \label{eq:senv-avg}
	\end{gather}
	Therefore,
	\begin{gather}
	\E_T( \varsigma_{T}) = \oldsigmaT, \label{eq:sprod-avg}
	\end{gather}
	and relation \eqref{eq:balance_eq_for_trajectories} reduces to the entropy balance equation \eqref{eq:balance} upon taking expectation with respect to~$\bP^F_{T}$.
\end{proposition}

\begin{proof}
	We start with the relation \eqref{eq:ssys-avg}. On one hand,
	\begin{align*}
	&\E_T\Big( \log \frac{\tr(\pi\init_{\ai} \sysstate\init)}{\dim \pi\init_{\ai}} \Big) \\
   &\qquad = \sum_{\ai} \log \frac{\tr(\pi\init_{\ai} \sysstate\init)}{\dim \pi\init_{\ai}} \sum_{\af, \vec{\imath}, \vec{\jmath}} \bP^F_{T}(\ai,\af, \vec{\imath}, \vec{\jmath}) \\
	&\qquad = \sum_{\ai} \log \frac{\tr(\pi\init_{\ai} \sysstate\init)}{\dim \pi\init_{\ai}} \sum_{\af, \vec{\imath}, \vec{\jmath}}  \tr\big( U_{T}\dotsm U_1   ( \pi\init_{\ai}\sysstate\init \pi\init_{\ai} \otimes \Pi_{\vec \imath}\,\Xi\, \Pi_{\vec \imath})  U_1^* \dotsm U_T^*  (\pi\fin_{\af}\otimes \Pi_{\vec \jmath})\big) \\
	&\qquad = \sum_{\ai} \log \frac{\tr(\pi\init_{\ai} \sysstate\init)}{\dim \pi\init_{\ai}} \sum_{ \vec{\imath}}  \tr\big( U_{T}\dotsm U_1   ( \pi\init_{\ai}\sysstate\init \pi\init_{\ai} \otimes \Pi_{\vec \imath}\,\Xi \,\Pi_{\vec \imath})  U_1^* \dotsm U_T^*  (\one \otimes \one)\big).
	\end{align*}
	Using that each $\envstate[k]$ is a function of $Y_k$, we have $\sum_{\vec{\imath}} \Pi_{\vec{\imath}} \,\Xi\, \Pi_{\vec{\imath}} = \Xi$, and
	\[
	\E_T\Big( \log \frac{\tr(\pi\init_{\ai} \sysstate\init)}{\dim \pi\init_{\ai}} \Big) = \sum_{\ai} \log \frac{\tr(\pi\init_{\ai} \sysstate\init)}{\dim \pi\init_{\ai}}  \tr\big( U_{T}\dotsm U_1   ( \pi\init_{\ai}\sysstate\init \pi\init_{\ai} \otimes \Xi )  U_1^* \dotsm U_T^* \big).
	\]
	As $\sysstate\init$ is a function of $\Ai$, and using the cyclicity of the trace, we have
	\begin{align*}
\E_T\Big( \log \frac{\tr(\pi\init_{\ai} \sysstate\init)}{\dim \pi\init_{\ai}} \Big)&= \sum_{\ai} \log \frac{\tr(\pi\init_{\ai} \sysstate\init)}{\dim \pi\init_{\ai}}  \tr\big(\pi\init_{\ai} \otimes \Xi \big) \frac{\tr(\pi\init_{\ai} \sysstate\init)}{\dim \pi\init_{\ai}}, \\
	&= \sum_{\ai}  \tr(\pi\init_{\ai} \sysstate\init) \log \frac{\tr(\pi\init_{\ai} \sysstate\init)}{\dim \pi\init_{\ai}}.
	\end{align*}
	Using the commutation relation $[\pi\init_{\ai},\sysstate\init] = 0$, the right-hand side is precisely $-S(\rho\init)$. The term involving $\sysstate\fin$ is treated similarly.

	We turn to \eqref{eq:senv-avg}:
	\begin{align*}
	&\E_T\Big( \sum_{k=1}^T \beta_k (E_{j_k}^{(k)} - E_{i_k}^{(k)}) \Big) \\
   &\qquad = \sum_{\vec{\imath}, \vec{\jmath}} \sum_{k=1}^T  \beta_k (E_{j_k}^{(k)} - E_{i_k}^{(k)}) \sum_{\ai,\af} \bP^F_{T}(\ai,\af, \vec{\imath}, \vec{\jmath}) \\
	\intertext{using $[\sysstate\init,\Ai] = 0$,}
	&\qquad = \sum_{\vec{\imath}, \vec{\jmath}} \sum_{k=1}^T  \beta_k (E_{j_k}^{(k)} - E_{i_k}^{(k)}) \tr\big( U_{T}\dotsm U_1   ( \sysstate\init  \otimes \Pi_{\vec \imath}\Xi \Pi_{\vec \imath})  U_1^* \dotsm U_T^*  (\one \otimes \Pi_{\vec \jmath})\big) \\
	&\qquad = \sum_{\vec{\jmath}} \sum_{k=1}^T  \beta_k E_{j_k}^{(k)} \tr\big( U_{T}\dotsm U_1   ( \sysstate\init  \otimes \Xi )  U_1^* \dotsm U_T^*  (\one \otimes \Pi_{\vec \jmath})\big) \\
	&\qquad\qquad{} - \sum_{\vec{\imath}} \sum_{k=1}^T  \beta_k E_{i_k}^{(k)} \tr\big( U_{T}\dotsm U_1   ( \sysstate\init  \otimes \Pi_{\vec \imath}\Xi \Pi_{\vec \imath})  U_1^* \dotsm U_T^*  (\one \otimes \one)\big) \\
	&\qquad= \sum_{k=1}^T \sum_{j_k} \beta_k  \tr\big( U_{T}\dotsm U_1   ( \sysstate\init  \otimes \Xi )  U_1^* \dotsm U_T^*  (\one \otimes \Pi_{j_k}^{(k)} E_{j_k}^{(k)})\big)\\
	&\qquad\qquad{} - \sum_{k=1}^T \sum_{i_k} \beta_k \tr\big(  \sysstate\init  \otimes \Xi\, \Pi_{i_{k}}^{(k)} E_{i_k}^{(k)}  \big).
	\end{align*}
	Using that each $\envstate[k]$ is a function of $Y_k$, we have $\sum_{i_k} \Pi_{i_{k}}^{(k)} E_{i_k}^{(k)} = \henv[k]$ and thus
	\[
	\E_T\Big( \sum_{k=1}^T \beta_k (E_{j_k}^{(k)} - E_{i_k}^{(k)}) \Big)= \sum_{k=1}^T \beta_k \tr(\envstate[k]\fin \henv[k]) - \sum_{k=1}^T \beta_k \tr(\envstate[k]\init \henv[k]). \qedhere
	\]
\end{proof}

Before we move on with our program, let us make a number of remarks on the choice of $\Ai = - \log \sysstate\init$ and $\Af = - \log \sysstate\fin_T$.
\begin{remarks}\label{rem:}\,\hfill
	\begin{itemize}
		\item We made the assumption above that the operator $\Af$ was uniformly bounded in $T$. The remarks below \Cref{cor:adiab_for_alpha=0} show that this is true for $\Af = - \log \sysstate\fin_T$ in the case in which the reduced dynamics are identical ($\cL\equiv \cL_k$) and irreducible, or in the case in which the $\cL_k$ are not all identical but are each irreducible and slowly vary in an adiabatic sense described below in \Cref{sec:adiabatic_limit}. 
		\item The observable $-\log\sysstate\init$ is the analogue of the information random variable in classical information theory.
		\item The observable $-\log\sysstate\fin$ has the same interpretation and is the initial condition for the backward process in \cite{EHM09,HorPar} but it might seem odd that the observer is expected to have access to $\sysstate\fin=\cL_T\circ\ldots\circ \cL_1(\sysstate\init)$. However, one can see that the reduced state of the probe after the forward experiment is random with $\bP^F_T$--expectation equal to $\sysstate\fin$. In addition, $\varsigma_T$ is a relative information random variable, and as such is relevant only to an observer who knows both distributions (here~$\bP^F_{T}$ and~$\bP^B_{T}$). Such an observer, knowing the possible outcomes for the random states after the experiment, and their distribution, would necessarily know their average,~$\sysstate\fin$.
	\end{itemize}
\end{remarks}

We are interested in the full statistics of the random variable $\varsigma_{T}$ that we will address through its cumulant generating functions in the limit $T\to\infty$. We will consider two cases: $\lim_{T\to\infty}\oldsigmaT < \infty$, and $\lim_{T\to\infty}\oldsigmaT =\infty$. The behaviour of this averaged quantity was investigated in~\cite{HJPR17} in an \emph{adiabatic} setting in which the maps $\cL_k$ do not need to be identical, but instead are required to slowly vary (in a particular sense). This setting is described next.

\section{Adiabatic limit of repeated interaction systems} \label{sec:adiabatic_limit}
\cite{HJPR17} analyzed the Landauer principle and its saturation in the framework of an adiabatic limit of RIS that we briefly recall here. It is in this framework in which we will analyze $\varsigma_T$.

We introduce the \emph{adiabatic parameter} $T \in \N$ and consider a repeated interaction process with $T$ probes, such that the parameters governing the $k$th probe and its interaction with $\sys$, namely $(\henv[k], \beta_k, v_k)$, are chosen by sampling sufficiently smooth functions, as described by the following assumption which we denote by the acronym ADRIS, short for \textbf{ad}iabatic \textbf{r}epeated \textbf{i}nteraction \textbf{s}ystem. Below, we say that a function $f$ is $C^2$ on $[0,1]$ if it is $C^2$ on $(0,1)$, and its first two derivatives admit limits at $0^+$, $1^-$.
\begin{description}
	\labitem{ADRIS}{ADRIS} We are given a family of RIS processes indexed by an adiabatic parameter~$T\in\nn$ such that there exist $C^2$ functions $s\mapsto \henv(s)$, $\beta(s)$, $v(s)$ on $[0,1]$ for which
	\begin{equation} \nonumber
	\henv[k] = \ham_\env(\tfrac {k}{T}), \qquad \beta_{k}=\beta(\tfrac {k}{T}), \qquad  v_{k} =v(\tfrac {k}{T})
	\end{equation}
	for all $k = 1, \dotsc, T$ when the adiabatic parameter has value~$T$.
\end{description}

In this case, we may define
\begin{equation}
\begin{aligned} \label{eq_versionscontinues}
U(s)&=\exp \big({-\i}\tau \big(\ham_\env(s) + \henv(s) +  v(s)\big)\big),\\
\cL(s)&=\tr_\env\big( U(s)\big(\cdot\otimes \,\envstate(s)\big)U(s)^*\big),
\end{aligned}
\end{equation}
where~$\envstate(s)$ is the Gibbs state at inverse temperature~$\beta(s)$ for the Hamiltonian~$\henv(s)$ and $\tau$ is kept constant. Then, $[0,1]\ni s \mapsto \cL(s)$ is a  $\cB(\cB(\H_\sys))$-valued $C^2$ function, and $\cL_{k}=\cL(\tfrac {k}{T})$ when the adiabatic parameter has value~$T$. Note that for each $s \in [0,1]$, the map $\cL(s)$ is completely positive (CP) and trace preserving (TP). For some results, we will need to make some extra hypotheses on the family $(\cL(s))_{s \in [0,1]}$. We introduce such conditions:
\begin{description}
	\labitem{Irr}{Irr} For each $s \in [0,1]$, the map~$\cL(s)$ is \emph{irreducible}, meaning that it has (up to a multiplicative constant) a unique invariant, which is a faithful state.
	\labitem{Prim}{Prim}  For each $s \in [0,1]$, the map~$\cL(s)$ is \emph{primitive}, meaning that it is irreducible and $1$ is its only eigenvalue of modulus one.
\end{description}
From \Cref{prop:decomp-irred}, we recall that the peripheral spectrum of an irreducible completely positive, trace-preserving map is a subgroup of the unit circle. We denote by $z(s)$ the order of that subgroup for $\cL(s)$.

The work \cite{HJPR17} used a suitable adiabatic theorem to characterize the large~$T$ behaviour of the total entropy production term~\eqref{eq:def_sigma_tot}, which monitors the saturation of the Landauer bound in the adiabatic limit (note that the terms in the sum~\eqref{eq:def_sigma_tot} are $T$-dependent through \ref{ADRIS}).  There, it was shown that for a RIS satisfying the assumptions \ref{ADRIS} and \ref{Prim}, the condition
\[
\limsup_{T\to\infty} \oldsigmaT < \infty
\]
can be shown to be equivalent to the identity $X(s) \equiv 0$, where
\begin{equation}
X(s) := U(s) \big(\sysstate\invar(s) \otimes \xi\init(s)\big) U(s)^* - \sysstate\invar(s) \otimes \xi\init(s), \label{eq:def-X}
\end{equation}
and $\sysstate\invar(s)$ is the unique invariant state of $\cL(s)$. If the assumption $X(s)\equiv 0$ does not hold, then $\lim_{T\to\infty}\oldsigmaT =\infty$. The same work also established that the condition $X(s)\equiv 0$ is equivalent to the existence of a family $(k_\sys(s))_{s \in [0,1]}$ of observables on $\H_\sys$ such that $[k_\sys(s)+h_{\env}(s),U(s)] \equiv 0$.

Let us briefly summarize the proof from \cite{HJPR17}, in high level terms. The goal is to estimate $\oldsigmaT = \sum_{k=1}^T \sigma_k$. From \eqref{eq:def_sigmak} we have 
\begin{equation} \label{eq:sigma_k-proof-sketch}
\sigma_k  \equiv \sigma_{k,T} = D\big(U(\tfrac{k}{T}) (\sysstate[k-1]\otimes \envstate(\tfrac{k}{T})) U(\tfrac{k}{T})^* \|  \cL(\tfrac{k}{T})(\sysstate[k-1])\otimes \envstate(\tfrac{k}{T})\big)
\end{equation}
with
\begin{equation}\label{eq:sysstate-proof-sketch}
\sysstate[k-1] = \cL(\tfrac{k-1}{T})\circ\dotsm\circ\cL(\tfrac{1}{T})(\rho\init).
\end{equation}
Note then that in $\sigma_k$, each quantity depends only on the current step, i.e.\@, on the various quantities defining the repeated interaction system evaluated at $s=\tfrac{k}{T}$, except for $\sysstate[k-1]$, which depends on all the previous timesteps $\{\tfrac{1}{T}, \tfrac{2}{T}, \dotsc, \tfrac{k-1}{T}\}$ as well. The proof therefore proceeds by estimating $\sysstate[k-1] = \rho\invar(\tfrac{k-1}{T}) + O(1/T) = \rho\invar(\tfrac{k}{T}) + O(1/T)$ where $\rho\invar(s)$ is the unique invariant state of $\cL(s)$, using a discrete adiabatic theorem, to remove the dependence on the previous timesteps. This is the main place where the assumptions \ref{ADRIS} and \ref{Prim} are used in the proof. Then, roughly speaking, a perturbation theory argument is used to control the relative entropy term $\sigma_k$ in terms of $\|X(k/T)\|$, yielding in the end that if $\|X(s)\|\equiv 0$, $\sigma^\text{tot}$ vanishes (under an assumption on the initial state), and otherwise diverges.

\smallskip

In this work, we aim to refine this analysis by investigating the entropy production at the level of trajectories, i.e.\@~study the random variable $\varsigma_{T}$ (which satisfies $\bE[\varsigma_{T}] = \oldsigmaT$ by \Cref{prop:averaged_balance_equation}). It turns out that by considering the moment-generating function of $\varsigma_{T}$, a similar proof technique to the one sketched above for $\oldsigmaT$ can be used.

\section{Moment generating functions and deformed CP maps}
First, we add an additional assumption about the observable $Y$ which is measured as part of the two-time measurement protocol.
When assuming \ref{ADRIS} and discussing measured observables $Y$, we will always assume the following:
\begin{description}
	\labitem{Comm}{Comm}
	There is a twice continuously differentiable $\cB(\H_\env)$-valued function $s \mapsto Y(s)$ on $[0,1]$ such that~$[Y(s), \henv(s)] = 0$ at all~$s \in [0,1]$ for which, when the adiabatic parameter has value~$T$,
	\[ Y_k = Y(\tfrac{k}{T}), \qquad k = 1, \dotsc, T. \]
\end{description}
The family of probe Hamiltonians themselves~$Y(s) = \henv(s)$ are suitable, but in our applications to Landauer's Principle, we will be particularly interested in~$Y(s) = \beta(s) \henv(s)$.

We recall that the quantities $\rvW$ and $\rvY$ are defined in \eqref{eq_defW} and \eqref{eq_defY}. We also recall that the \emph{moment generating function} of a real-valued random variable $V$ (with respect to the probability distribution $\bP^F_{T}$, which will always be implicit in this chapter) is defined as the map $M_V:\alpha\mapsto \E_T\big(\e^{\alpha V}\big)$, and the moment generating function of a pair $(V_1,V_2)$ as the map $M_{(V_1,V_2)}:(\alpha_1,\alpha_2)\mapsto \E_T\big(\e^{\alpha_1 V_1+\alpha_2 V_2}\big)$. When $V$ or $(V_1,V_2)$ are given by the random variables $\rvY$, $\rvW$, the above functions $M_{V}$ (resp.\ $M_{(V_1,V_2)}$) are defined for all $\alpha\in\cc$ (resp.\ for all $(\alpha_1,\alpha_2)\in\cc^2$). For relevant properties of moment generating functions we refer the reader to Sections 21 and 30 of \cite{Bill}.

Our main tool to study these moment generating functions is the following proposition:
\begin{proposition} \label{prop:phiTalpha}
	For $\alpha\in\cc$, define an analytic deformation of~$\cL(s)$ by the complex parameter~$\alpha$ corresponding to the observable~$Y(s)$:
	\begin{align}
	\cL_Y^{(\alpha)}(s) :  \cB(\H_\sys) &\to \cB(\H_\sys) \nonumber \\
	 \eta &\mapsto \tr_\env\big(\e^{ \alpha Y(s) }  U(s)   (\eta \otimes \xi(s))\e^{-\alpha Y(s)} U(s)^*\big). \label{eq_defLk_alpha}
   \end{align}
	Under assumption \textup{\ref{Comm}}, the moment generating function of $\rvY$ is given by
	\[
	M_{\rvY}(\alpha) = \tr_\sys \big( \cL_Y^{(\alpha)}(\tfrac{T}{T})  \dotsm \cL_Y^{(\alpha)}(\tfrac{1}{T}) (\sum_{\ai} \pi\init_{\ai}\rho\init\pi\init_{\ai})\big).
	\]
	If in addition $[\Ai,\sysstate\init]=0$, then the moment generating function of the pair $(\rvY, \rvW)$
	is given~by
	\begin{gather*}
	M_{(\rvY,\rvW)}(\alpha_1,\alpha_2) = \tr \big(\Exp{-\alpha_2 \Af} \cL_Y^{(\alpha_1)}(\tfrac{T}{T})  \dotsm \cL_Y^{(\alpha_1)}(\tfrac{1}{T}) (\Exp{+\alpha_2 \Ai}\rho\init)\big).
	\end{gather*}
	so that in particular the moment generating function of $\rvZ$ is given by
	\begin{equation}\label{eq:mgf_Z}
	M_{\rvZ}(\alpha) = \tr_\sys \big(\Exp{+\alpha \Af}  \cL_Y^{(\alpha)}(\tfrac{T}{T})  \dotsm \cL_Y^{(\alpha)}(\tfrac{1}{T})(\Exp{-\alpha \Ai}\rho\init)\big).
	\end{equation}
\end{proposition}
\begin{remarks}\,\hfill
\begin{itemize}
	\item  The complex deformation of the map~$\cL(s)$ given here is similar to the deformations introduced in \cite{HMO07} for hypothesis testing on spin chains, and to the complex deformation of Lindblad operators introduced in \cite{JPW14} suited to the study of entropy fluctuations for continuous time evolution.
	\item With the choices $Y(s) = \beta(s)\henv(s)$, $A\init = -\log \rho\init$, and $A\fin = -\log\rho\fin_T$, the balance equation \eqref{eq:decomp-entropy-prod-traj} yields $\varsigma_T = \rvZ$. Thus, in this case \eqref{eq:mgf_Z} provides a formula for the moment generating function of $\varsigma_T$ in terms of the composition $\cL_Y^{(\alpha)}(\tfrac{T}{T})  \dotsm \cL_Y^{(\alpha)}(\tfrac{1}{T})(\Exp{-\alpha \Ai}\rho\init)$. In particular, the dependence of $\varsigma_T$ on the different timesteps of the evolution is captured entirely in that composition, just as for the averaged quantity $\sigma_k$ in \eqref{eq:sigma_k-proof-sketch}. This is the key step that allows an adiabatic theorem to approximate this composition in order to analyze $\varsigma_T$ in the limit $T\to\infty$.
\end{itemize}
\end{remarks}
\begin{proof}
	\Ref{Comm} yields that $[Y(s),\xi(s)]=0$ for all $s$. Then by the expression \eqref{eq:def_P_F^T},
		\begin{align*}
		&\ee\big(\e^{\alpha \rvY}\big)\\
&\quad=
		\sum_{\vec\imath,\vec\jmath}\sum_{\ai,\af} \exp \Big(\alpha \sum_{k=1}^Y (y_{j_k}^{(k)}- y_{i_k}^{(k)})\Big)\, \pp^F_{T}(\ai,\af,\vec\imath,\vec\jmath)\\
		&\quad= \sum_{\vec\imath,\vec\jmath} \tr\!\Big( U_{T}\dotsm U_1   \Big(\sum_{\ai} \pi\init_{\ai} \rho\init\pi\init_{\ai}  \otimes \prod_{k=1}^T \e^{-\alpha y_{i_k}^{(k)}}\Pi_{i_k}\,\Xi\Big)
U_1^* \dotsm U_T^*  \Big(\sum_b\pi\fin_{\af}\otimes \prod_{k=1}^T \e^{\alpha y_{j_k}^{(k)}}\Pi_{j_k}\Big)\!\Big)\\
		&\quad= \tr\!\Big( U_{T}\dotsm U_1   \Big(\sum_{\ai} \pi\init_{\ai} \rho\init\pi\init_{\ai}  \otimes \prod_{k=1}^T\e^{-\alpha Y_k}\,\Xi\Big)
U_1^* \dotsm U_T^*  \,\Big(\id\otimes  \prod_{k=1}^T\e^{\alpha Y_k}\Big)\Big)\\
		&\quad= \tr \!\big(\cL_Y\ealpha(\tfrac TT)\circ\ldots\circ \cL_Y\ealpha(\tfrac1T)(\sum_{\ai} \pi\init_{\ai} \rho\init\pi\init_{\ai})\big).
	\end{align*}
	Assume in addition that  $[\Ai,\sysstate\init]=0$. Then similarly, for $\alpha_1,\alpha_2$ in $\rr$,
	\begin{align*}
		&\ee\big(\e^{\alpha_1 \rvY+\alpha_2 \rvW}\big)\\
		&\quad =
		\sum_{\vec\imath,\vec\jmath}\sum_{\ai,\af} \exp \Big(\alpha_1 \sum_{k=1}^Y (y_{j_k}^{(k)}- y_{i_k}^{(k)})\Big)\,\e^{\alpha_2(\ai-\af)}\, \pp^F_{T}(\ai,\af,\vec\imath,\vec\jmath)\\
		&\quad = \sum_{\vec\imath,\vec\jmath} \tr\Big( U_{T}\dotsm U_1   \big(\sum_{\ai} \e^{+\alpha_2 \ai}\pi\init_{\ai} \rho\init  \otimes \prod_{k=1}^T \e^{-\alpha_1 y_{i_k}^{(k)}}\Pi_{i_k}\,\Xi\big)
         \\ & \quad\qquad\qquad\qquad
         U_1^* \dotsm U_T^*  (\sum_b \e^{-\alpha_2 \af}\pi\fin_{\af}\otimes \prod_{k=1}^T \e^{\alpha_1 y_{j_k}^{(k)}}\Pi_{j_k})\Big)\\
		&\quad =  \tr\Big( U_{T}\dotsm U_1   \big(\e^{+\alpha_2 \Ai} \rho\init  \otimes \prod_{k=1}^T\e^{-\alpha_1 Y_k}\,\Xi\big) U_1^* \dotsm U_T^*  (\e^{-\alpha_2 \Af}\otimes \prod_{k=1}^T \e^{\alpha_1 Y_k})\Big)\\
		&\quad = \tr \big(\e^{-\alpha_2 \Af}\cL_Y^{(\alpha_1)}(\tfrac TT)\circ\ldots\circ \cL_Y^{(\alpha_1)}(\tfrac1T)( \e^{+\alpha_2 \Ai} \rho\init )\big).\qedhere
	\end{align*}
\end{proof}
Section 3.2 \cite{HJPR18} establishes several properties of the maps $\cL_Y\ealpha(s)$ for $s\in[0,1]$ and $\alpha \in \bC$, including that they are each completely positive and irreducible (under the assumption \ref{Irr}), although in general not trace-preserving. Appendix A of the same work goes on to analyze the decomposition of irreducible maps given in \Cref{prop:decomp-irred} in the presence of the deformation parameter $\alpha \in \bR$. The following summarizes the results of that work.

\begin{proposition}\label{prop:irred-properties-deformation}
Let $s\mapsto \cL(s)$ be a family of CPTP maps satisfying \textnormal{\ref{Irr}}, and define $\cL_Y\ealpha(s)$ by \eqref{eq_defLk_alpha}. Then there exist maps~$\lambda_Y\ealpha(s)$,~$\invalpha_Y(s)$ and~$\rho\ealpha_Y(s)$ from $[0,1]\times \rr \ni(s,\alpha)$ to, respectively, $\rr_{>0}$, the set of positive-definite operators, and the set of faithful states of $\H_\sys$; and maps $z(s), u(s)$ from $[0,1]$ to, respectively, $\nn$ and the set of unitary operators, with the following properties:
\begin{itemize}
	\item the identities  $[u(s),\invalpha_Y(s)]=[u(s),\rho\ealpha_Y(s)]=0$, and $u(s)^{z(s)}=\id$ hold;
	\item the peripheral spectrum of $\cL_Y\ealpha(s)$ is $\lambda\ealpha_Y(s) S_{z(s)}$, where
	\[
	S_{z}=\{\theta^m  :  \theta=\e^{2\i \pi /z}, {m=0,\ldots,z-1}\};
	\]
	\item the spectral decomposition $u(s) = \sum_{m=1}^{z(s)} \e^{2\i \pi m/z(s)} p_m(s)$ holds;
	\item the map $\eta\mapsto \tr( \invalpha_Y(s) u(s)^{-m}  \eta) \rho\ealpha_Y(s) u(s)^{m}$ is the spectral projector of $\cL_Y\ealpha(s)$ associated with  $\lambda_Y\ealpha(s) \,\e^{2\i \pi m/ z(s)}$;
	\item the unitary $u(s)$ and cardinal $z(s)$ of the peripheral spectrum of $\cL_Y\ealpha(s)$ do not depend on~$\alpha$ or $Y$.
\end{itemize}
\end{proposition}
Note that since $\cL^{(0)}(s) = \cL(s)$, we have $\lambda^{(0)}_Y(s)=1$, $\mathrm{I}^{(0)}_Y(s)=\id$ and $\rho^{(0)}_Y(s)=\rho\invar(s)$ for all $Y$ and $s$. As mentioned above, the case  $Y=\beta h_\env$ will be particularly relevant to the discussion of the Landauer's Principle. We therefore drop the indices $Y$, and simply denote by $\lambda\ealpha(s)$, $\invalpha(s)$ and $\rho\ealpha(s)$ the above quantities, in the case where $Y=\beta h_\env$.
We also define the rescaled map
\[\tilde\cL_Y\ealpha(s)=\big(\lambda\ealpha_Y(s)\big)\inv \cL_Y\ealpha(s)\]
which has spectral radius 1.

\section{An adiabatic theorem} \label{sec:adiabatic_theorem}

The following proposition establishes an adiabatic theorem for compositions of the form $\tilde \cL_Y\ealpha(\tfrac kT) \ldots  \tilde \cL_Y\ealpha(\tfrac 1T) \rho\init$ in terms of the quantities given in \Cref{prop:irred-properties-deformation}. See Proposition 3.12 in \cite{HJPR18} for the proof, which is based on the proof of the adiabatic theorem given in \cite{HJPR17}.
 \begin{proposition}\label{prop_adiabaticlemma}
	Consider an \textup{\ref{ADRIS}} with the family $(\cL(s))_{s \in [0,1]}$ satisfying \textup{\ref{Irr}} with $z(s) \equiv z$. Then, there exist continuous functions $\rr \ni \alpha \mapsto \ell'(\alpha)  \in (0,1)$ and $\rr \ni \alpha \mapsto C(\alpha)  \in \rr_+$,
   and a function $\alpha\mapsto T_0(\alpha)\in\nn$ that is bounded on any compact set of $\rr$, such that for all $\alpha \in \rr$, $T\geq T_0(\alpha)$, and~$k\leq T$,
	\begin{align*}
		&\Big\|\tilde \cL_Y\ealpha(\tfrac kT) \ldots  \tilde \cL_Y\ealpha(\tfrac 1T) \rho\init -z\e^{-\vartheta\ealpha_Y} \sum_{m=0}^{z-1} \tr\big(\invalpha(0) p_m(0) \rho\init\big)  \rho\ealpha_Y(\tfrac kT) p_{m - k}(\tfrac kT) \Big\|\\
		&\hspace{0.5\textwidth} \leq  \frac{C(\alpha) }{T(1-\ell'(\alpha) )}+ C(\alpha)\ell'(\alpha)^k
	\end{align*}
	where the index of the spectral projector $p_{m-k}(\tfrac kT)$ is interpreted modulo $z$, and
	$$
		\vartheta\ealpha_Y:=\int_0^{k/T} \tr\big(\invalpha_Y(s) \,\frac\partial{\partial s}\rho\ealpha_Y(s)\big)\d s.
	$$
\end{proposition}

By taking $\alpha=0$, this result allows adiabatic approximation of the state of $\sys$ under the physical evolution $\cL(\frac kT) \dotsm   \cL(\frac1T)$ after $k$ steps of an irreducible RIS. This corresponds to a generalization of the results of \cite{HJPR17}, which could only treat the primitive case, i.e.\@ $z=1$.
\begin{corollary} \label{cor:adiab_for_alpha=0}
Consider an \textup{\ref{ADRIS}} with the family $(\cL(s))_{s \in [0,1]}$ satisfying \textup{\ref{Irr}} with $z(s) \equiv z$. Then, there exists $\ell' < 1$, $C>0$,  and $T_0>0$ such that for all $T\geq T_0$, and $k\leq T$,
\[
\Big\| \cL(\frac kT) \dotsm   \cL(\frac1T) \, \rho\init -\rhoadiab(k,T)
		\Big\| \leq  \frac{C }{T(1-\ell')}+ C\ell'^k
\]
	where
\begin{equation}
	\rhoadiab(k,T):= z \sum_{n=0}^{z-1} \tr\big( p_n(0) \rho\init\big) \rho\invar(\tfrac kT)  p_{n - k}(\tfrac kT) \label{eq:def_rho_adiab}
\end{equation}
	is a state, and the index of the spectral projector $p_{n-k}(\tfrac kT)$ is interpreted modulo $z$. Moreover, if $\rho\init$ is faithful, we have the uniform bound
	\[
	\inf_{T>1} \inf_{k\leq T} \inf \spec\rhoadiab(k,T) \geq  z \Big(\min_{1\leq j \leq z}  \tr\big( p_j(0) \rho\init\big) \Big) \inf_{s\in [0,1]} \inf \spec\rho\invar(s) > 0.
	\]
\end{corollary}
\begin{proof}
We apply Proposition~\ref{prop_adiabaticlemma} for $\alpha=0$, and use that $I^{(0)}(s)\equiv \id$, $\rho_Y^{(0)}= \rho\invar$, and $\vartheta\ealpha_Y = 0$ which follows from $\tr \rho^{(0)}_Y(s)\equiv 1$. Next, we check the formula $\tr(\rho\invar(\tfrac kT) p_\ell(\tfrac kT)) =\frac{1}{z}$  for each $\ell=0,\dotsc,z-1$. We drop the argument $\tfrac kT$ in what follows, and write $\cL_{k/T}(\cdot) = \sum_i V_i \cdot V_i^*$ the Kraus decomposition.
Recalling that $p_\ell V_i = V_i p_{\ell+1}$ for all $i$ and $\ell$ via \eqref{eq:irred-kraus-intertwine},
\begin{align*}
   \tr(\rho\invar p_\ell) &= \tr(\cL(\rho\invar) p_\ell) = \sum_{i} \tr(V_i \rho\invar V_i^* p_\ell) \\
      &= \sum_i \tr(V_i \rho\invar p_{\ell+1} V_i^*) = \tr(\cL(\rho\invar p_{\ell+1})),
\end{align*}
so $\tr(\rho\invar p_\ell) =\tr( \rho\invar p_{\ell+1})$ using that $\cL$ is trace-preserving.
As $\sum_\ell \tr(\rho\invar p_\ell) = \tr \rho\invar =1$, we must have $\tr(\rho\invar p_\ell) = \frac{1}{z}$. Therefore,
\begin{align*}
   \tr (\rhoadiab(k,T)) &= z \sum_{n} \tr\big( p_n(0) \rho\init\big)\tr( \rho\invar(\tfrac kT)  p_{n - k}(\tfrac kT))= \sum_{n} \tr\big( p_n(0) \rho\init\big) \\
      &= \tr \rho\init = 1.
\end{align*}
Moreover, given a normalized vector $\psi \in \H$, we have
\begin{align*}
\braket{\psi,\rhoadiab(k,T) \psi} &= z \sum_{n} \tr\big( p_n(0) \rho\init\big) \braket{\psi,\rho\invar(\tfrac kT)  p_{n - k}(\tfrac kT)\psi}\\
&=z \sum_{n} \tr\big( p_n(0) \rho\init\big) \braket{p_{n - k}(\tfrac kT)\psi,\rho\invar(\tfrac kT)  p_{n - k}(\tfrac kT)\psi}.
\end{align*}
using $[\rho\invar(\tfrac kT),p_{n - k}(\tfrac kT)]=0$. Since $\rho\invar(\tfrac kT) > 0$ and $ \tr\big( p_n(0) \rho\init\big) > 0$, each term in the sum is non-negative, and we have
\begin{align*}
\braket{\psi,\rhoadiab(k,T) \psi} &\geq z  \big(\min_{1\leq j \leq z}  \tr\big( p_j(0) \rho\init\big) \big) \sum_n \braket{\psi,\rho\invar(\tfrac kT)  p_{n - k}(\tfrac kT)\psi} \\
&= z  \big(\min_{1\leq j \leq z}  \tr\big( p_j(0) \rho\init\big) \big) \braket{\psi,\rho\invar(\tfrac kT) \psi}\\
&\geq z \big(\min_{1\leq j \leq z}  \tr\big( p_j(0) \rho\init\big) \big) \inf_{s\in [0,1]} \inf \spec\rho\invar(s).\qedhere
\end{align*}
\end{proof}
\begin{remarks}\,\hfill
	\begin{itemize}
		\item Given an \textup{\ref{ADRIS}} the family $(\cL(s))_{s \in [0,1]}$ satisfying \textup{\ref{Irr}}, for faithful $\rho\init$
		the state
		$
		\rho\fin_T = \cL_{T} \dotsm \cL_{1} \rho\init
		$ is faithful for each $T > 1$ using \Cref{prop:irred-preserve-faithful}. Corollary \ref{cor:adiab_for_alpha=0} and Weyl's inequalities (see Section III.2 in \cite{Bha97}) give the stronger result $\inf_{T>1} \inf\spec\rho\fin_T > 0$.
		In particular, we may make the choice $\Af = - \log \rho\fin_T$ which is bounded uniformly in $T$.
		\item If we assume, in the notation of \Cref{prop:decomp-irred}, that $P_n\rho\init=0$ for $n>0$ (i.e.\@\ $\rho\init$ has no components corresponding to the peripheral eigenvalues of $\cL(0)$ other than $1$), then one can check that $\rhoadiab(k,T)=\rho\invar(\frac kT)$.
	\end{itemize}
\end{remarks}
 
\section{The statistics of the entropy production at the level of trajectories}
In this section, we summarize the statistical properties of $\zeta_T$ which are derived in \cite{HJPR18} using the adiabatic theorem described in \Cref{sec:adiabatic_theorem}. For the sake of brevity, most proofs will be omitted.

First, the adiabatic theorem can be used to directly compute the limiting moment generating function in the special case in which the reduced dynamics is primitive and $X(s) \equiv 0$.
\begin{theorem}\label{prop:mgf_X_zero}
 Under the assumptions \textup{\ref{ADRIS}} and~\textup{\ref{Prim}}, and with $X(s) \equiv 0$, the limiting moment generating function $M_\varsigma(\alpha):=\lim_{T\to\infty}M_{\varsigma_{T}}(\alpha)$ satisfies
\[
M_\varsigma(\alpha)=Q_{-\alpha}(\sysstate\invar(0) \| \sysstate\init),\]
for $Q_\alpha(\eta \| \zeta) :=  \tr(\eta^\alpha \zeta^{1-\alpha})$
which can be related to the R\'enyi relative entropy via $D_\alpha(\eta\|\zeta) = \frac{1}{1-\alpha}\log Q_\alpha(\eta\|\zeta)$.
\end{theorem}
\begin{remark}
In particular, if $\rho\init$ is chosen as the initial invariant state $\rho\invar(0)$, then $M_\varsigma(\alpha) \equiv 1$, and hence $\varsigma_T \to 0$ weakly. This greatly strengthens the results of \cite{HJPR17} which showed that under the assumptions of the proposition, if $\rho\init=\rho\invar(0)$ then $\oldsigmaT=\bE[\varsigma_T] \to 0$.
\end{remark}

Next, we remove the assumptions of primitivity and that $X(s)\equiv 0$, and use the adiabatic theorem to compute the exponential rate of change of the moment generating function with $T$ as $T\to\infty$.
\begin{lemma} \label{lemma_cvgCGF}
	Under the assumptions \textup{\ref{ADRIS}} and \textup{\ref{Irr}} with $z(s) \equiv z$, for any faithful initial state~$\sysstate\init > 0$, for any~$\alpha \in \rr$, the moment generating function of the random variable~$\rvY$ with respect to~$\pp^F_{T}$ satisfies
	\begin{equation} \label{eq_defLambdaalpha}
		\lim_{T \to \infty} \frac{1}{T} \log M_{\Delta a_T + \rvY}(\alpha) = \lim_{T \to \infty} \frac{1}{T} \log M_{\rvY}(\alpha) = \int_0^1 \log \lambda_Y^{(\alpha)}(s) \d s=: \Lambda_Y(\alpha).
	\end{equation}
\end{lemma}

\begin{proof}\label{lem:log-conv}
	Lemma 5.1 of \cite{HJPR18} uses the adiabatic theorem \Cref{prop_adiabaticlemma} to show that
	$$
		\lim_{T \to \infty} \frac{1}{T} \log \tr\big(\tilde \cL_Y\ealpha(\tfrac TT) \dotsb \tilde \cL_Y\ealpha(\tfrac 1T)(\rho\init)\big) = 0.
	$$
	But the moment generating function reads
	\begin{align*}
		M_{\rvY}(\alpha) &= \tr\big( \cL_Y\ealpha(\tfrac TT) \dotsb \cL_Y\ealpha(\tfrac 1T)\rho\init\big) \\
			&= \Big( \prod_{k=1}^T \lambda_Y\ealpha(\tfrac{k}{T}) \Big) \tr\big(\tilde \cL_Y\ealpha(\tfrac TT) \dotsb \tilde \cL_Y\ealpha(\tfrac 1T)(\rho\init)\big)
	\end{align*}
	by definition of $\tilde \cL_Y\ealpha(s)$. Hence, the result follows from the Riemann sum convergence
	\begin{align*}
		\lim_{T \to \infty} \frac{1}{T} \log \Big( \prod_{k=1}^T \lambda_Y\ealpha(\tfrac{k}{T}) \Big)
		&= \lim_{T \to \infty} \sum_{k=1}^T(\tfrac{k}{T} - \tfrac{k-1}{T}) \log  \lambda_Y\ealpha(\tfrac{k}{T}) \\
		&= \int_0^1 \log \lambda_Y^{(\alpha)}(s) \d s.
	\end{align*}
	The same holds for $M_{\Delta a_T + \rvY}(\alpha)$ because $\rvY(\omega)$ and~$\rvZ(\omega)$ only differ by the uniformly bounded term~$-\Delta a_T(\omega) + \Delta y_T^\text{tot}(\omega)$.
\end{proof}

Many results can be derived from the convergence of the moment generating function of $\varsigma_T$ shown in \Cref{lemma_cvgCGF}. First, we formulate a large deviations principle; see e.g.\@~\cite{Ell06} for more on large deviations.
 In the statement below we denote the interior of a set $E\subseteq\rr$ by $\inte E$ and the closure by $\clos E$.
\begin{theorem} \label{theo_ldp}
	Assume \textup{\ref{ADRIS}} and \textup{\ref{Irr}}, and that the initial state~$\sysstate\init$ is faithful. Let $\Lambda_Y$ be defined by relation \eqref{eq_defLambdaalpha} and denote by $\Lambda_Y^*$ the Fenchel--Legendre transform of $\Lambda_Y$, i.e.\@\ for $x\in\rr$ let
	\[\Lambda_Y^*(x)=\sup_{\alpha\in\rr}\big(\alpha x - \Lambda_Y(\alpha)\big).\]
Then $\Lambda_Y^*(x)=+\infty$ for $x\not\in [\nu_{Y,-},\nu_{Y,+}]$, and for any Borel set $E$ of $\rr$ one has
\begin{align*}
   -\inf_{x\in \inte E} \Lambda_Y^*(x) &\leq \liminf_{T\to\infty} \frac1T \log \bP^F_{T}\big(\frac{\rvY}T \in \inte E\big) \\
      &\leq \limsup_{T\to\infty} \frac1T \log \bP^F_{T}\big(\frac{\rvY}T \in \clos E\big)\leq -\inf_{x\in \clos E} \Lambda_Y^*(x).
\end{align*}
The same statement holds with $\rvZ$ in place of $\rvY$. In particular, for $Y=\beta h_\env$, one has
\begin{align*}
   -\inf_{x\in \inte E} \Lambda^*(x) &\leq \liminf_{T\to\infty} \frac1T \log \bP^F_{T}\big(\frac{\varsigma_T}T \in \inte E\big) \\
      &\leq \limsup_{T\to\infty} \frac1T \log \bP^F_{T}\big(\frac{\varsigma_T}T \in \clos E\big)\leq -\inf_{x\in \clos E} \Lambda^*(x)
\end{align*}
and the same statement holds with $\Delta s_{\env,T}$ in place of $\varsigma_T$.
\end{theorem}

Let us explore two special cases of this result with the choice $Y(s)=\beta(s)h_\env(s)$.
First, if $X(s)\equiv 0$, one can compute\footnote{See Section 4 of \cite{HJPR18} for the proof.} that $\lambda\ealpha(s)=1$ for all $\alpha\in\rr$ and $s\in[0,1]$. In that case $\Lambda(\alpha)\equiv 0$ and
\[\Lambda^*(x) = \left\{
\begin{array}{cl}
0 & \mbox{ if } x=0,\\
+\infty & \mbox{ otherwise.}
\end{array}
\right.\]
The above large deviation statement therefore gives a concentration of~$\tfrac 1T \varsigma_{T}$ or $\tfrac 1T {\Delta s_{\env,T}}$ at zero which is faster than exponential.

For the second case, let us make the following assumption.
\begin{description}
	\labitem{TRI}{TRI} We say that a repeated interaction system satisfying \ref{ADRIS} satisfies \emph{time-reversal invariance} if for every $s\in[0,1]$ there exist two antiunitary involutions $C_\sys(s) : \H_\sys \to \H_\sys$ and $C_\env(s) : \H_\env \to \H_\env $ such that if $C(s)=C_\sys(s)\otimes C_\env(s)$ one has for all $s\in[0,1]$
	\[  [\hsys,C_\sys(s)] = 0, \qquad [\henv(s),C_\env(s)] = 0,  \qquad [v(s),C(s)] = 0 .\]
\end{description}
This holds for example if each $\hsys, \henv$ and $v$ are real valued matrices in the same basis, and $C_\sys$, $C_\env$ are complex conjugation in the corresponding basis.

In this case, one finds that for all $s\in[0,1]$,
\begin{equation} \label{eq_symlambda}
	\lambda\ealpha(s)=\lambda^{(-1-\alpha)}(s),
\end{equation}
or equivalently that the function $\alpha\mapsto \lambda\ealpha(s)$ is symmetric about $\alpha=-1/2$ for all $s\in[0,1]$. This is shown in Lemma 3.11 of \cite{HJPR18}.

The symmetry \eqref{eq_symlambda} is central to the Gallavotti--Cohen Theorem which relates in a parameter free formulation the probabilities of observing entropies of opposite signs. Indeed, it implies that $\Lambda$ is symmetric about the $\alpha=-1/2$ axis. A direct computation shows that
	\begin{equation} \label{eq_relationGC}
		\Lambda^*(x)=x+\Lambda^*(-x).
	\end{equation}
	A consequence of Theorem \ref{theo_ldp} together with this equality is that if e.g.\@\ $\Lambda''(0)\neq 0$,
	\begin{equation}\label{eq:senv_pos_vs_neg}
	\lim_{\delta\to 0}\lim_{T\to\infty} \frac1T \log \frac{\pp^F_{T}(+\Delta s_{\env,T}\in[s-\delta,s+\delta])}{\pp^F_{T}(-\Delta s_{\env,T}\in[s-\delta,s+\delta])} = -s.
	\end{equation}
	Roughly speaking, the probability of obtaining a sequence of measurements corresponding to an increase in the entropy of the chain is exponentially smaller than the probability of observing a decrease in entropy. In light of the balance equation \eqref{eq:decomp-entropy-prod-traj}, since $\Delta s_{\sys, T}(\omega)$ is bounded as $T\to\infty$ (i.e.\@~the finite-size system can only add or remove a finite amount of entropy), this corresponds to an exponentially larger probability for $\zeta_T$ to be positive than negative. This is a more refined statement than the fact that $\bE[\zeta_T] = \oldsigmaT \geq 0$.
	\eqref{eq:senv_pos_vs_neg} is obtained by observing that in the present case, $\Lambda$ is analytic in a neighbourhood of the real axis, and so is $\Lambda^*$ if $\Lambda$ is strictly convex.  See e.g.\@\ \cite{EHM09,JOPP12,CJPS18} for more information on the role of symmetries such as \eqref{eq_relationGC}.

The next consequence of \Cref{theo_ldp} is a result similar to a law of large numbers for~$\rvY$:
\begin{corollary} \label{coro_LLN}
	Under the same assumptions as in Theorem \ref{theo_ldp}, for all $\epsilon>0$ there exists $r_\epsilon>0$ such that for $T$ large enough
	\begin{equation}  \label{eq_expcvg}
	\pp^F_{T}\big(|\frac1T \,\rvY - \Lambda_Y'(0)|>\epsilon\big)\leq \exp \left(-r_\epsilon T\right).
	\end{equation}
\end{corollary}
\begin{remark} \label{remark_expcvg}
	Such a result is sometimes called exponential convergence. If one could replace~$\pp^F_{T}$ by a $T$-independent probability measure $\pp^F$ in~\eqref{eq_expcvg} then the Borel--Cantelli lemma would imply that $\frac1T \rvY$ converges $\pp^F$-almost-surely to $\Lambda_Y'(0)$.
   It implies, however, that $\lim_{T\to\infty} \frac1T \ee(\rvY)=\Lambda_Y'(0)$. In the case $Y=\beta h_\env$, the positivity of $\oldsigmaT$ implies $\Lambda'(0)\geq 0$. The formula
   \[
  \Lambda'_Y(0) =\int_0^1\beta(s)\,\tr\Big(X(s) \big(\id\otimes h_\env(s)\big)\Big)\d s,
   \]
   valid for $Y=\beta h_\env$, shows that $\Lambda'(0)=0$ if $X(s)\equiv 0$. This expression can be rewritten in a more illuminating way \cite{BCJP}, using the following relation in which the dependence on $s$ is suppressed:
   \begin{align*}	
   D(\rho\invar \otimes \xi \| U^* (\rho\invar \otimes \xi) U) &= -S(\rho\invar \otimes \xi) - \tr[(\rho\invar \otimes \xi) \log (U^* \rho\invar \otimes \xi U )] \\
   &=-S(\rho\invar) - S(\xi) - \tr[U(\rho\invar \otimes \xi) U^* \log (\rho\invar \otimes \xi )] \\
   &=-S(\rho\invar) - S(\xi) - \tr[U(\rho\invar \otimes \xi) U^* ( \log (\rho\invar) \otimes \one )] \\
   &\qquad - \tr[U(\rho\invar \otimes \xi) U^* ( \one \otimes \log(\xi) )] \\
   &= -S(\rho\invar) - S(\xi) + S(\rho\invar) - \tr[U(\rho\invar \otimes \xi) U^* ( \one \otimes \log(\xi) )] \\
   &= \tr[ (\xi - \xi\fin) \log \xi ] = \beta\tr[ (\xi\fin - \xi) \henv ]
   \end{align*}
   where $\xi\fin := \tr_\sys [U (\rho\invar \otimes \xi) U^*]$. Hence, restoring the dependence on $s$,
   \[
   \beta(s)\,\tr\Big(X(s) \big(\id\otimes h_\env(s)\big)\big)\Big) = D\Big(\rho\invar(s) \otimes \xi(s) \Big\| U^*(s) (\rho\invar(s) \otimes \xi(s)) U(s) \Big)
   \]
   and thus
   \begin{equation*}
   \Lambda'_Y(0) = \int_0^1D\Big(\rho\invar(s) \otimes \xi(s) \Big\| U^*(s) (\rho\invar(s) \otimes \xi(s)) U(s) \Big) \d s.
   \end{equation*}
   By the invariance of the relative entropy under joint unitary conjugations, we may rewrite this formula more naturally as
\begin{equation}
   \Lambda'_Y(0) = \int_0^1D\Big(U(s) (\rho\invar(s) \otimes \xi(s))U^*(s) \Big\|  \rho\invar(s) \otimes \xi(s) \Big) \d s.
   \end{equation}
   In particular, $\Lambda'_Y(0)=0$ if and only if $U^*(s)(\rho\invar(s) \otimes \xi(s)) U(s)=  \rho\invar(s) \otimes \xi(s)$ for all $s\in [0,1]$, and hence $X(s) = 0$ if and only if $\Lambda'_Y(0)=0$.
\end{remark}

One can also obtain a central limit-type result by a slight improvement of the results in Theorem~\ref{theo_ldp}.
\begin{theorem} \label{theo_clt}
	Under the same assumptions as in Theorem \ref{theo_ldp} we have
	\[ \frac1{\sqrt T}\big(\rvY- T \,\Lambda_Y'(0)\big)\underset{T\to\infty}\to \mathcal N\big(0,\Lambda_Y''(0)\big)\]
	in distribution.
\end{theorem}

\begin{remark} \label{remark_dvgoldsigmaT}
	In the case $Y=\beta h_\env$, \Cref{theo_clt} and the remark following \Cref{coro_LLN}  show that if $\Lambda''(0)\neq 0$ (which is generically expected) then $\oldsigmaT\to\infty$ as $T\to\infty$.
\end{remark} 
\section{Examples}\label{sec:RIS_examples}
Let us recall the simplest non-trivial RIS, which was already considered in \Cref{ex:RWA}. Here, we change notation to call the reduced dynamics $\cL$ instead of $\Phi$. In this example, the system and probes are 2-level systems, with $\cH_\sys = \cH_\env = \bC^2$, and we choose Hamiltonians $h_\sys := E a^*a$ and $\henv[k] \equiv \henv  := E_0 b^*b$ where $a/a^*$ (resp. $b/b^*$) are the Fermionic annihilation/creation operators for $\sys$ (resp. $\env$), with $E, E_0 > 0$  constants with units of energy. As matrices in the (ground state, excited state) bases $\{|0\rangle, |1\rangle\}$ for $\sys$ and $\env$, we write
\[
a=  b= \begin{pmatrix}
0 & 1 \\ 0 & 0
\end{pmatrix}, \quad a^* = b^* = \begin{pmatrix}
0 & 0 \\ 1 & 0
\end{pmatrix}, \qquad a^*a=b^*b = \begin{pmatrix}
0 & 0 \\ 0 &1
\end{pmatrix}.
\]
As in \Cref{ex:RWA}, we consider a constant potential $v_\text{RW}\in \cB(\cH_\sys \otimes \cH_\env)$,
\[
v_\text{RW} = \frac{\mu_1}{2}(a^* \otimes b + a \otimes b^*)
\]
where $\mu_1 =1$ with units of energy. Given $s\mapsto \beta(s) \in [0,1]$ a $C^2$ curve of inverse probe temperatures, an interaction time $\tau>0$ and coupling constant $\lambda>0$, we let
\[U = \exp\big(-\i \tau (\hsys + \henv + \lambda v_\text{RW})\big).\]
Then $\spec\cL(s)$ is independent of $s$, with $1$ as a simple eigenvalue with eigenvector $$\rho\invar(s) = \exp(-\beta^*(s) \hsys)/\tr(\exp(-\beta^*(s) \hsys))$$ for $\beta^*(s) = \frac{E_0}{E}\beta(s)$.

With $\nu := \sqrt{(E-E_0)^2 + \lambda^2}$,  the assumption $\nu\tau\not \in 2\pi \Z$ yields that $\cL(s)$ is primitive, and moreover, the fact that $[v_\text{RW}, a^*a +  b^*b] = 0$ yields $X(s)\equiv 0$. Here, we may take $k_\sys \equiv \frac{E_0}{E}\hsys$ independently of $s$, which satisfies $[k_\sys + \henv, U ] \equiv 0$.

We choose an initial system state $\rho\init >0$, and
set $Y(s) := \beta(s) \henv$, $\Ai := \log \rho\init$, $\Af := \log \rho\fin_T$, for $\rho\fin_T := \cL(\tfrac TT) \dotsm \cL(\tfrac 1T) \rho\init$. 
Then \Cref{prop:mgf_X_zero} yields the asymptotic moment generating function of $\varsigma_{T}$:
if $\rho\init$ has spectral decomposition $\rho\init=r_0 |v_0\rangle\langle v_0|+r_1 |v_1\rangle\langle v_1|>0$, then
\begin{equation}
\begin{split}
\lim_{T\to\infty}M_{\varsigma_{T}}(\alpha) &=
(1+e^{-\beta(0)E_0})^\alpha\\
   &\qquad\qquad \Big(r_0^{1+\alpha}(|\langle 0|v_0\rangle|^2+|\langle 1|v_0\rangle|^2e^{\alpha\beta(0)E_0})\alpha\\
      &\qquad\qquad\qquad +r_1^{1+\alpha}(|\langle 0|v_1\rangle|^2+|\langle 1|v_1\rangle|^2e^{\alpha\beta(0)E_0})\Big).
\end{split}
\end{equation}
Now, let us consider the same example with a different interaction potential. Instead of $v_\text{RW}$, we consider the so-called full-dipole interaction potential $v_\text{FD}\in \cB(\cH_\sys \otimes \cH_\env)$ given by
\[
v_\text{FD} := \frac{\mu_1}{2} (a+a^*) \otimes (b +  b^*),
\]
This example was considered in \cite[Section 7.1]{HJPR17}, where it was shown that \ref{Prim} is satisfied, and that $\sigma_T \to \infty$ with a finite and nonzero rate $\lim_{T\to\infty} \frac{1}{T}\sigma_T$ for generic choices of parameters $\{E,E_0,\tau\}$.

\begin{figure}[ht]

 \centering

  \includegraphics{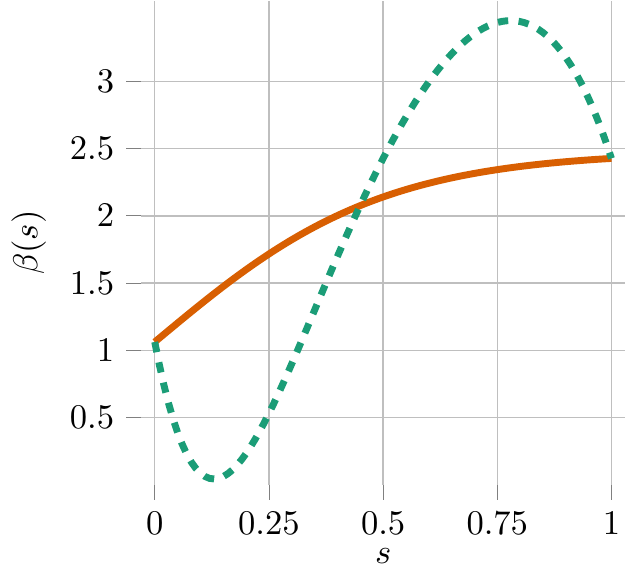}

\caption{Two choices of curves $s\mapsto \beta(s)$. The solid orange line, $\beta(s)= \beta_1(s)$, is given by \eqref{eq:beta1}, and in dashed green line, $\beta(s) = \beta_2(s)$, is given by \eqref{eq:beta2}. These curves of inverse temperatures start and end at the same points, and have the same integral. The differences in $\Lambda$ and $\Lambda^*$ shown in \Cref{fig:FD_Lambda} below are due to the differences in $\beta_1(s)$ and $\beta_2(s)$ shown here.  \label{fig:FDbetas}}
\end{figure}

We take $Y(s) =\beta(s) \henv$ before. Introducing the shorthand $\eta := \sqrt{(E_0+E)^2 + \lambda^2}$, we compute a matrix expression for $\cL\ealpha_s$ by  identifying $\cB(\cH_\sys) \cong \operatorname{Mat}_{2\times 2}(\bC) \cong \bC^4$ via $\begin{psmallmatrix}
\eta_{11} & \eta_{12}\\
\eta_{21} & \eta_{22}
\end{psmallmatrix} \mapsto \begin{psmallmatrix}
\eta_{11} \\ \eta_{12} \\ \eta_{21} \\ \eta_{22}
\end{psmallmatrix}$.
Working in the (ground state, excited state) basis for~$\sys$, we obtain
$$
   \cL\ealpha(s)
      =
      \begin{pmatrix}
         a & 0 & 0 & d \\
         0 & b & c & 0 \\
         0 & c & e & 0 \\
         f & 0 & 0 & g
      \end{pmatrix},
$$
where
\begin{align*}
   a &= \frac{\left(2 (E_0+E)^2+\lambda ^2+\lambda ^2 \cos (\eta
     \tau )\right)}{2 \left(1+\Exp{E_0 \beta(s) }\right)\eta ^2\Exp{-E_0 \beta(s) } }+\frac{2 (E_0-E)^2+\lambda ^2+\lambda ^2 \cos (\nu  \tau
     )}{2 \left(1+\Exp{E_0 \beta(s) }\right)\nu ^2}, \\
   d &= \lambda ^2 \left(-\frac{2
   \Exp{-E_0 \alpha  \beta(s) } (\cos (\eta  \tau )-1)}{4 \left(1+\Exp{E_0 \beta(s) }\right)\eta ^2}-\frac{2 \Exp{E_0 (\alpha +1)
   \beta(s) } (\cos (\nu  \tau )-1)}{4 \left(1+\Exp{E_0 \beta(s) }\right)\nu ^2}\right), \\
   c &= \frac{\lambda ^2 \cosh \left(E_0 \beta(s)(\frac{1}{2}+\alpha) \right)
      \text{sech}\left(\frac{E_0 \beta(s) }{2}\right) \sin \left(\frac{\eta  \tau }{2}\right) \sin
      \left(\frac{\nu  \tau }{2}\right)}{\sqrt{E_0^4+2 \left(\lambda ^2-E^2\right)
      E_0^2+\left(E^2+\lambda ^2\right)^2}}, \\
   b &= \frac{\left(\mathrm{i} \eta  \cos \left(\frac{\eta  \tau }{2}\right)+(E_0+E) \sin
     \left(\frac{\eta  \tau }{2}\right)\right) \left((E_0-E) \sin \left(\frac{\nu  \tau
     }{2}\right)-\mathrm{i} \nu  \cos \left(\frac{\nu  \tau }{2}\right)\right)}{\sqrt{E_0^4+2
     \left(\lambda ^2-E^2\right) E_0^2+\left(E^2+\lambda ^2\right)^2}}, \\
   e &= \frac{
      \left(-\Exp{\mathrm{i} \nu  \tau } E_0+E_0-E+\nu +\Exp{\mathrm{i} \nu  \tau } (E+\nu
      )\right) \left(\eta  \cos \left(\frac{\eta  \tau }{2}\right)+\mathrm{i} (E_0+E) \sin
      \left(\frac{\eta  \tau }{2}\right)\right)}{2 \eta  \nu \Exp{\frac{1}{2} \mathrm{i} \nu  \tau }}, \\
   f &= \frac{\Exp{-E_0 \alpha  \beta(s) } \lambda ^2 }{4
     \left(1+\Exp{E_0 \beta(s) }\right)}\left(\frac{2-2 \cos (\nu  \tau )}{\nu ^2}-\frac{2
     \Exp{E_0 (2 \alpha +1) \beta(s) } (\cos (\eta  \tau )-1)}{\eta ^2}\right), \\
   g &= \frac{ \left(2
   (E_0+E)^2+\lambda ^2+\lambda ^2 \cos (\eta  \tau )\right)}{2 \left(1+\Exp{-E_0
   \beta(s) }\right)\eta ^2 \Exp{E_0 \beta(s) }}+\frac{2
   (E_0-E)^2+\lambda ^2+\lambda ^2 \cos (\nu  \tau )}{2 \left(1+\Exp{-E_0
   \beta(s) }\right)\nu ^2},
\end{align*}
which depend on $s$ through $\beta(s)$. The computation was performed with Mathematica, using \cite{RIS_mathematica_code}. We make a particular choice of parameters, $\lambda = 2$, $\tau=0.5$, $E_0 = 0.8$, $E=0.9$, and two choices of $[0,1]\ni s \mapsto \beta(s)$:
\begin{equation}\label{eq:beta1}
\beta_1(s) = \frac{2(3+4\tanh(2s))}{3 + 2 \log(\cosh(2))}
\end{equation}
and
\begin{equation}\label{eq:beta2}
\beta_2(s) =a_1 \tanh(2 s) -a_2 \tanh\big(\frac{s}{2}\big) - a_3 s^3 + a_4 s^2 -a_5 s + a_6
\end{equation}
for $a_1 = 35.483$, $a_2=141.929$, $a_3=42.945$, $a_4 = 93.5$, $a_5=17.808$, $a_6 = 1.061$. We have $\beta_1(0) = \beta_2(0) = 1.06$, and $\beta_1(1) = \beta_2(1) = 2.43$, as well as $\int_0^1 \beta_1(s) \d s = \int_0^1 \beta_2(s) \d s = 2$. These are plotted in Figure~\ref{fig:FDbetas}.

\begin{figure}[ht]
\centering
  \includegraphics{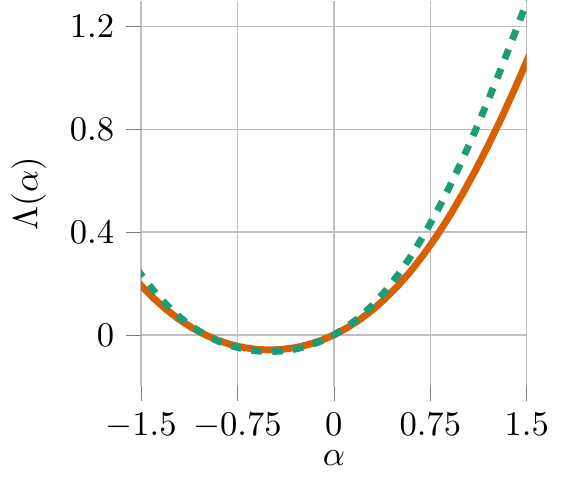}
\quad
  \includegraphics{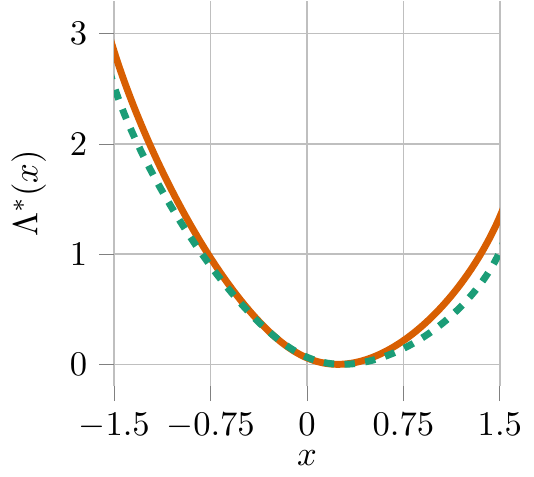}

\caption{
Left: The function $\Lambda(\alpha)$ for $Y = \beta \henv$ with $v_k = v_\text{FD}$, $\lambda = 2$, $\tau=0.5$, $E_0 = 0.8$,  $E=0.9$, plotted for each choice of $\beta(s)$.
Right: The rate function $\Lambda^*(\alpha)$, for the same setup.
In each plot, the solid orange line corresponds to the choice $\beta(s) = \beta_1(s)$, defined in \eqref{eq:beta1}, and the dashed green line corresponds to $\beta(s) = \beta_2(s)$, defined in \eqref{eq:beta2}.  \label{fig:FD_Lambda}}
\end{figure}

\begin{figure}[ht]
\centering
   \includegraphics[width=.8\textwidth]{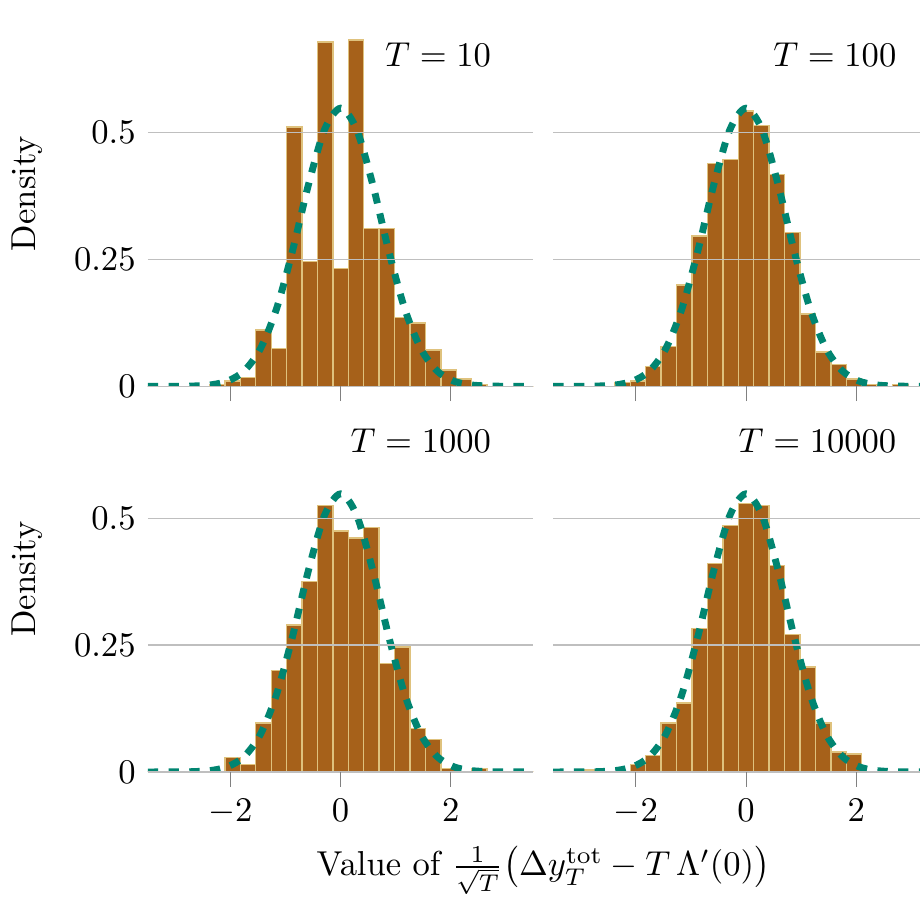}
   \caption{Convergence of $ \frac1{\sqrt T}\big(\rvY- T \,\Lambda'(0)\big)$ to a normal distribution, where $\Lambda'(0) \approx 0.240$, with $\beta(s)=\beta_1(s)$ given by~\eqref{eq:beta1}. Each plot was generated by simulating the two-time measurement protocol in 2,000 instances of the repeated interaction system with the parameters given in \Cref{fig:FD_Lambda}. The value of $ \frac1{\sqrt T}\big(\rvY- T \,\Lambda'(0)\big)$ was calculated for each instance and plotted in a histogram in orange, with bar heights normalized to yield total mass 1. In green, the probability density function of $\mathcal{N}(0,\Lambda''(0))$ is plotted, where $\Lambda''(0) \approx 0.530$. As~$T$ increases, one sees qualitatively the convergence of $ \frac1{\sqrt T}\big(\rvY- T \,\Lambda'(0)\big)$ to the normal distribution, as guaranteed by Theorem~\ref{theo_clt}.  \label{fig:CLT_FD1}}
\end{figure}

\begin{figure}[ht]
\centering
   \includegraphics[width=.8\textwidth]{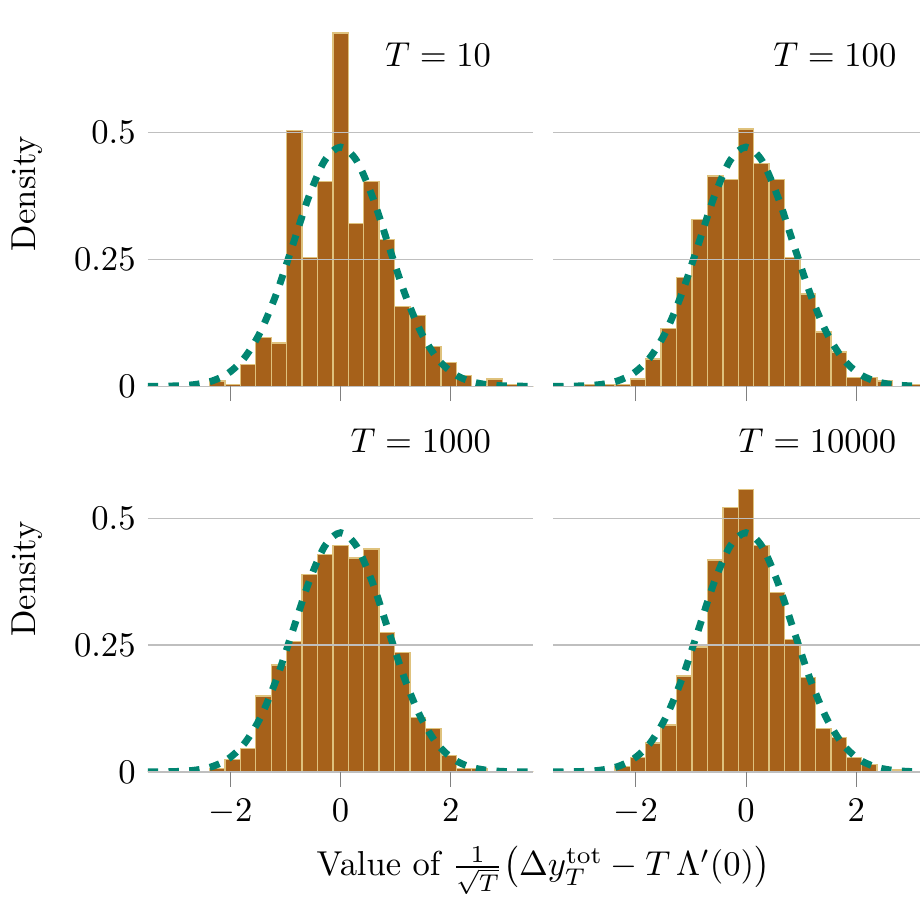}
   \caption{The same setup as Figure~\ref{fig:CLT_FD1}, with $\beta(s)=\beta_2(s)$ given by~\eqref{eq:beta2}. Here, $\Lambda'(0) \approx 0.275$, and $\Lambda''(0) \approx 0.716$. \label{fig:CLT_FD2}}
\end{figure}

We compute numerically the function $\Lambda(\alpha)$ for each choice of $s\mapsto \beta(s)$, as shown in Figure~\ref{fig:FD_Lambda}. Figures~\ref{fig:CLT_FD1} and \ref{fig:CLT_FD2} shows the convergence described by Theorem \ref{theo_clt} by simulating 2,000 instances of this repeated interaction system at four values of $T$.

\vspace{2em}

\FloatBarrier  

\chapter{Open questions from Part II}

\begin{enumerate}
	\item In \Cref{cor:pptfulleeb} it is established that faithful PPT quantum channels are eventually entanglement-breaking. Can this result be extended to the non-faithful case? Moreover, can a universal finite number $n\in \mathbb{N}$ be found such that any PPT quantum channel $\Phi$ is such that its $n$-fold composition $\Phi^n$ is entanglement-breaking? The PPT$^2$ conjecture is that such $n=2$ suffices (even in the non-trace-preserving case).
	\item In \Cref{chap:RIS}, a two-time measurement protocol is used to provide a detailed analysis of the entropy production of repeated interaction systems, and has similarly been used in other studies in quantum statistical mechanics. Can such a protocol play a role in an information-theoretic analysis as well?
\end{enumerate}

\begin{spacing}{0.9}

\printbibliography[heading=bibintoc, title={References}]

\end{spacing}

\begin{appendices} 

\chapter{Mathematica code}\label{sec:cx_mathematica}
The following is Mathematica code to perform the symbolic manipulations used in the proof of \Cref{prop:symm-concave-cx}. This code was run on Mathematica version 12.0.0.0.

\mmaSet{leftmargin=3.5em}

\begin{mmaCell}[addtoindex=0,moredefined={g},morepattern={x1_, x2_,x3_, x4_, x1, x2, x3, x4}]{Input}
  g[x1_,x2_,x3_,x4_] := Log[E^(x1 + 2*x2 + 3*x3 + 4*x4)
  + E^(x4 + 2*x3 + 3*x2 + 4*x1)]
\end{mmaCell}

\begin{mmaCell}[moredefined={f, g},morepattern={x1_, x2_, x3_, x4_,x1, x2, x3, x4}]{Input}
  f[x1_,x2_,x3_,x4_] := -1*(g @@@  Permutations[\{x1,x2,x3,x4\}]
   //Total )/Factorial[4]
\end{mmaCell}

\begin{mmaCell}[moredefined={f}]{Input}
  f[x1,x2,x3,x4]
\end{mmaCell}

\begin{mmaCell}[morelst={basicstyle=\small}]{Output}
  \mmaFrac{1}{24} (-2 Log[\mmaSup{e}{x1+3 x2+4 x3+2 x4}+\mmaSup{e}{4x1+2 x2+x3+3 x4}]-2 Log[\mmaSup{e}{3 x1+x2+4 x3+2 x4}+\mmaSup{e}{2x1+4 x2+x3+3x4}]
  -2 Log[\mmaSup{e}{x1+4 x2+3 x3+2 x4}+\mmaSup{e}{4 x1+x2+2 x3+3x4}]-2 Log[\mmaSup{e}{4 x1+x2+3 x3+2 x4}+\mmaSup{e}{x1+4 x2+2 x3+3x4}]
  -2 Log[\mmaSup{e}{3x1+4 x2+x3+2 x4}+\mmaSup{e}{2 x1+x2+4 x3+3 x4}]-2 Log[\mmaSup{e}{4x1+3 x2+x3+2 x4}+\mmaSup{e}{x1+2 x2+4 x3+3 x4}]
  -2 Log[\mmaSup{e}{2x1+3 x2+4 x3+x4}+\mmaSup{e}{3x1+2 x2+x3+4 x4}]-2 Log[\mmaSup{e}{3 x1+2 x2+4 x3+x4}+\mmaSup{e}{2x1+3 x2+x3+4 x4}]
  -2 Log[\mmaSup{e}{2 x1+4 x2+3 x3+x4}+\mmaSup{e}{3x1+x2+2 x3+4x4}]-2 Log[\mmaSup{e}{4 x1+2 x2+3 x3+x4}+\mmaSup{e}{x1+3 x2+2 x3+4x4}]
  -2 Log[\mmaSup{e}{3 x1+4 x2+2 x3+x4}+\mmaSup{e}{2 x1+x2+3 x3+4x4}]-2 Log[\mmaSup{e}{4x1+3 x2+2 x3+x4}+\mmaSup{e}{x1+2 x2+3 x3+4 x4}])
\end{mmaCell}

\begin{mmaCell}[moredefined={substitute}]{Input}
  substitute = \{x1 \(\pmb{\to}\) 5/10, x2 \(\pmb{\to}\) 3/10, x3 \(\pmb{\to}\) 2/10, x4 \(\pmb{\to}\) 0\}
\end{mmaCell}

\begin{mmaCell}{Output}
  \{x1 \(\to\)\mmaFrac{1}{2},x2 \(\to\)\mmaFrac{3}{10},x3 \(\to\) \mmaFrac{1}{5},x4\(\to\)0\}
\end{mmaCell}

\begin{mmaCell}[moredefined={f, substitute}]{Input}
  (D[f[x1,x2,x3,x4], x2, x4] + D[f[x1,x2,x3,x4], x3, x1]
-D[f[x1,x2,x3,x4], x3, x4] -D[f[x1,x2,x3,x4], x2, x1] )
      /. substitute // Simplify
\end{mmaCell}

\begin{mmaCell}[uselistings=false]{Output}
  -((\mmaSup{(-1+\mmaSup{e}{1/5})}{2} \mmaSup{e}{1/5} (3-4\mmaSup{e}{1/5}+13 \mmaSup{e}{2/5}-13 \mmaSup{e}{3/5}+36\mmaSup{e}{4/5}-32 e+84 \mmaSup{e}{6/5}-82
\mmaSup{e}{7/5}+157 \mmaSup{e}{8/5}-127 \mmaSup{e}{9/5}+235\mmaSup{e}{2}-202 \mmaSup{e}{11/5}+382 \mmaSup{e}{12/5}-290\mmaSup{e}{13/5}+500 \mmaSup{e}{14/5}-382
\mmaSup{e}{3}+685 \mmaSup{e}{16/5}-513 \mmaSup{e}{17/5}+817\mmaSup{e}{18/5}-574 \mmaSup{e}{19/5}+970 \mmaSup{e}{4}-680\mmaSup{e}{21/5}+1052 \mmaSup{e}{22/5}-691
\mmaSup{e}{23/5}+1114 \mmaSup{e}{24/5}-756 \mmaSup{e}{5}+1114\mmaSup{e}{26/5}-691 \mmaSup{e}{27/5}+1052 \mmaSup{e}{28/5}-680\mmaSup{e}{29/5}+970
\mmaSup{e}{6}-574 \mmaSup{e}{31/5}+817 \mmaSup{e}{32/5}-513\mmaSup{e}{33/5}+685 \mmaSup{e}{34/5}-382 \mmaSup{e}{7}+500\mmaSup{e}{36/5}-290 \mmaSup{e}{37/5}+382
\mmaSup{e}{38/5}-202 \mmaSup{e}{39/5}+235 \mmaSup{e}{8}-127\mmaSup{e}{41/5}+157 \mmaSup{e}{42/5}-82 \mmaSup{e}{43/5}+84\mmaSup{e}{44/5}-32 \mmaSup{e}{9}+36
\mmaSup{e}{46/5}-13 \mmaSup{e}{47/5}+13 \mmaSup{e}{48/5}-4\mmaSup{e}{49/5}+3 \mmaSup{e}{10}))/(3\mmaSup{(1+\mmaSup{e}{1/5})}{2} \mmaSup{(1+\mmaSup{e}{2/5})}{2}\mmaSup{(1-\mmaSup{e}{1/5}+\mmaSup{e}{2/5})}{2}
\mmaSup{(1+\mmaSup{e}{4/5})}{2}\mmaSup{(1-\mmaSup{e}{2/5}+\mmaSup{e}{4/5})}{2}\mmaSup{(1-\mmaSup{e}{1/5}+\mmaSup{e}{2/5}-\mmaSup{e}{3/5}+\mmaSup{e}{4/5}e-e+\mmaSup{e}{6/5})}{2}\mmaSup{(1+\mmaSup{e}{8/5})}{2}))
\end{mmaCell}

\begin{mmaCell}{Input}
  N[\%5, 20]
\end{mmaCell}

\begin{mmaCell}{Output}
  -0.15621966760667033050
\end{mmaCell} 

\end{appendices}

\printthesisindex 

\end{document}